%% file: ms.tex
\documentclass[useAMS,usenatbib]{mn2e}
\usepackage{subfigure}
\usepackage{graphicx}
\usepackage{amsmath}
\usepackage{amssymb}
\usepackage{pdflscape}
\usepackage{xcolor}
\usepackage[normalem]{ulem}

\title[The nuclear and extended mid-infrared emission of Seyfert galaxies]{The nuclear and extended mid-infrared emission of Seyfert galaxies}
\author[I. Garc\'ia-Bernete et al.]
{\parbox{\textwidth}{I. Garc\'ia-Bernete$^{1,2}$\thanks{E-mail: igarcia@iac.es}, C. Ramos Almeida$^{1,2}$, J.A. Acosta-Pulido$^{1,2}$,A. Alonso-Herrero$^{3,4,5}$,O. Gonz\'alez-Mart\'in$^{6}$, A. Hern{\'a}n-Caballero$^{7,8}$, M. Pereira-Santaella$^{9}$, N. A. Levenson$^{10}$, C. Packham$^{5,11}$, E. S. Perlman$^{12}$, K. Ichikawa$^{11}$, P. Esquej$^{13}$ and T. D\'iaz-Santos$^{14}$.
}\vspace{0.4cm}\\
\parbox{\textwidth}{$^{1}$Instituto de Astrof\'isica de Canarias, Calle V\'ia L\'actea, s/n, E-38205, La Laguna, Tenerife, Spain\\
$^{2}$Departamento de Astrof\'isica, Universidad de La Laguna, E-38206, La Laguna, Tenerife, Spain\\
$^{3}$Centro de Astrobiolog\'ia, CSIC-INTA, ESAC Campus, E-28692, Villanueva de la Ca\~nada, Madrid, Spain\\
$^{4}$Department of Physics, University of Oxford, Oxford OX1 3RH, UK\\
$^{5}$Department of Physics and Astronomy, University of Texas at San Antonio, One UTSA Circle, San Antonio, TX 78249, USA\\
$^{6}$Centro de Radioastronom\'ia y Astrof\'isica (CRyA-UNAM), 3-72 (Xangari), 8701, Morelia, Mexico\\
$^{7}$Instituto de F\'isica de Cantabria, CSIC-Universidad de Cantabria, E-39005, Santander, Spain\\
$^{8}$Departamento de Astrof\'isica y CC. de la Atm\'osfera, Facultad de CC. F\'isicas, Universidad Complutense de Madrid, E-28040 Madrid, Spain\\
$^{9}$Centro de Astrobiolog\'ia, CSIC-INTA, Ctra de Torrej\'on a Ajalvir, km 4, E-28850, Torrej\'on de Ardoz, Madrid, Spain\\
$^{10}$Gemini Observatory, Casilla 603, La Silla, Chile\\
$^{11}$National Astronomical Observatory of Japan, 2-21-1 Osawa, Mitaka, Tokyo 181-8588, Japan\\
$^{12}$Department of Physics \& Space Sciences, Florida Institute of Technology, 150 W. University Blvd.,Melbourne, FL 32901, USA\\
$^{13}$European Space Astronomy Centre (ESAC)/ESA, E-28691, Villanueva de la Ca\~nada, Madrid, Spain\\
$^{14}$N\'ucleo de Astronom\'ia de la Facultad de Ingenier\'ia, Universidad Diego Portales, Av. Ejercito Libertador 441, Santiago, Chile\\
}
}
\begin{document}
\date{}
\pagerange{\pageref{firstpage}--\pageref{lastpage}} \pubyear{2016}
\maketitle
\label{firstpage}
\begin{abstract}

We present subarcsecond resolution mid-infrared (MIR) images obtained with 8-10~m-class ground-based telescopes of a complete volume-limited (D$_L<$40\,Mpc) sample of 24 Seyfert galaxies selected from the {\textit{Swift/BAT}} nine month catalog. We use those MIR images to study the nuclear and circumnuclear emission of the galaxies. Using different methods to classify the MIR morphologies on scales of $\sim$400 pc, we find that the majority of the galaxies (75-83\%) are extended or possibly extended and 17-25\% are point-like. This extended emission is compact and it has low surface brightness compared with the nuclear emission, and it represents, on average, $\sim$30\% of the total MIR emission of the galaxies in the sample. We find that the galaxies whose circumnuclear MIR emission is dominated by star formation show more extended emission (650$\pm$700\,pc) than AGN-dominated systems (300$\pm$100\,pc). In general, the galaxies with point-like MIR morphologies are face-on or moderately inclined (b/a$\sim$0.4-1.0), and we do not find significant differences between the morphologies of Sy1 and Sy2. We used the nuclear and circumnuclear fluxes to investigate their correlation with different AGN and SF activity indicators. We find that the nuclear MIR emission (the inner $\sim$70~pc) is strongly correlated with the X-ray emission (the harder the X-rays the better the correlation) and with the [O\,IV]~$\lambda$25.89~$\mu$m emission line, indicating that it is AGN-dominated. We find the same results, although with more scatter, for the circumnuclear emission, which indicates that the AGN dominates the MIR emission in the inner $\sim$400 pc of the galaxies, with some contribution from star formation.

\end{abstract}


\section{Introduction}
\label{intro}
Active galactic nuclei (AGNs) are powered by accretion of material onto supermassive black holes (SMBHs), which release enormous quantities of energy in the form of radiation and/or mechanical outflows to the host galaxy interstellar medium. This feedback is fundamental to the formation and evolution of the host galaxies \citep{Hopkins10}. Seyfert galaxies are intermediate-luminosity AGNs, characterized by a bright unresolved nucleus generally hosted by a spiral galaxy \citep{Adams77}. Seyfert galaxies are classified by the presence of broad lines (type\,1) or otherwise (type\,2) in the optical spectrum \citep{Khachikian71,Khachikian74}, and these types depend on orientation, according to the unified model \citep{Antonucci1993}. This scheme proposes that there is dust surrounding the active nucleus distributed in a toroidal geometry, which obscures the central engines of type 2, and allows a 
direct view in the case of type 1 sources. The dusty torus absorbs the AGN radiation and, then, reprocesses it to emerge in the infrared (IR), peaking in the mid-IR (MIR; $\sim$5-30~$\mu$m), according to torus models (e.g. \citealt{Pier92,Efstathiou95,Schartmann05,Honig06,Nenkova08,bNenkova08,Stalevski2012,Siebenmorgen2015}).

MIR observations of the nuclear regions of active galaxies allow to study the emission of dust heated by the AGN, but also by star-formation (SF) when present (e.g. \citealt{Radomski2003,Packham05,Sales13,Esquej14,Herrero14,Ramos14,Ruschel14,Bernete2015}). 
Prominent features in the MIR spectrum of AGN are the 9.7~$\mu$m silicate feature, and the
Polycyclic Aromatic Hydrocarbon (PAH) emission bands, although the latter can be diluted by the bright AGN continuum (see e.g. \citealt{Esquej14, Herrero14, Ramos14, Bernete2015}). The PAH features are often used as indicators of the star formation rate (SFR) of galaxies (see e.g. \citealt{Peeters04,Wu05,Diamond12,Esquej14}), together with low ionization potential (IP) MIR emission lines such as [Ne\,II]$\lambda$12.81~$\mu$m \citep{Spinoglio92,Ho07,Pereira-Santaella2010,Spinoglio12}. 

The unprecedented angular resolution achieved by 8-10~m-class ground-based telescopes in the MIR is crucial to correctly separate the nuclear emission from the foreground galaxy emission. As the MIR-emitting torus is compact (r$<$10~pc; see e.g. \citealt{Tristram09,Burtscher13}), this angular resolution is fundamental to isolate its emission from other emitting sources at larger scales, as well as to disentangle the heating source of the diffuse circumnuclear MIR emission. However, our understanding about the dominant heating source of the dust on these physical scales (inner kpcs) remains unclear, because of the paucity of ground-based MIR instruments and the limited size of the samples studied to date \citep{Horst08,Gandhi09,Levenson09,Mason12,Gonzalez-Martin2013}. A major step forward was attained with  
the publication of the subarcsecond MIR imaging atlas of local AGN \citep{Asmus2014}. These authors presented a compilation of subarcsecond MIR imaging for 253 AGNs (204 with nuclear component detected), and found that a large fraction of the galaxies present extended MIR morphologies. They also found that the lower angular resolution data are significantly affected by non-AGN emission, and that the subarcsecond resolution MIR fluxes are generally less than half compared to {\textit{Spitzer/IRS}} data in a significant fraction of the sample (31\%).

The aim of this work is to study for the first time the nuclear and circumnuclear MIR emission of a complete and volume limited sample of X-ray selected Seyfert galaxies (see Section \ref{sample}) to obtain statistically significant results. Therefore, here we used the nine month {\textit{Swift/Burst Alert Telescope}} (BAT; \citealt{Tueller2008}) AGN catalog, which is a very hard X-ray survey (14-195~keV), to select our AGN sample.

The paper is organized as follows. Section \ref{sample} and Section \ref{observations} describe the sample selection and the observations, respectively. The main results on the MIR emission are presented in Section \ref{MIR_emission}. Section \ref{morphology}  describes de MIR morphological analysis, and in Section \ref{correlations}  we study different MIR correlations with AGN and SF indicators. Finally, in Section \ref{Conclusions} we summarize the main conclusions of this work. 

Throughout this paper we assumed a cosmology with H$_0$=73 km~s$^{-1}$~Mpc$^{-1}$, $\Omega_m$=0.27, and $\Omega_{\Lambda}$=0.73, and a velocity-field corrected using the \citet{Mould00} model, which includes the influence of the Virgo cluster, the Great Attractor, and the Shapley supercluster.

\section{Sample Selection}
\label{sample}
Previous ultraviolet (UV), optical and IR surveys are often incomplete, since UV and optical surveys are missing obscured sources, and IR surveys introduce a bias against dust-free AGNs. A complete sample would be designed to select AGN by an isotropic property, such as the hard X-ray emission, which is commonly used as an isotropic indicator of AGN luminosity (see e.g. \citealt{Mulchaey94, Melendez08, Rigby09, Diamond09}). 

The very hard 14-195 keV band of the Swift/BAT catalog is far less sensitive to the effects of obscuration than optical or softer X-ray wavelengths. Indeed, this AGN catalog is one of the most complete to date (see e.g. \citealt{Winter2009,Winter2010,Weaver2010,Ichikawa2012,Ueda15}).

The sample studied here consists of 24 Seyfert galaxies selected from the nine month catalog observed with {\textit{Swift/BAT}} \citep{Tueller2008}, which is flux limited in the very hard 14-195~keV X-ray band (153 sources). The Swift/BAT sources were selected based on a detection at 4.8$\sigma$ or higher. Note that this catalog is sensitive over 80\% of the sky to a flux threshold of 3.5$\times$10$^{-11}$~erg\,cm$^{-2}$\,s$^{-1}$ in the 14-195~keV band and covers one third of the sky near the ecliptic poles at 2.5$\times$10$^{-11}$~erg\,cm$^{-2}$\,s$^{-1}$. 

We chose the nine month catalog \citep{Tueller2008} for selecting our sample since the sources are bright and nearby, and most of them had archival MIR data when we started this work. Since we are interested in the study of the nuclear and circumnuclear emission of Seyfert galaxies, we selected all the Seyfert galaxies in the nine month catalog with luminosity distances D$_L<$40\,Mpc. Considering the average angular resolution of 8-10~m-class ground-based telescopes ($\sim$0.3\arcsec ~in the N-band), this left us with a sample of 24 Seyfert galaxies for which we have a resolution $\leqslant$50\,pc in the MIR (hereafter BAT Complete Seyfert sample at D$_L<$40\,Mpc; BCS$_{40}$ sample). Here we present high angular resolution MIR imaging observations obtained with 8-10~m-class ground-based telescopes of these galaxies (public and proprietary). See Section \ref{subarcsecond} for further details on the subarcsecond resolution observations.

\begin{figure}
\centering
\par{
\includegraphics[width=8.0cm]{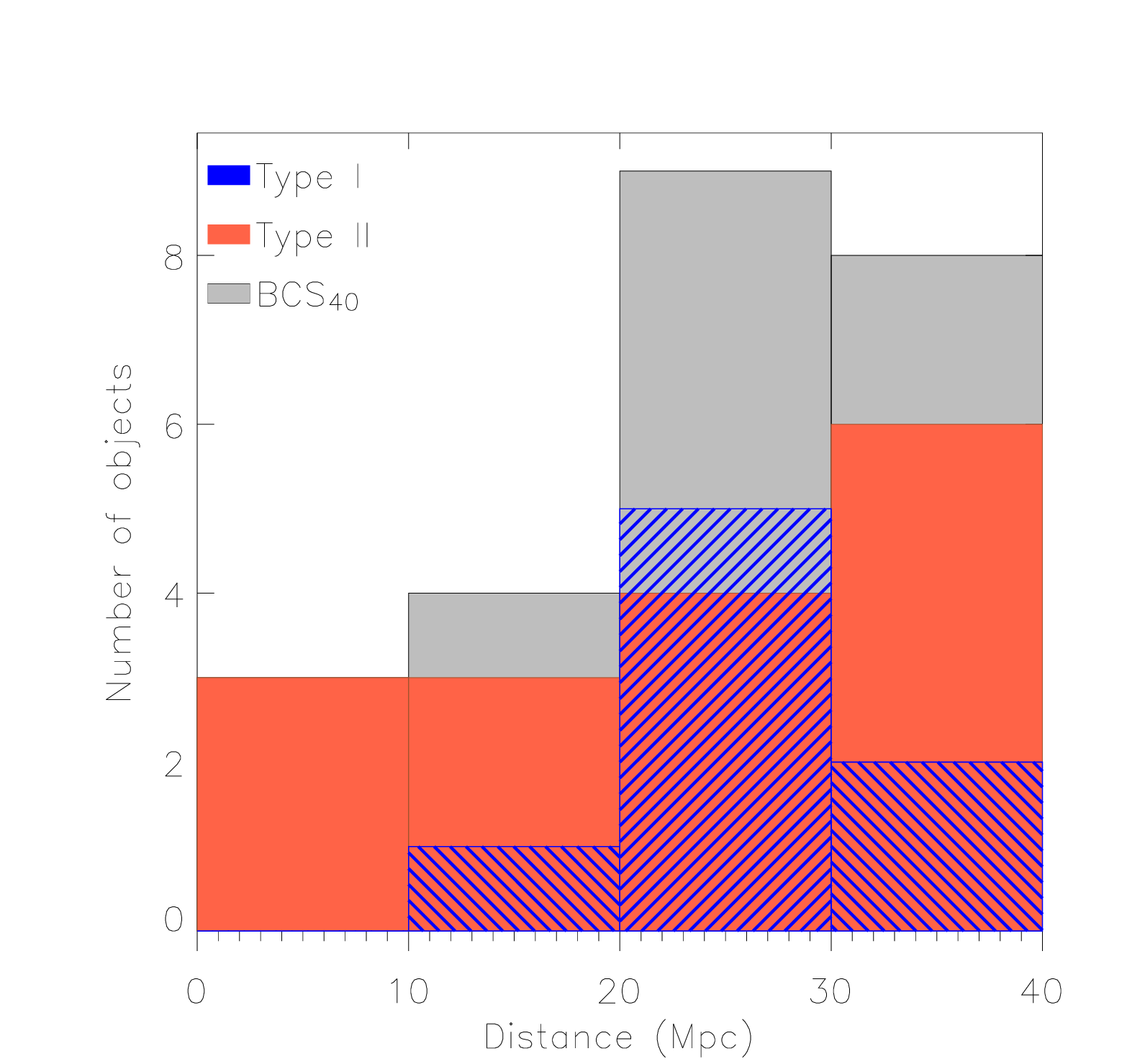}
\includegraphics[width=8.0cm]{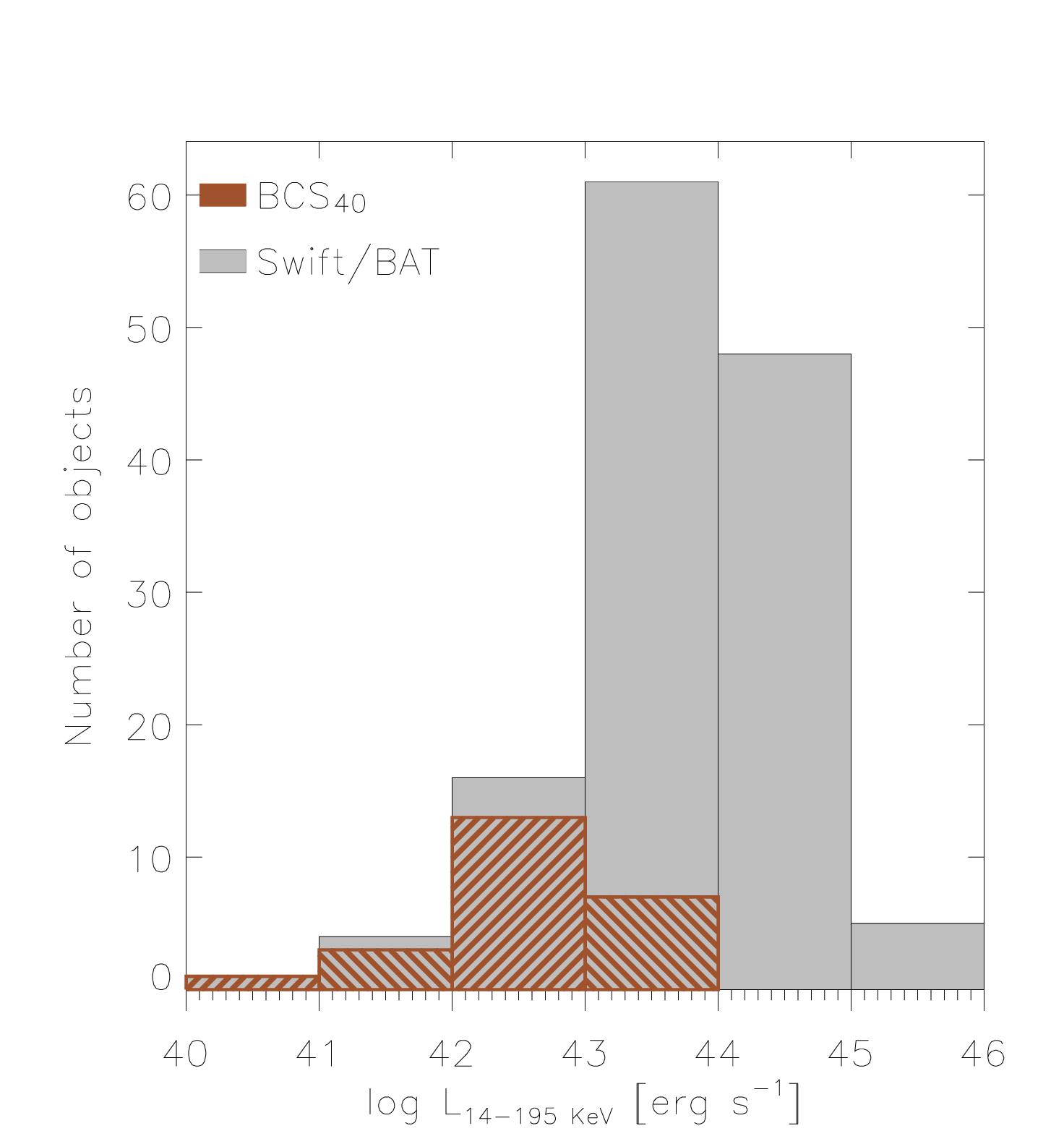}
\par}
\caption{Top panel: Luminosity distance distribution of the BCS$_{40}$ sample. The blue hatched and red filled histograms are the distribution of Sy1 and Sy2 galaxies, respectively. The grey filled histograms correspond to the total BCS$_{40}$ sample. Bottom panel: 14-195~keV X-ray luminosity distribution of the total nine month {\textit{Swift/BAT}} AGN catalog (excluding beamed sources$-$BL Lac objects and blazars$-$) and the BCS$_{40}$ sample (grey filled and brown hatched histograms, respectively).}
\label{fig1}
\end{figure}

The BCS$_{40}$ sample sorted by luminosity distance is shown in Table \ref{tab1}. We present the luminosity distance distribution of the sample in the top panel of Fig. \ref{fig1}. Our sample contains 8 Sy1 and 16 Sy2, with the majority of the galaxies in the 20-40\,Mpc range (71\% of the sample). The median distance of the BCS$_{40}$ sample is 24.1\,Mpc. Hereafter the Seyfert types are referred as Sy1 (Seyferts 1, 1.2 and 1.5) and Sy2 (Seyferts 1.8, 1.9 and 2). The closest galaxies in the sample are Sy2, although the median values for Sy1 and Sy2 are relatively similar, 25.8\,Mpc and 22.1\,Mpc, respectively. In the bottom panel of Fig.\ref{fig1} we present the comparison between the nine month {\textit{Swift/BAT}} catalog and BCS$_{40}$ sample 14-195~keV X-ray luminosity. The latter corresponds to the lowest luminosity region of the nine month {\textit{Swift/BAT}} sample.

\input{tab1.tex}

\section{Observations}
\label{observations}
In this section we describe all the MIR observations analyzed in this work, which we divide in subarcsecond and arcsecond resolution data.

\subsection{Subarcsecond resolution data}
\label{subarcsecond}
Here we describe the subarcsecond resolution MIR imaging data, which are from 8-10~m-class ground-based telescopes. For all subarcsecond resolution MIR observations the standard MIR chopping and nodding technique was used to remove the time-variable sky background and the thermal emission from the telescope. In addition to the target observations, an image of a point spread function (PSF) standard star was obtained immediately after or before the science target for accurately sampling the image quality, and to allow flux calibration.

In Fig. \ref{figA1} of Appendix \ref{A} we present the N-band images of the BCS$_{40}$ sample, and Table \ref{tab2} summarizes the details of the observations, as well as the derived galaxy measurements. In the following we consider that the nucleus is resolved when the FWHM of the galaxy is larger than the FWHM of the PSF standard star. In Table \ref{tab2} we also present the classification of the MIR images of the galaxies by comparing the FWHM of the PSF standard star and the galaxy in each band.

\input{tab2.tex}

\subsubsection{Gran Telescopio CANARIAS/CanariCam observations}
\label{CC}

We obtained MIR images of three Seyfert galaxies (NGC\,4388, NGC\,3227 and UGC\,6728) and their corresponding PSF standard star with the filter Si2 ($\lambda_c=$8.7~$\mu$m) using the instrument CanariCam (CC; \citealt{Telesco03}) on the 10.4~m~GTC. NGC\,4388 and NGC\,3227 were observed as part of an ESO/GTC large programme (182.B-2005; \citealt{Herrero15}), and UGC\,6728 as part of proposal GTC43-15A (PI: I. Garc\'ia-Bernete). CC is an MIR (7.5-25~$\mu$m) imager with spectroscopic, coronagraphic and polarimetric capabilities and uses a Si:As Impurity Band Conduction (IBC) detector, which covers a FOV of 26 arcsec $\times$ 19 arcsec on the sky and it has a pixel scale of 0.0798 arcsec (hereafter 0.08 arcsec). The chopping and nodding throws were 15 arcsec. 

The data reduction was carried out with the \textit{RedCan} pipeline \citep{Gonzalez-Martin2013}, which performs sky subtraction, stacking of individual observation, rejection of the bad frames and flux calibration. The spatial resolution of the image was determined by measuring the full width at half-maximum (FWHM) of the PSF standard star. See Table \ref{tab2} for further details on the observations. 

\subsubsection{Gemini/T-ReCS observations}
\label{T-ReCS}
We retrieved MIR images of ten Seyfert galaxies and their corresponding PSF standard stars per filter and science object from the Gemini archive. The images were taken with the Si2 ($\lambda_c=$8.7~$\mu$m) and Qa ($\lambda_c=$18.3~$\mu$m) filters using the instrument Thermal-Region Camera Spectrograph (T-ReCS; \citealt{Telesco1998}) on the 8.1~m Gemini-South Telescope. T-ReCS is a MIR (8-25~$\mu$m) imager and long-slit spectrograph and uses a Raytheon Si:As IBC detector, which covers a FOV of 28.5 arcsec $\times$ 21.4 arcsec, providing a pixel scale of 0.089 arcsec. See Table \ref{tab2} for further details on the observations.

We reduced the Gemini/T-ReCS data using the \textit{RedCan} pipeline \citep{Gonzalez-Martin2013}, as described in Section \ref{CC}.

\subsubsection{Gemini/MICHELLE observations}
\label{Michelle}
We compiled data for seven Seyfert galaxies obtained with the instrument MICHELLE \citep{Glasse97} on the 8.1~m Gemini-North Telescope.  MICHELLE is an MIR (7-26~$\mu$m) imager and spectrograph, which uses a Si:As IBC detector, covering a FOV of 32 arcsec $\times$ 24 arcsec on the sky. Its pixel scale is 0.1005 arcsec. The images were taken with different filters, namely N$'$ ($\lambda_c=$11.2~$\mu$m), Si5 ($\lambda_c=$11.6~$\mu$m), Si6 ($\lambda_c=$12.5~$\mu$m) or Qa ($\lambda_c=$18.1~$\mu$m). See Table \ref{tab2} for further details on the observations.

The Gemini/MICHELLE data reduction was carried out with the Gemini {\textit{IRAF}}\footnote{IRAF is distributed by the National Optical Astronomy Observatory, which is operated by the Association of Universities for Research in Astronomy (AURA) under cooperative agreement with the National Science Foundation.} packages, particularly with the {\textit{MIDIR}} \citep{tody86} reduction task. The Gemini {\textit{IRAF}} packages include sky subtraction,
stacking of individual images and rejection of bad images. The flux calibration was carried out using the {\textit{PHOT}} {\textit{IRAF}} task and the image of the PSF standard star to work out the relation between count and flux. 

\subsubsection{Gemini/OSCIR observations}
\label{OSCIR}
Two images of the galaxy NGC\,4151 were taken with the N ($\lambda_c=$10.75~$\mu$m) and IHW18 ($\lambda_c=$18.17~$\mu$m) filters using the University of Florida MIR camera/spectrometer OSCIR on the 8.1~m Gemini-North telescope. OSCIR uses a Rockwell 128$\times$128 Si:As IBC detector. On Gemini North, OSCIR has a plate scale of 0.089 arcsec pixel$^{-1}$, corresponding to a FOV of 11.4 arcsec $\times$ 11.4 arcsec. The chopping and nodding throws were 15 arcsec. See Table \ref{tab2} for further details on the observations.

We took the fully reduced and flux calibrated images from \citet{Radomski2003}, also presented in \citet{Ramos09}. The data were reduced using in-house-developed IDL routines (see \citealt{Radomski2003} for further details on the data reduction).

\subsubsection{Very Large Telescope/VISIR observations}
\label{VISIR}
We finally compiled MIR images for the rest of the sample (4 Seyfert galaxies) taken with different filters (see Table \ref{tab2}) with the instrument VISIR \citep{Lagage2004} on the Unit 3 of the 8.2~m Very Large Telescope (VLT) telescope (Melipal). VISIR is a MIR (16.5-24.5~$\mu$m) imager and spectrograph, which uses 256$\times$256 Si:As IBC detector, covering a FOV of 19.2 arcsec $\times$ 19.2 arcsec and its pixel scale is 0.075 arcsec. 

We downloaded the fully reduced and calibrated science data of the galaxies from the {\textit{Subarcsecond mid-infrared atlas of local AGN}}\footnote{http://dc.zah.uni-heidelberg.de/sasmirala/q/cone/form} \citep{Asmus2014}. In addition, we retrieved PSF standard stars for each observation from the ESO archive\footnote{http://archive.eso.org/eso/eso$\_$archive$\_$main.html}. We reduced the PSF standard stars with the VISIR\footnote{http://www.eso.org/observing/dfo/quality/VISIR/pipeline/pipe$\_$gen.html} pipeline, which performs flat-fielding correction, bad pixel removal, source alignment, and co-addition of frames are executed to produce a combined image for each filter. See Table \ref{tab2} for further details on the observations.

\subsection{Arcsecond resolution data}
\label{Arcsecond}
Here we describe the arcsecond resolution MIR data, which correspond to observations taken with the {\textit{Spitzer Space Telescope}} and the {\textit{Wide-field Infrared Survey Explorer}} (\textit{WISE};  \citealt{Wright10}), both with lower spatial resolution but higher sensitivity than the ground-based observations described in Section \ref{subarcsecond}.
\subsubsection{Spitzer Space Telescope observations}
\label{Spitzer}
We compiled arcsecond resolution MIR data of 18 galaxies obtained with the instrument Infrared Array Camera ({\textit{IRAC}}; \citealt{Fazio04}) using the 8~$\mu$m channel (angular resolution $\sim$1.9\arcsec). The {\textit{IRAC}} FOV is 5.2 arcmin $\times$ 5.5 arcmin on the sky and its pixel scale is 1.2 arcsec. We downloaded the reduced and calibrated mosaiked data from the {\textit{Spitzer}} Heritage Archive (SHA). Note that these mosaics are resampled to a pixel size of 0.6 arcsec. The 8~$\mu$m {\textit{IRAC}} data are shown in Fig. \ref{figA2} of Appendix \ref{A}. Besides, in Fig. \ref{figA1} of Appendix \ref{A}, we also show the comparison between the subarcsecond resolution N-band data and the arcsecond resolution 8~$\mu$m {\textit{Spitzer}} images for the nuclear region ($\leqslant$650~pc) of the galaxies. 

In addition, low-resolution MIR spectra were retrieved for the whole sample from the Cornell Atlas of {\textit{Spitzer/IRS}} Source (CASSIS\footnote{http://cassis.astro.cornell.edu/atlas/} v4; \citealt{Lebouteiller11}). The spectra were obtained using the InfraRed Spectrograph (IRS; \citealt{Houck04}). The bulk of the observations were obtained in staring mode using the low-resolution (R$\sim$60-120) IRS modules: the short-low (SL; 5.2-14.5~$\mu$m) and the long-low (LL; 14-38~$\mu$m). The spectra were reduced with the CASSIS software, using the optimal extraction to get the best SNR. We only needed to apply a small offset to stitch together the different modules, taking the shorter wavelength module (SL2; 5.2-7.6~$\mu$m) as the basis, which has associated a slit width of 3.6\arcsec. The spectra are shown in Appendix \ref{B}. Note that for NGC\,4138 there is no low-resolution staring mode spectrum. Therefore, we have extracted a spectrum in a 7.7\arcsec aperture diameter from the spectral mapping observations available in the SHA.

\subsubsection{WISE observations}
\label{WISE}
For these galaxies in our sample that do not have {\textit{IRAC}} images available and for cases whose {\textit{IRAC}} images are saturated, we downloaded MIR images from the {\textit{WISE}} All-Sky Data Release\footnote{http://irsa.ipac.caltech.edu/Missions/wise.html}, taken in the 12~$\mu$m band (angular resolution $\sim$6.5\arcsec). The {\textit{WISE}} FOV is 47 arcmin x 47 arcmin on the sky and its pixel scale is 2.75 arcsec. We downloaded the reduced fully calibrated data. We show the 12~$\mu$m {\textit{WISE}} images in Fig. \ref{figA2} of Appendix \ref{A}. Besides, in Fig. \ref{figA1} of Appendix \ref{A}, we also include the arcsecond resolution 12~$\mu$m {\textit{WISE}} images in the comparison between the subarcsecond and arcsecond resolution N-band images.

\section{MIR emission}
\label{MIR_emission}
In this section we study in detail the properties of the inner regions (few hundred parsecs) of the BCS$_{40}$ sample. 

\subsection{Subarcsecond resolution nuclear fluxes: subtraction of scaled PSFs}
\label{nuclear_fluxes}

\begin{figure}
\centering
\includegraphics[width=7.8cm]{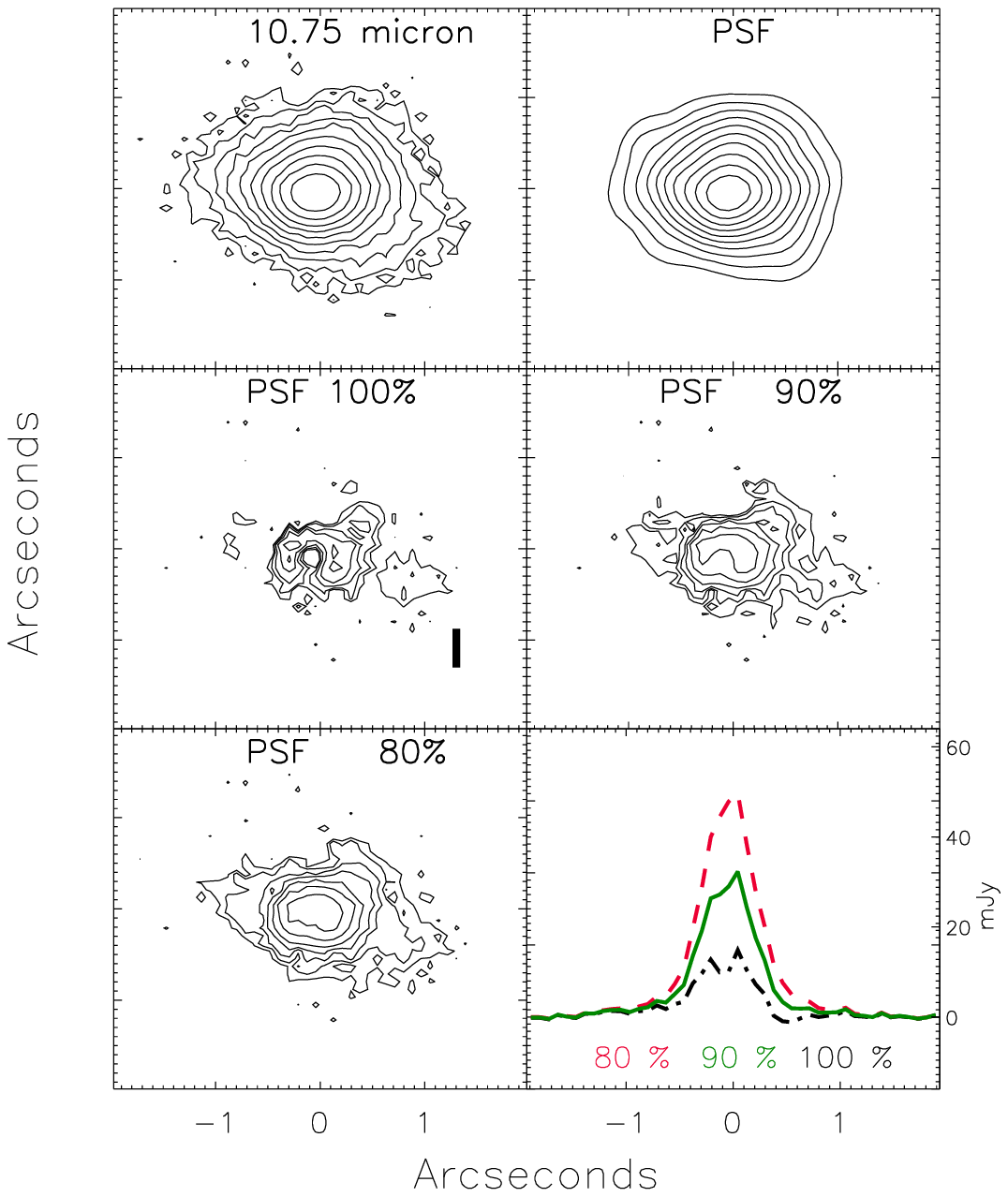}
\caption{MIR ($\lambda_c=$10.75~$\mu$m) GEMINI/OSCIR contours at $\geq$3$\sigma$ of NGC\,4151, the PSF standard star and, 
scaled PSF subtraction at different levels, and 1-D profiles of the residuals at 100, 90 and 80 percent levels (black dot-dashed, solid green and red dashed lines, respectively). The best subtraction is 90\% according to the flat galaxy profile shown in the bottom-right panel. North is up, and east to left.}
\label{fig2}
\end{figure}

Taking advantage of the angular resolution provided by the ground-based instruments described in Section \ref{subarcsecond}, we obtained nuclear fluxes for all the galaxies in our sample. We used the PSF standard stars to obtain these nuclear fluxes. Fig. \ref{fig2} shows an example of PSF subtraction at various levels (in $\geq$3$\sigma$ contours) for the MIR ($\lambda_c=$10.75~$\mu$m) GEMINI/OSCIR image of NGC\,4151. The method consists in the following: we first matched the PSF-star image (see top right panel of Fig. \ref{fig2}) to the peak of the galaxy emission, that is, at a 100\% level. Then we subtracted the scaled PSF-star from the galaxy image (see top left panel of Fig. \ref{fig2}) at different percentage peak levels until we obtained a non-oversubtracted residual image. This is something that we determine by looking at the 1-D profile shown in the bottom right panel of Fig. \ref{fig2} but also at the contour plots shown in the middle and bottom-left panels (100, 90 and 80\% in the case of Fig. \ref{fig2}). The contour plot at 100\% subtraction appears clearly oversubtracted in the galaxy nucleus, and the 1-D profile as well. The 90\% subtraction is the best fit according to the previous criteria. Finally, we measured the unresolved component by integrating the emission in a 1~arcsec radius aperture\footnote{Note that we carried out the sky subtraction using a concentric ring wide enough to contain a good estimate of the sky background.} on the scaled PSF-star image. The host galaxy contribution corresponds to the total galaxy emission minus the scaled PSF (i.e. the residual of the subtraction). This method has been widely tested in ground-based MIR images (e.g. \citealt{Soifer00,Radomski2002,Radomski2003,Levenson09,Ramos09,Ramos2011,Ramos14,Mason12,Bernete2015,Martinez-Paredes2015}).

The MIR nuclear fluxes calculated using this method are listed in Table \ref{tab2}. The estimated uncertainty of the fluxes determined using PSF subtraction is $\sim$15\%, which includes also the photometric calibration uncertainty (see \citealt{Herrero14,Herrero15} for further details).

Once we obtained the subarcsecond resolution MIR nuclear fluxes (hereafter nuclear fluxes), we can quantify the circumnuclear MIR emission of the galaxies by subtracting the nuclear emission from the total fluxes. To determine the apertures containing the total flux for each galaxy, we used increasing apertures in order to construct a photometric profile (flux versus aperture radius) and chose the aperture which contains $\sim$95\% of the maximum of the profile. Table \ref{tab2} shows these total fluxes.

As we have MIR observations obtained with different filters (see Table \ref{tab2}), we used the nuclear fluxes calculated as described above and the {\textit{Spitzer/IRS}} spectra of the galaxies to obtain homogeneous nuclear fluxes at  8~$\mu$m\footnote{We chose 8~$\mu$m as our wavelength of reference since this is the wavelength of the {\textit{Spitzer}} images.}. The process involves using spectral decomposition in AGN, PAH and stellar emission components to estimate the AGN contribution of the IRS spectra. To do so, we used the DeblendIRS\footnote{http://www.denebola.org/ahc/deblendIRS/} routine \citep{Hernan-caballero2015} and the MIR nuclear fluxes in the various filters as priors to constrain the flux of the AGN component. Once we derived the AGN contribution, we used it to extrapolate our nuclear fluxes to 8~$\mu$m for all the galaxies. In Table \ref{tab3} we list these nuclear fluxes. See Appendix \ref{B} for further details on the spectral decomposition. 

\input{tab3.tex}

\subsection{Comparison with arcsecond resolution MIR data}
\label{arcsecond_fluxes}

In order to compare the subarcsecond and arcsecond resolution MIR data, we calculate 8~$\mu$m arcsecond resolution nuclear fluxes. To do that, we also used the PSF subtraction method to try to recover the nuclear emission from the lower angular resolution {\textit{Spitzer/IRAC}} images at 8~$\mu$m\footnote{For the galaxies whose {\textit{IRAC}} images are saturated we calculate arcsecond resolution nuclear fluxes from the IRS spectra (see Table \ref{tab3}).}. 

In the case of the {\textit{IRAC}} data, we used the core Point Response Functions (PRFs)\footnote{http://irsa.ipac.caltech.edu/data/SPITZER/docs/}. Note that the core PRFs are the most adequate for the PSF-fitting photometry due to the faithful size and structure of the PRF center. However, these PRFs do not include the extended region of the PRF wings. Therefore, to take into account the wings we have used scaled core PRFs to derive the nuclear fluxes and then multiplied them by the factor between the core and the extended PRFs fluxes to obtain realistic values of the nuclear emission. We list the arcsecond resolution nuclear fluxes calculated using this method in Table \ref{tab3} (see Section \ref{nuclear_fluxes} for further details). 

\begin{figure*}
\centering
\par{
\includegraphics[width=8.8cm]{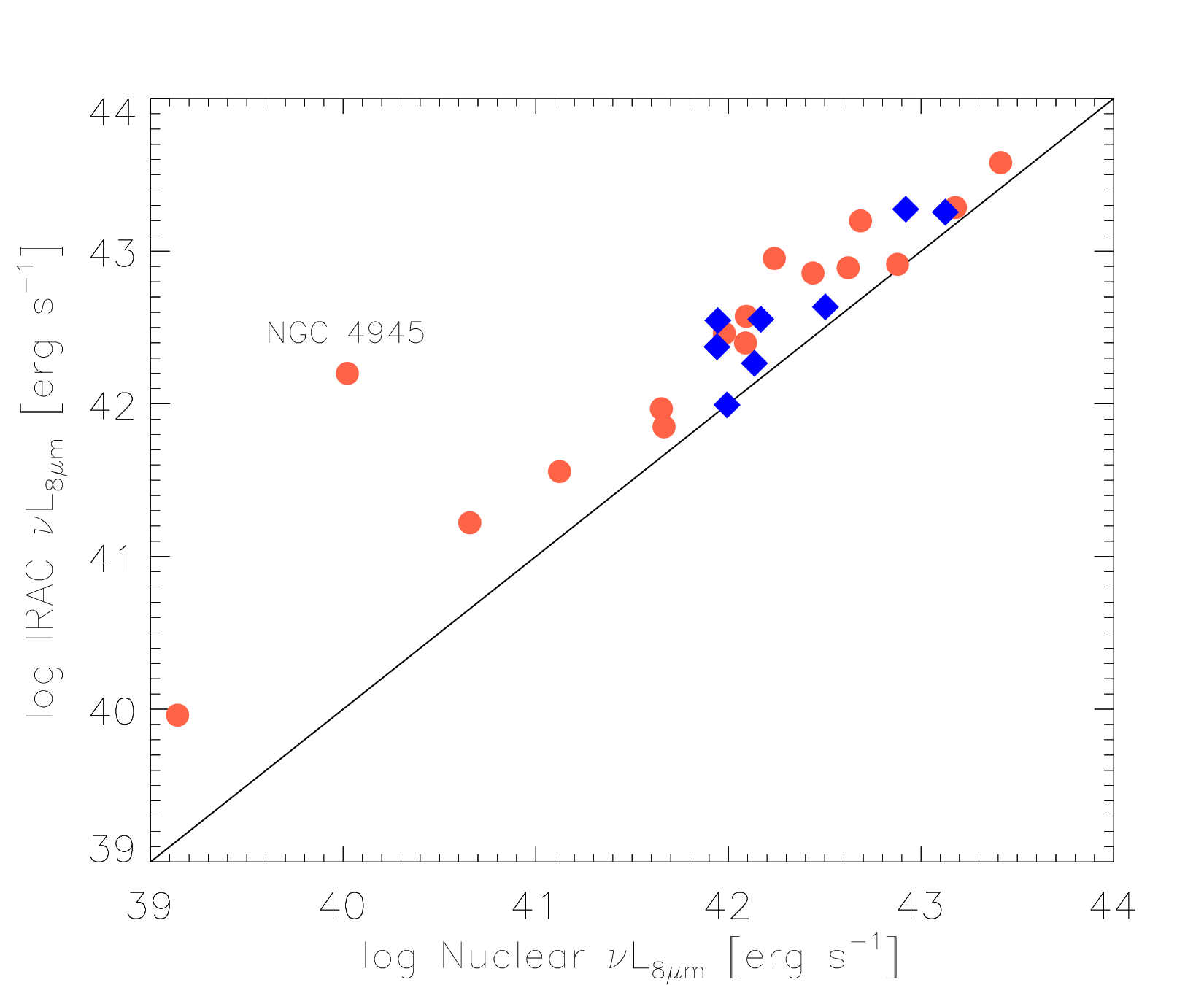}
\includegraphics[width=8.8cm]{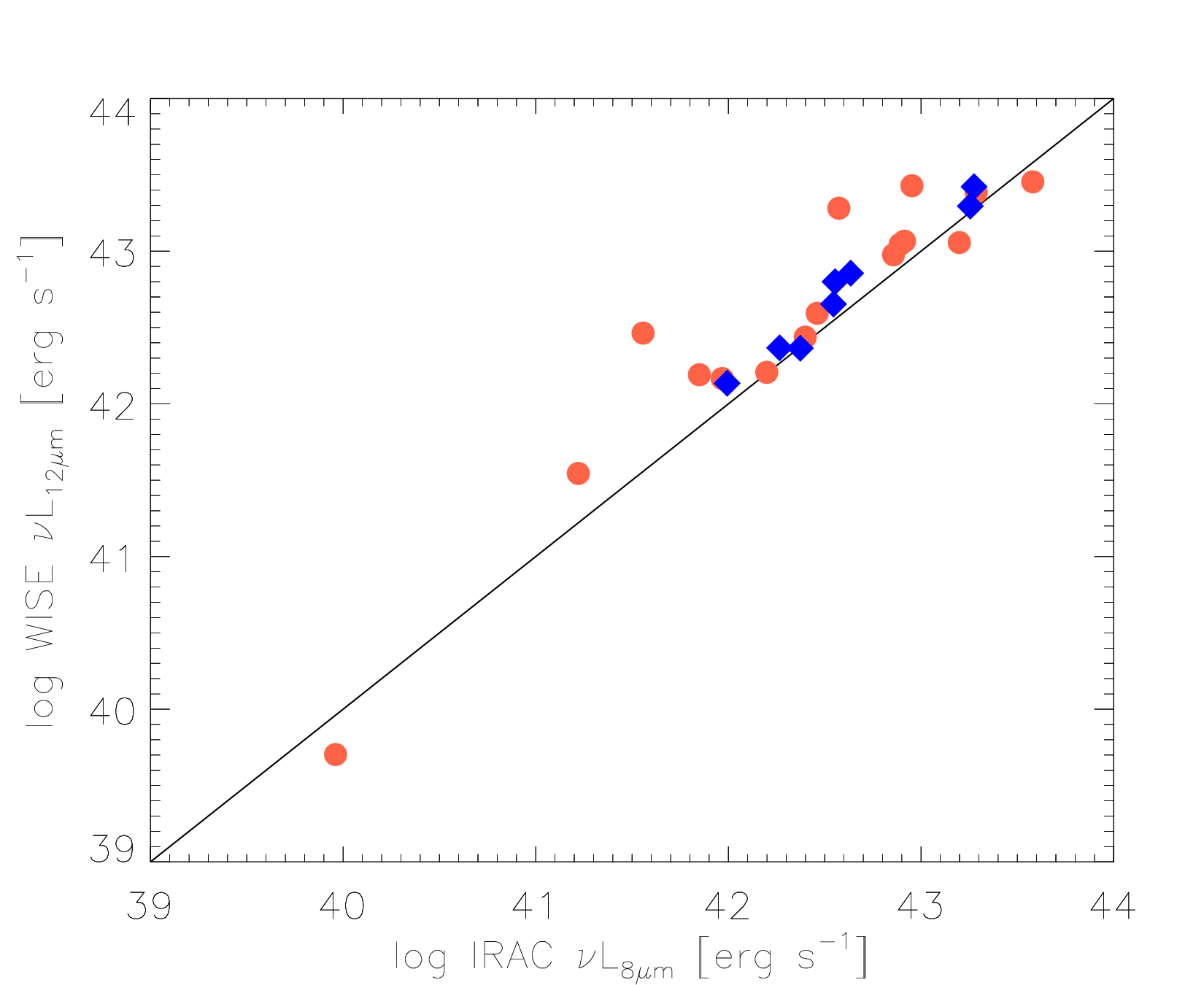}
\par} 
\caption{Left panel: arcsecond resolution 8~$\mu$m {\textit{IRAC}} nuclear luminosities versus subarcsecond resolution nuclear luminosities at 8~$\mu$m. Right panel: arcsecond resolution 8~$\mu$m {\textit{IRAC}} nuclear luminosities versus arcsecond resolution 12~$\mu$m {\textit{WISE}} luminosities. Blue diamonds and red circles are Sy1 and Sy2 galaxies, respectively. We plot the 1:1 line for comparison.}
\label{fig3}
\end{figure*}

In the left panel of Fig. \ref{fig3} we show the comparison between the arcsecond and subarcsecond resolution nuclear 8~$\mu$m luminosities. As expected, the lower angular resolution nuclear fluxes are generally larger than the subarcsecond resolution nuclear fluxes, being the median value of the ratio between subarcsecond and arcsecond resolution nuclear emission 0.44$\pm$0.05. Note that the median value of this ratio is smaller for Sy2 (0.38$\pm$0.06) than for Sy1 (0.74$\pm$0.09) galaxies. We only recover the subarcsecond nuclear fluxes, using the lower angular resolution MIR data, for the galaxies UGC\,6728 and NGC\,2110. These are the only galaxies in the BCS$_{40}$ sample with a $>$90\% AGN contribution to the {\textit{Spitzer/IRS}} spectrum (see Appendix \ref{B}). On the other hand, the point deviating the most from the 1:1 line is NGC\,4945, which has the most extended MIR emission of the BCS$_{40}$ sample.

We checked the relation between the arcsecond resolution 8~$\mu$m {\textit{IRAC}} nuclear emission and the {\textit{WISE}} photometry (see right panel of Fig. \ref{fig3}). We retrieved the {\textit{WISE}} photometry from the {\textit{WISE}} Source Catalog\footnote{http://irsa.ipac.caltech.edu/cgi-bin/Gator/nph-dd}, which were performed using a point source profile-fitting (angular resolution $\sim$6.5\arcsec~at 12~$\mu$m). We found that the majority of the {\textit{WISE}} fluxes are larger than the {\textit{IRAC}} fluxes, likely due to the larger scales probed by {\textit{WISE}} and to the possible contribution of the 11.3~$\mu$m PAH feature.

We note that we are not taking into account variability effects in the previous comparisons. However, the variability is not expected to be very important in the MIR, according to simulations and observations. Simulations predict longer variability time-scales in the MIR than in the optical (e.g. \citealt{Honig11}) and, using {\textit{Spitzer}} data, \citet{Garcia-Gonzalez15} found that the contribution of MIR-variable AGN to the general AGN population is small.

\section{Morphological analysis}
\label{morphology}

\subsection{Parsec-scale morphologies}
\label{nuclear_morph}
In Fig. \ref{figA1} of Appendix \ref{A} we show the subarcsecond resolution MIR data for the BCS$_{40}$ sample. We used different methods, which we describe below, to classify the MIR morphologies on scales of $\sim$400 pc (average value of the sample) using the ground-based MIR images for each galaxy.

First, we classified by eye the extension of the MIR emission following the same method as in \citet{Asmus2014}: a) point-like nucleus with no extended emission; b) possibly extended emission, marginal extension; and c) extended emission, with significant and consistent extension. For further details on the high angular resolution morphologies, see Fig. \ref{fig4} and Fig. \ref{figA1} of Appendix \ref{A}. A large fraction of the sample show extended or possibly extended MIR morphologies (83\%; see Table \ref{tab4}), with a variety of structures and features. In Appendix \ref{A} we also present the comparison between the arcsecond and subarcsecond resolution data in the N-band. In some cases we can identify similar structures and orientations of the extended MIR emission (ESO\,005-G004, NGC\,2992, NGC\,3227, NGC\,3783, NGC\,4945 and NGC\,7172). However, for the majority of the targets, the {\textit{IRAC}} PSF is larger than the extent of the subarcsecond resolution MIR emission. Therefore, the high angular resolution data are crucial to study the circumnuclear emission of Seyfert galaxies.

\input{tab4.tex}

\begin{figure}
\centering
\includegraphics[width=8.0cm]{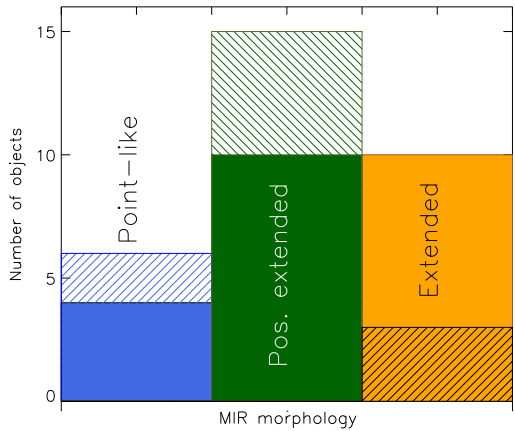}
\caption{Classification of the BCS$_{40}$ morphologies from the MIR subarcsecond resolution images. 
The filled and hatched histograms represent the visual and quantitative classifications, respectively.
Blue, green and orange histograms correspond to point-like, possibly extended and extended morphologies.}
\label{fig4}
\end{figure}

\begin{figure}
\centering
\includegraphics[width=8.8cm]{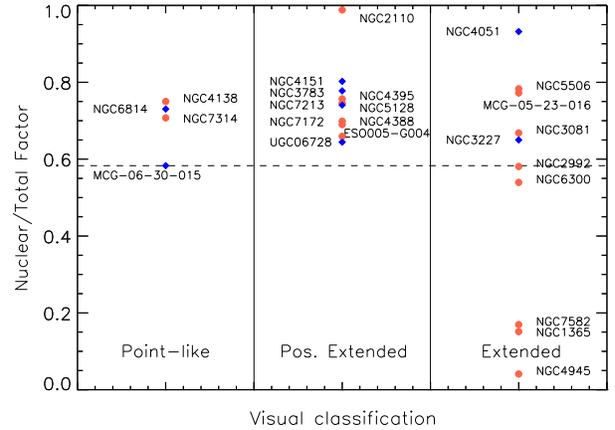}
\caption{Comparison between the morphological classifications by eye and using the nuclear versus high angular resolution total flux ratio. Blue diamonds and red circles are Sy1 and Sy2 galaxies, respectively. The horizontal black dashed line corresponds to the minimum value of the compactness factor for point-like morphologies.}
\label{fig5}
\end{figure}

Second, we used the high angular resolution nuclear and total fluxes to evaluate the strength of the nuclear emission against the total emission. To do so, we calculated the compactness factor as the ratio between the nuclear and total flux. Since we have heterogeneous photometry, we calculated the compactness factor for each galaxy using the average ratio of all the bands available. Using this factor and comparing it with the results from the visual classification, we find that our sample is dominated by galaxies whose circumnuclear emission
has low surface brightness compared with the nuclear emission (see Fig. \ref{fig5}). The horizontal black dashed line in Fig. \ref{fig5} corresponds to the minimum value of the compactness factor for point-like morphologies ($\sim$0.6). The overall majority of the extended and possibly extended galaxies are above this value.
This is due to the low surface brightness of the extended emission (e.g. MCG-05-23-016, NGC\,2992, NGC\,3081, NGC\,3227, NGC\,4051 and NGC\,5506). Therefore, in addition to quantitative methods, it is important to perform visual classifications in order to detect low surface brightness extended MIR emission when present. On the other hand, there are few galaxies with bright extended emission (see Table \ref{tab4} and Fig. \ref{fig5}). 

Third, we use the residuals from the PSF subtraction method described in Section \ref{nuclear_fluxes} and we found that most of the galaxies ($\sim$75\%) show residuals at $\geq$3$\sigma$, which correspond to the circumnuclear emission (see Table \ref{tab4}).

Finally, using the two quantitative methods described above, we can calculate the strength of the extended emission and compare it with the visual classification. Thus, by using the nuclear to total flux ratio and the 3$\sigma$ residuals, we consider the MIR morphologies: a) point-like, when there is $\geq$50\%  nuclear contribution to the total flux and no 3$\sigma$ residuals; b) possibly extended, when there is $\geq$50\%  nuclear contribution to the total flux and 3$\sigma$ residuals; and c) extended, when there is $\leq$50\%  nuclear contribution to the total flux and 3$\sigma$ residuals. The result of this classification is shown in Table \ref{tab4} and in Fig. \ref{fig4}.

Using different methods to classify the MIR morphologies we found, from visual classification, that the majority of the sample show extended or possibly extended morphologies (83\%) and 17\% are point-like. From the quantitative classification, we found that most of the galaxies present extended or possibly extended emission (75\%) and 25\% are point-like. Therefore, there is a good agreement between the results obtained using the qualitative and quantitative methods (see Fig. \ref{fig4}). This extended emission represents, on average, $\sim$30\% of the total emission of the BCS$_{40}$ sample ($\sim$25\% for Sy1 and $\sim$30\% for Sy2).

We found that 87\% of the Sy2 in the BCS$_{40}$ show extended MIR morphologies, versus 75\% for Sy1 galaxies. We used the Fisher's exact test method and we found that this difference is not significant.

The percentage of point-like morphologies measured for the BCS$_{40}$ sample is in agreement with the results reported by \citet{Asmus2014} for a sample of 204 AGNs detected in the MIR with ground-based instruments (19\%). However, we found a larger contribution of extended or possibly extended MIR morphologies (75-83\% of the BCS$_{40}$ sample) than the 47\% (21\% extended and 26\% possibly extended) reported by \citet{Asmus2014}\footnote{The remaining 34\% of the sample studied in \citet{Asmus2014} corresponds to galaxies with unknown extension due to insufficient data.}. The differences between the MIR morphologies of the two samples could be related with a distance effect, since the median value of the distance for the sample analyzed by \citet{Asmus2014} is 71.7\,Mpc, whereas for our sample is 24.1\,Mpc. 

Previous studies of LIRGs and ULIRGs based on ground-based MIR data as those presented here concluded that AGNs are, in general, less extended than SF-dominated systems (e.g. \citealt{Diaz-santos10}, \citealt{Diaz-santos11}, \citealt{Imanishi11}). \citet{Diaz-santos10} found that, on average, the MIR continuum becomes more compact than the PAH emission as the MIR is increasingly dominated by the AGN. In \citet{Imanishi11}, the authors studied a sample of 18 ULIGRs and found that SF-dominated galaxies have more extended emission, with low surface brightnesses, than AGN-dominated galaxies, which show more compact MIR morphologies. \citet{Soifer01} found different sizes (ranging from 100 pc to 1.5 kpc) for the MIR emission using ground-based data of a sample of starburst galaxies. These sizes are in good agreement with our results, as we found that the extended MIR emission in AGN-dominated systems is more compact (300$\pm$100\,pc) than in SF-dominated systems (650$\pm$700\,pc) and composite galaxies (350$\pm$500\,pc). The classification of the galaxies as AGN-dominated, SF-dominated and composite comes from the spectral decomposition of the {\textit{Spitzer/IRS}} spectra (see Appendix \ref{B}).

\begin{figure}
\centering
\par{
\includegraphics[width=8.0cm]{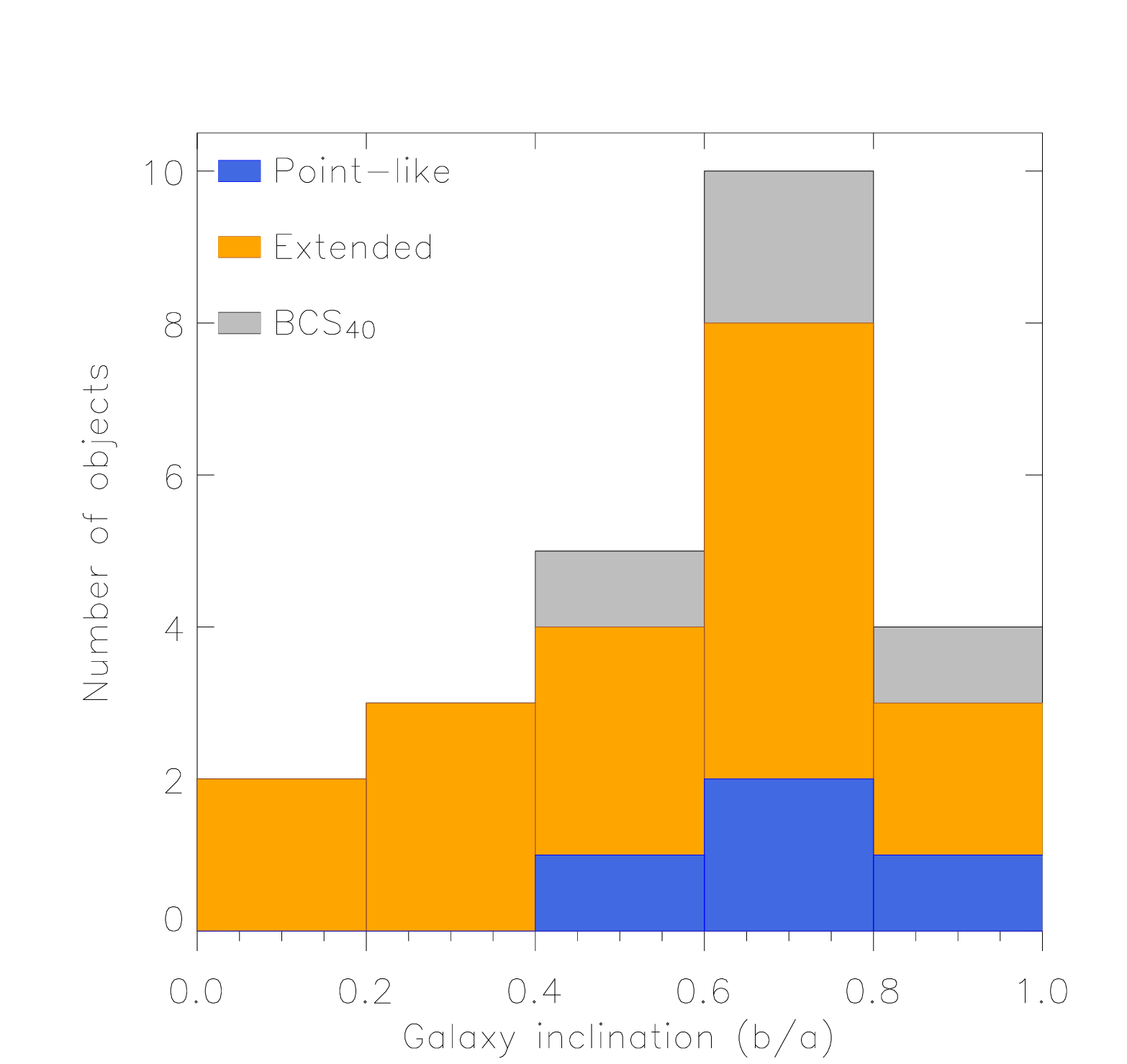}
\includegraphics[width=8.0cm]{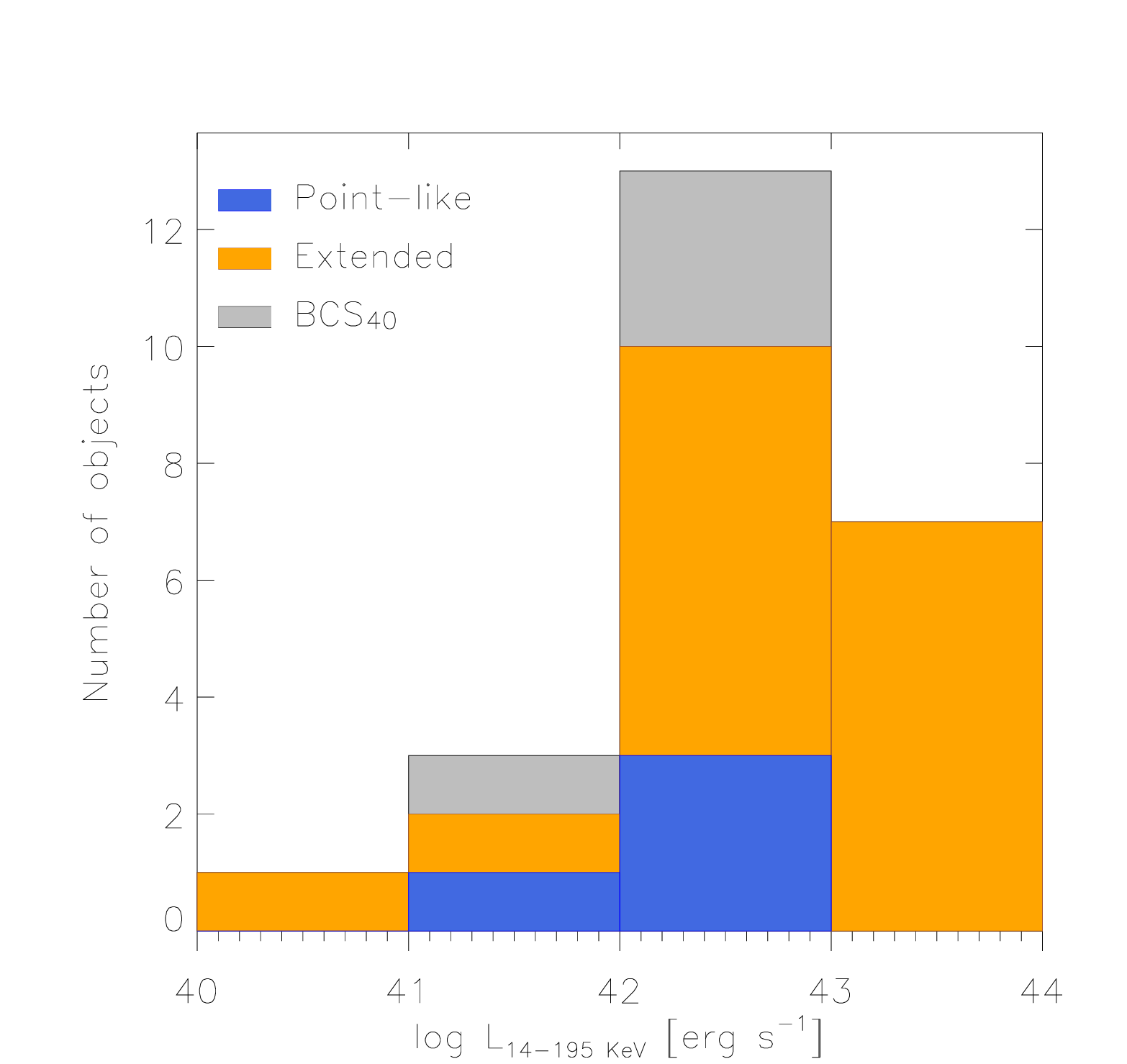}
\par}
\caption{Top panel: Galaxy inclination (b/a) distribution of the BCS$_{40}$ sample.  Bottom panel: 14-195~keV X-ray luminosity distribution of the sample. The blue and orange filled histograms are the distributions of point-like and extended MIR morphologies, respectively. The grey filled histograms correspond to the total BCS$_{40}$ sample.}
\label{fig6}
\end{figure}

We finally checked whether the different MIR morphologies that we find are related to galaxy inclination and/or luminosity. Using the visual classification method, we found that the galaxies with point-like MIR morphologies are face-on or moderately inclined (b/a$\sim$0.4-1.0). On the other hand, the galaxies which are extended in the MIR have different values of b/a, from edge-on to face-on (see top panel of Fig. \ref{fig6}). Regarding the AGN luminosity, we found the galaxies with point-like morphologies having intermediate luminosities within the range covered by the galaxies showing extended morphologies (see bottom panel of Fig. \ref{fig6}).

\subsection{Kpc-scale morphologies}
\label{extended}

\begin{figure}
\centering
\par{
\includegraphics[width=3.2cm, angle=90]{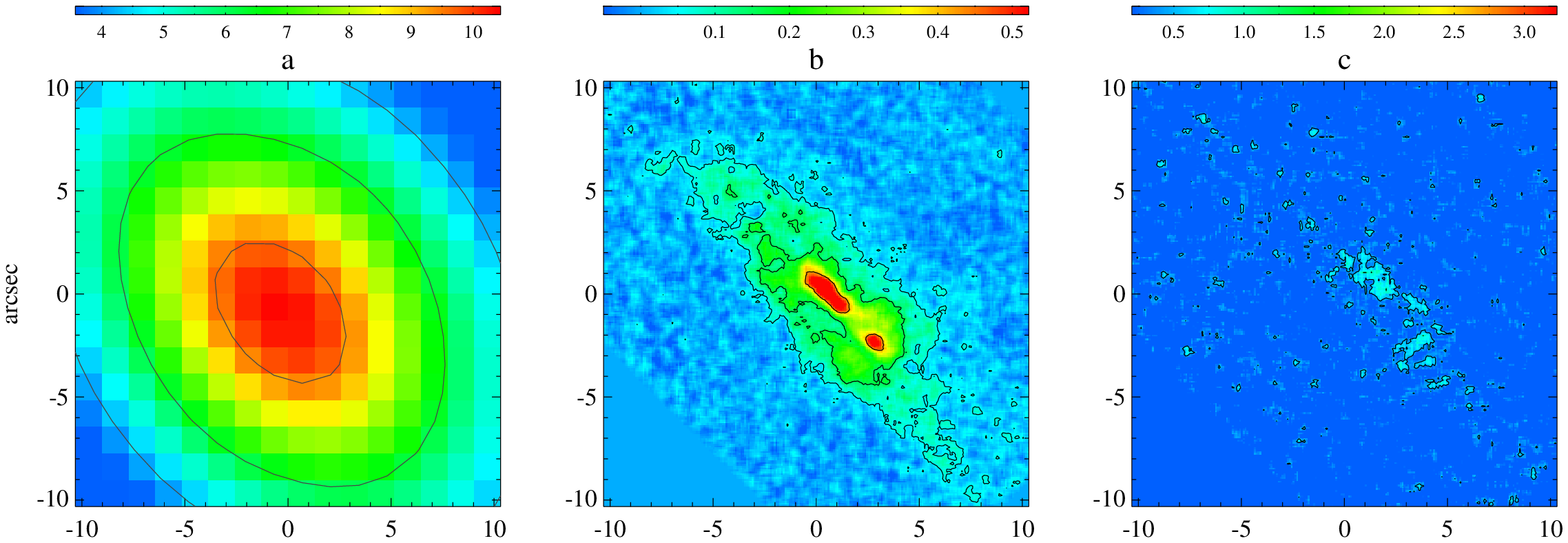}
\includegraphics[width=3.2cm, angle=90]{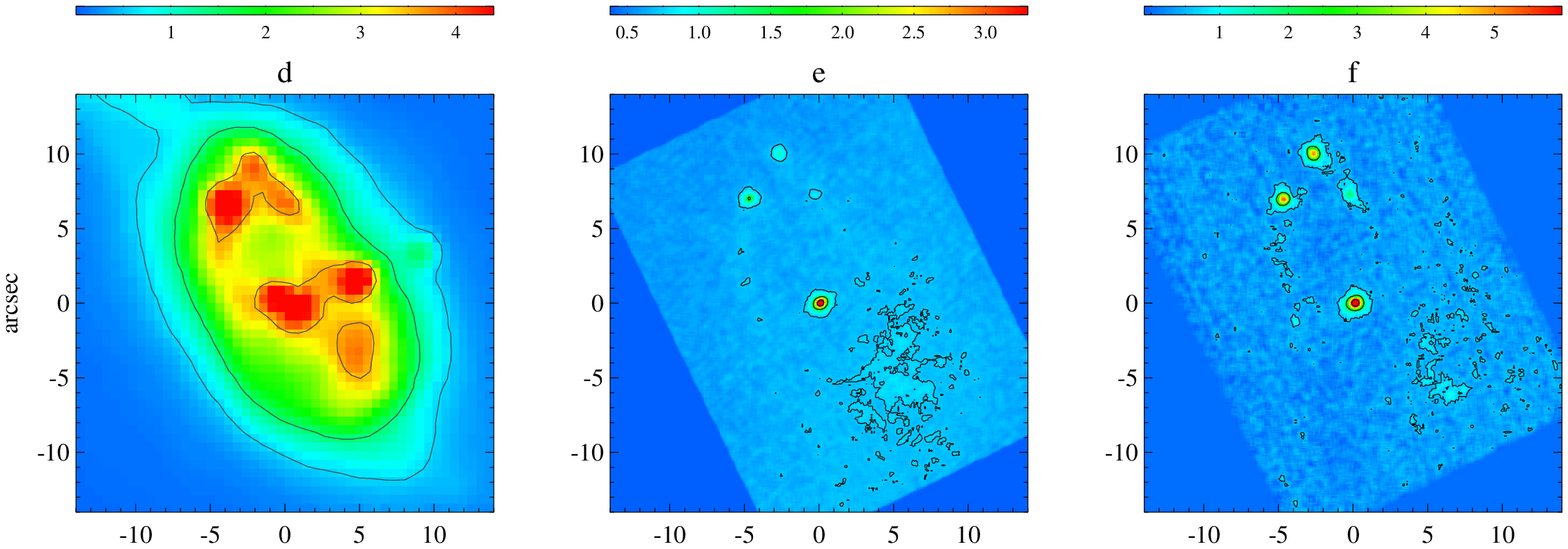}
\includegraphics[width=3.2cm, angle=90]{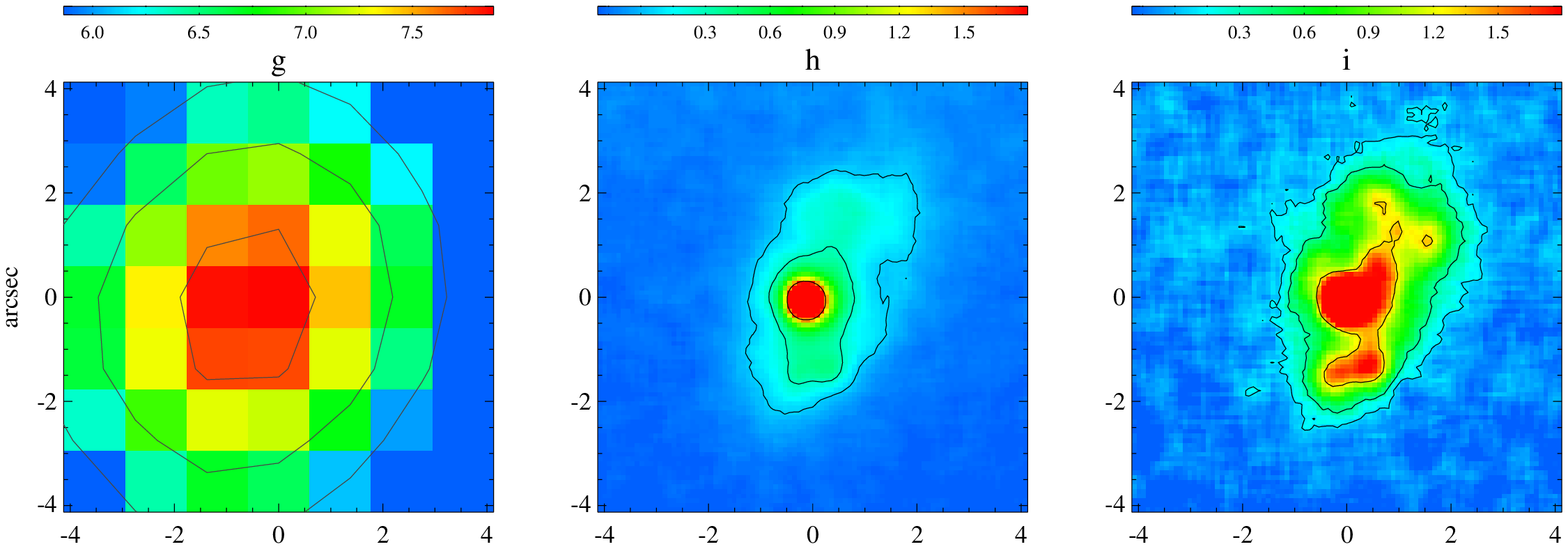}
\includegraphics[width=3.2cm, angle=90]{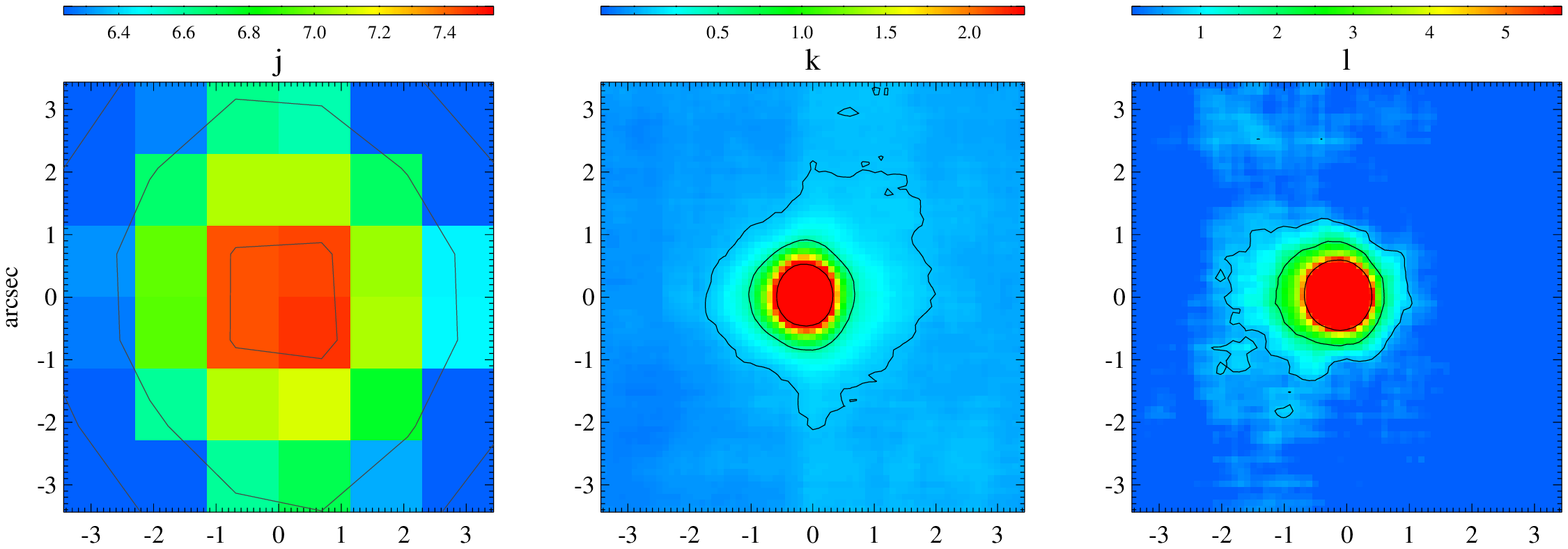}
\includegraphics[width=3.2cm, angle=90]{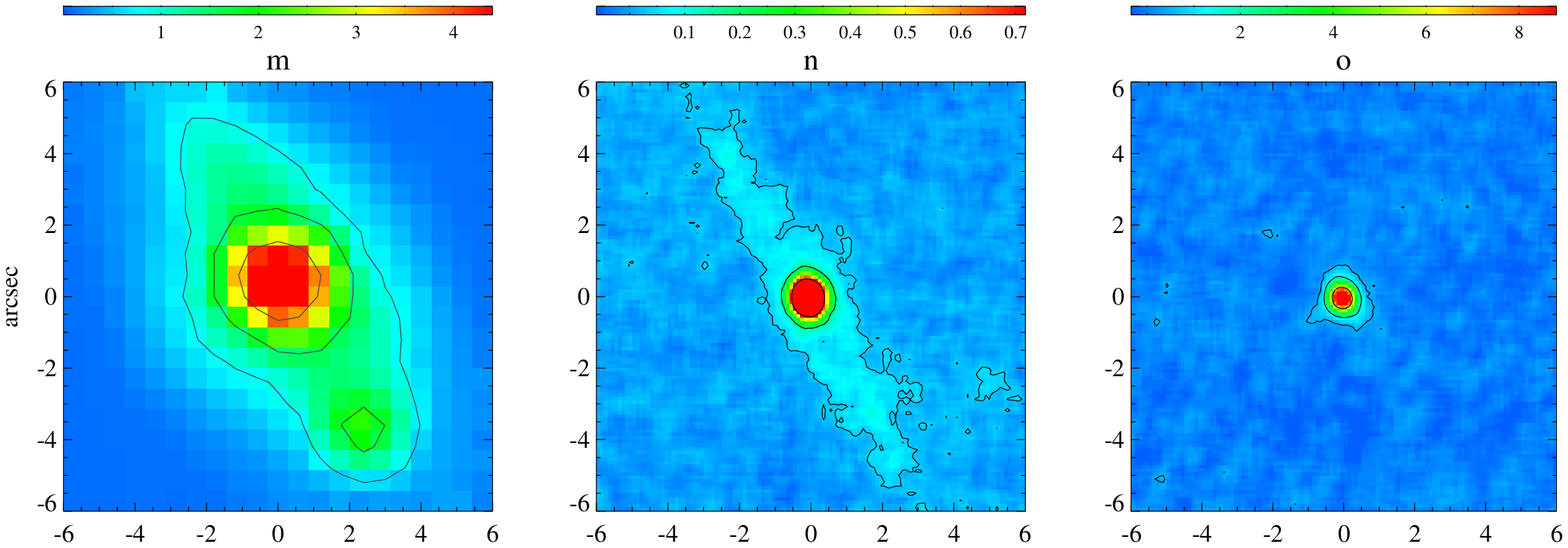}
\includegraphics[width=3.2cm, angle=90]{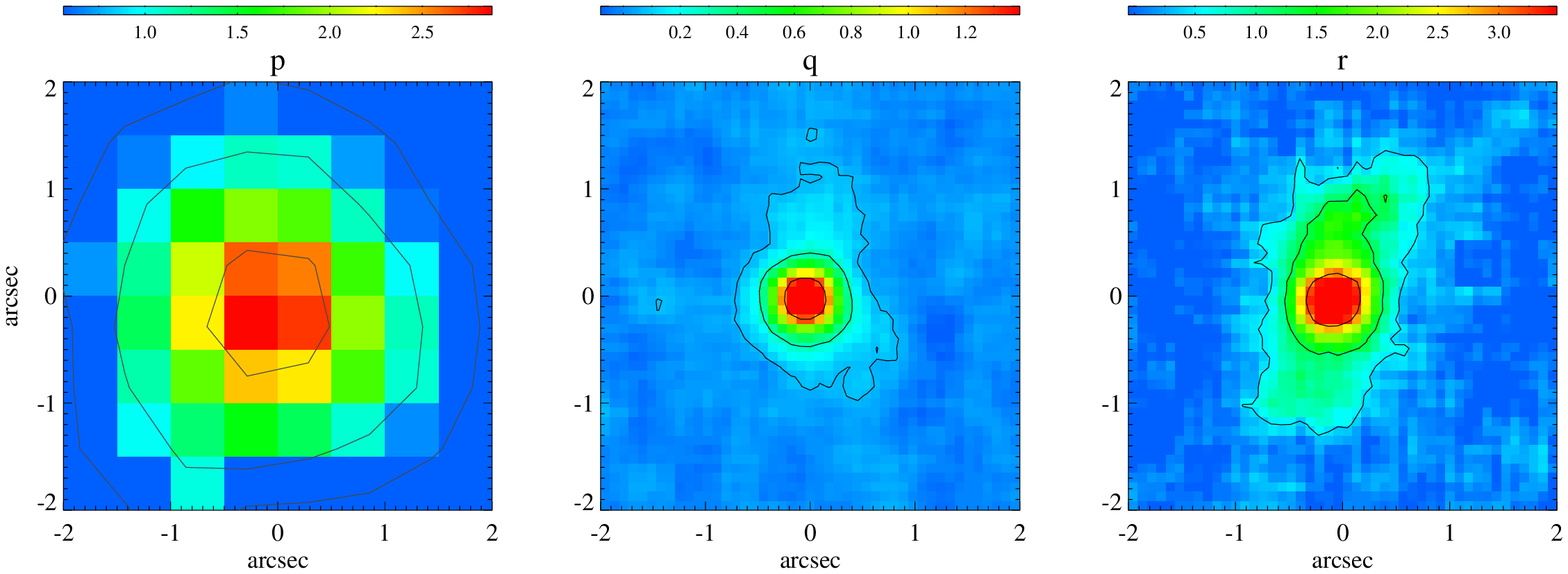}
\par}
\caption{Galaxies showing extended MIR emission on scales larger than 400\,pc in the high angular resolution data. From left to right: arcsecond resolution data, N- and Q-band subarcsecond resolution data. (a)-(c) NGC\,4945; (d)-(f) NGC\,1365; (g)-(i) NGC\,7582; (j)-(l) NGC\,5506; (m)-(o) NGC\,2992; (p)-(r) NGC\,3081. All images have been smoothed (3 pixel box). Colour bars correspond to fluxes in mJy/pixel units. North is up, and east to the left.}
\label{fig7}
\end{figure}

Despite the fact that a large fraction of the sample (75-83\%) show extended morphologies, most of this extended emission is compact and concentrated close to the nuclear region, and it has low surface brightness compared with the nuclear emission. Only six Sy2 galaxies present extended MIR emission at large-scales ($>$400 pc, which is the sample average value) in the high angular resolution MIR data. These galaxies were previously studied in the MIR: a) NGC\,4945 \citep{Imanishi11}; b) NGC\,1365 \citep{Herrero12}; c) NGC\,7582 \citep{Wold06,WoldGaliano06}; d) NGC\,5506 \citep{Roche07}; e) NGC\,2992 \citep{Bernete2015}; and f) NGC\,3081 \citep{Ramos11c}. Here we make a comparison between this large-scale MIR emission as detected in the arcsecond and subarcsecond resolution data. In Fig. \ref{fig7} we show the arcsecond resolution 8~$\mu$m {\textit{IRAC}} or 12$~\mu$m {\textit{WISE}} images and the subarcsecond resolution N- and Q-band images of these galaxies. The extended emission detected in the high angular resolution data has a similar structure and orientation than those detected in the {\textit{IRAC}} or {\textit{WISE}} images on the same scales (see Fig. \ref{fig7}). For 4/6 of these galaxies (NGC\,1365, NGC\,2992, NGC\,4945 and NGC\,7582) the bulk of this extended emission is due to SF activity, while for NGC\,3081 and NGC\,5506 is mainly produced by the AGN activity. Although these four galaxies show strong PAH features in their IR spectra, there are other galaxies in the sample with strong PAH features as well which do not show extended emission at large-scales (e.g. NGC\,7172 and ESO\,005-G004). Perhaps the circumnuclear regions of these six galaxies are dustier than the rest. See Appendix \ref{c} for details on the extended emission of these six objects.

We also used the {\textit{IRAC}} images to compare the luminosities of the kpc-scale MIR extended emission of Seyfert galaxies with those of normal galaxies. To do so, we calculated {\textit{IRAC}} 8~$\mu$m total fluxes\footnote{To determine the total fluxes we used the same method as in Section \ref{nuclear_fluxes}.} for the galaxies in our sample that have IRAC data and we found a median value of log($\nu$L$_{8\mu m}$)$=$43.17$\pm$0.06~erg\,s$^{-1}$. The median value reported by \citet{Munoz-Mateos09} for 70 nearby galaxies in the SINGs sample \citep{Kennicutt03} is log($\nu$L$_{8\mu m}$)$=$42.52$\pm$0.11~erg\,s$^{-1}$. As expected, the total MIR luminosity of Seyfert galaxies is higher than in normal galaxies, due to the AGN contribution. If we subtract the {\textit{IRAC}} nuclear luminosities from our total luminosities, we measured log($\nu$L$_{8\mu m}$)$=$43.06$\pm$0.10~erg\,s$^{-1}$ for the extended emission, which remains larger than the value reported for the SINGs sample. This is likely due to the extra-contribution of AGN-heated dust in the circumnuclear region of the galaxies in our sample.

Finally, using the arcsecond resolution data, we also classified the sample morphologies as: a) irregular; b) point-like; c) elliptical; and d) spiral.  We found that $\sim$60\% of the sample present spiral morphologies (see Table \ref{tab5}). In Table \ref{tab5} we have also indicated whether or not the galaxies show a compact nucleus. Using this classification, we found that all the galaxies without a compact nucleus have extended morphologies according to our visual classification of the subarcsecond resolution data. 

\input{tab5.tex}

\section{MIR correlations with AGN and SF indicators}
\label{correlations}

Dust grains in the nuclear region of AGN are heated mainly by nuclear activity, although other heating sources can be SF and/or jets. This radiation is re-emitted in the IR range, peaking in the MIR. On the other hand, due to the high energies involved in the accretion process, X-rays are good tracers of the AGN power. For this reason, the X-ray MIR correlation has been widely used in the literature to investigate the origin of the MIR emission in different types of AGN \citep{Lutz04,Ramos07,Horst08,Gandhi09,
Levenson09,Ichikawa2012,Mason12,Matsuta2012,Sazonov12,Asmus2015,Mateos15}.

Another AGN tracer commonly used is [O\,IV]$\lambda$25.89~$\mu$m (IP$\sim$55~eV; \citealt{Spinoglio92,Spinoglio12}), which correlates well with both the hard X-rays \citep{Melendez08,Rigby09,Diamond09} and the soft X-rays \citep{Prieto02}. The [O\,IV] emission line has proved to be an accurate indicator of the AGN luminosity (see \citealt{Weaver2010} and references therein), and particularly in the case of Seyfert galaxies \citep{Pereira-Santaella2010}.

On the other hand, low IP MIR emission lines such as [Ne\,II]$\lambda$12.81~$\mu$m (IP$\sim$21\,eV) are used to quantify the SFR of Seyfert galaxies \citep{Spinoglio92,Ho07,Pereira-Santaella2010,Spinoglio12}. However, for luminous AGNs there is a significant contribution of the AGN to the [Ne\,II]~$\lambda$12.81~$\mu$m emission \citep{Pereira-Santaella2010}. We refer the reader to \citet{Roche91,Spinoglio92,Ho07,Pereira-Santaella2010,Dasyra11,Spinoglio12} for further discussion on the MIR diagnostics described above. 

It is interesting then to correlate different AGN and SF indicators with the MIR 
nuclear and circumnuclear emission of Seyfert 
galaxies to investigate the dominant heating source of the dust. With this aim, we compiled X-ray luminosities (the intrinsic 2-10~keV and 14-195~keV) and integrated MIR emission line luminosities ([O\,IV]$\lambda$25.89~$\mu$m and [Ne\,II]$\lambda$12.81~$\mu$m) for our 
sample (see Table \ref{tab3}) and we performed linear regressions in log$-$log space (see Table \ref{tab6}). We note that, although hereafter we will refer to luminosity-luminosity correlations only, we confirmed that the results hold in flux-flux space (see Table \ref{tabD1} of Appendix \ref{D}).

\input{tab6.tex}

\begin{figure*}
\centering
\par{
\includegraphics[width=8.8cm]{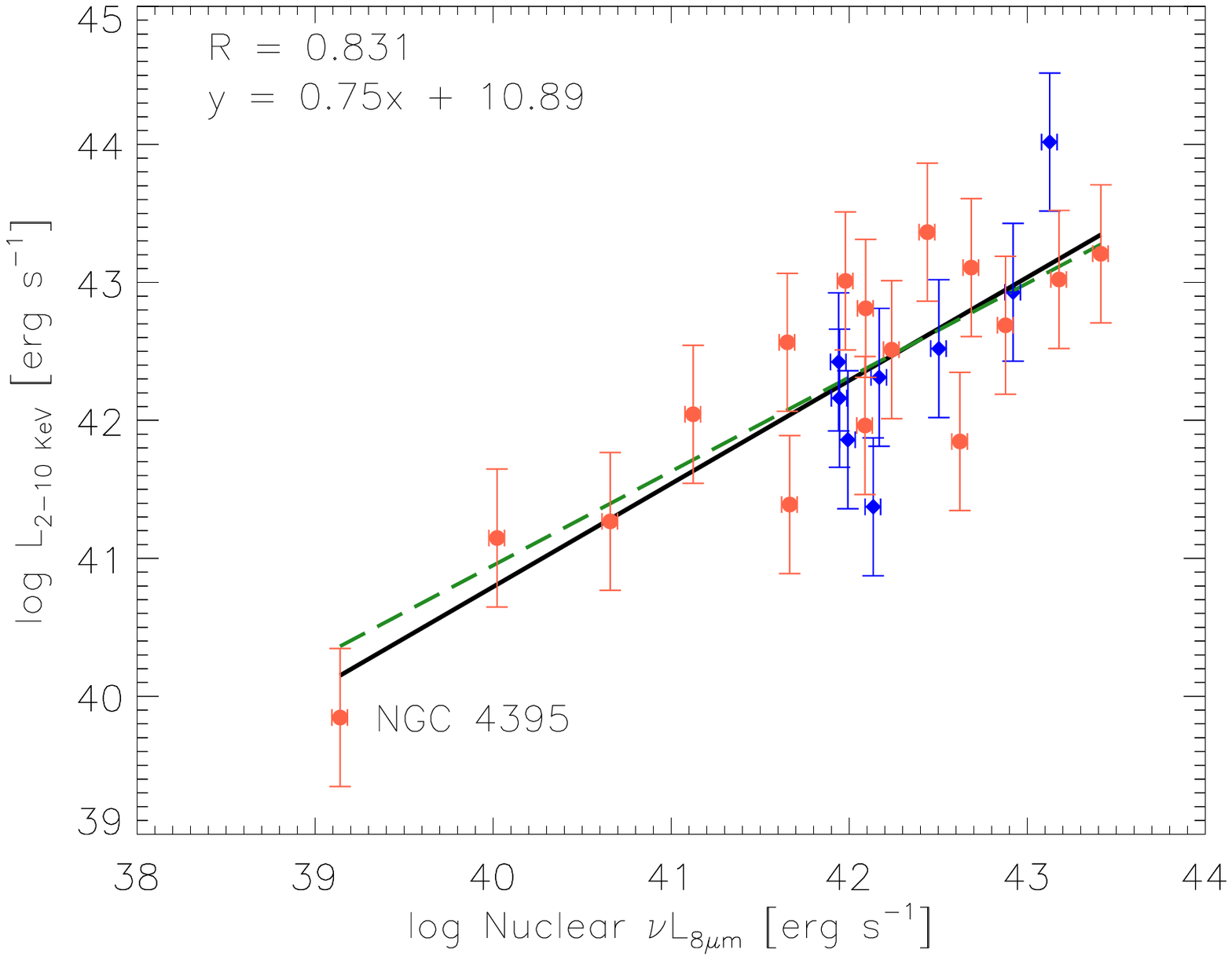}
\includegraphics[width=8.8cm]{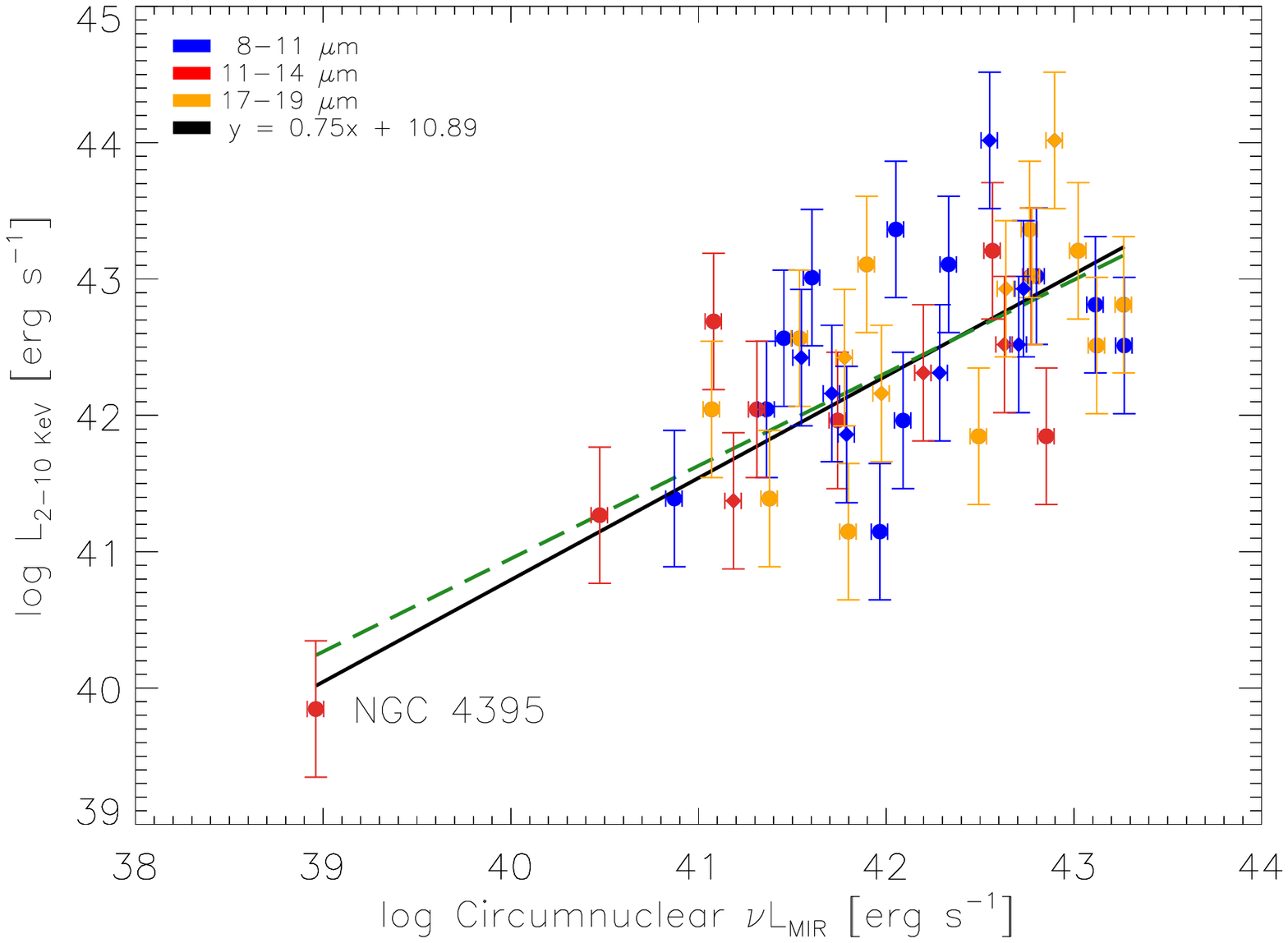}
\includegraphics[width=8.8cm]{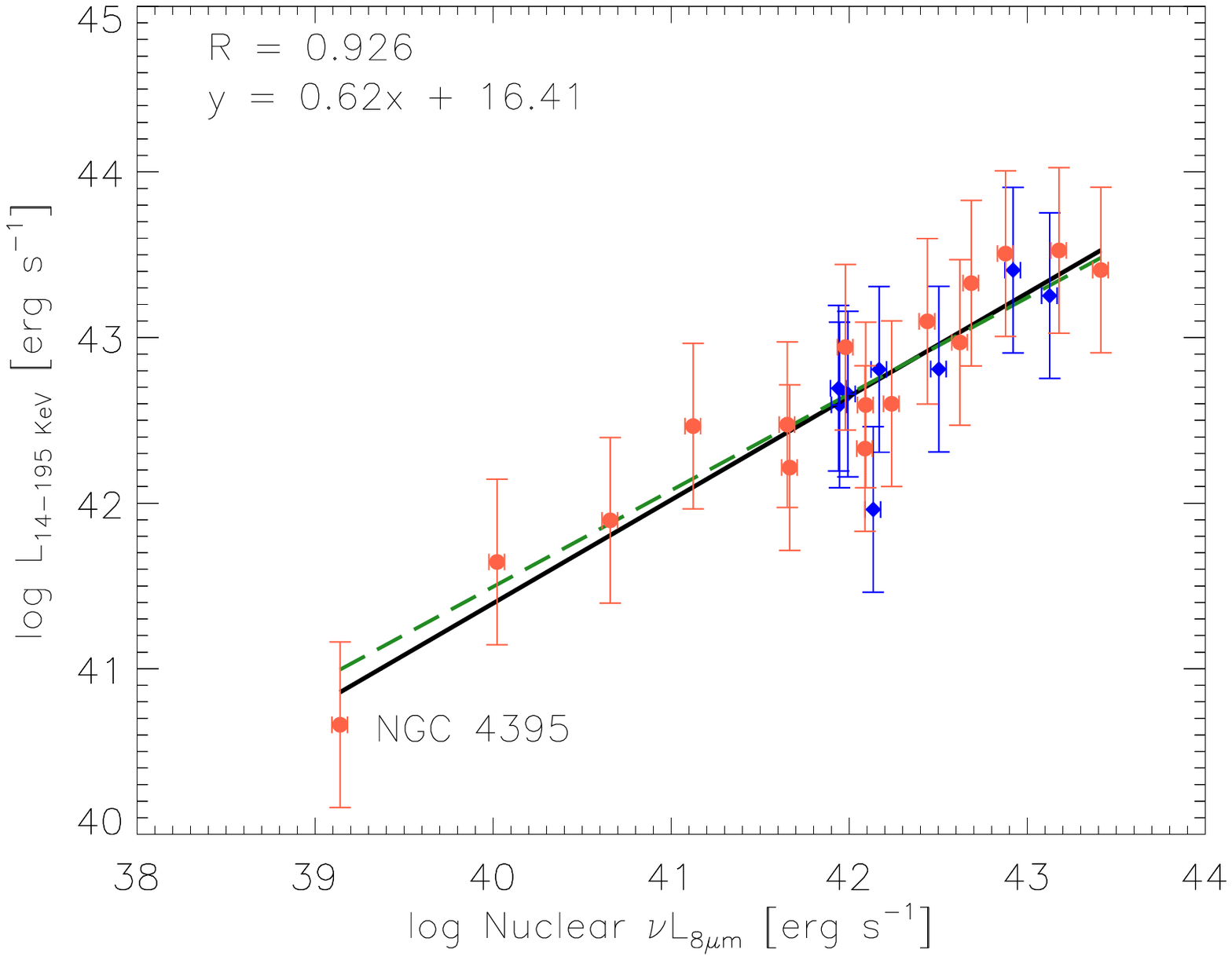}
\includegraphics[width=8.8cm]{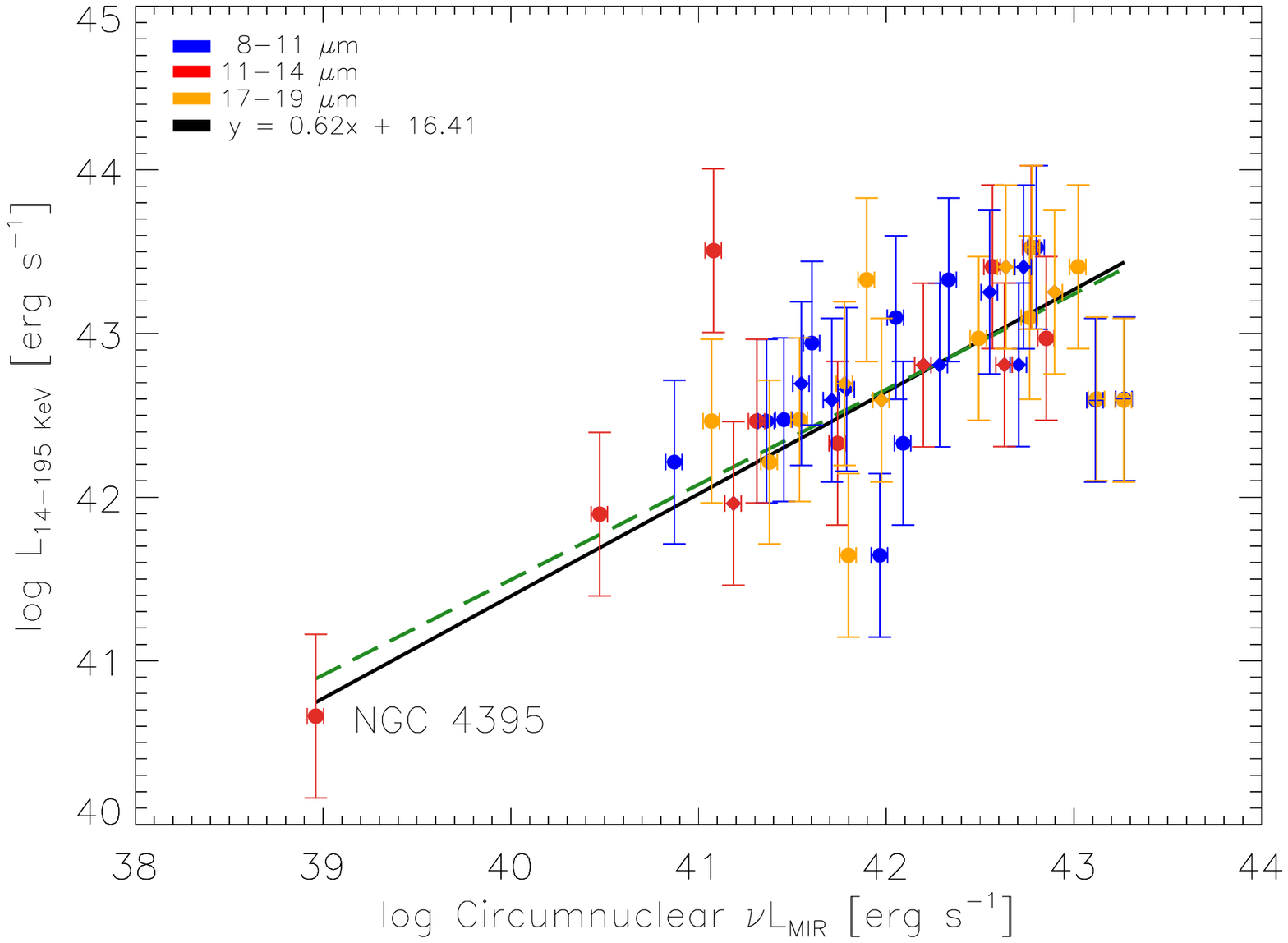}
\par} 
\caption{High angular resolution MIR--X-ray luminosity correlations. Top left panel: 8~$\mu$m nuclear 
luminosity versus intrinsic 2-10~keV X-ray luminosity. Bottom left panel: same as in the top left panel, 
but for the 14-195~keV X-ray luminosity. Right panels: Same as in the left panels, but for the circumnuclear MIR luminosities,
with different colors indicating different wavelengths. The black lines correspond to the nuclear correlation results from 
the fits shown in the left panels. The green dashed lines correspond to the same fit, but excluding NGC\,4395. Blue diamonds and red 
circles are Sy1 and Sy2 galaxies, respectively. The vertical error bars correspond to one order of magnitude, which is the uncertainty associated to multi-epoch X-ray measurements. The horizontal error bars correspond to the photometry uncertainty reported in Section \ref{nuclear_fluxes}.}
\label{fig8}
\end{figure*}

In the left panels of Fig. \ref{fig8} we show the relation between the high angular resolution MIR nuclear 
emission and the intrinsic 2-10~keV and 14-195~keV X-ray emission. As can be seen from the 
top left panel of Fig. \ref{fig8}, there is a tight correlation between the MIR nuclear luminosities and the intrinsic 2-10~keV X-ray emission (Pearson's correlation coefficient R=0.83). This correlation 
improves when we use the harder 14-195~keV X-ray luminosity (R=0.93; bottom left panel of Fig. \ref{fig8} and Table \ref{tab6}). The tight correlations shown in the left panels of Fig. \ref{fig8} suggest that our nuclear MIR fluxes are AGN-dominated. Points deviating from this correlation indicate the presence of other heating sources (i.e. star formation and/or jets) on the scales probed by the data (the inner $\sim$70 pc). 

The slope of the nuclear 8~$\mu$m--2-10~keV correlation is 0.75$\pm$0.11, in agreement with previous 
studies of Seyfert galaxies using similar MIR wavelengths (6 to 12~$\mu$m; e.g. \citealt{Lutz04,Ramos07,Horst08,Gandhi09,Levenson09,Sazonov12}). For example, \citet{Ramos07} found 
a slope of 0.8 by using nuclear fluxes obtained from ISOCAM data at 6.75~$\mu$m. Using 
high angular resolution 12 $\mu$m fluxes from VISIR, \citet{Horst08} and \citet{Gandhi09} found slopes of $\sim$1 and 0.9, respectively. Finally, \citet{Levenson09} reported a value between 0.7-0.9 for this slope by employing different methods to obtain nuclear fluxes at 8--12 $\mu$m using data from the Gemini telescopes. 

We also find good nuclear MIR--hard X-ray correlations when we perform the fits for the Sy1 and Sy2. In 
the case of the MIR--2-10~keV correlation, we measure slopes of 1.4$\pm$0.4 and 0.7$\pm$0.1 for Sy1 and Sy2 (R=0.83 and 0.86 respectively). This difference practically disappears when we use the 14--195~keV
luminosity (see bottom left panel of Fig. \ref{fig8}). In this case we measure slopes of 0.7$\pm$0.3 and 0.6$\pm$0.1 for Sy1 and Sy2, with R=0.75 and 0.96 respectively. The less significant correlation that we found  
for the Sy1 is likely a consequence of the small number of objects. We conclude that the fits to the Sy1 and Sy2 galaxies are consistent with each other, indicating that the nuclear 8~$\mu$m emission of Seyfert 
galaxies is essentially independent of the viewing angle and line-of-sight obscuration and nearly isotropic on the small scales probed here (the inner $\sim$70 pc). We have also compared the nuclear MIR luminosity distributions of Sy1 and Sy2 galaxies using the KS test and we do not find significant differences between the two Seyfert types. This is in agreement with the predictions from clumpy torus models and with observational results: the nuclear 8--18 $\mu$m SEDs of Sy1 and Sy2 galaxies are almost identical (e.g. \citealt{Nenkova08,bNenkova08,Ramos2011}). 

As we have quantified the circumnuclear emission of the galaxies, we can do the same exercise to investigate the heating source of this MIR extended emission. However, in the case of the circumnuclear emission we have heterogeneous photometry (i.e., fluxes measured in different filters), and therefore we cannot perform linear fits as we did for the nuclear fluxes\footnote{We derived nuclear fluxes at 8 $\mu$m by combining the high angular resolution data and spectral decomposition of the Spitzer spectra, but unfortunately, we cannot do the same for the circumnuclear emission.}. Thus, we separated the circumnuclear luminosities in three bands (8--11~$\mu$m, 11--14~$\mu$m and 17--19~$\mu$m) and plotted them in the right panels of Fig. \ref{fig8}, with different colors indicating different wavelengths, versus the hard X-ray luminosities. We show the results from the nuclear fits (those in the left panels of Fig. \ref{fig8}) for comparison, and in order to investigate the scatter of the circumnuclear luminosities, we measured the mean absolute deviations of this emission from the nuclear fits (see Table \ref{tab6}).

In general, there is also a correlation between the circumnuclear MIR and hard X-ray luminosity (at 2-10~keV and 14-195~keV), although with more scatter than that of the nuclear MIR emission. The mean absolute deviations from the nuclear fits are 0.63 and 0.56 for the 2-10~keV and 14-195~keV luminosities, respectively. Indeed, some of the points deviating more from the nuclear fits correspond to the galaxies whose extended MIR emission is mainly produced by SF (see Appendices \ref{B} and \ref{c}). Therefore, for the majority of the galaxies, we find that the AGN is the main contributor to their circumnuclear emission (the inner $\sim$400 pc), with some contribution from SF. Other AGN components such as jets and narrow emission line clouds, apart from the dusty torus, emit in the MIR, and this emission can be detected at kpc-scales. For example, in the case of the Sy2 galaxy NGC\,1068, \citet{Mason06} detected low surface brightness MIR emission from dust in the ionization cones extending to hundreds of parsecs.

Another interesting result from the right panels of Fig. \ref{fig8} is that the correlation seems to be independent of the MIR wavelength chosen. The distribution of the points in the MIR--X-ray plots is similar for the three ranges considered (8--11 $\mu$m, 11--14 $\mu$m, and 17--19 $\mu$m), which is in agreement with the results reported by \citet{Ichikawa2012} using lower angular resolution {\textit{AKARI}} data. Moreover, the circumnuclear MIR luminosities of the two Seyfert types are almost identical according to the KS test, as we also found for the nuclear luminosities. This is also compatible with an AGN-dominated circumnuclear emission.

\begin{figure*}
\centering
\par{
\includegraphics[width=8.8cm]{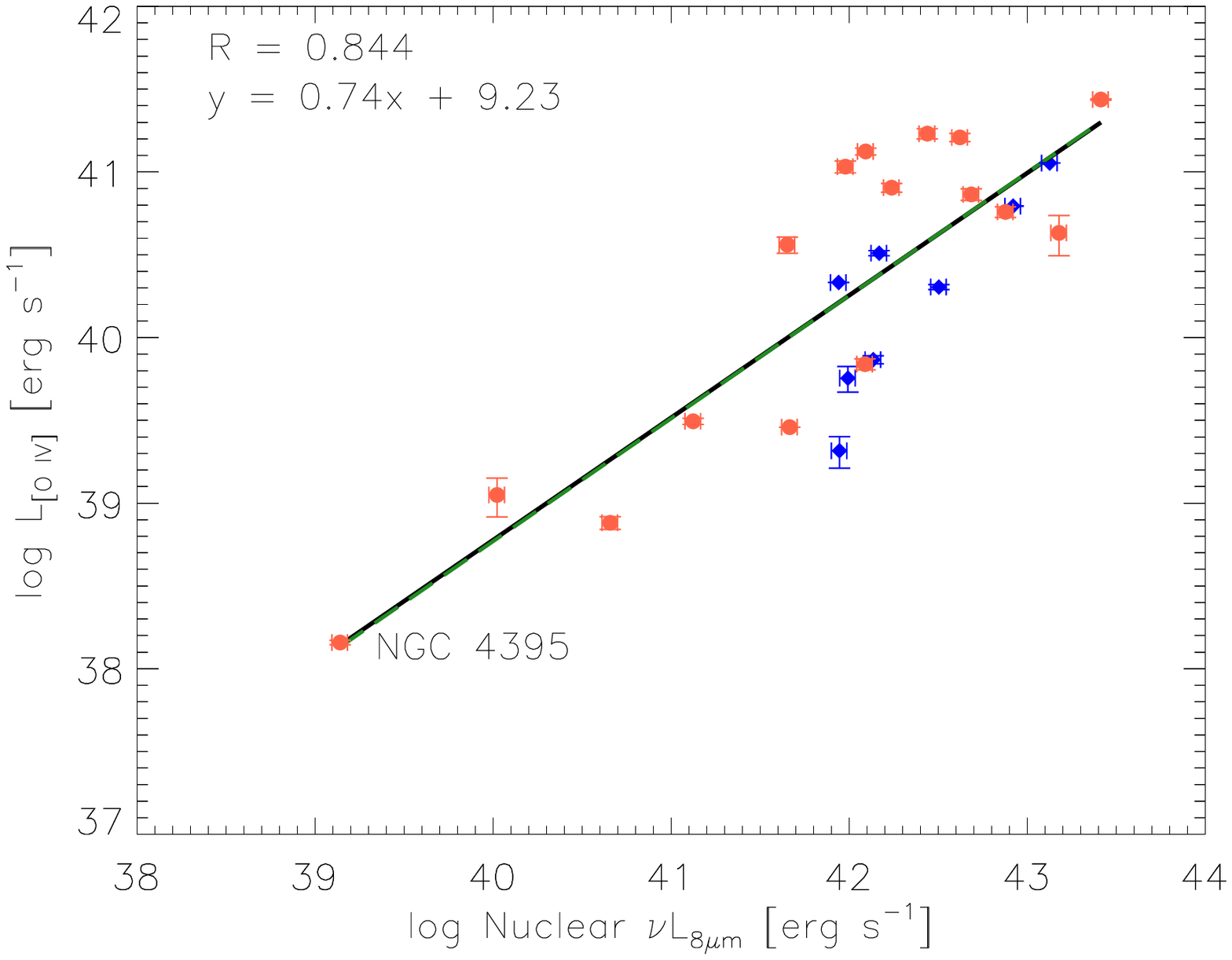}
\includegraphics[width=8.8cm]{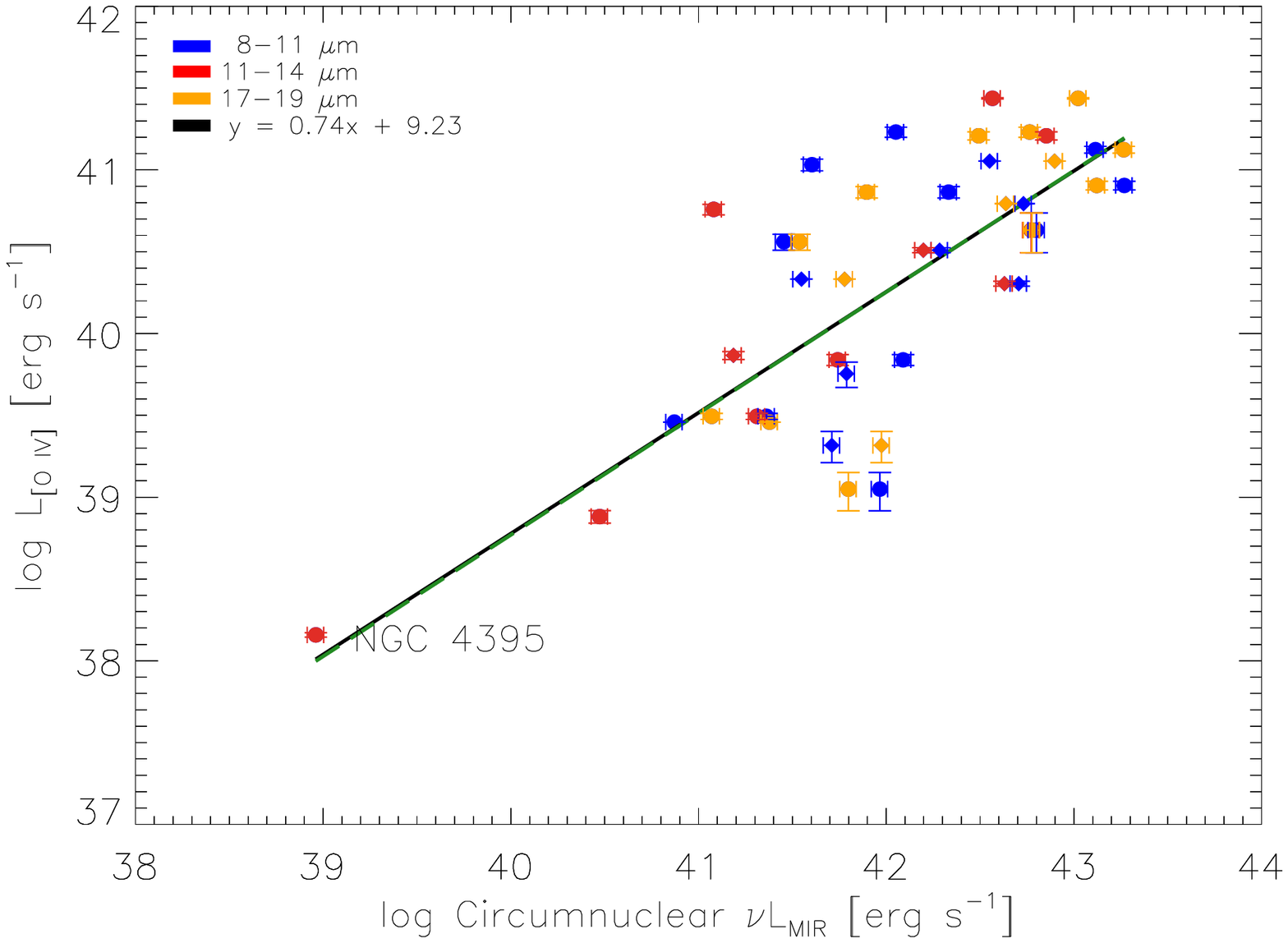}
\includegraphics[width=8.8cm]{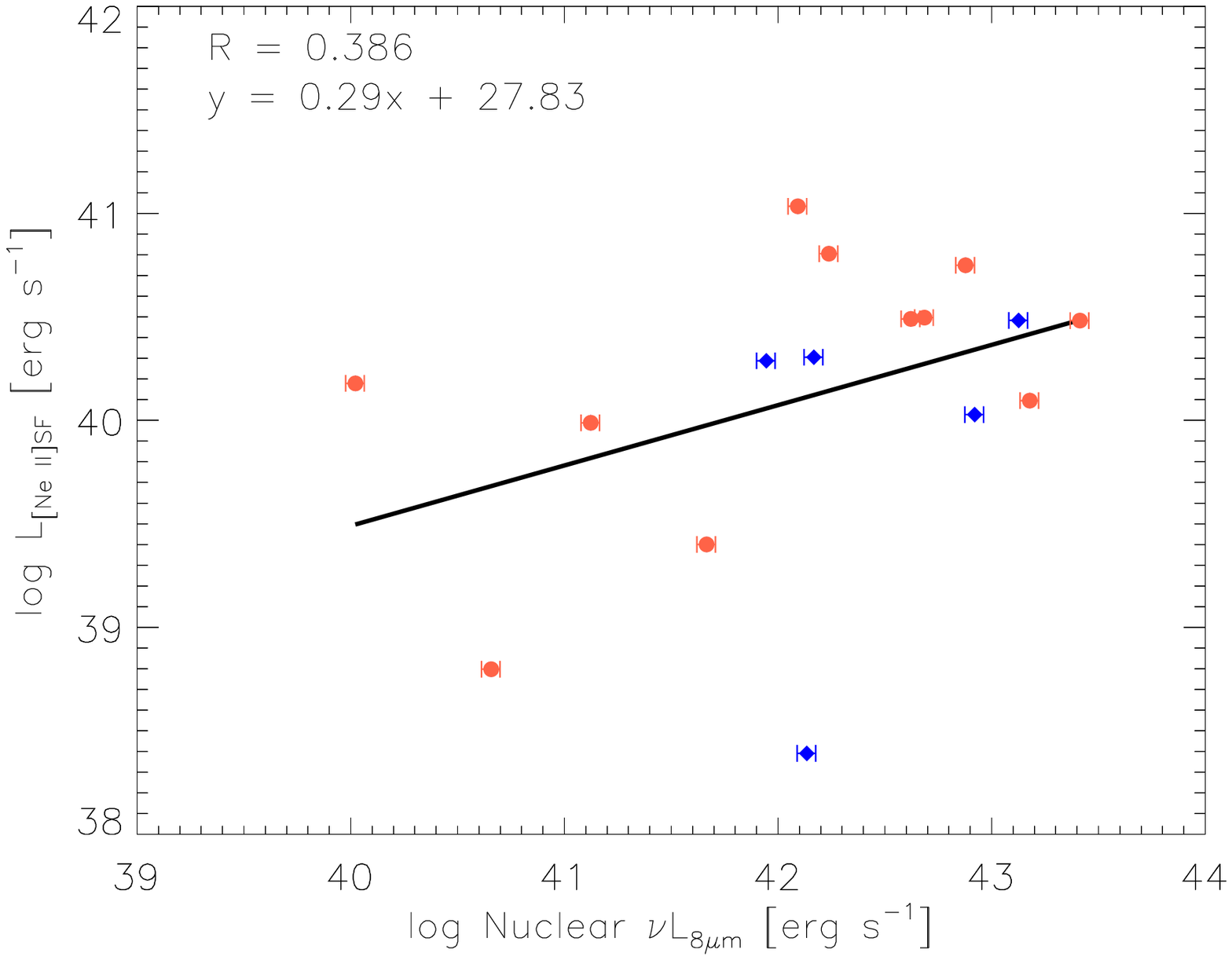}
\includegraphics[width=8.8cm]{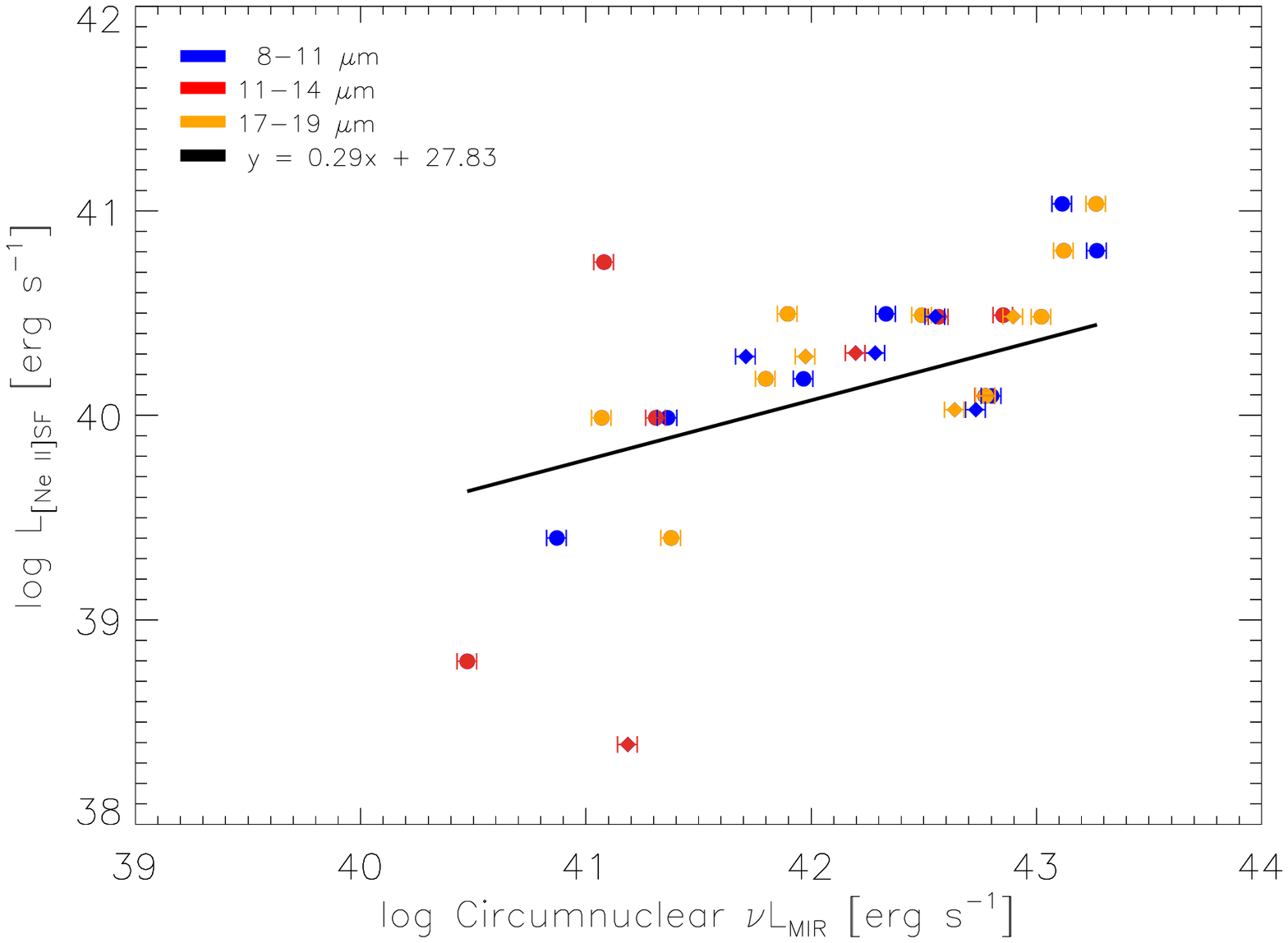}
\par} 
\caption{
Top left panel: 8~$\mu$m nuclear luminosity versus integrated [O\,IV]~$\lambda$25.89~$\mu$m 
emission line luminosity. Bottom left panel: same but for the integrated [Ne\,II]$_{SF}\lambda$12.81~$\mu$m 
emission luminosity. Right panels: same as in the left panels, but for the circumnuclear MIR luminosities. The black lines 
correspond to the nuclear correlation results from the fits shown in the left panels. The green dashed lines 
correspond to the same fit, but excluding NGC\,4395. Blue diamonds and red circles are Sy1 and Sy2 galaxies, respectively. The vertical error bars correspond to the uncertainties listed in Table \ref{tab3}, and the horizontal error bars correspond to the photometry uncertainty reported in Section \ref{nuclear_fluxes}.}
\label{fig9}
\end{figure*}

To further investigate the origin of the nuclear and circumnuclear MIR emission of the sample, in 
Fig. \ref{fig9} we show the same correlations as in Fig. \ref{fig8}, but now using the MIR emission lines 
[O\,IV]$\lambda$25.89~$\mu$m and [Ne\,II]$\lambda$12.81~$\mu$m instead of X-ray luminosities. 
We find a tight correlation between the nuclear MIR and [O\,IV]~$\lambda$25.89~$\mu$m luminosities. Indeed, the slope and Pearson's correlation coefficient are practically identical to those measured for the nuclear 8 $\mu$m--2-10 keV fit (top left panel of Fig. \ref{fig8} and Table \ref{tab6}). We also measure a steeper slope for Sy1 galaxies (1.0$\pm$0.3 with R=0.81) than for Sy2 (0.8$\pm$0.1 with R=0.89), but the values are consistent within the errors. Again, the tight correlation between our nuclear MIR luminosities and the [O\,IV] emission of the galaxies indicates that our nuclear 8 $\micron$ fluxes are AGN-dominated. 

In order to have a reliable SF activity indicator, we estimated the AGN and SF contributions to the [Ne\,II]$\lambda$12.81~$\mu$m emission (hereafter [Ne\,II]$_{AGN}$ and [Ne\,II]$_{SF}$, respectively), following the same method as in \citet{Melendez08b}. We use the [Ne\,II]--[O\,IV] correlation\footnote{We found the same correlation log([Ne\,II]$_{AGN}$)=(0.81$\pm$0.16)$\cdot$log([O\,IV])+7.24 as \citet{Melendez08b}, within the errors.} for the AGN-dominated systems in our sample (see Appendix \ref{B}) to estimate [Ne\,II]$_{AGN}$ fluxes for all the galaxies. We then subtracted these [Ne\,II]$_{AGN}$ fluxes from the total [Ne\,II] emission and derived [Ne\,II]$_{SF}$. From the bottom left panel of Fig. \ref{fig9} we see that the correlation between nuclear MIR emission and [Ne\,II]$_{SF}\lambda$12.81~$\mu$m is less significant (R=0.39) and it has larger scatter than those discussed before. This also indicates that there is little or no contribution of SF to our nuclear MIR fluxes.

Considering the low X-ray and MIR luminosities of the galaxy NGC\,4395, which is labelled in Figs. \ref{fig8} and \ref{fig9}, we have checked that the correlations shown in these figures are not due to the presence of this galaxy in the sample (see Table \ref{tab6}). We find almost identical results if we exclude NGC\,4395 from the fits (green dashed lines in the previously mentioned figures).

Finally, in the right panels of Fig. \ref{fig9} we show the same correlations, but for the circumnuclear MIR emission, as in the right panels of Fig. \ref{fig8}. We also found a good correlation with the [O\,IV]~$\lambda$25.89~$\mu$m emission line luminosity, with the mean absolute deviation being identical to those measured for the hard X-rays (0.56). For the [Ne\,II]$_{SF}\lambda$12.81~$\mu$m line, the distribution of points in the bottom right panel of Fig. \ref{fig9} is similar to the nuclear fluxes, with the mean deviation from the nuclear fit being 1.18.

Summarizing, the tightness of the correlations between the nuclear MIR emission and both the X-rays and 
[O\,IV], as well as the less significant correlation with [Ne\,II]$_{SF}$ confirm that the 8 $\micron$ emission of the inner $\sim$70 pc of the BCS$_{40}$ sample is AGN-dominated. We find practically the same correlations, although with larger scatter, for the circumnuclear emission. This suggests that the AGN dominates the MIR emission in the inner $\sim$400 pc of the galaxies, although with some contribution from SF for the galaxies deviating more from the nuclear correlations.

\section{Conclusions}
\label{Conclusions}

In this work we present the first detailed study of the nuclear and circumnuclear MIR emission of a complete sample of Seyfert galaxies (24 galaxies; BCS$_{40}$ sample) selected in the X-rays using high angular resolution images from 8-10 m-class ground-based telescopes. We also used {\textit{Spitzer}} and/or {\textit{WISE}} arcsecond resolution MIR imaging in order to compare the MIR morphologies and nuclear fluxes. Finally, we investigated the relationship of the MIR nuclear and circumnuclear emission with the intrinsic 2-10~keV, the 14-195~keV X-ray emission and different MIR emission lines. The main results are as follows: \\ 

\textbullet \ \  Using different methods to classify the MIR nuclear morphologies, we found, from visual classification, that the majority (83\%) of the sample show extended or possibly extended morphologies, whereas 17\% are point-like. From the quantitative classification, we found that most of the galaxies present extended or possibly extended emission (75\%) and 25\% are point-like.

\textbullet \ \ This extended MIR emission is compact and it has low surface brightness compared with the nuclear emission: it represents, on average, $\sim$30\% of the total emission of the BCS$_{40}$ sample ($\sim$25\% for Sy1 and $\sim$30\% for Sy2).

\textbullet \ \ We find that the extended MIR emission in AGN-dominated systems is more compact (300$\pm$100\,pc) than in SF-dominated systems (650$\pm$700\,pc) and composite galaxies (350$\pm$500\,pc).

\textbullet \ \ Using the visual classification method, we find that the galaxies with point-like MIR morphologies are face-on or moderately inclined (b/a$\sim$0.4-1.0). On the other hand, the galaxies which are extended in the MIR have different values of b/a, from edge-on to face-on.

\textbullet \ \ We find that the MIR emission is practically the same for Sy1 and Sy2, at nuclear and circumnuclear scales. This result is in agreement with the predictions from clumpy torus models if the main heating source of the circumnuclear emission is nuclear activity.

\textbullet \ \ The tightness of the correlations between the nuclear MIR emission and both the X-rays and 
[O\,IV], and the less significant correlation with [Ne\,II]$_{SF}$ confirm that the 8 $\micron$ emission of the inner $\sim$70 pc of the BCS$_{40}$ sample is AGN-dominated.

\textbullet \ \ We find practically the same correlations, although with slightly larger scatter, for the circumnuclear emission. This suggests that the AGN dominates the MIR emission in the inner $\sim$400 pc of the galaxies, although with some contribution from SF for the galaxies deviating more from the nuclear correlations.

\section*{Acknowledgments}

IGB acknowledges financial support from the Instituto de Astrof\'isica de Canarias through Fundaci\'on La Caixa. This research was partly supported by a Marie Curie Intra European Fellowship within the 7th European Community Framework Programme (PIEF-GA-2012-327934). CRA and IGB acknowledge financial support from the Spanish Ministry of Science and Innovation (MICINN) through project PN AYA2013-47742-C4-2-P. CRA also acknowledges the Ram\'on y Cajal Program of the Spanish Ministry of Economy and Competitiveness. A.A.-H. and AHC acknowledges financial support from the Spanish Ministry of Economy and Competitiveness through grant AYA2012-31447 which is party funded by the FEDER program. A.A.-H. also acknowledges AYA2015-64346-C2-1-P, which is partly funded by the FEDER programme. AHC also acknowledges funding by the Spanish Ministry of Economy and Competitiveness under grants AYA2015-70815- ERC and AYA2012-31277. OGM acknowledges to the PAPIIT project IA100516.

This work is based on observations made with the Gran Telescopio de CANARIAS (GTC), installed in the Spanish Observatorio del Roque de los Muchachos of the Instituto de Astrof\'isica de Canarias, in the island of La Palma. The GTC/CC programs under which the data were obtained are GTC43-15A and an ESO/GTC large programme (182.B-2005). 

The Gemini programs under which the data were obtained are GS-2004A-Q-41, GS-2005A-Q-6,, GN-2006-Q-11, GN-2006-Q-30, GS-2006-Q-62, GS-2007A-DD-7, GN-2007A-Q-49, GS-2009B-Q-43, GN-2010A-C-7, GS-2010B-Q-71, GS-2011B-Q-20 and GS-2012-Q-43. The VLT/VISIR programs under which the data were obtained are 076.B-0599(A), 077.B-0137(A), 078.B-0303(A), 080.B-0860(A), 084.B-0366(E), 086.B-0349(C) and 086.B-0349(D).

This work is also based on observations made with the \textit{Spitzer Space Telescope}, which is operated by the Jet Propulsion Laboratory, Caltech under NASA contract 1407.

The Cornell Atlas of Spitzer/IRS Sources (CASSIS) is a product of the Infrared Science Center at Cornell University, supported by NASA and JPL.

This publication makes use of data products from the Wide-field Infrared Survey Explorer, which is a joint project of the University of California, Los Angeles, and the Jet Propulsion Laboratory/California Institute of Technology, funded by the National Aeronautics and Space Administration.

Finally, we are extremely grateful to the GTC staff for their constant and enthusiastic support, and to the anonymous referee for useful comments.

\appendix

\section{PC- and kPC-scale morphologies}
\label{A}
Here we present the comparison between the arcsecond and the subarcsecond resolution MIR images of the BCS$_{40}$ sample. We have used the ground-based N-band images, which are closer in wavelength to the {\textit{IRAC}} and {\textit{WISE}} data. A large fraction of the galaxies in the BCS$_{40}$ sample have 8$\mu$m {\textit{IRAC}} data, but we have used 12$\mu$m {\textit{WISE}} images when the central part of the {\textit{IRAC}} image is saturated or when there is no 8$\mu$m {\textit{IRAC}} data available. In the left panels of Fig. \ref{figA1} we show the N-band high angular resolution images of the sample (central 3.6\arcsec region). In the central panels we show the same region as in the left panels, but for the arcsecond resolution data, and in the right panels, the central 6\arcsec region for comparison. In some cases we can identify similar structures and orientations in the subarcsecond and arcsecond resolution MIR data, as e.g. in ESO\,005-G004, NGC\,2992, NGC\,3227, NGC\,3783, NGC\,4945 and NGC\,7172.

\begin{figure}
\centering
\par{
\includegraphics[width=3.0cm, angle=90]{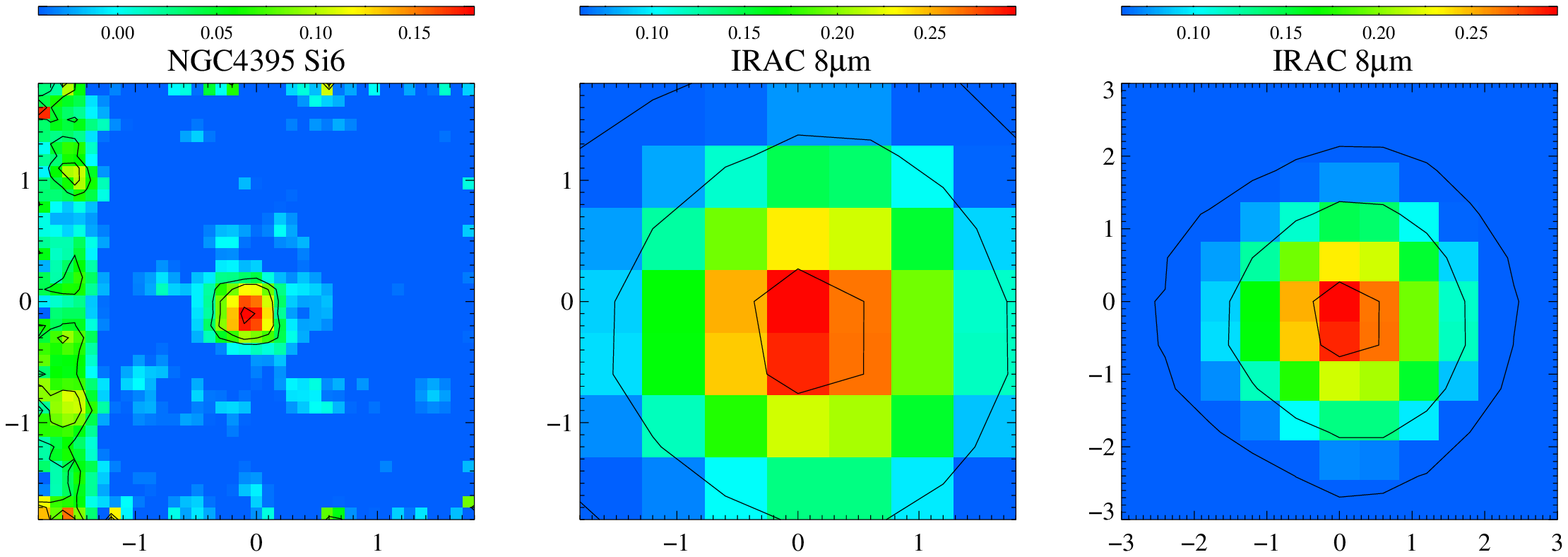}
\includegraphics[width=3.0cm, angle=90]{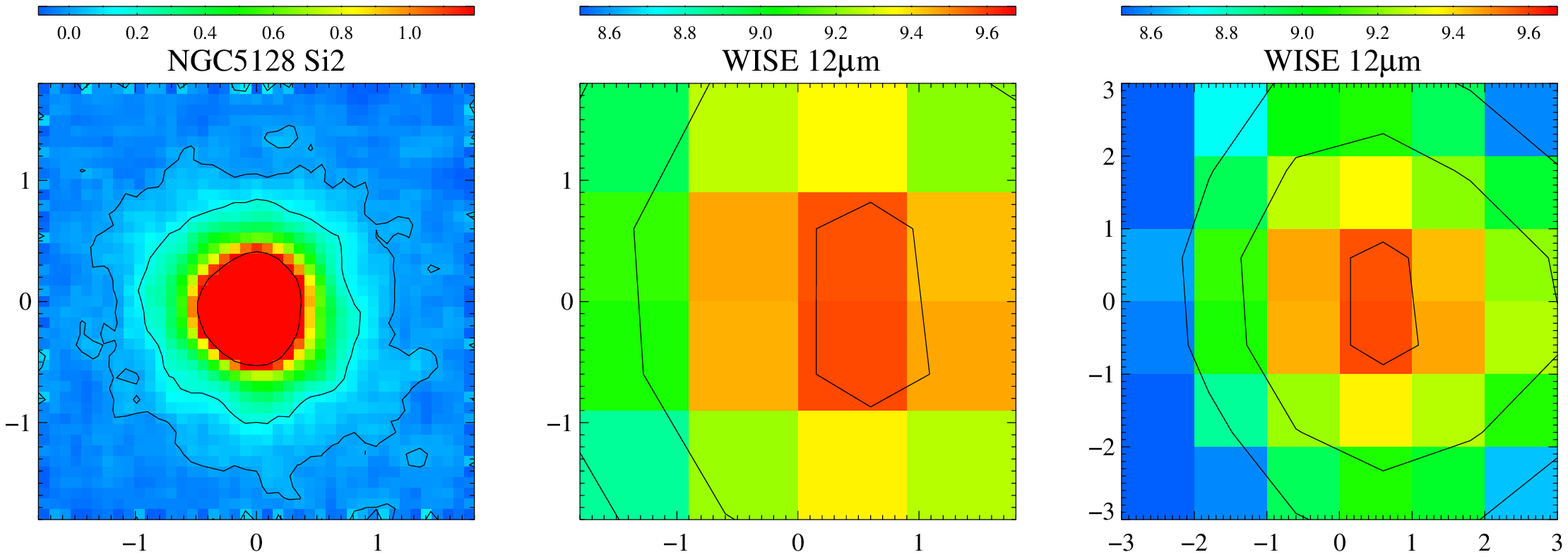}
\includegraphics[width=3.0cm, angle=90]{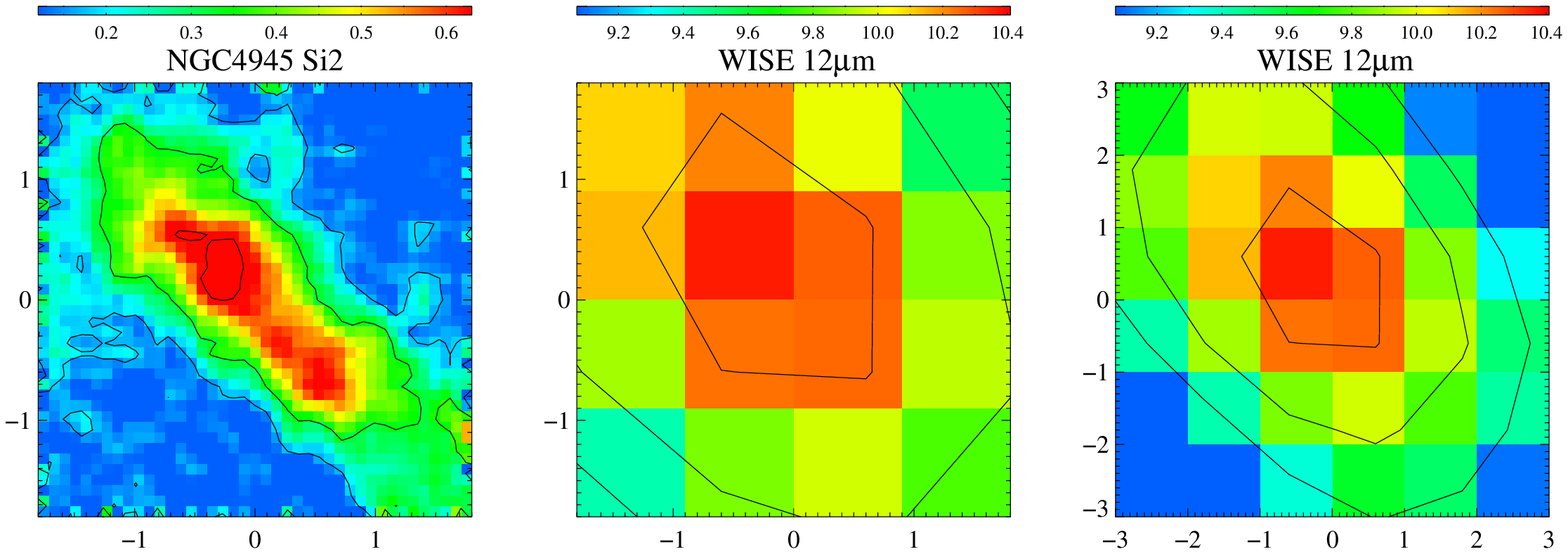}
\includegraphics[width=3.0cm, angle=90]{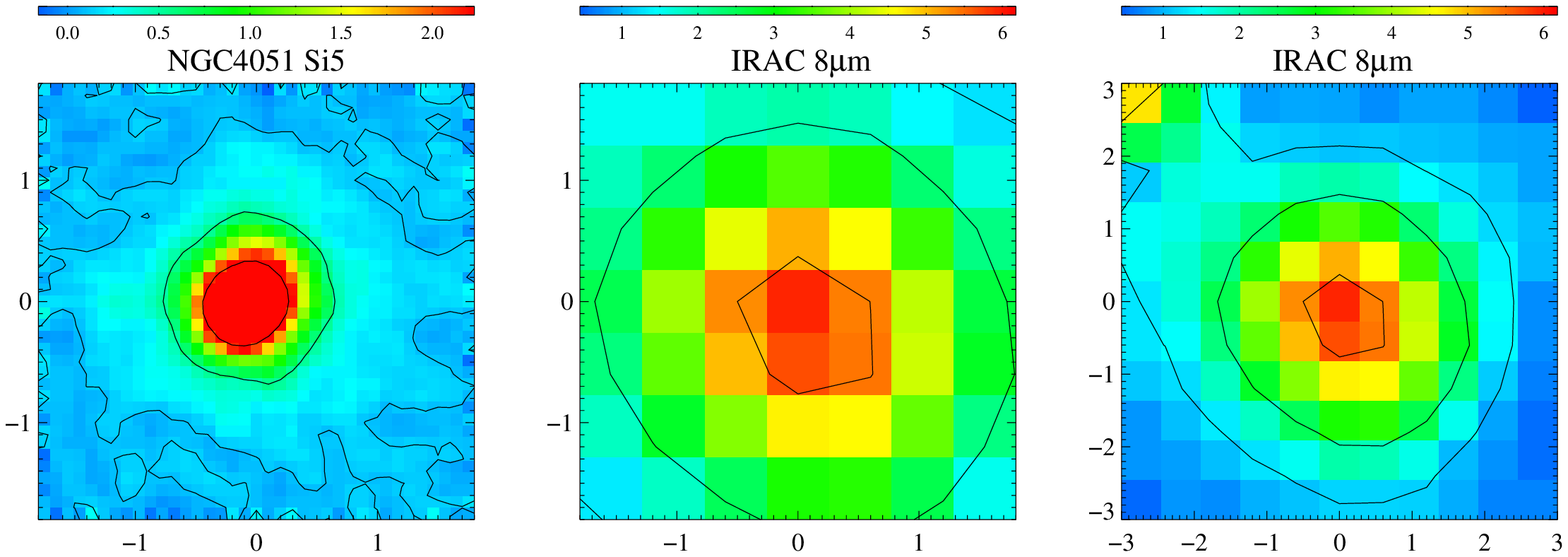}
\includegraphics[width=3.0cm, angle=90]{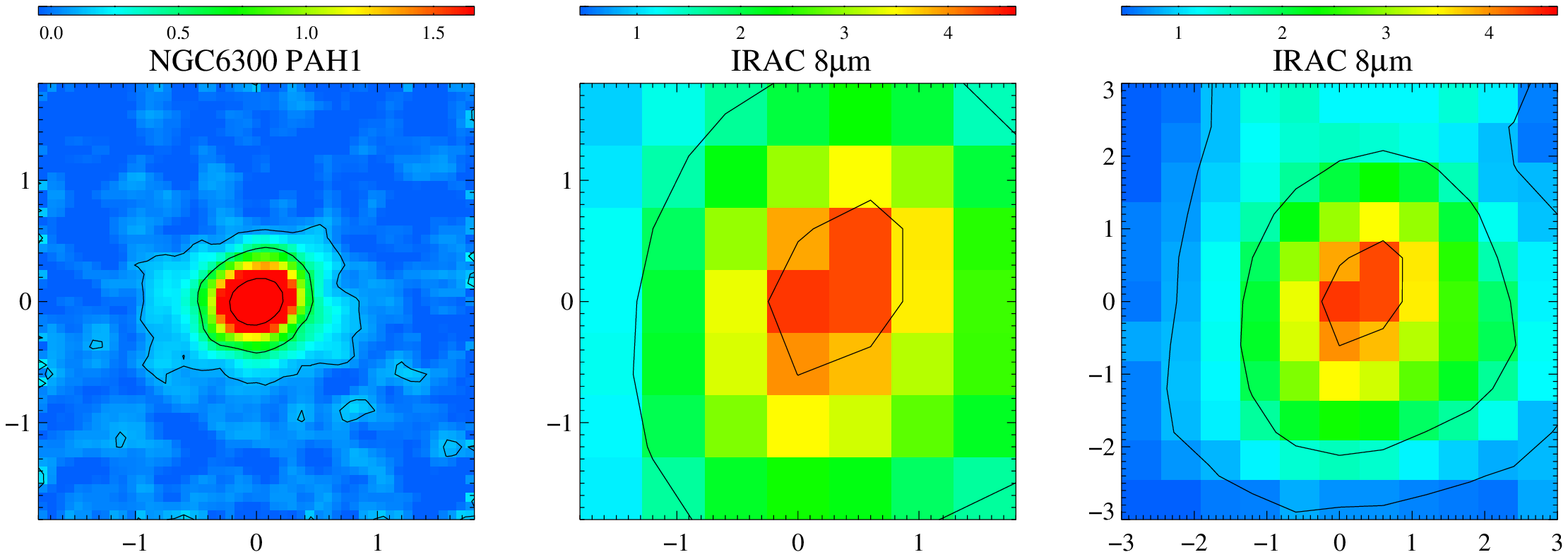}
\includegraphics[width=3.0cm, angle=90]{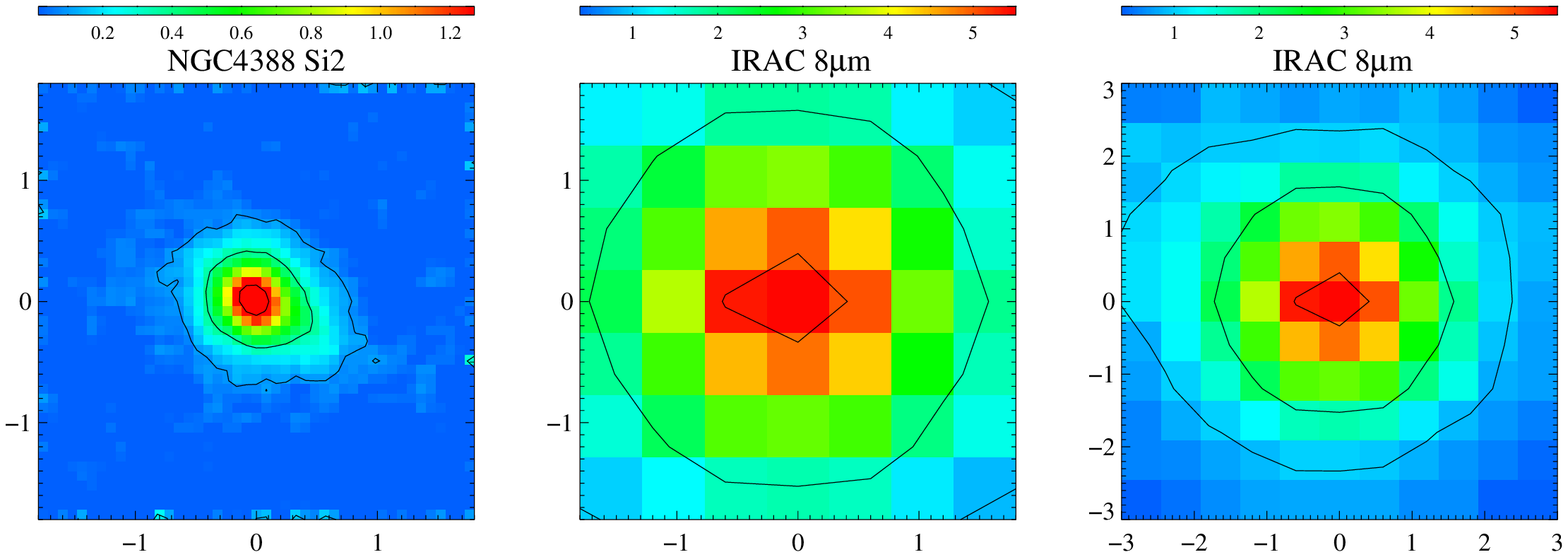}
\includegraphics[width=3.0cm, angle=90]{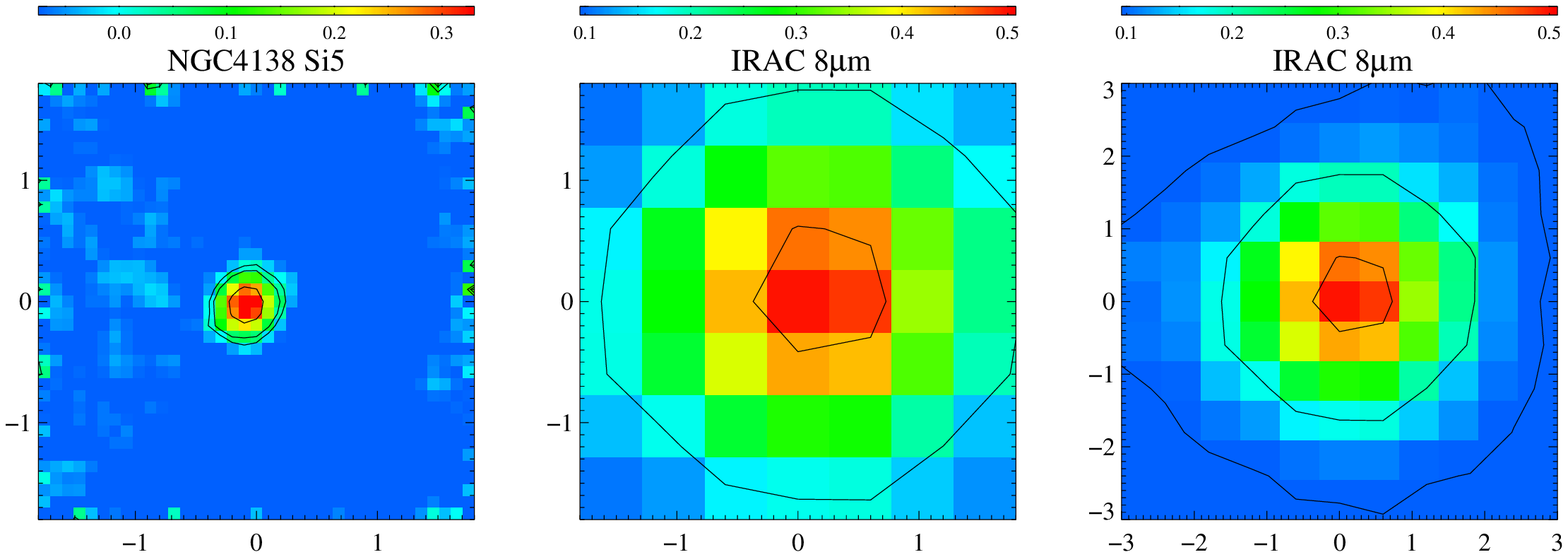}
\par}
\caption{Comparison between the arcsecond and subarcsecond resolution MIR images of the BCS$_{40}$ sample. Left panels: N-band high angular resolution images (central 3.6\arcsec region). Central panels: 8~$\mu$m {\textit{IRAC}} or 12~$\mu$m {\textit{WISE}} arcsecond resolution data (central 3.6\arcsec region). Right panels: same as in the central panels, but on a larger scale (central 6\arcsec region). All images are smoothed (box of 3 pixels) and have their own contours overlaid (in black). All images have been smoothed (3 pixel box). Colour bars correspond to fluxes in mJy/pixel units. North is up, and east to the left.}
\label{figA1}
\end{figure}

\begin{figure}
\contcaption
\centering
\par{
\includegraphics[width=3.0cm, angle=90]{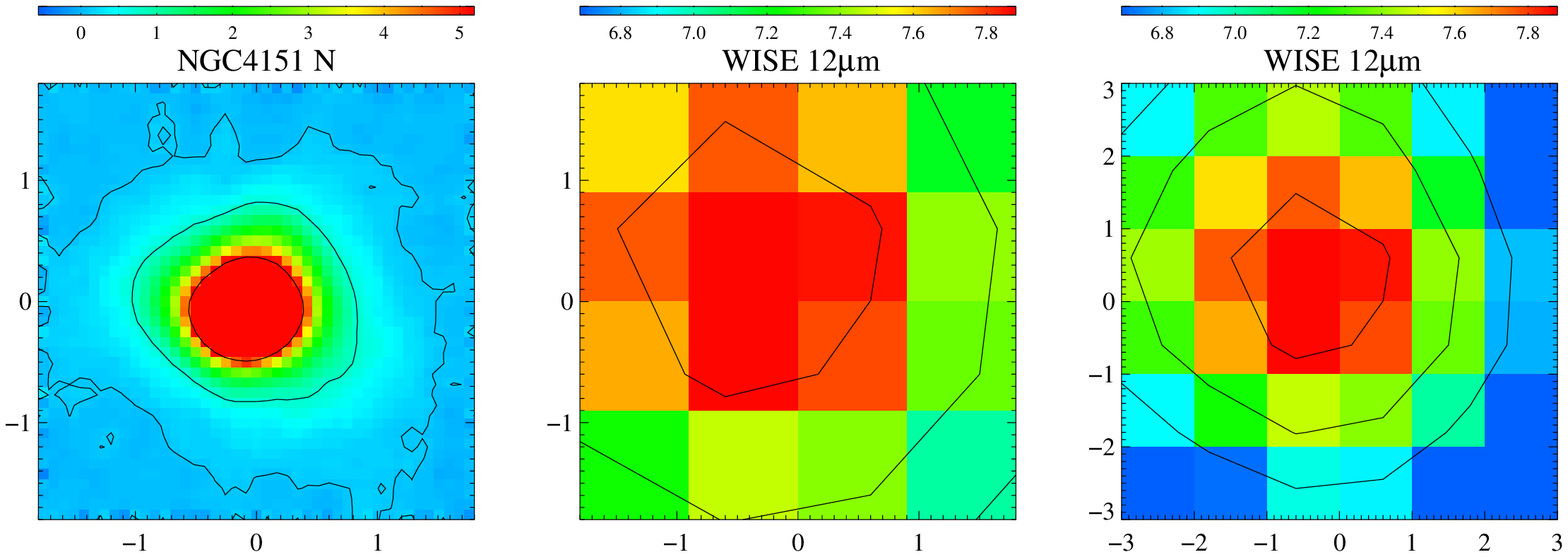}
\includegraphics[width=3.0cm, angle=90]{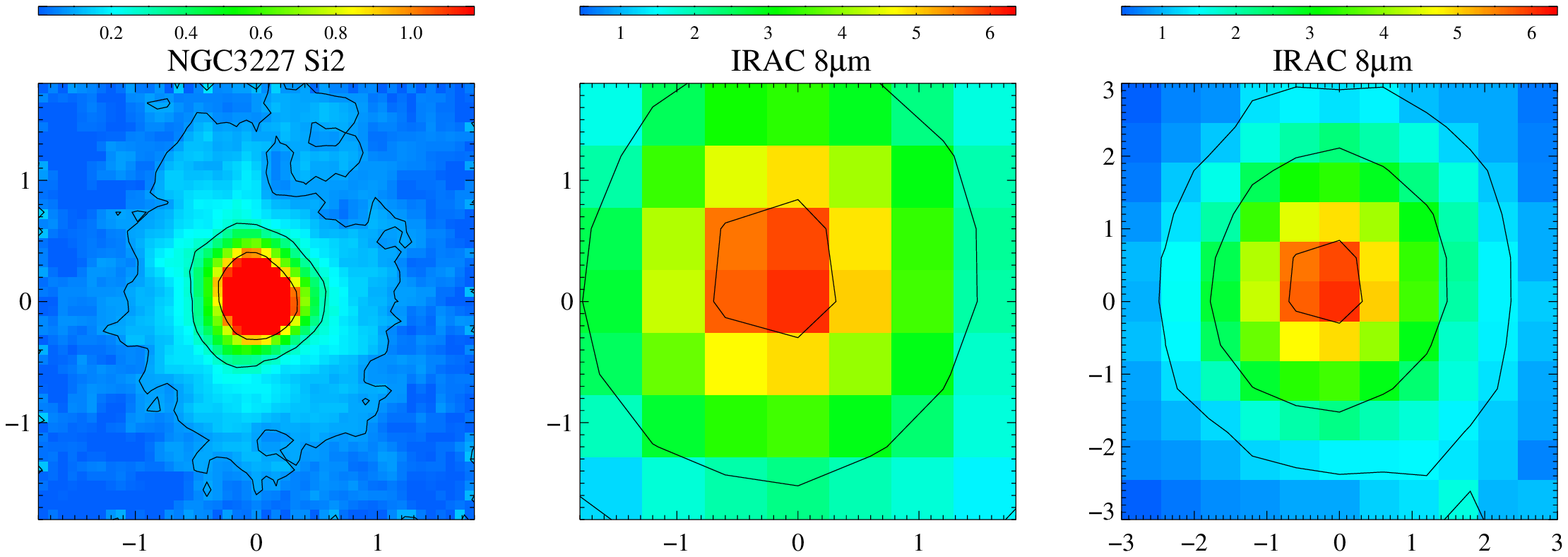}
\includegraphics[width=3.0cm, angle=90]{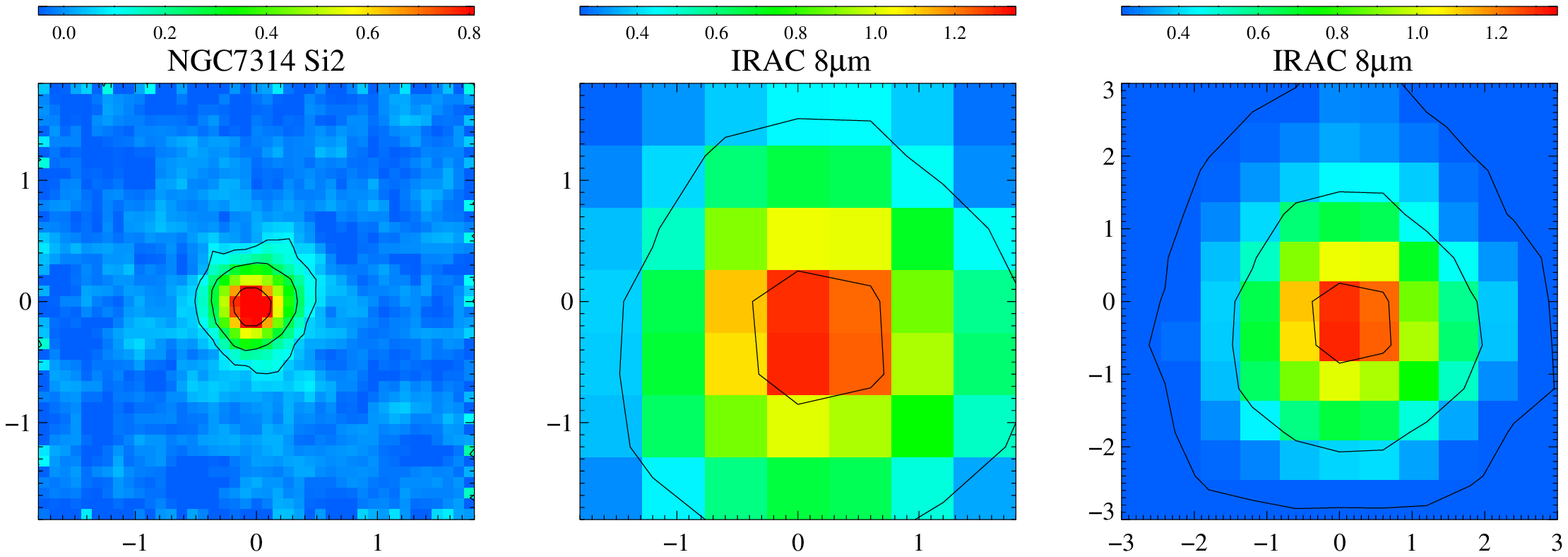}
\includegraphics[width=3.0cm, angle=90]{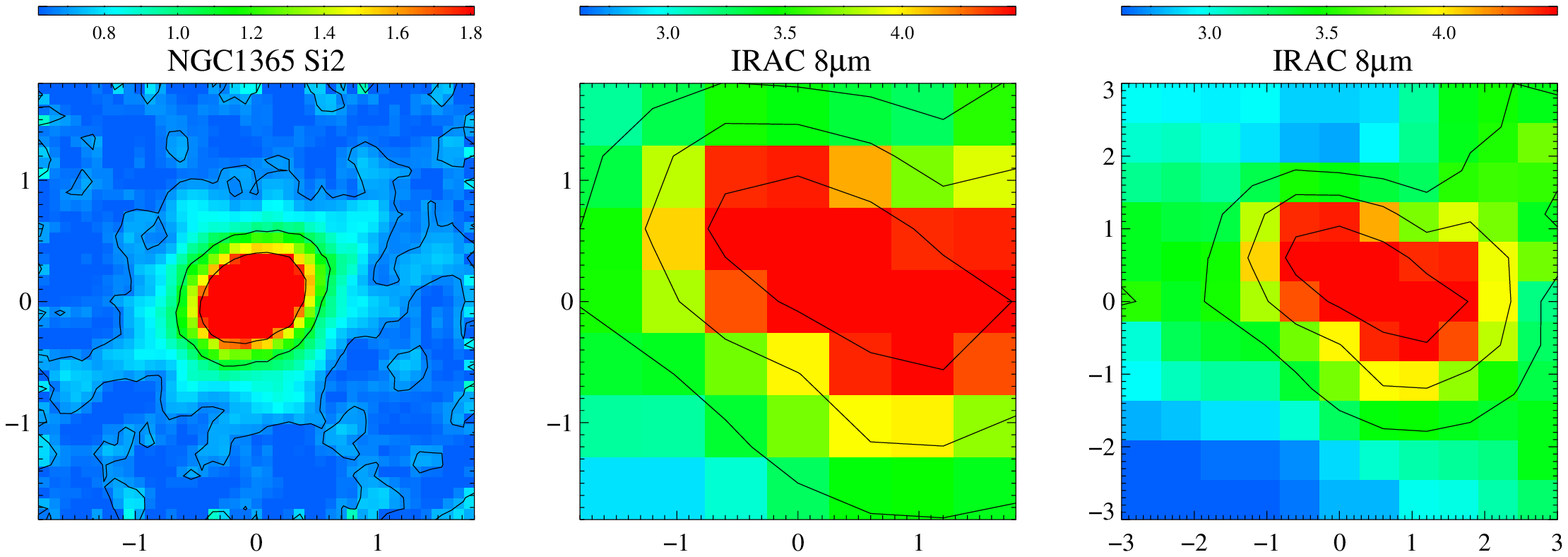}
\includegraphics[width=3.0cm, angle=90]{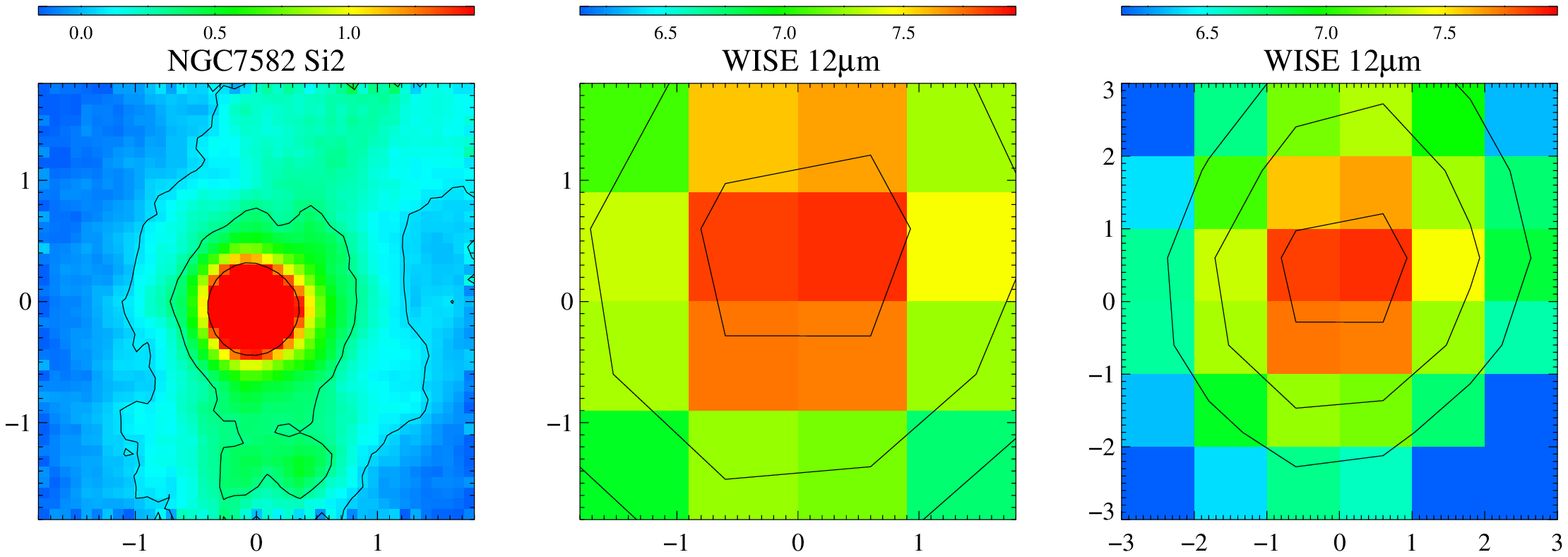}
\includegraphics[width=3.0cm, angle=90]{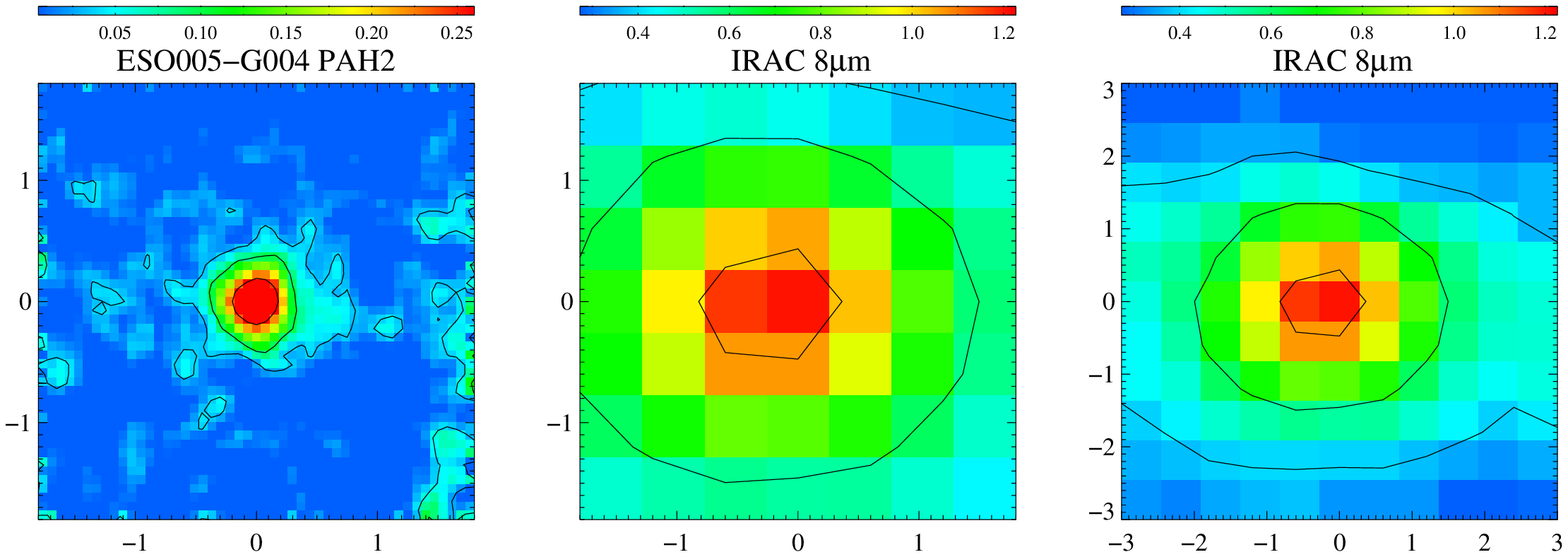}
\includegraphics[width=3.0cm, angle=90]{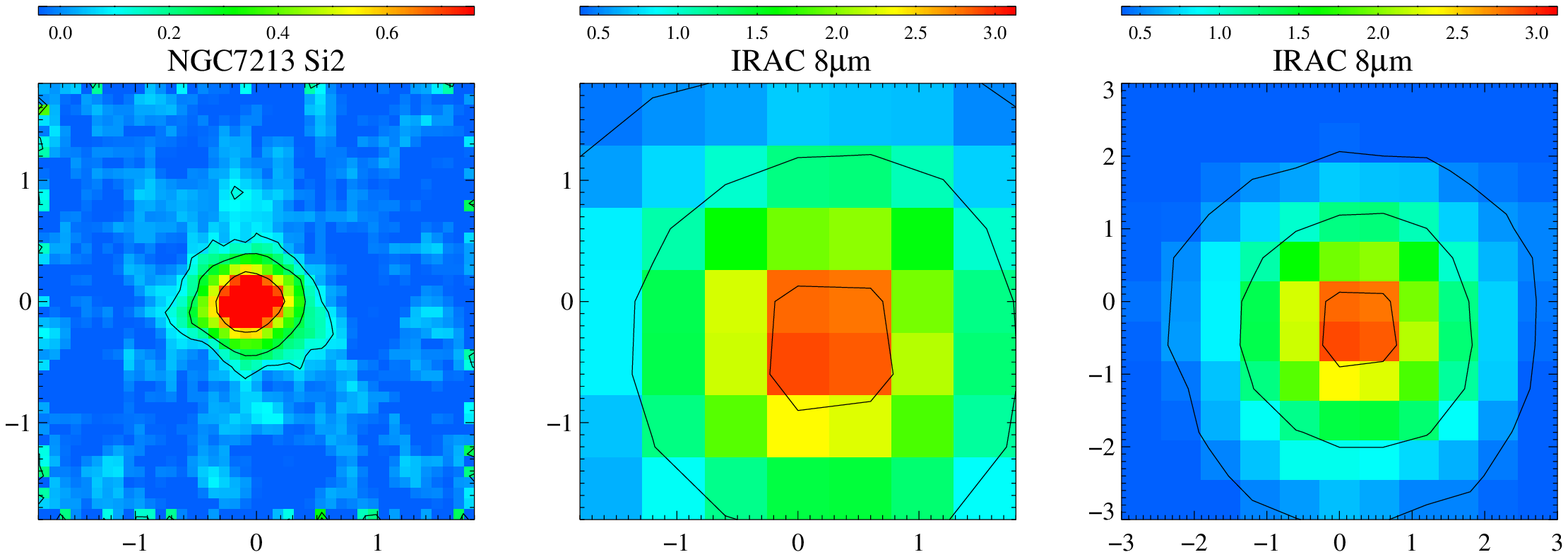}
\par}
\end{figure}

\begin{figure}
\contcaption
\centering
\par{
\includegraphics[width=3.0cm, angle=90]{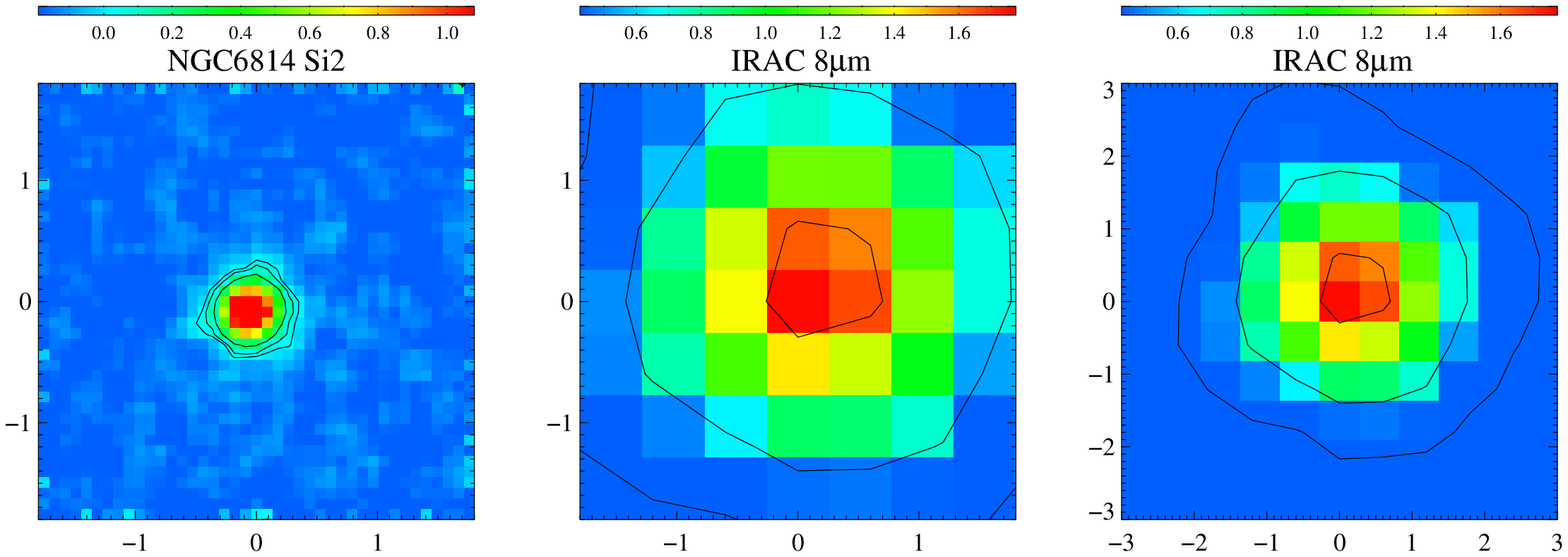}
\includegraphics[width=3.0cm, angle=90]{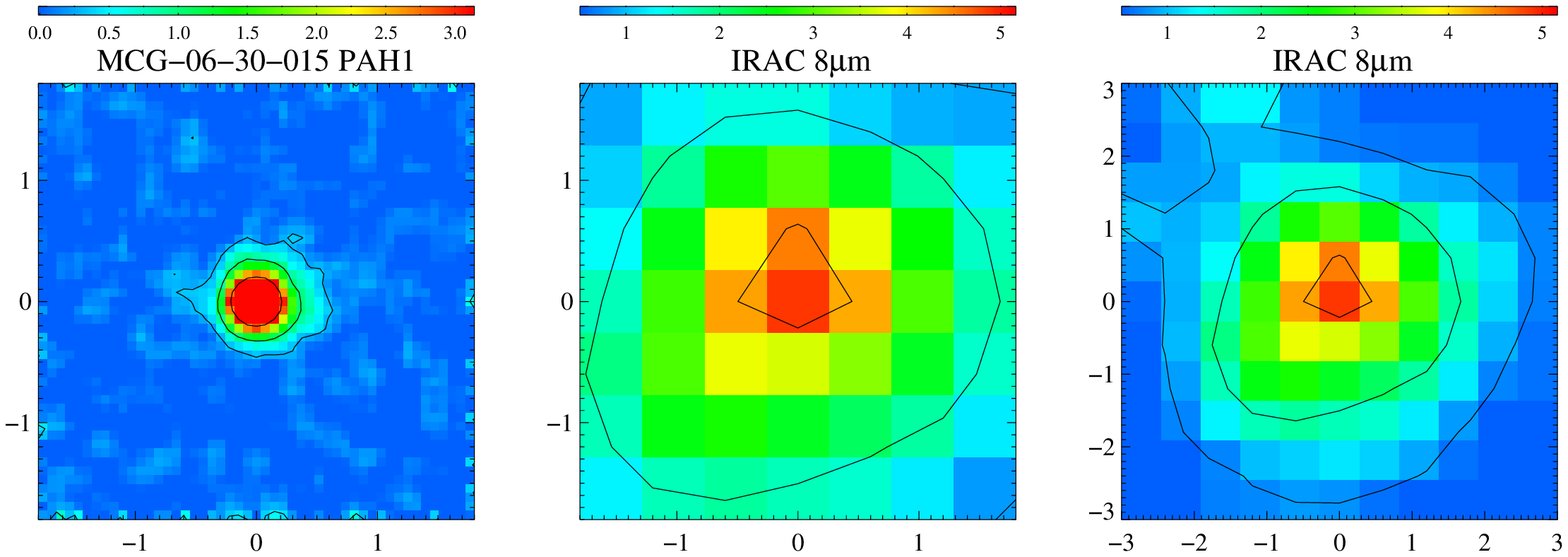}
\includegraphics[width=3.0cm, angle=90]{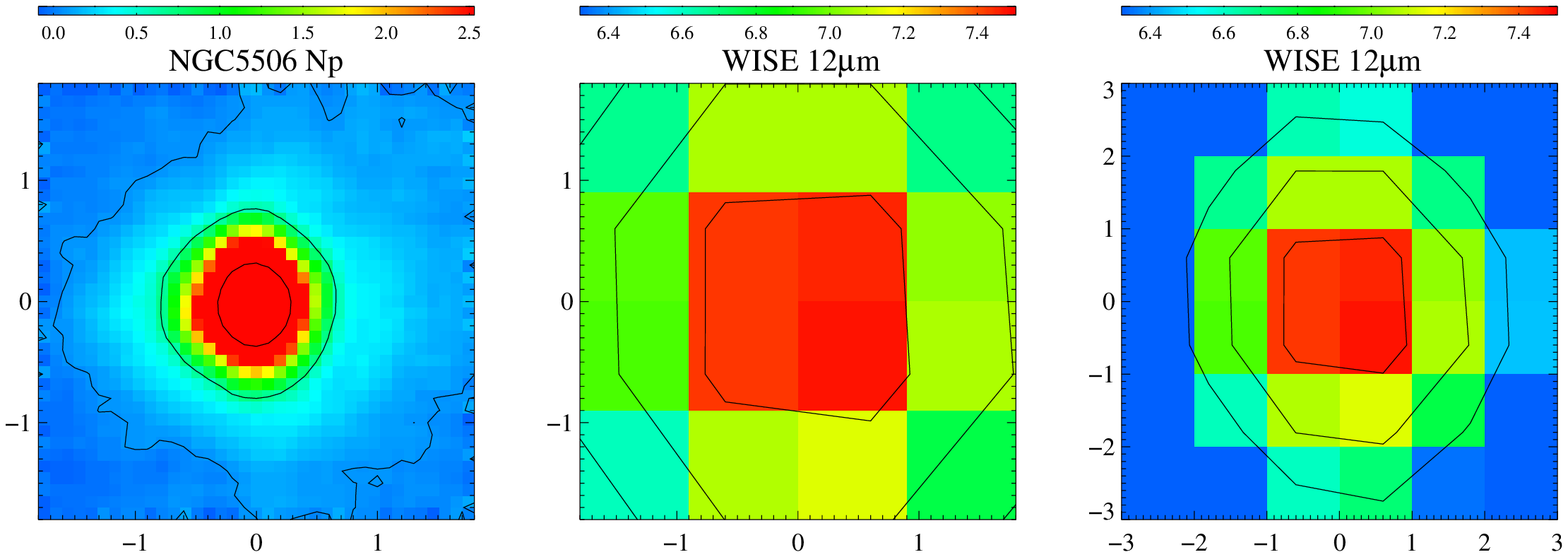}
\includegraphics[width=3.0cm, angle=90]{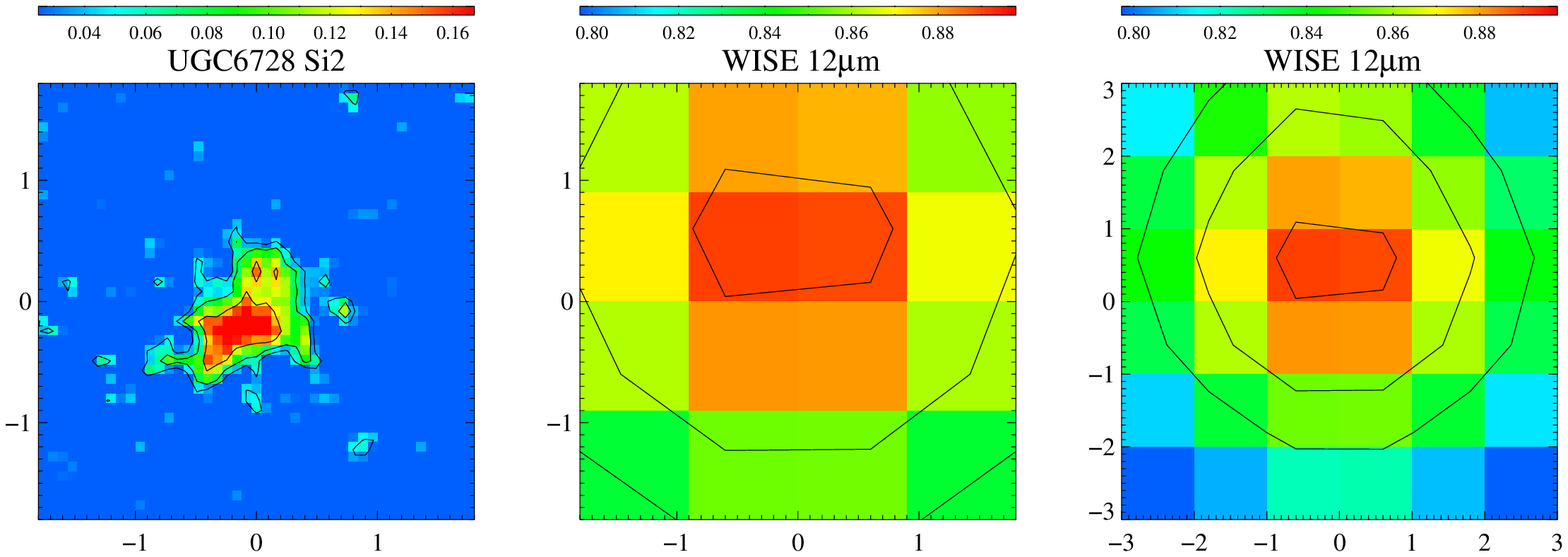}
\includegraphics[width=3.0cm, angle=90]{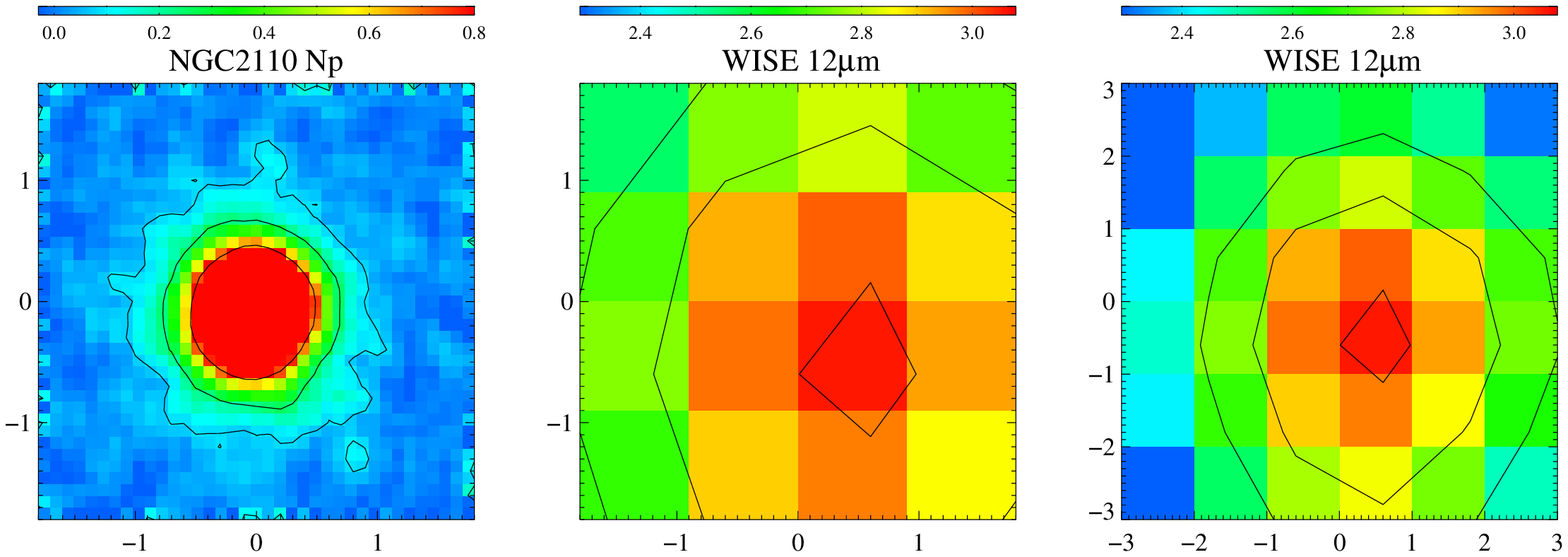}
\includegraphics[width=3.0cm, angle=90]{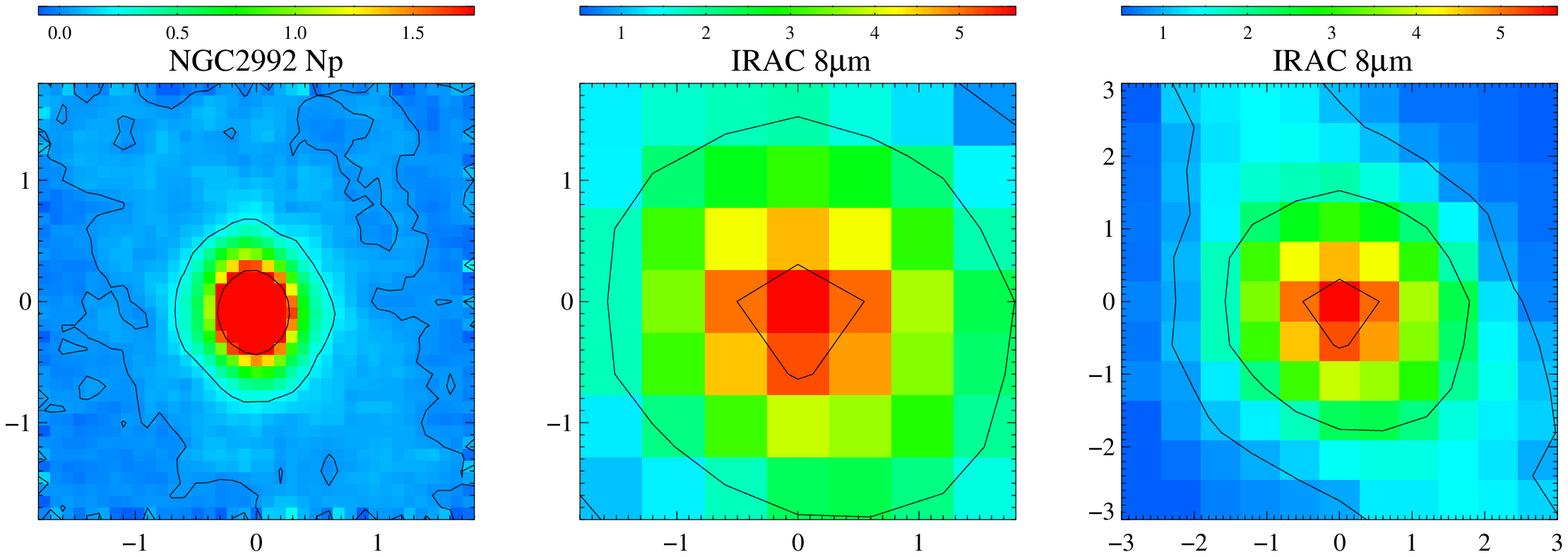}
\includegraphics[width=3.0cm, angle=90]{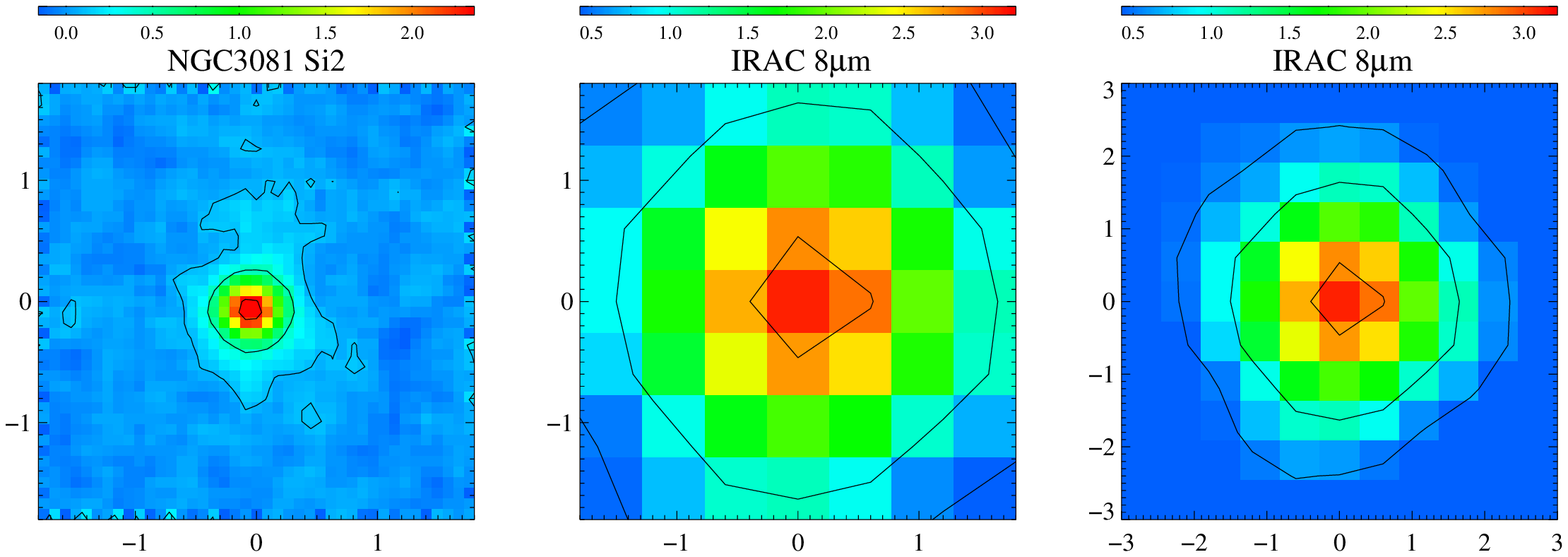}
\par}
\end{figure}

\begin{figure}
\contcaption
\centering
\par{
\includegraphics[width=3.0cm, angle=90]{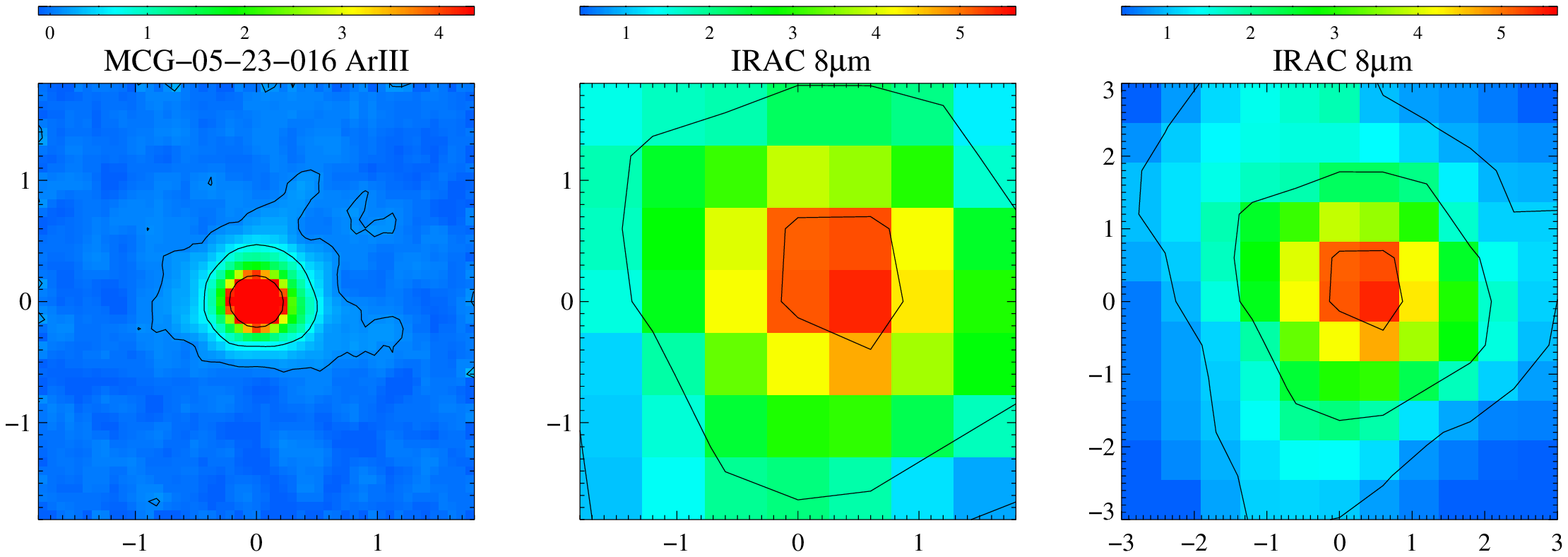}
\includegraphics[width=3.0cm, angle=90]{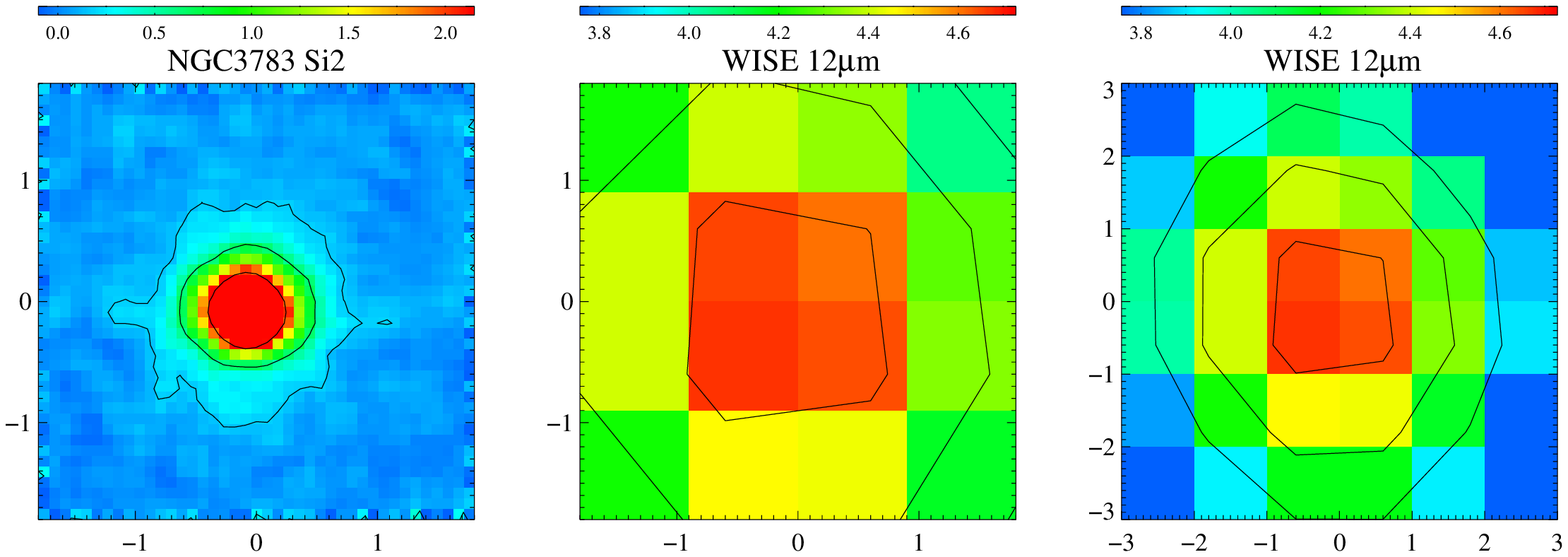}
\includegraphics[width=3.0cm, angle=90]{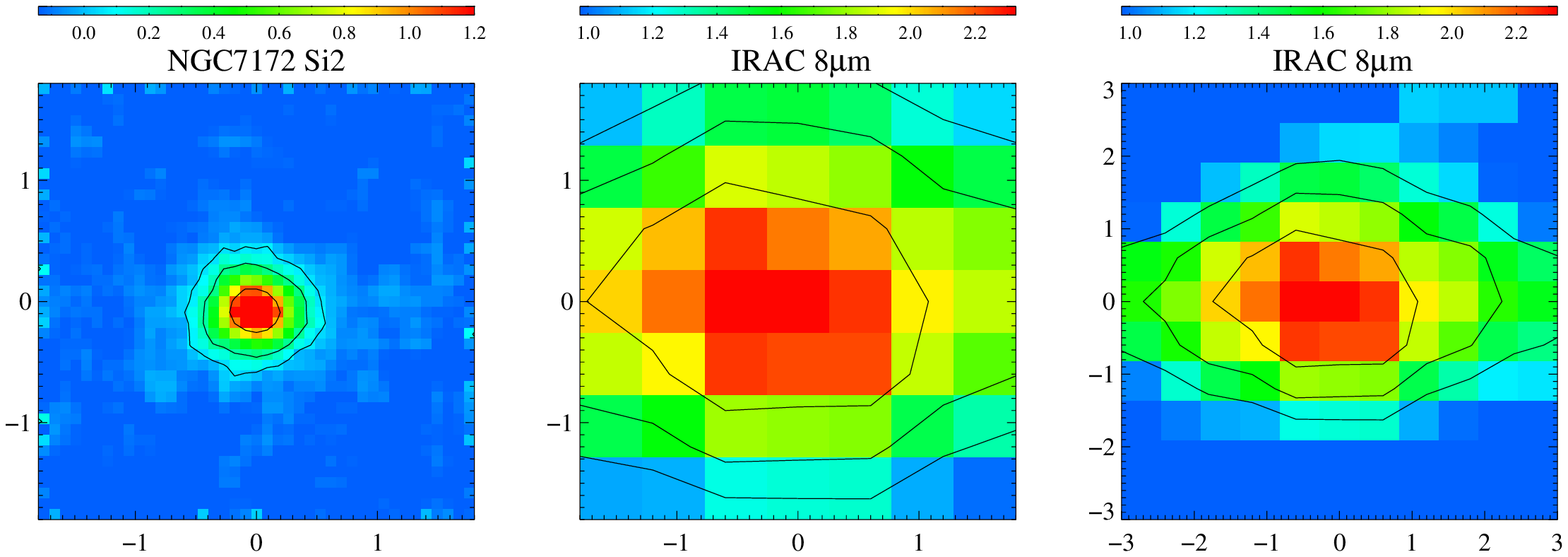}
\par}
\end{figure}

We also show the arcsecond resolution MIR images of the BCS$_{40}$ sample in Fig. \ref{figA2}, which we used in the classification reported in Section \ref{extended}.

\begin{figure*}
\centering
\par{
\includegraphics[width=4.3cm]{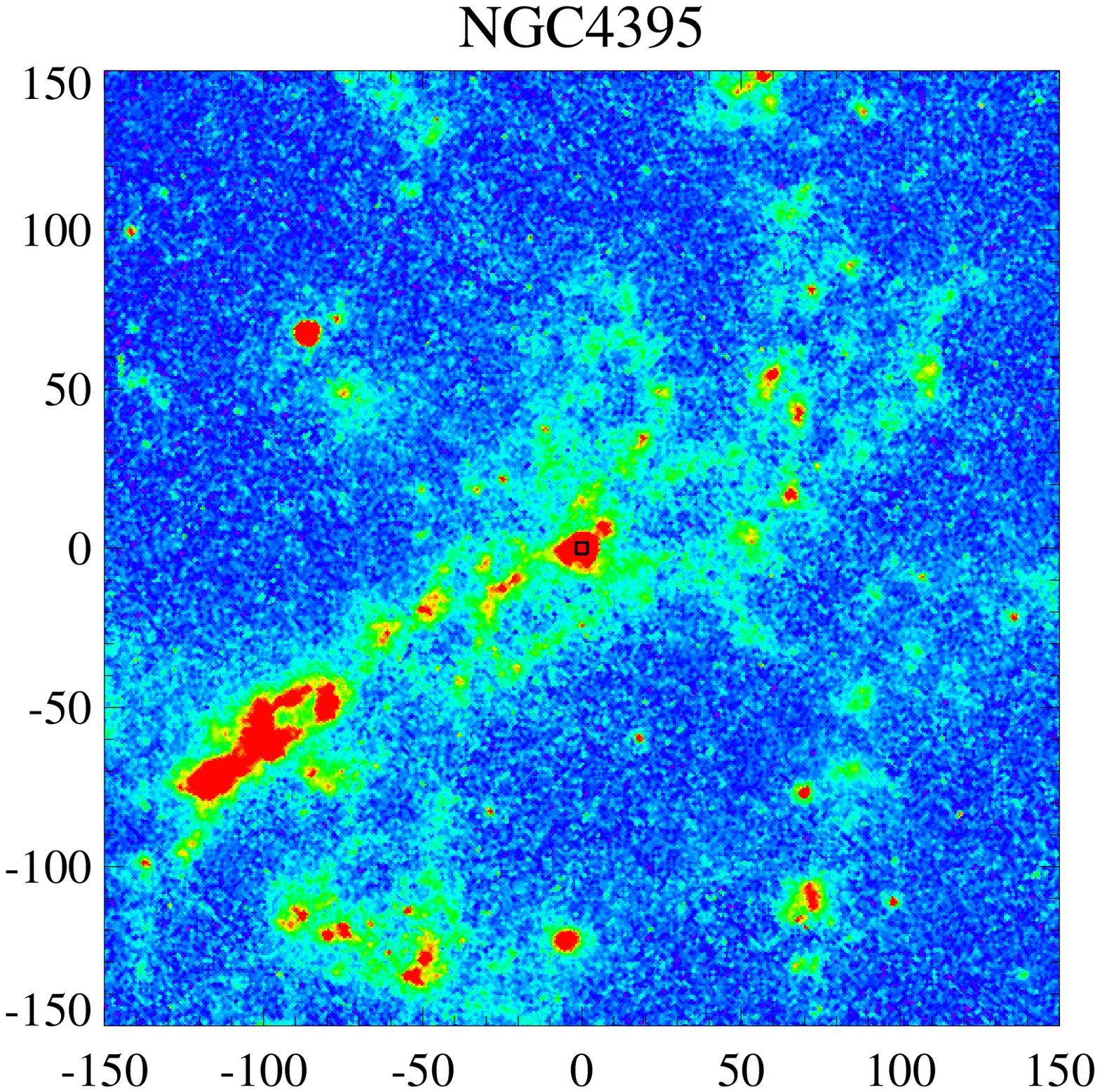}
\includegraphics[width=4.3cm]{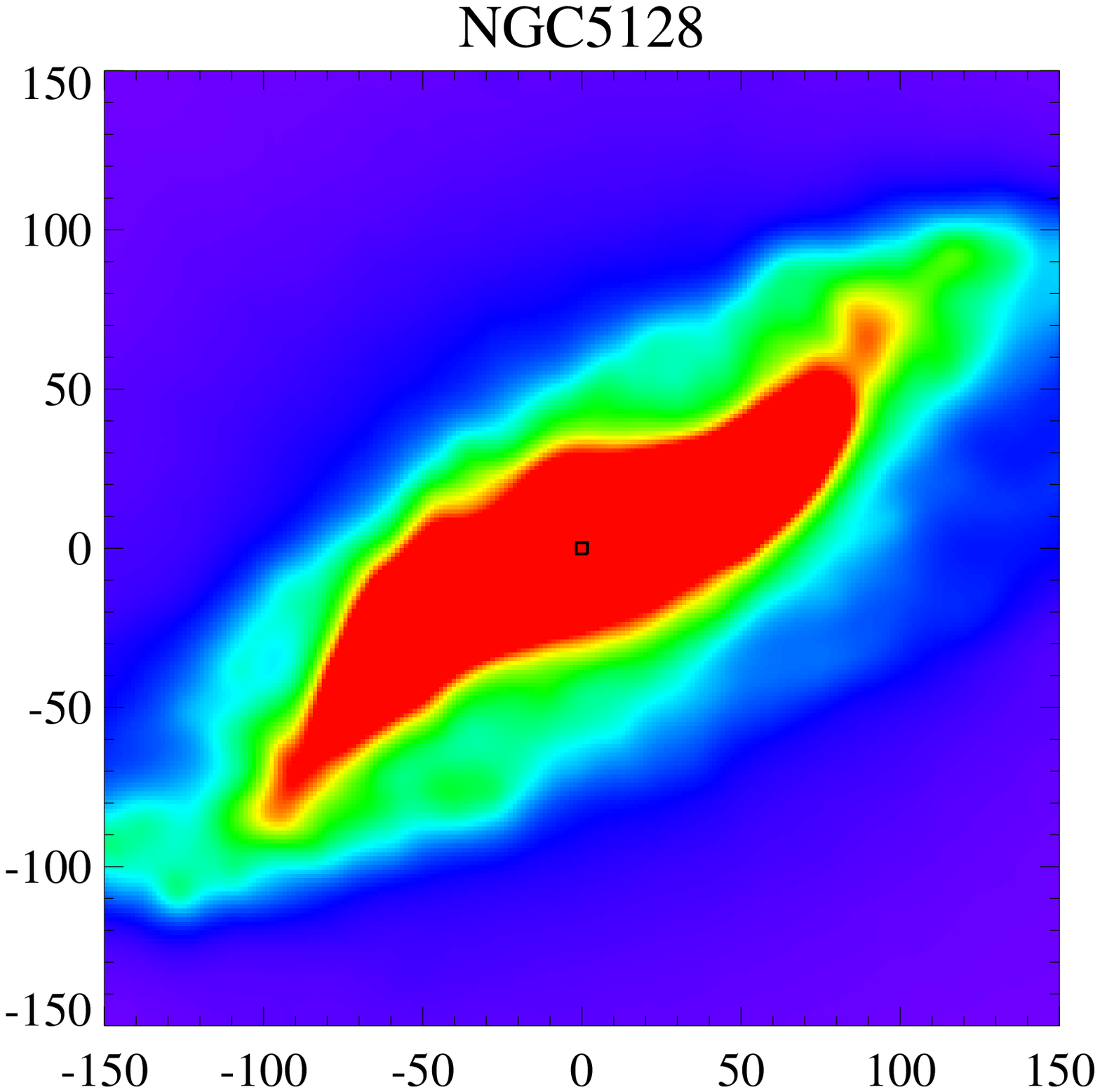}
\includegraphics[width=4.3cm]{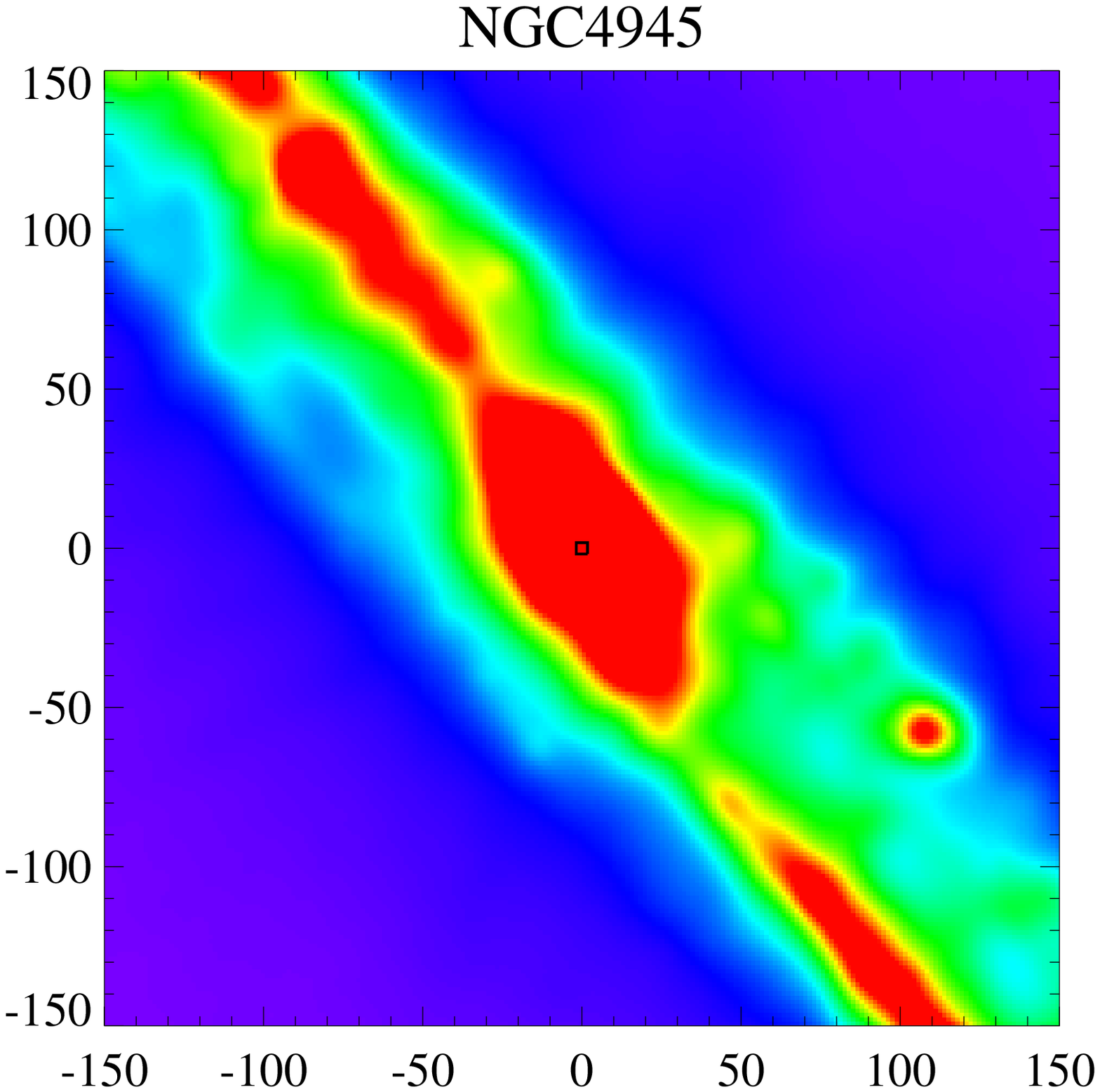}
\includegraphics[width=4.3cm]{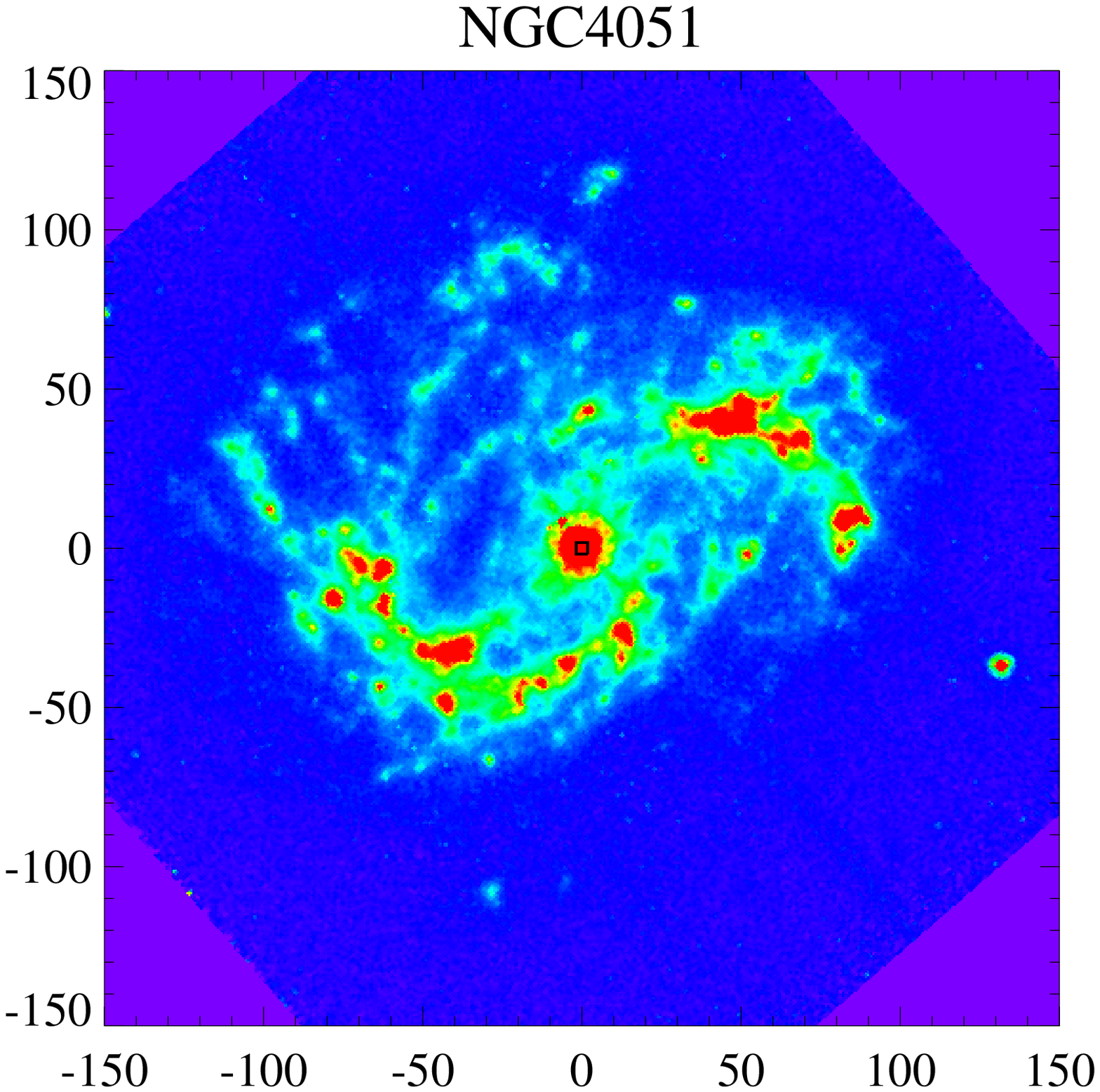}
\includegraphics[width=4.3cm]{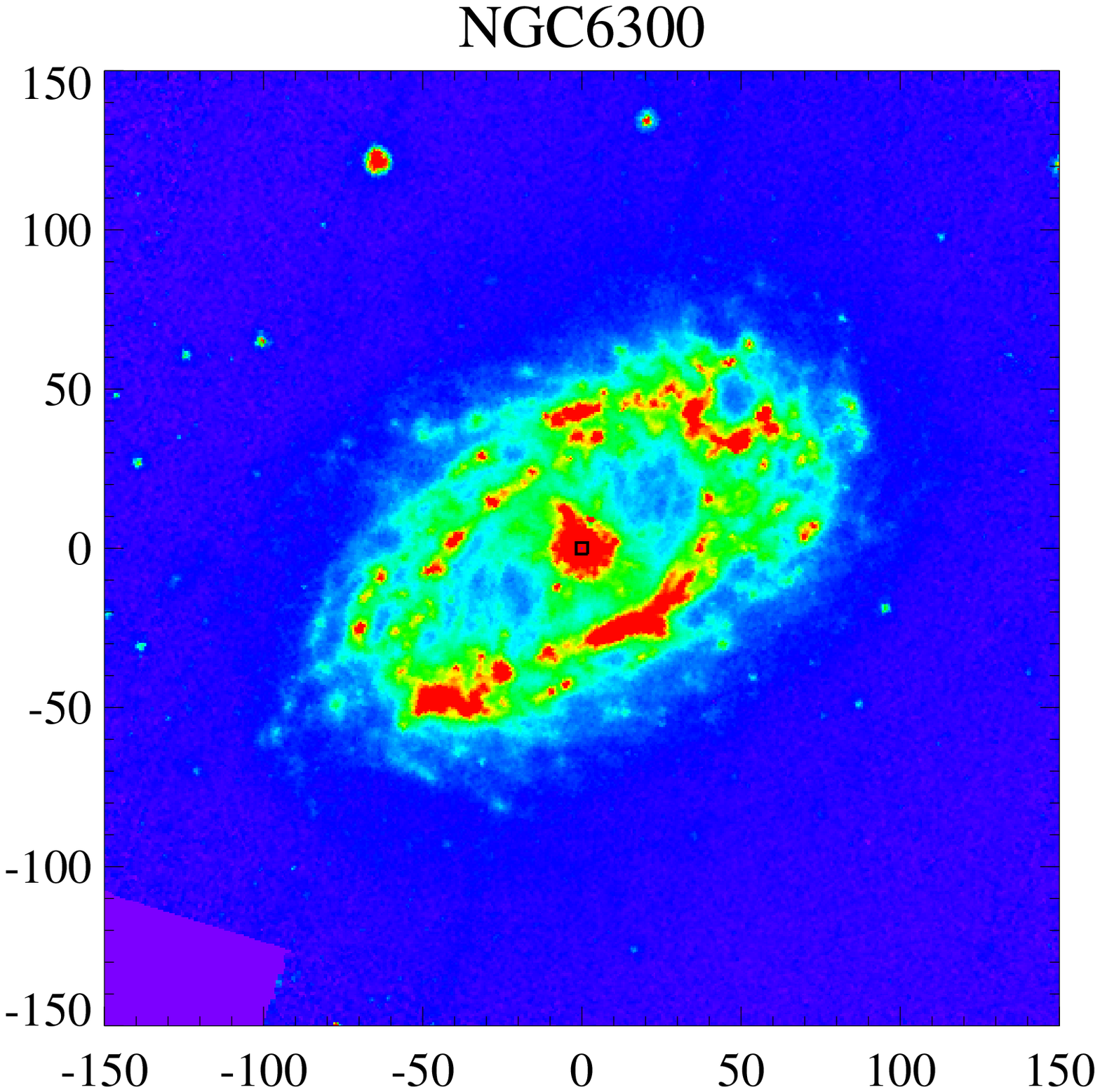}
\includegraphics[width=4.3cm]{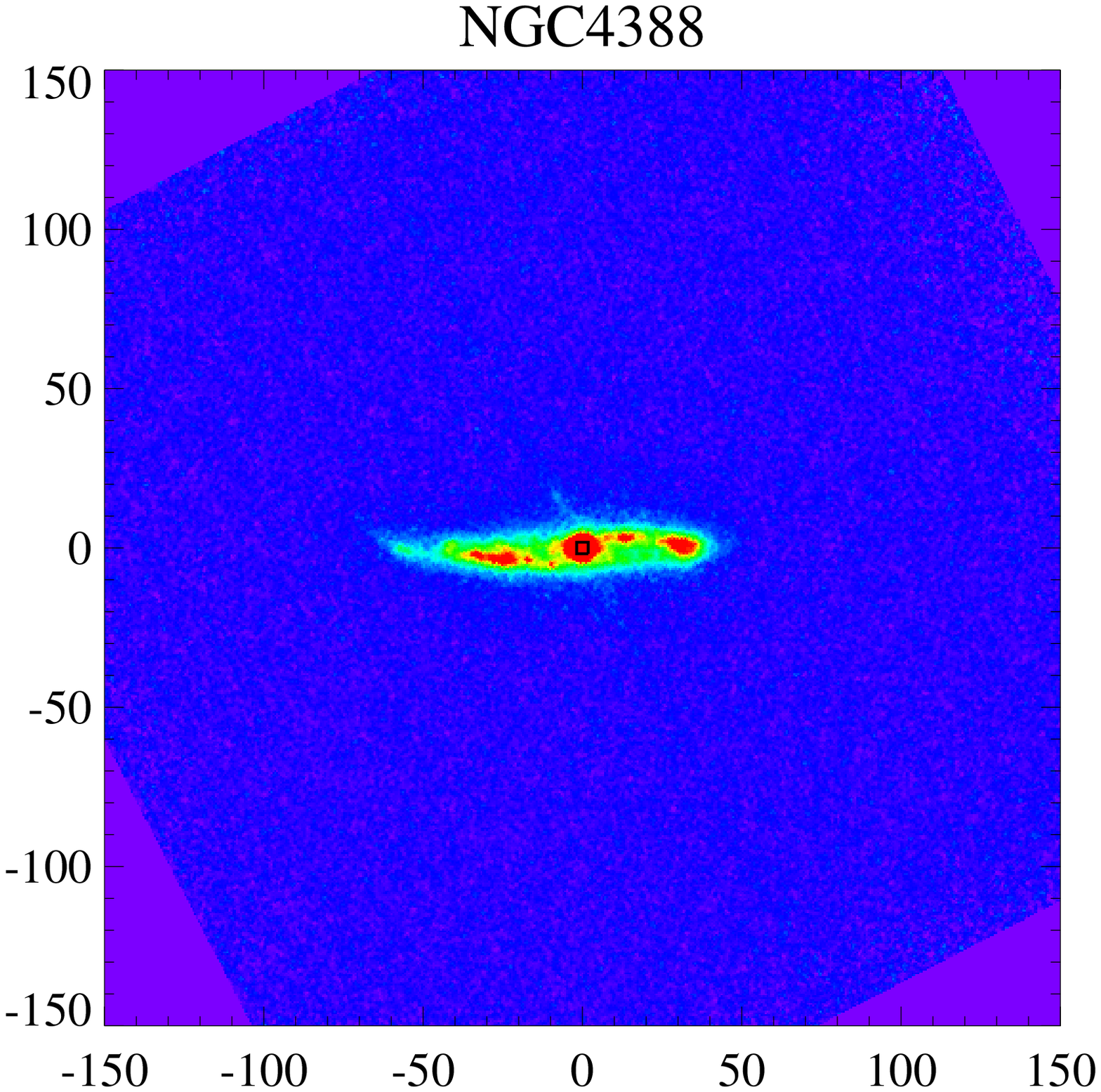}
\includegraphics[width=4.3cm]{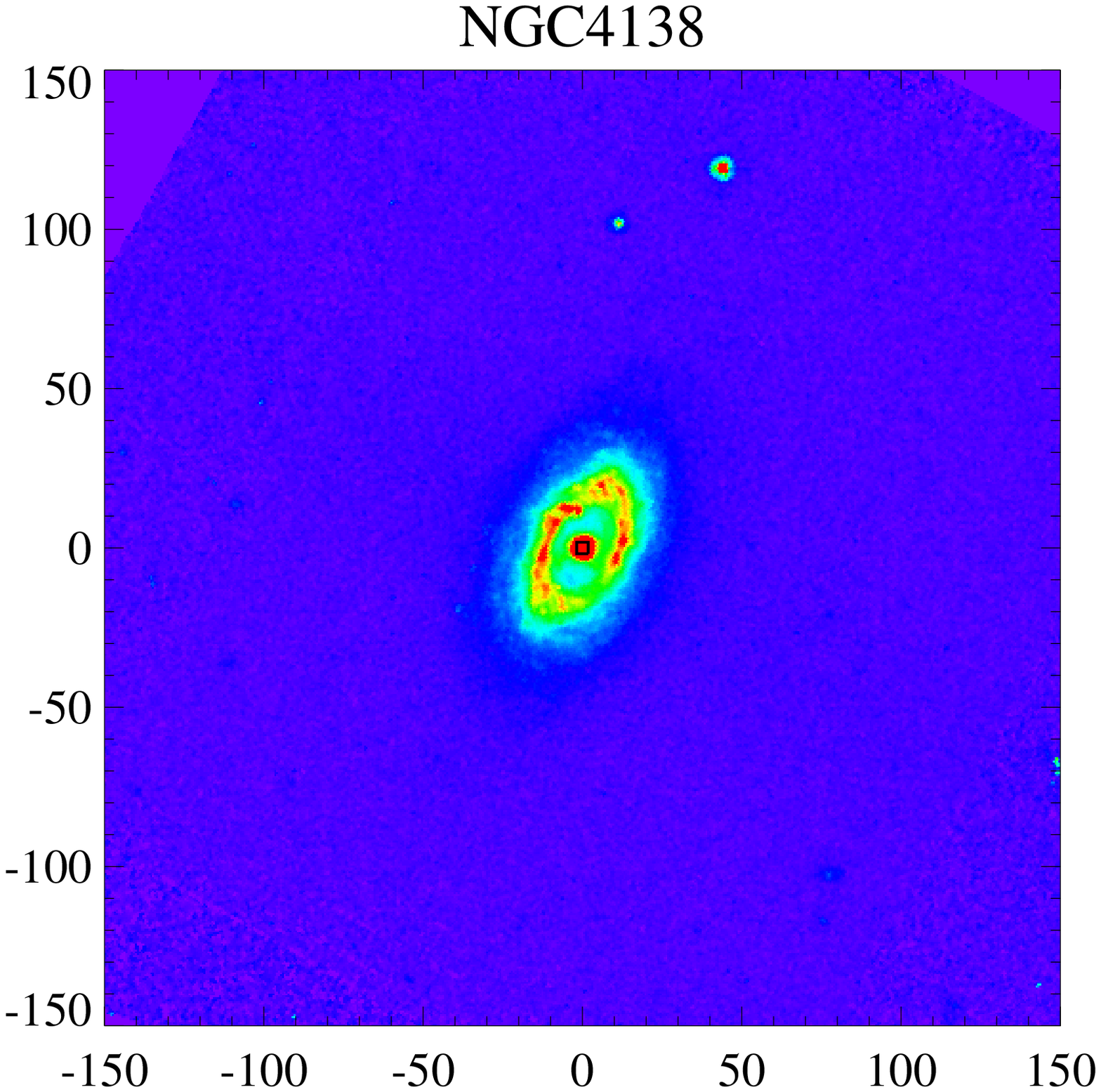}
\includegraphics[width=4.3cm]{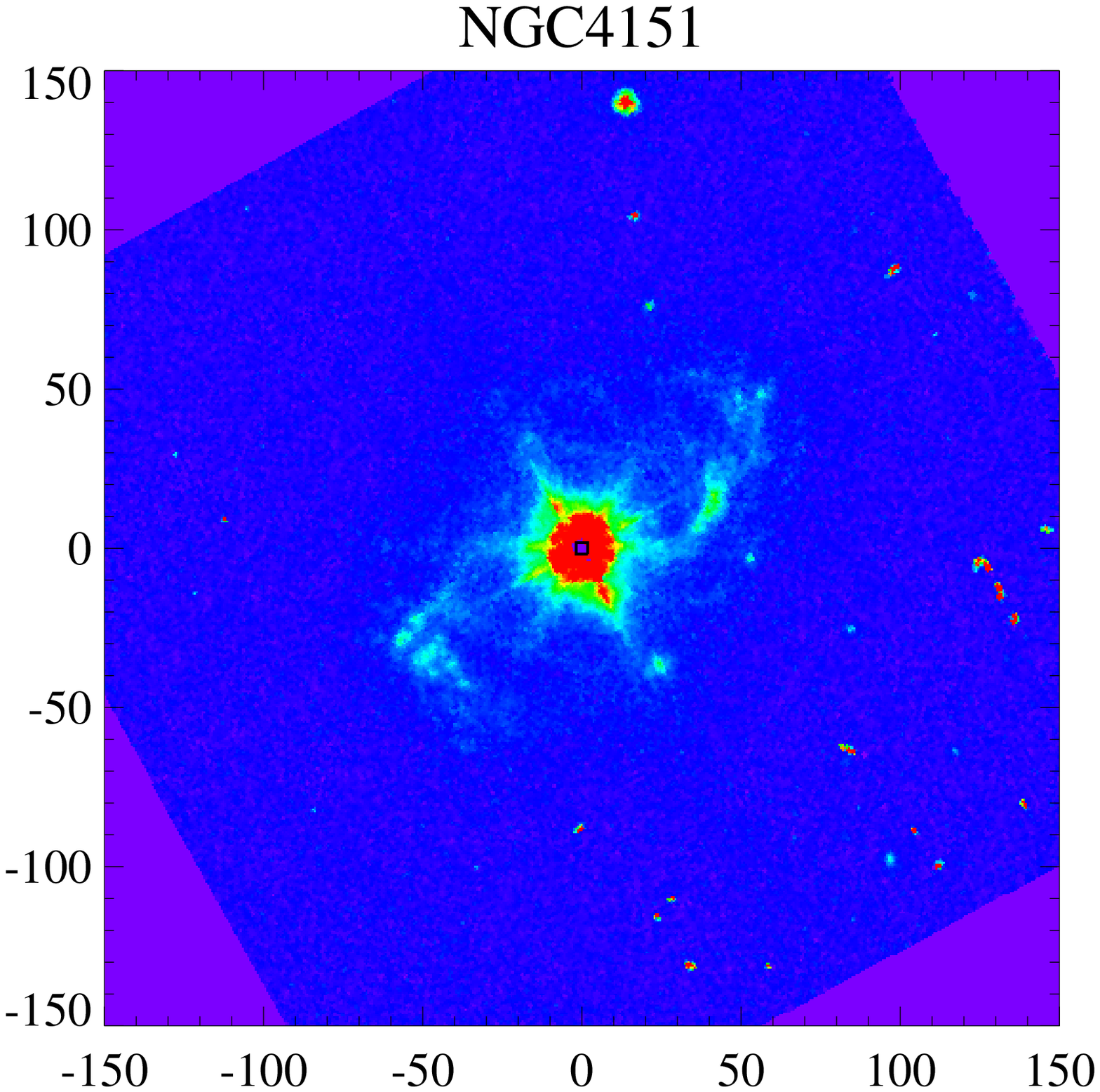}
\includegraphics[width=4.3cm]{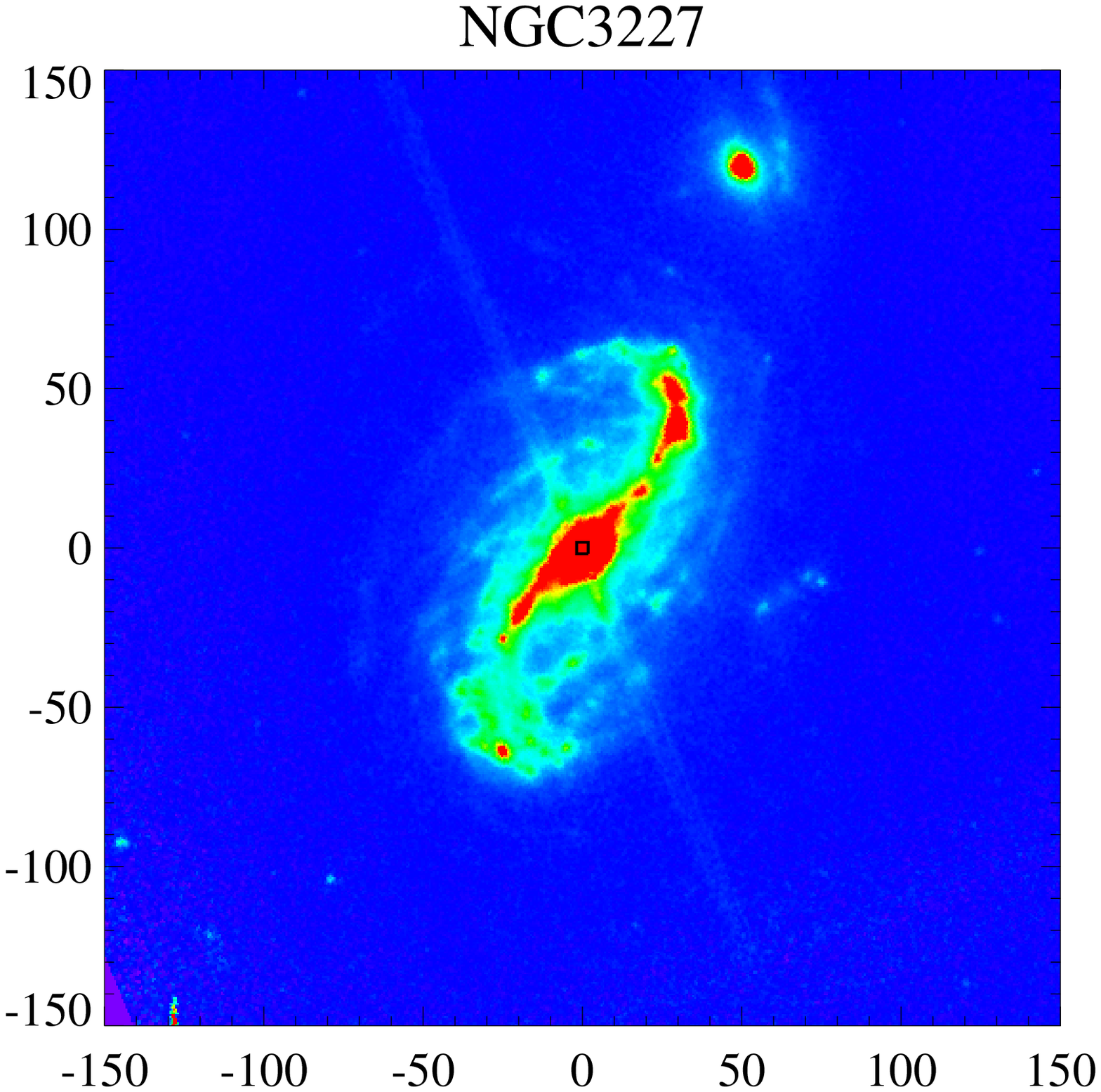}
\includegraphics[width=4.3cm]{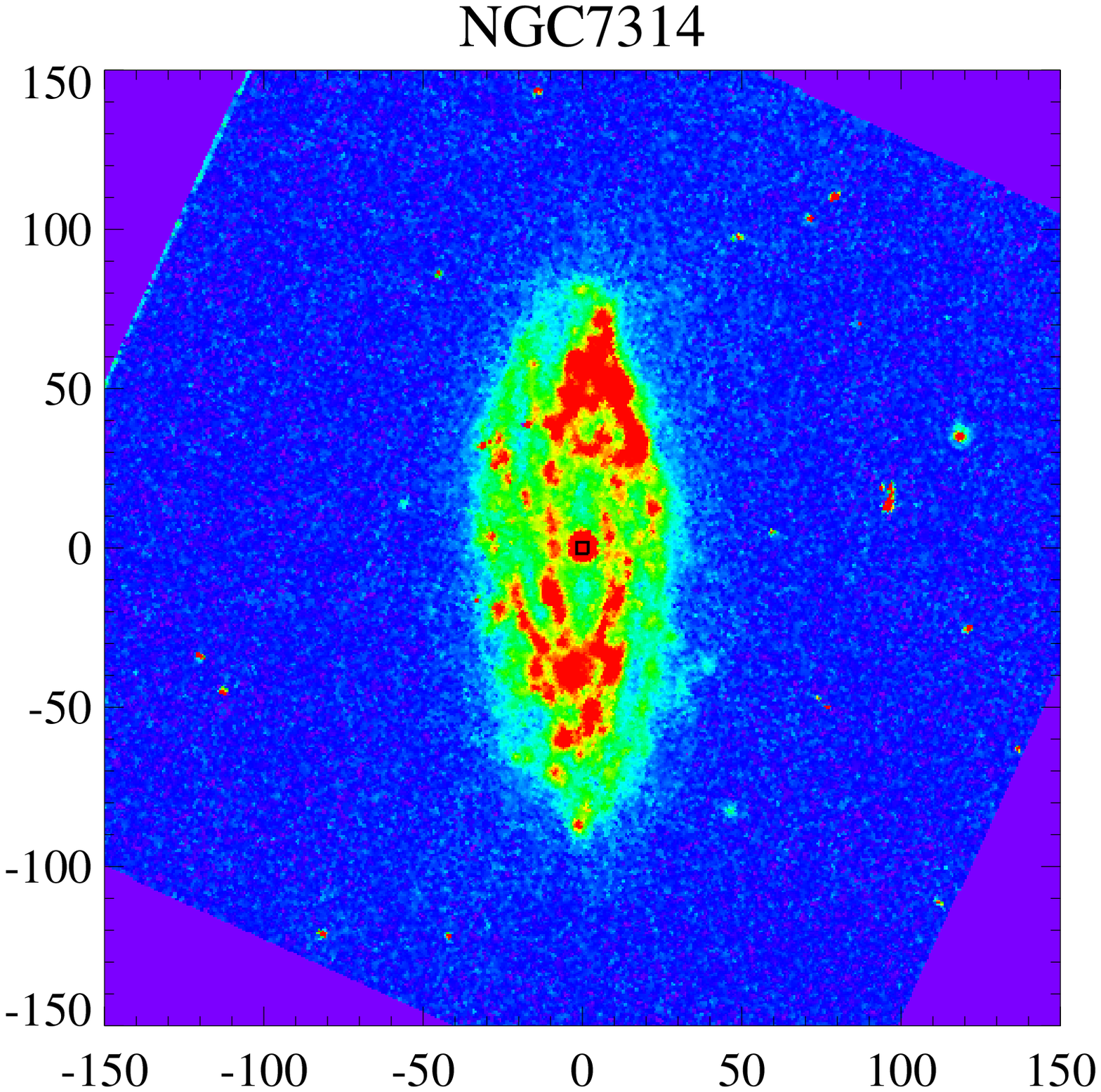}
\includegraphics[width=4.3cm]{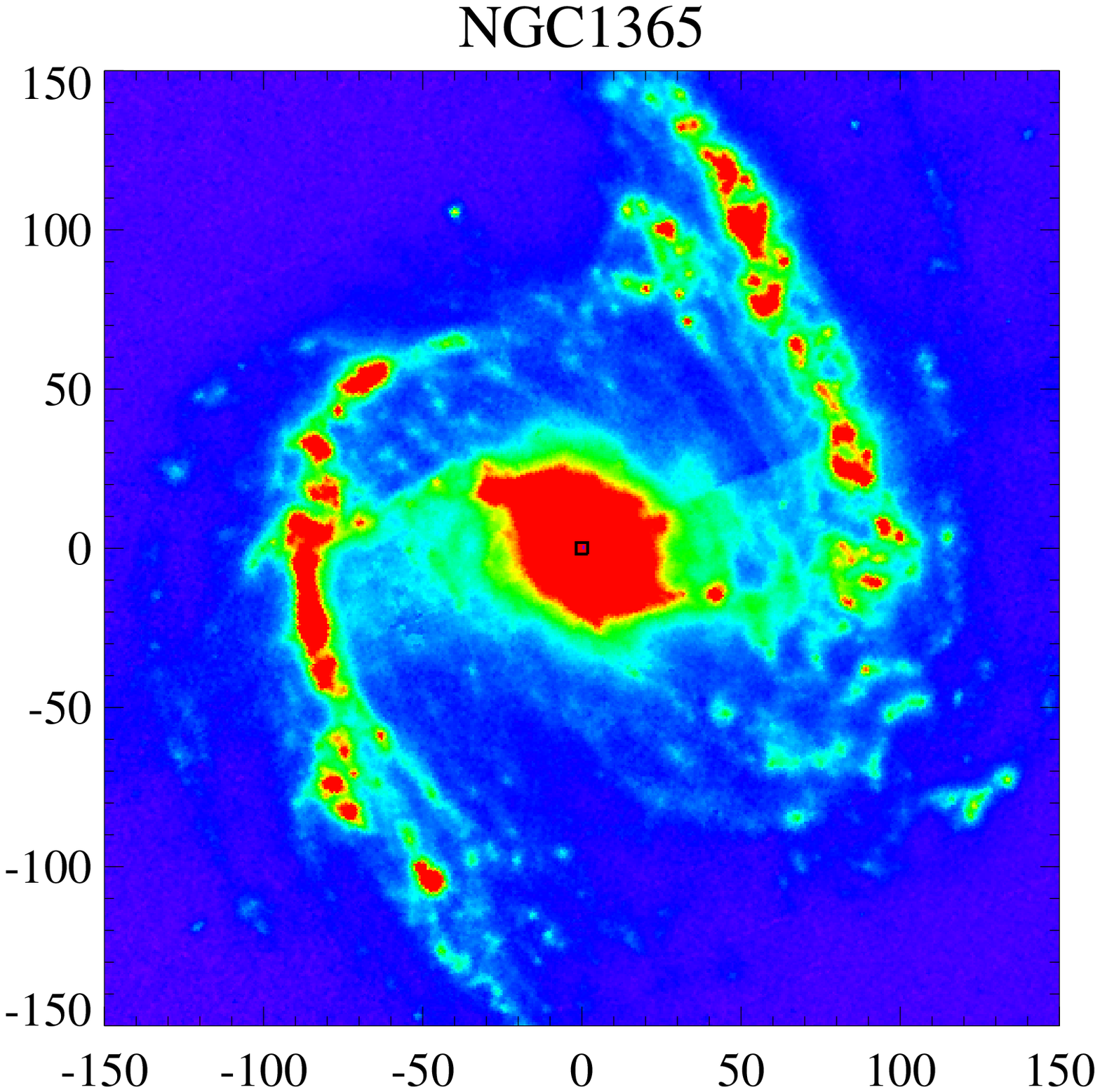}
\includegraphics[width=4.3cm]{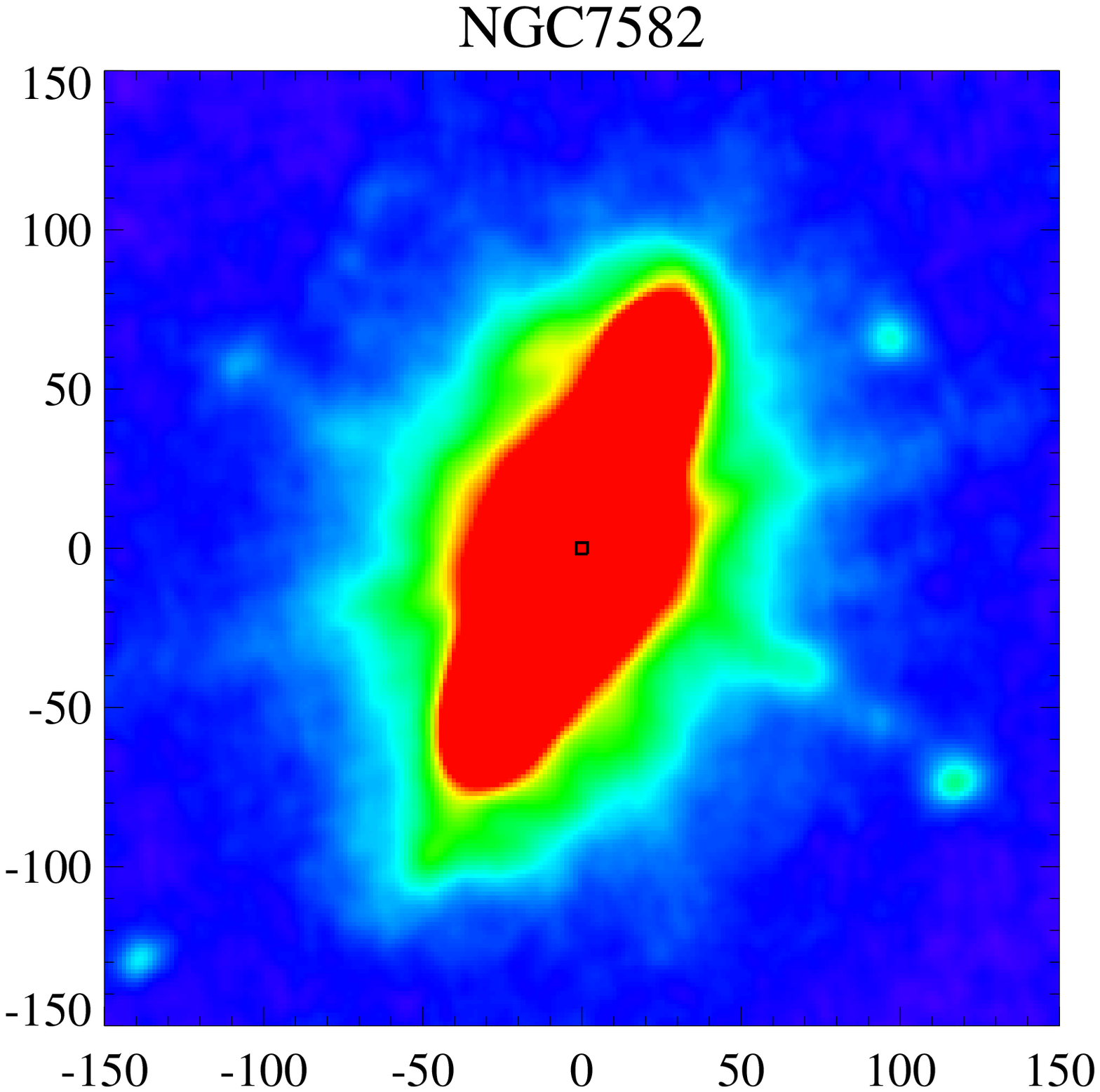}
\includegraphics[width=4.3cm]{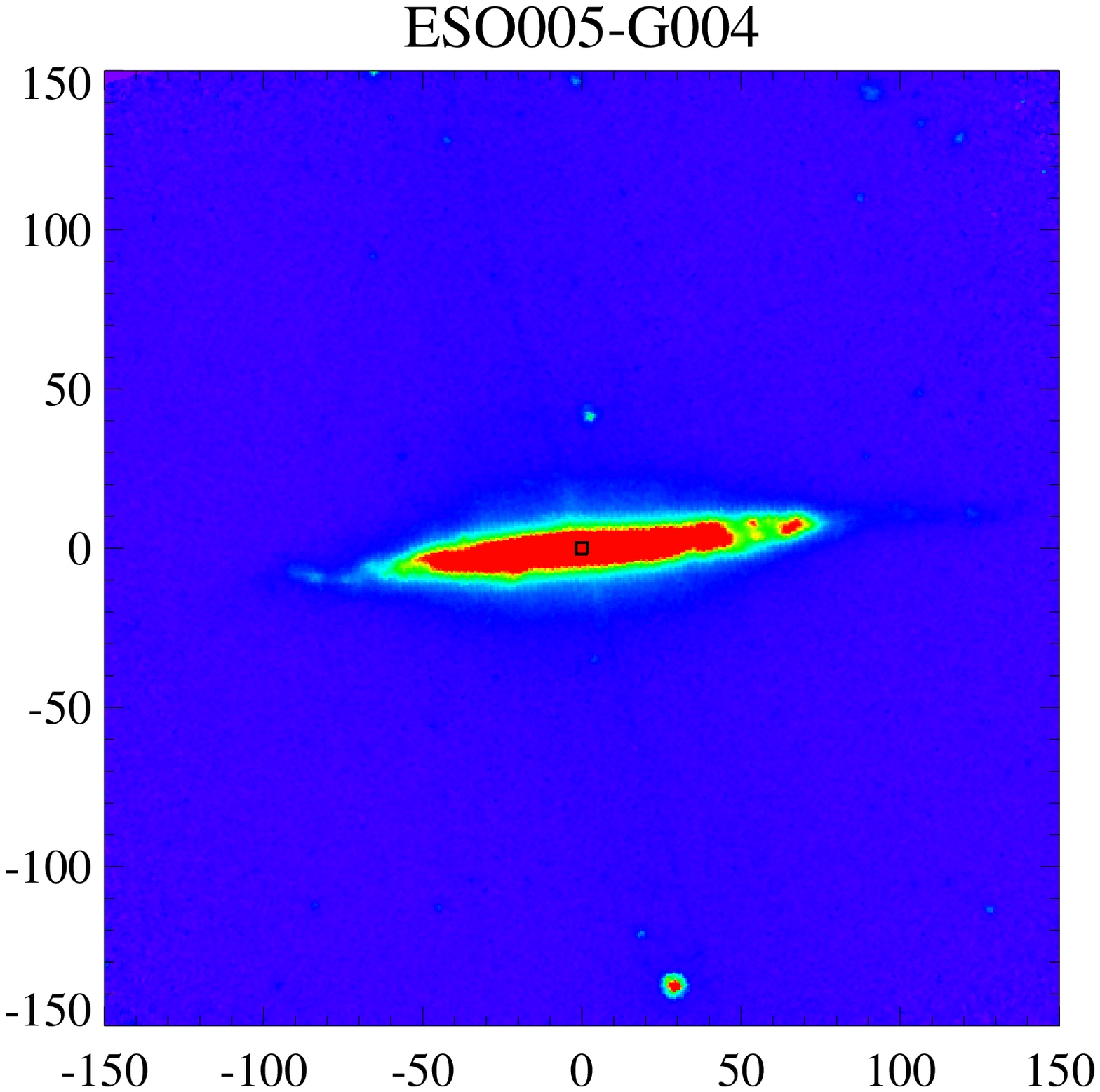}
\includegraphics[width=4.3cm]{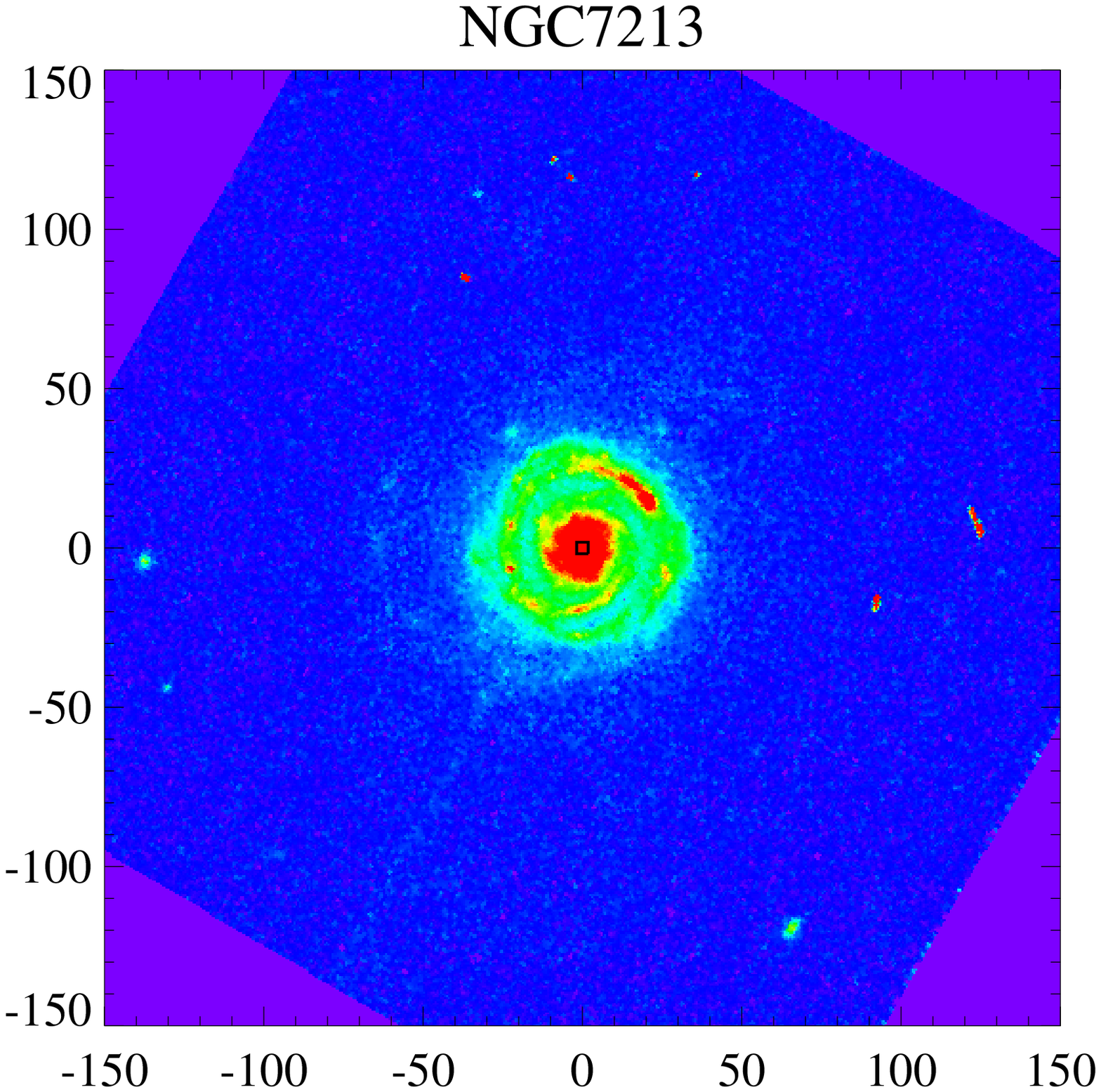}
\includegraphics[width=4.3cm]{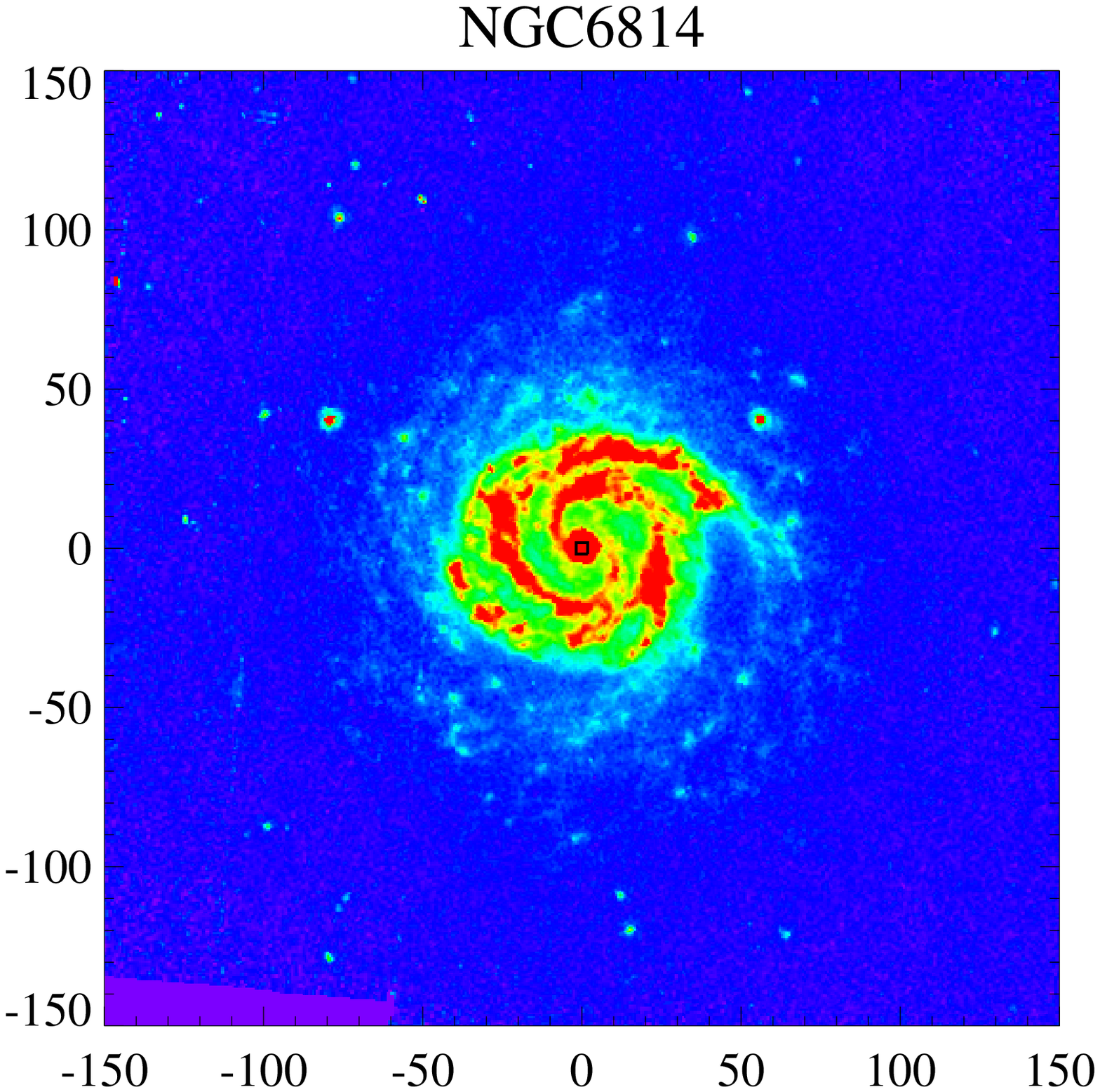}
\includegraphics[width=4.3cm]{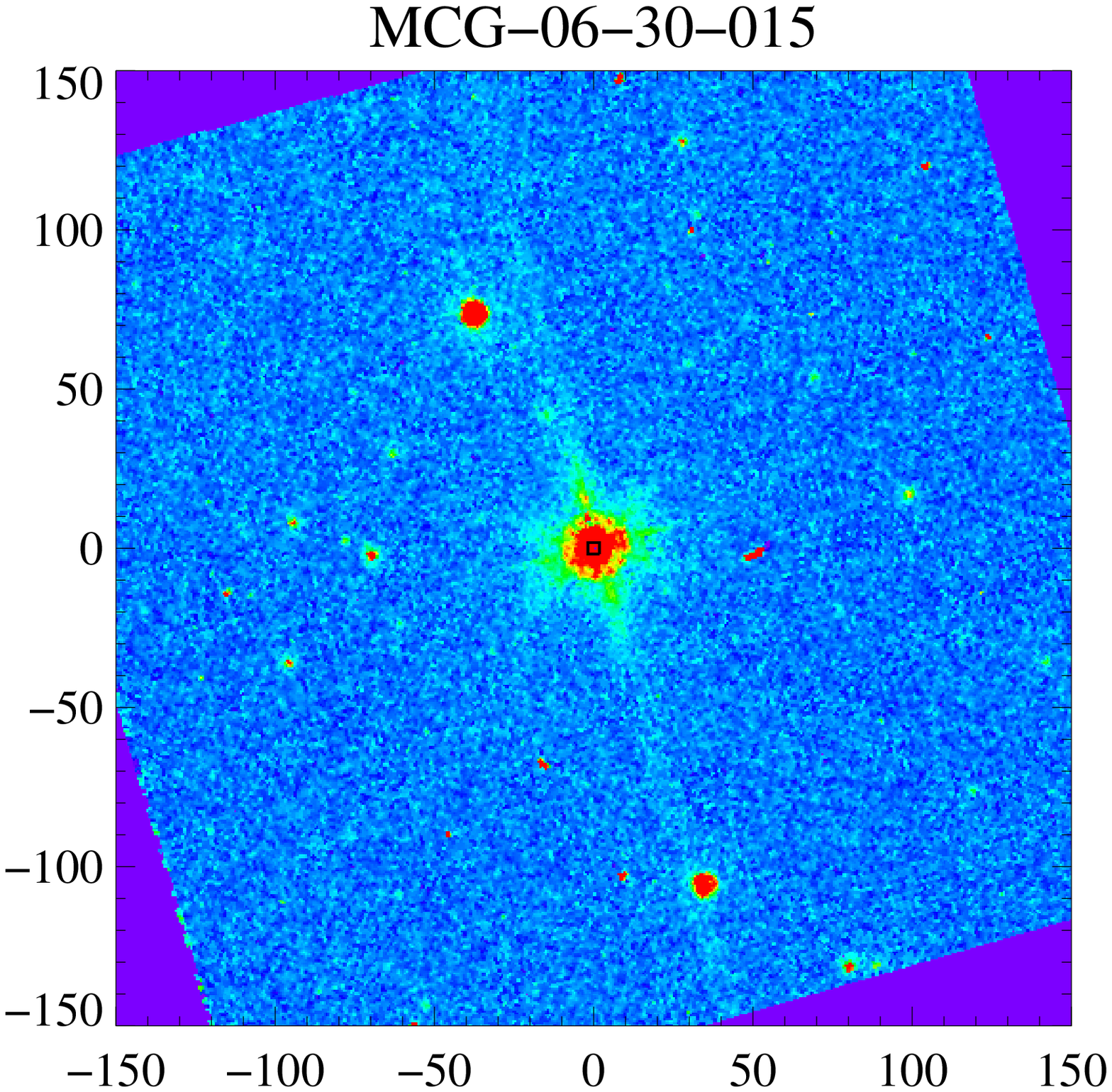}
\includegraphics[width=4.3cm]{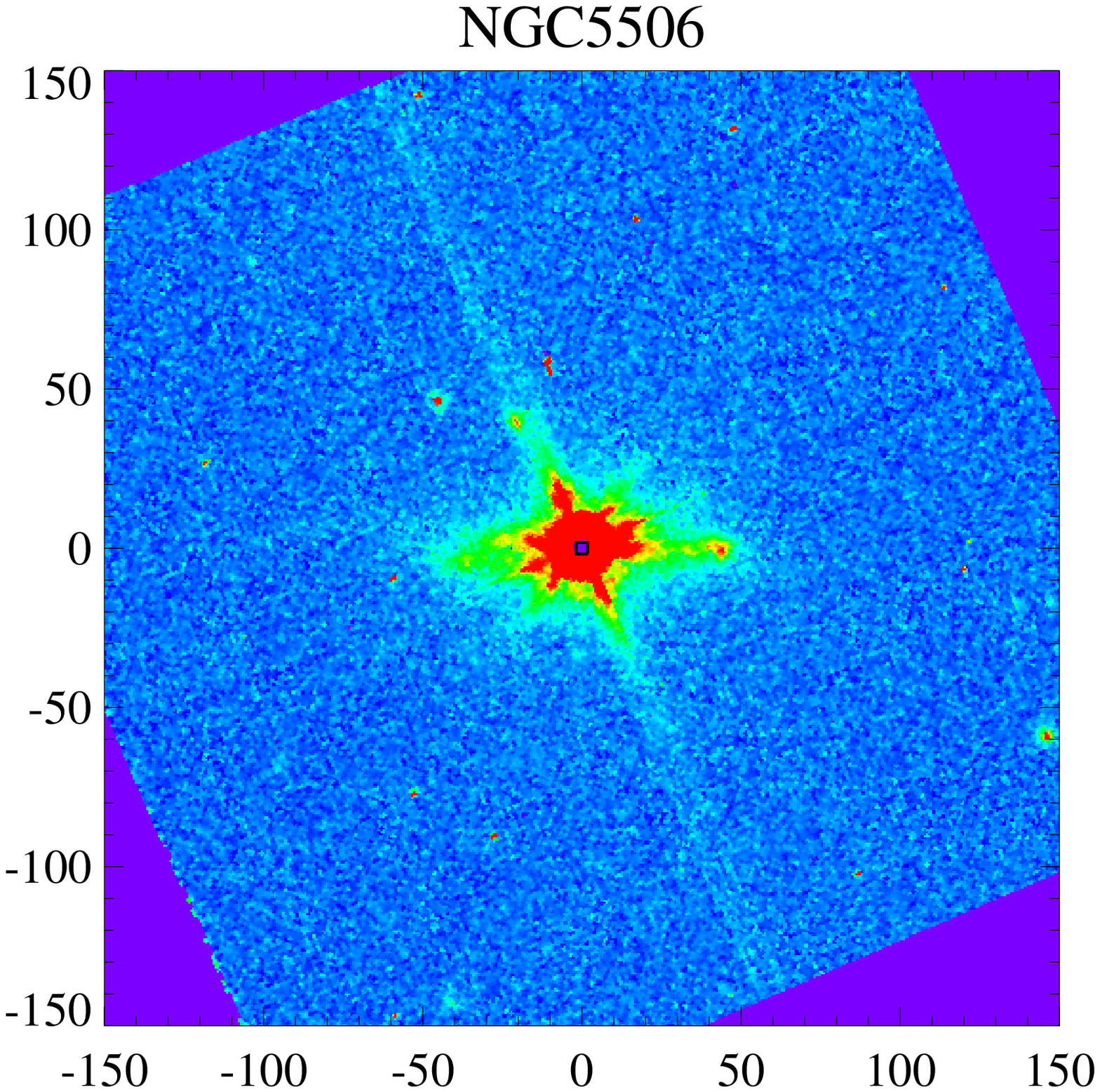}
\includegraphics[width=4.3cm]{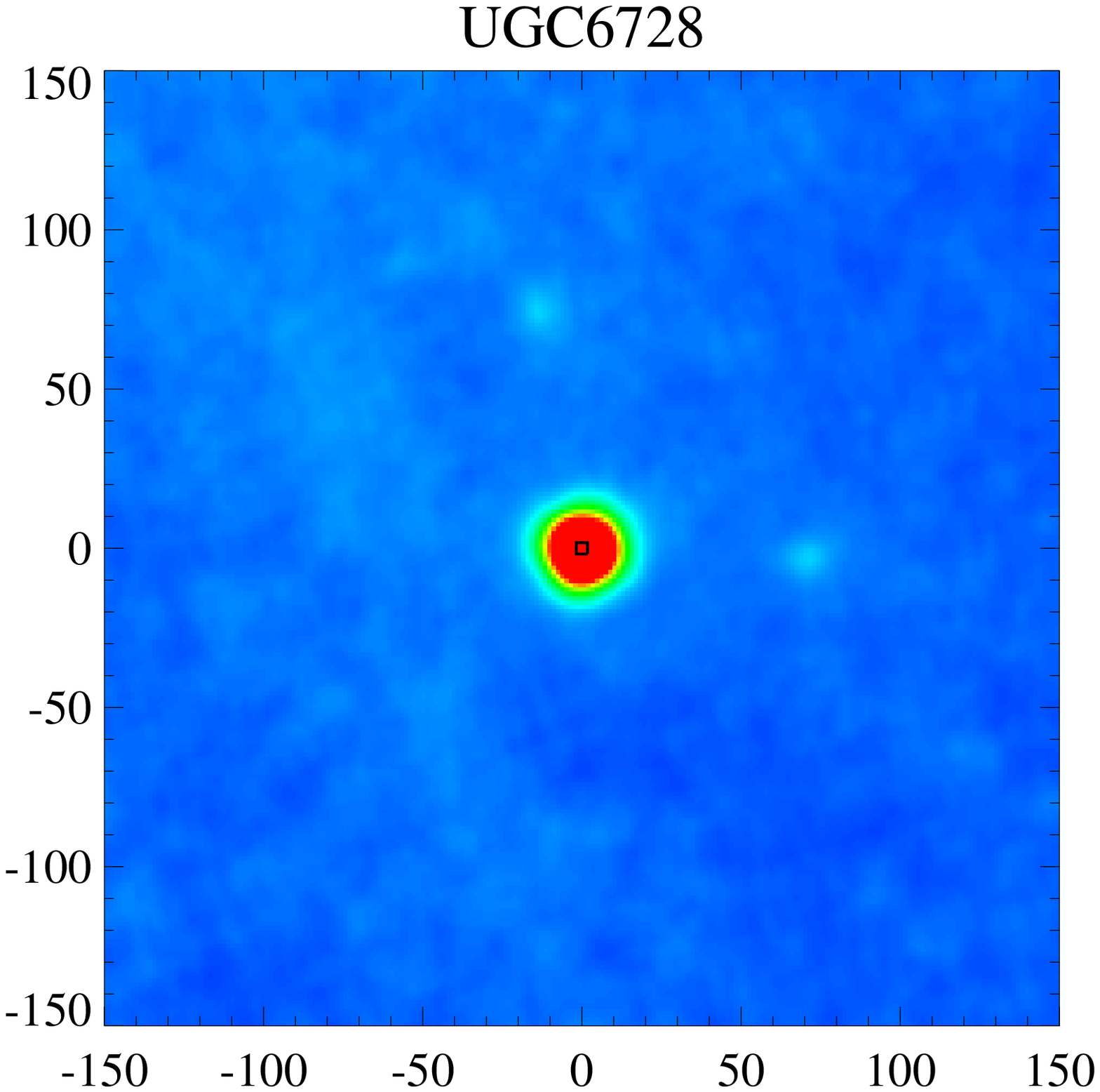}
\includegraphics[width=4.3cm]{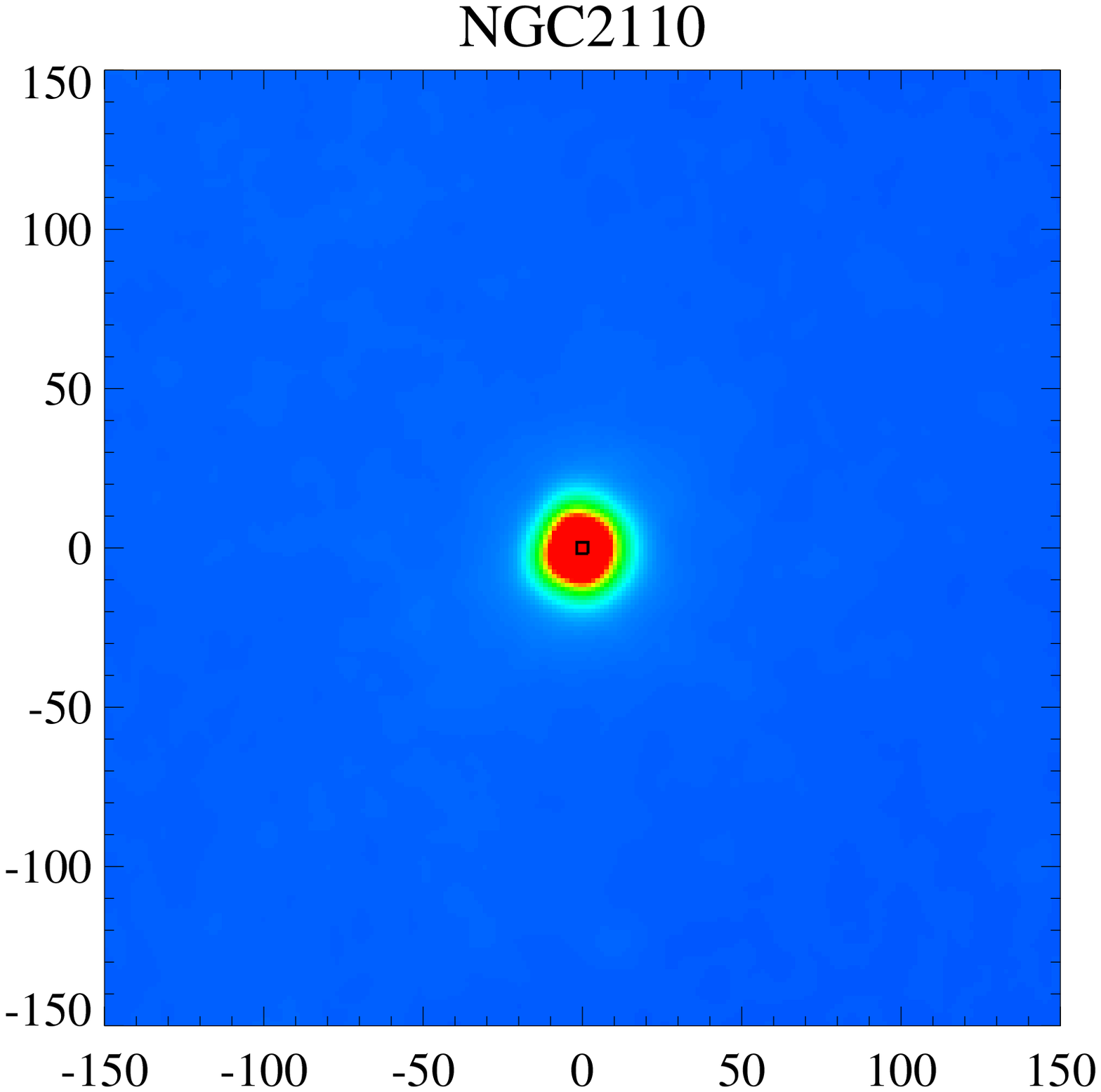}
\includegraphics[width=4.3cm]{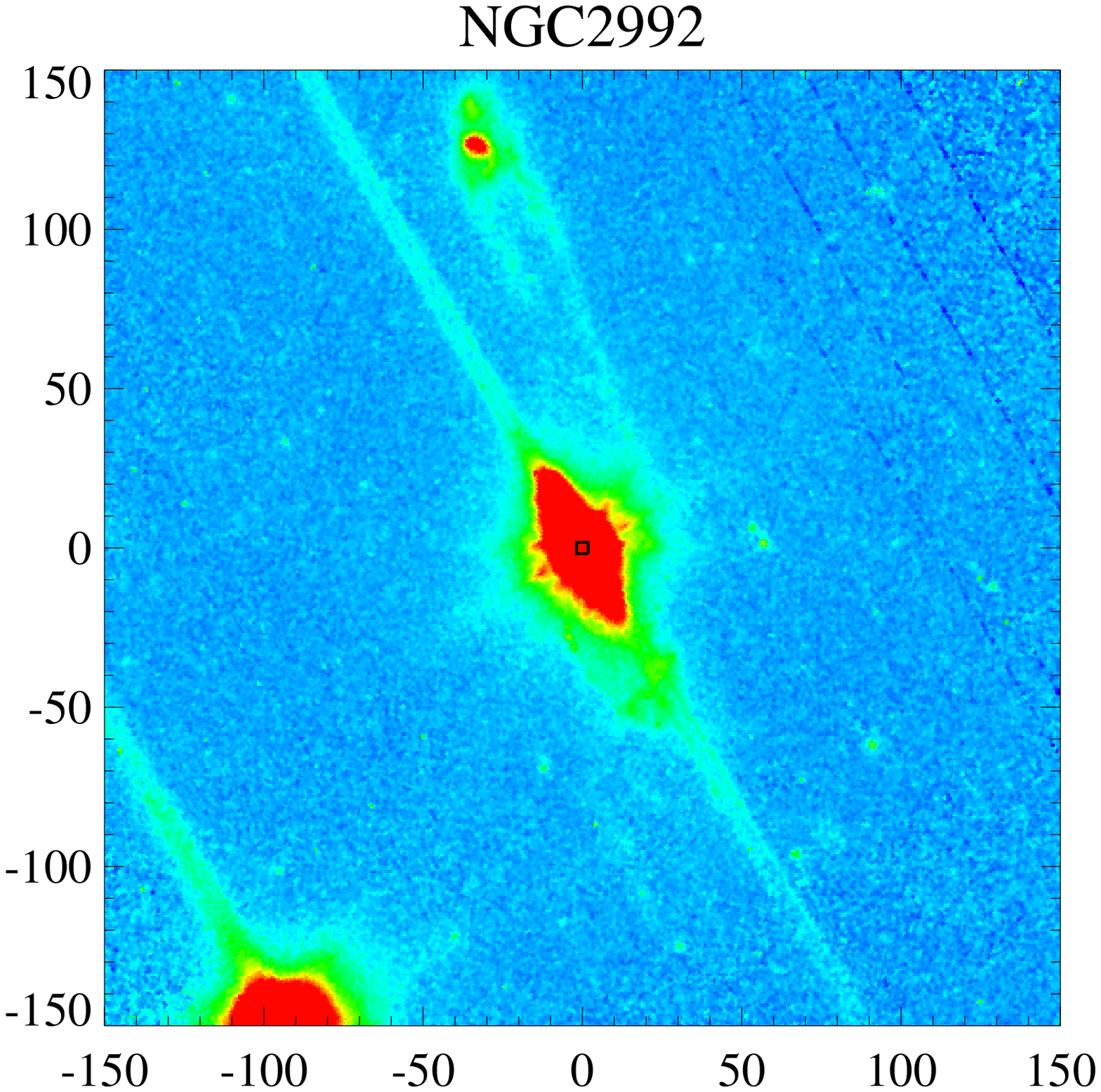}
\par}
\caption{8$\mu$m {\textit{IRAC}} or 12$\mu$m {\textit{WISE}} images of the BCS$_{40}$ sample. The images of NGC\,5128, NGC\,4945, NGC\,7582, UGC\,6728 and NGC\,2110 correspond to 12$\mu$m {\textit{WISE}} data. The black square boxes correspond to a size of 3.6 arcsec (the FOV shown in left and central panels of Fig. \ref{figA1}). North is up and east to the left.}
\label{figA2}
\end{figure*}

\begin{figure*}
\centering
\contcaption
\par{
\includegraphics[width=4.3cm]{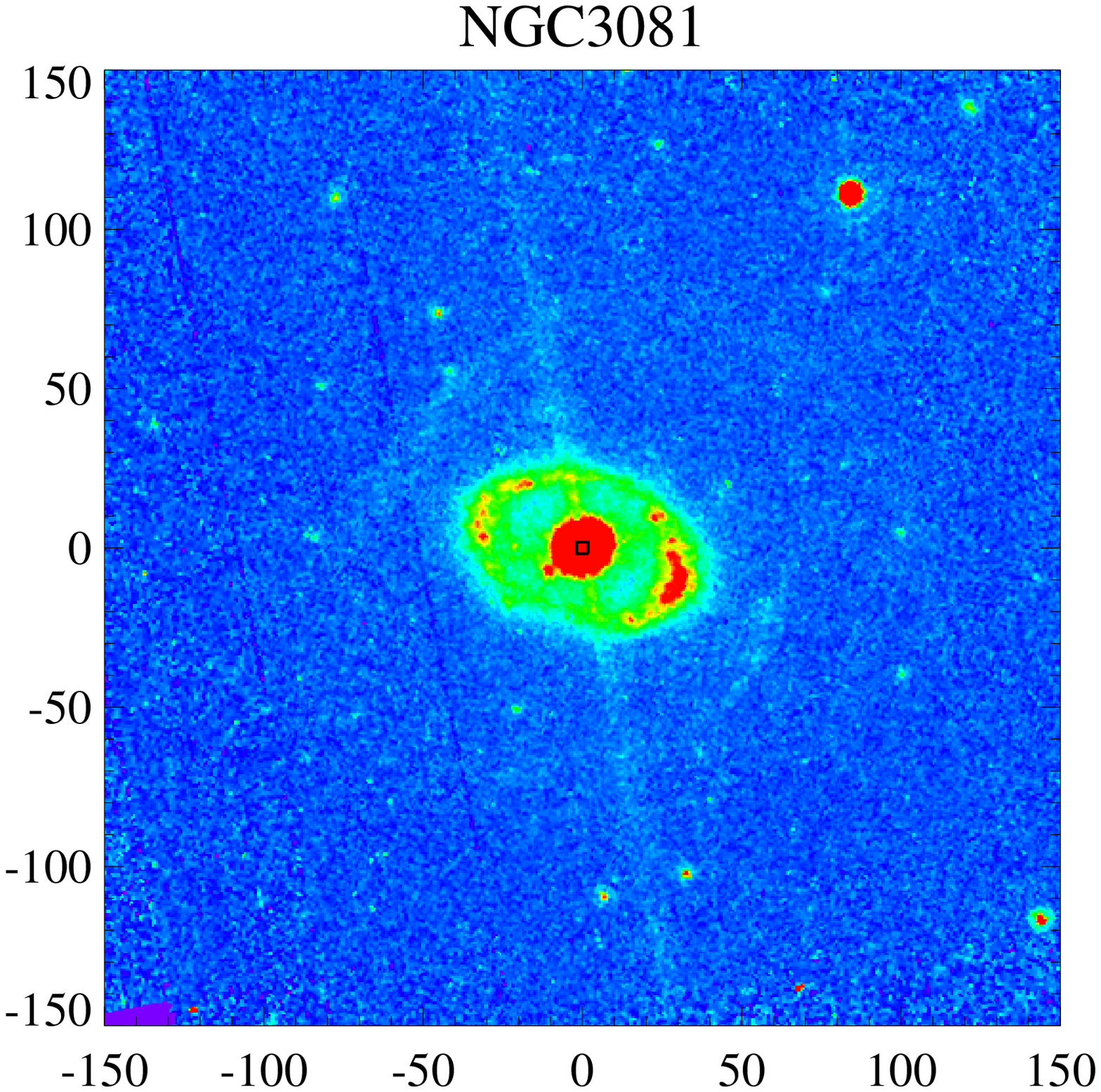}
\includegraphics[width=4.3cm]{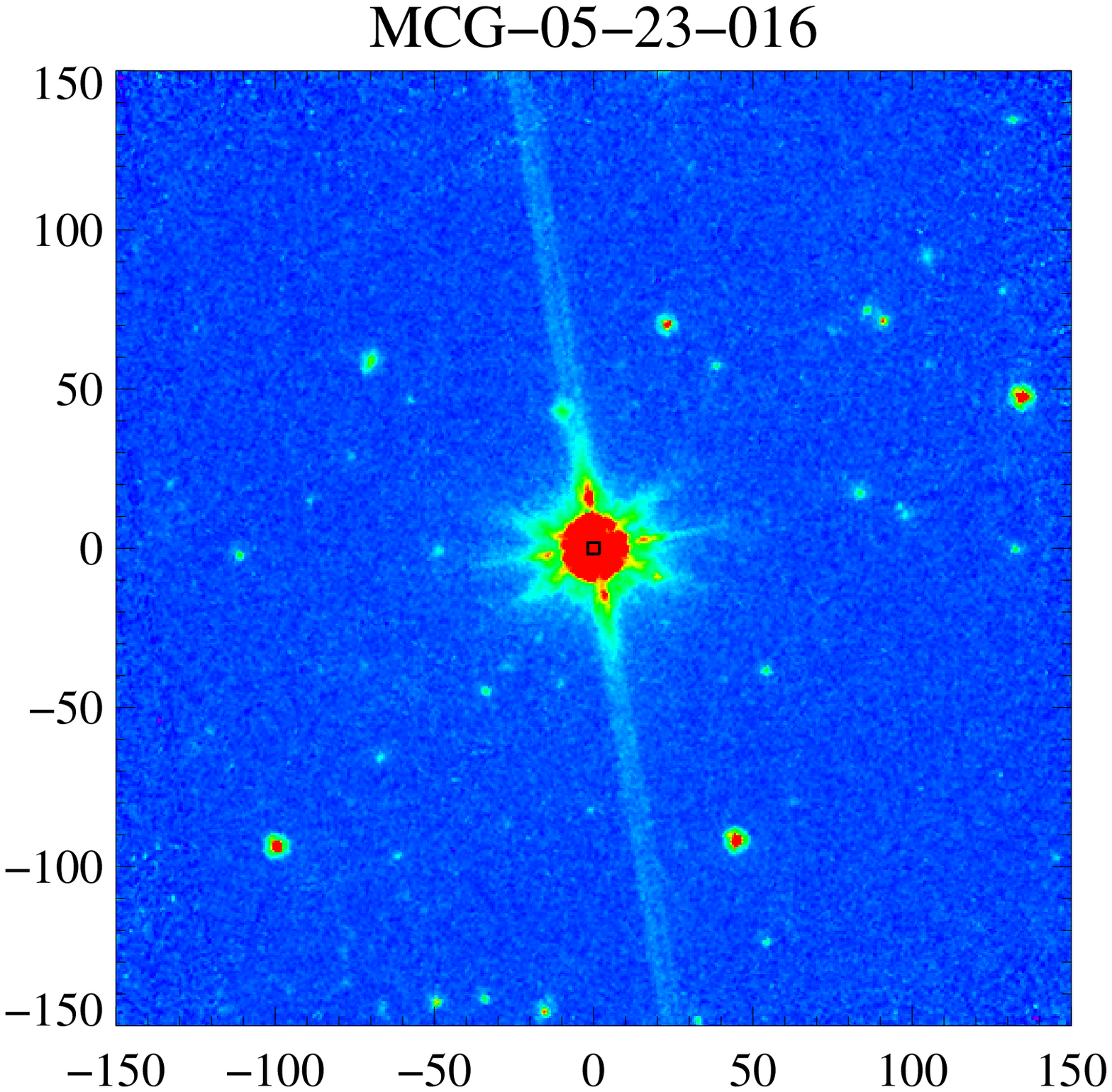}
\includegraphics[width=4.3cm]{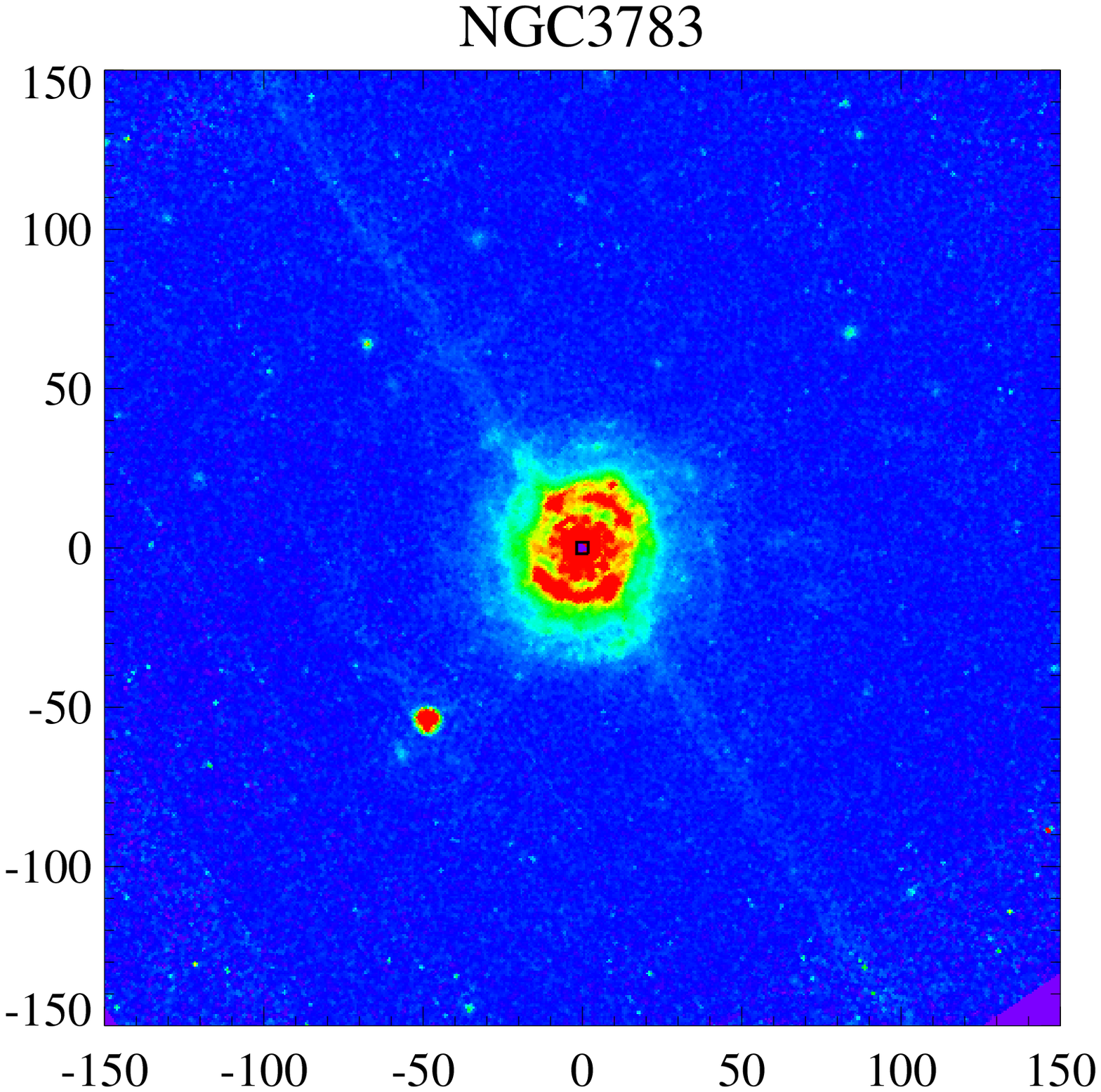}
\includegraphics[width=4.3cm]{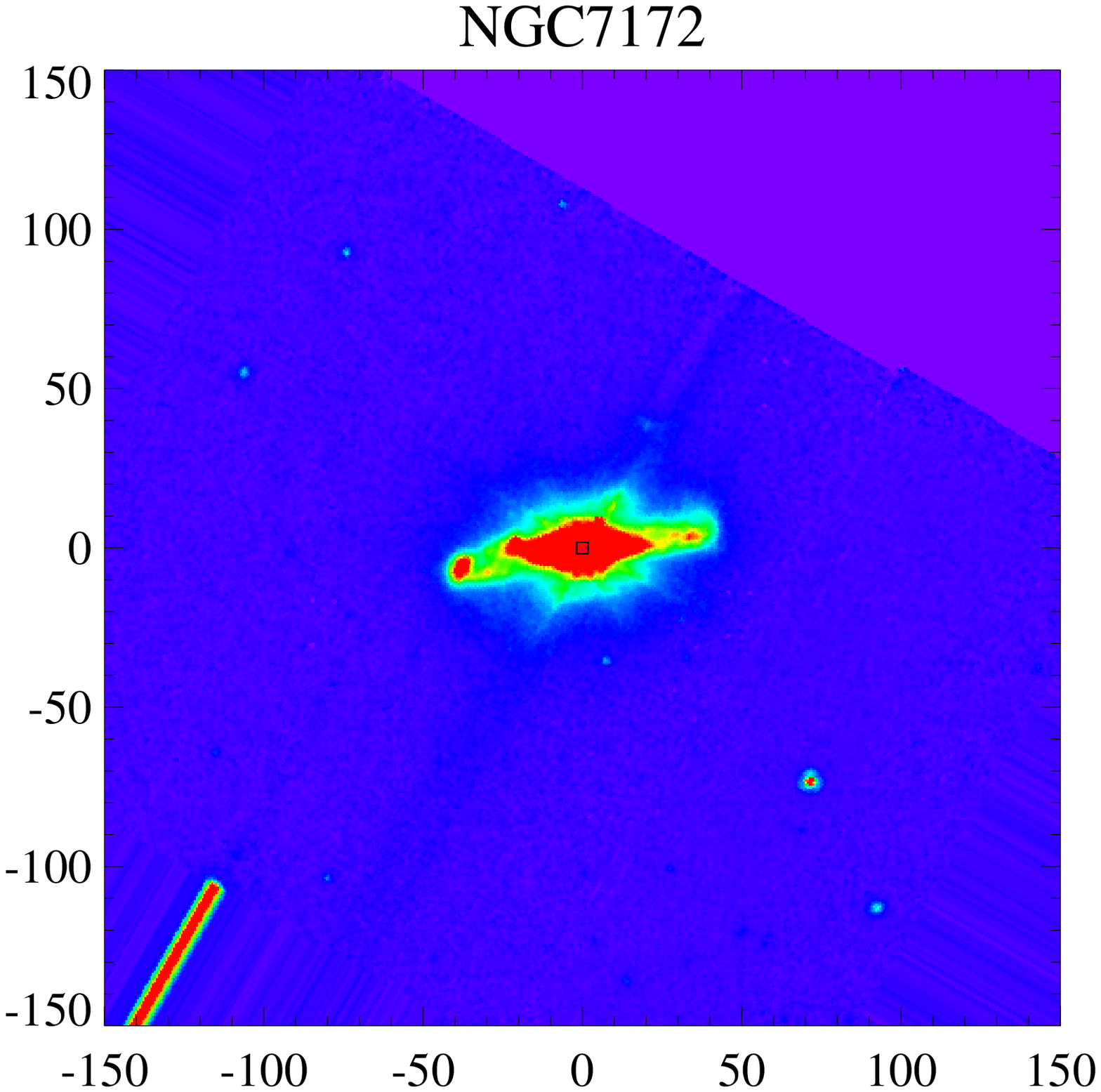}
\par}
\end{figure*}

\section{Spectral Decomposition}
\label{B}

Here we show the {\textit{Spitzer/IRS}} spectra of the BCS$_{40}$ sample, which were retrieved from the CASSIS atlas \citep{Lebouteiller11}. For NGC\,4138 there is no low-resolution staring mode spectrum, and instead, we have extracted an spectrum in a 7.7\arcsec aperture diameter from the mapping mode data cube available in the {\textit{Spitzer}} Heritage Archive.

Considering the spatial scales probed by the {\textit{Spitzer/IRS}} spectra of the whole sample ($\leqslant$650~pc), we expect contribution from both AGN and SF. To estimate the AGN contribution to the {\textit{Spitzer/IRS}} spectra, we use the DeblendIRS routine \citep{Hernan-caballero2015}, that decomposes MIR spectra using a linear combination of three spectral components: AGN, PAH and stellar emission. We used the high angular resolution MIR nuclear fluxes in the various filters reported in Table \ref{tab2} as priors to better constrain the AGN component. A detailed description of the method is given in \citet{Hernan-caballero2015}. 

We present the results of the {\textit{Spitzer/IRS}} spectral decomposition of the BCS$_{40}$ sample in Fig. \ref{figB1}. We used the AGN contribution to obtain homogeneous nuclear fluxes at 8~$\mu$m using a 1~$\mu$m window. These fluxes are presented in Section \ref{nuclear_fluxes}. In Table \ref{tabB1} we present the main properties derived from this spectral decomposition.

\input{tabB1.tex}

We found that the majority of the Sy1 have a higher contribution of the AGN than the Sy2 galaxies. The median values of the fractional contribution of the AGN to the {\textit{Spitzer/IRS}} spectra for Sy1 and Sy2 are 0.82 and 0.62, respectively. 

Using the AGN and PAH contributions to the MIR spectra (see Table \ref{tabB1}) we can classify the systems as AGN-dominated, when there is $\geqslant$70\% AGN contribution to the MIR spectrum and it does not show PAH emission; and SF-dominated, when there is $\leqslant$50\% AGN contribution to the MIR spectrum and it shows PAH emission. The rest of the galaxies are composite objects. See Table \ref{tabB1} for further details on the systems classification.

In the case of NGC\,7213 we cannot obtain a good fit, as the modelling does not reproduce the 9.7~$\mu$m silicate emission feature. However, the nuclear MIR fluxes of NGC\,7213 are in agreement with the spectral shape of the AGN component derived from the fit.

\begin{figure*}
\centering
\par{
\includegraphics[width=8.0cm]{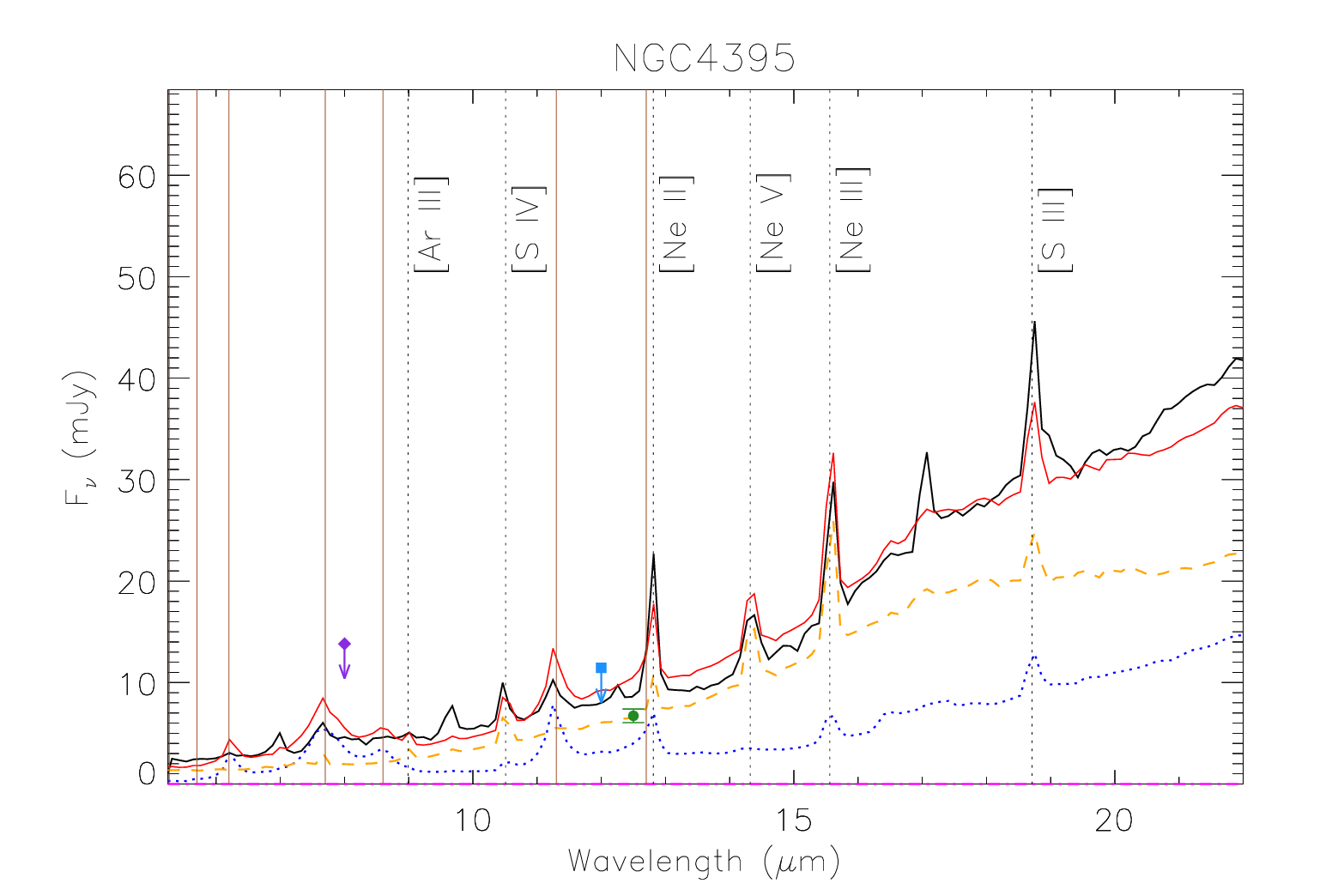}
\includegraphics[width=8.0cm]{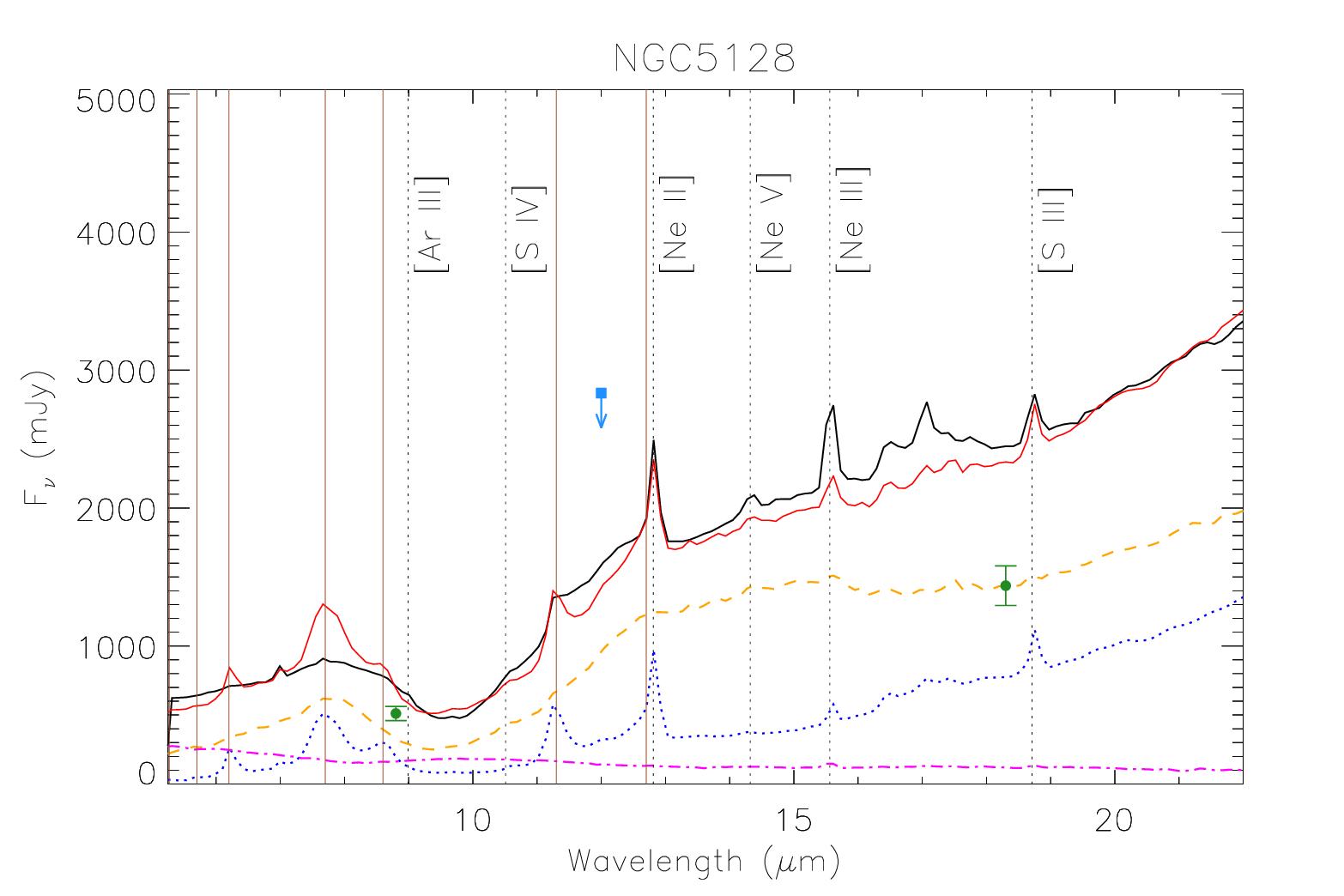}
\includegraphics[width=8.0cm]{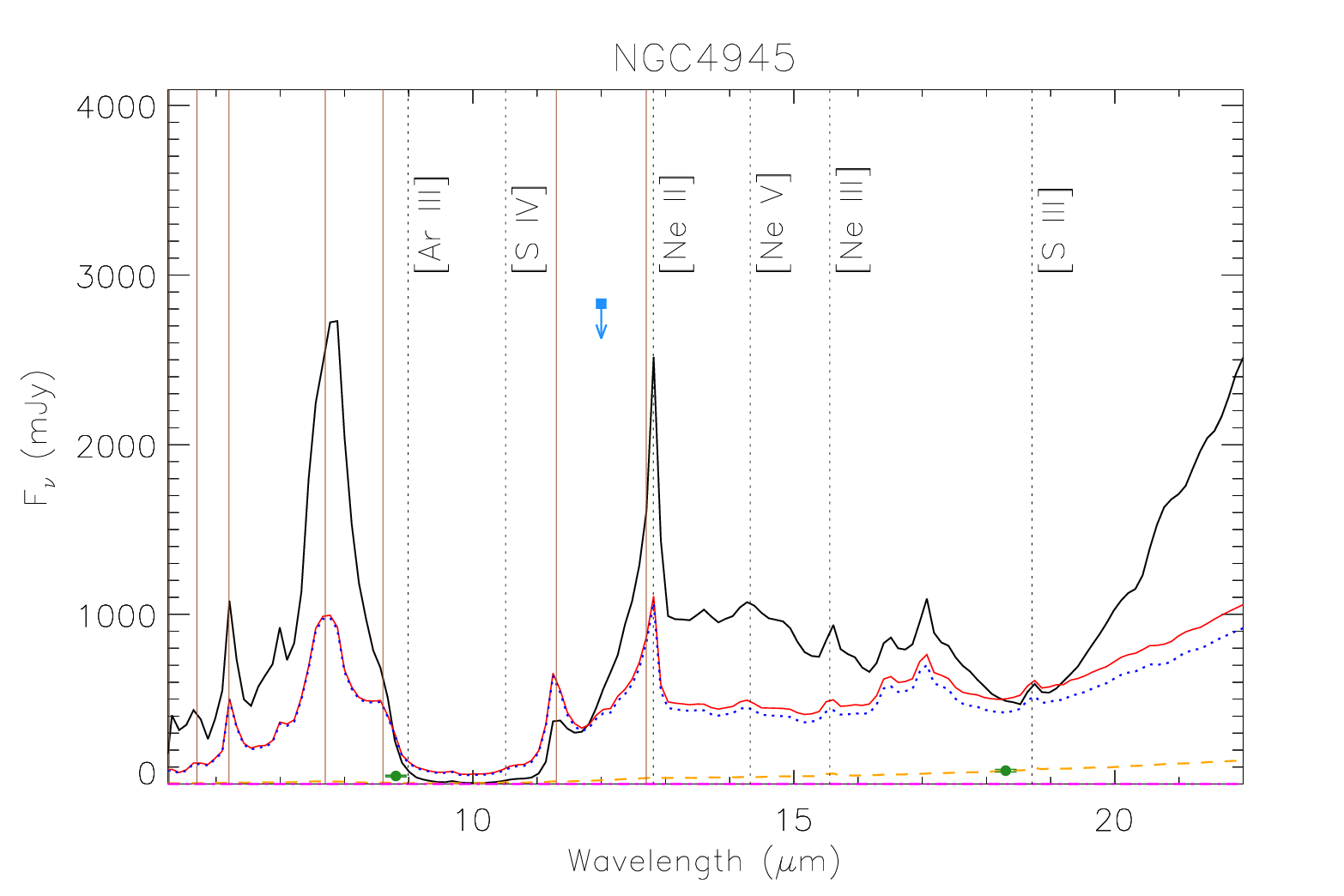}
\includegraphics[width=8.0cm]{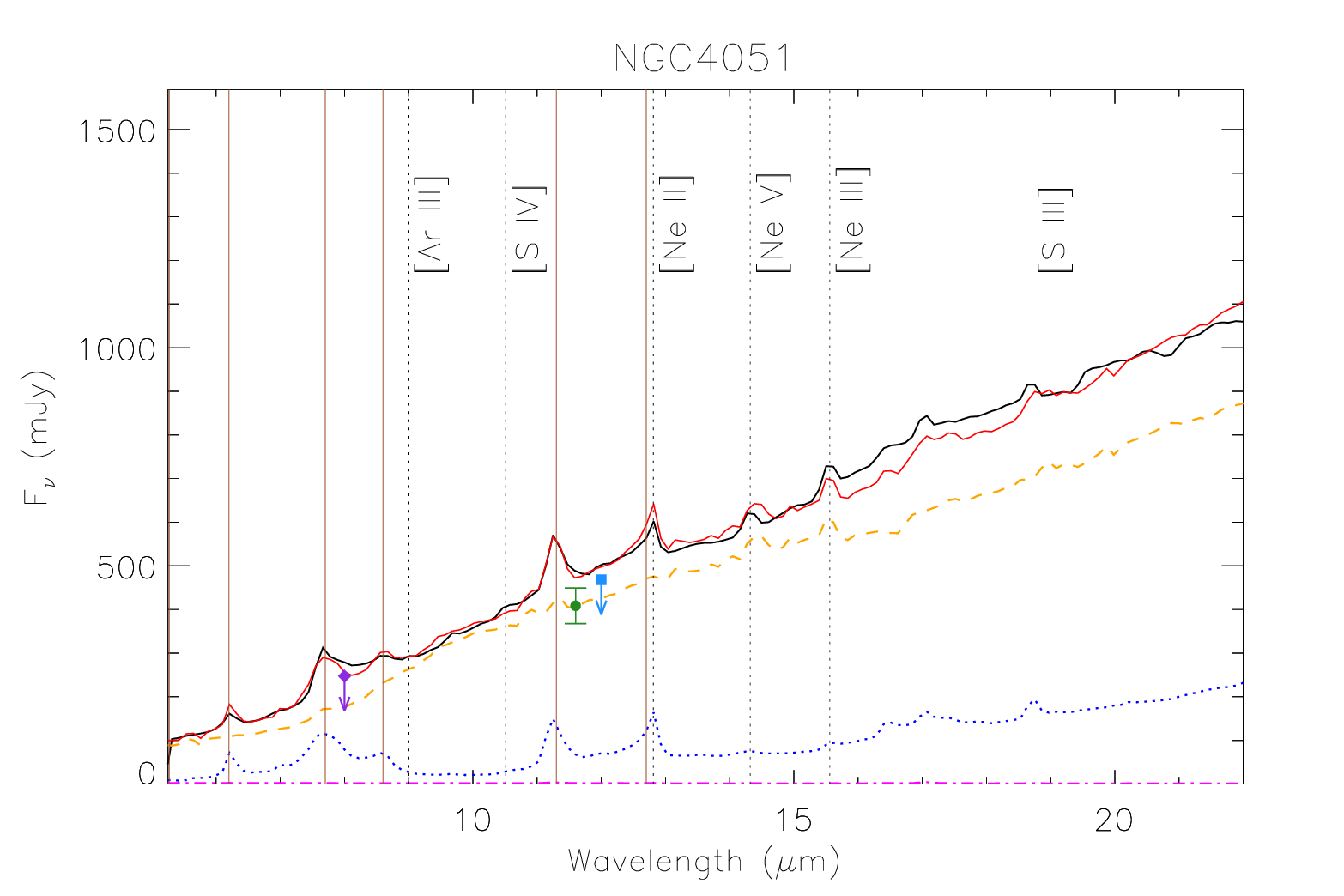}
\includegraphics[width=8.0cm]{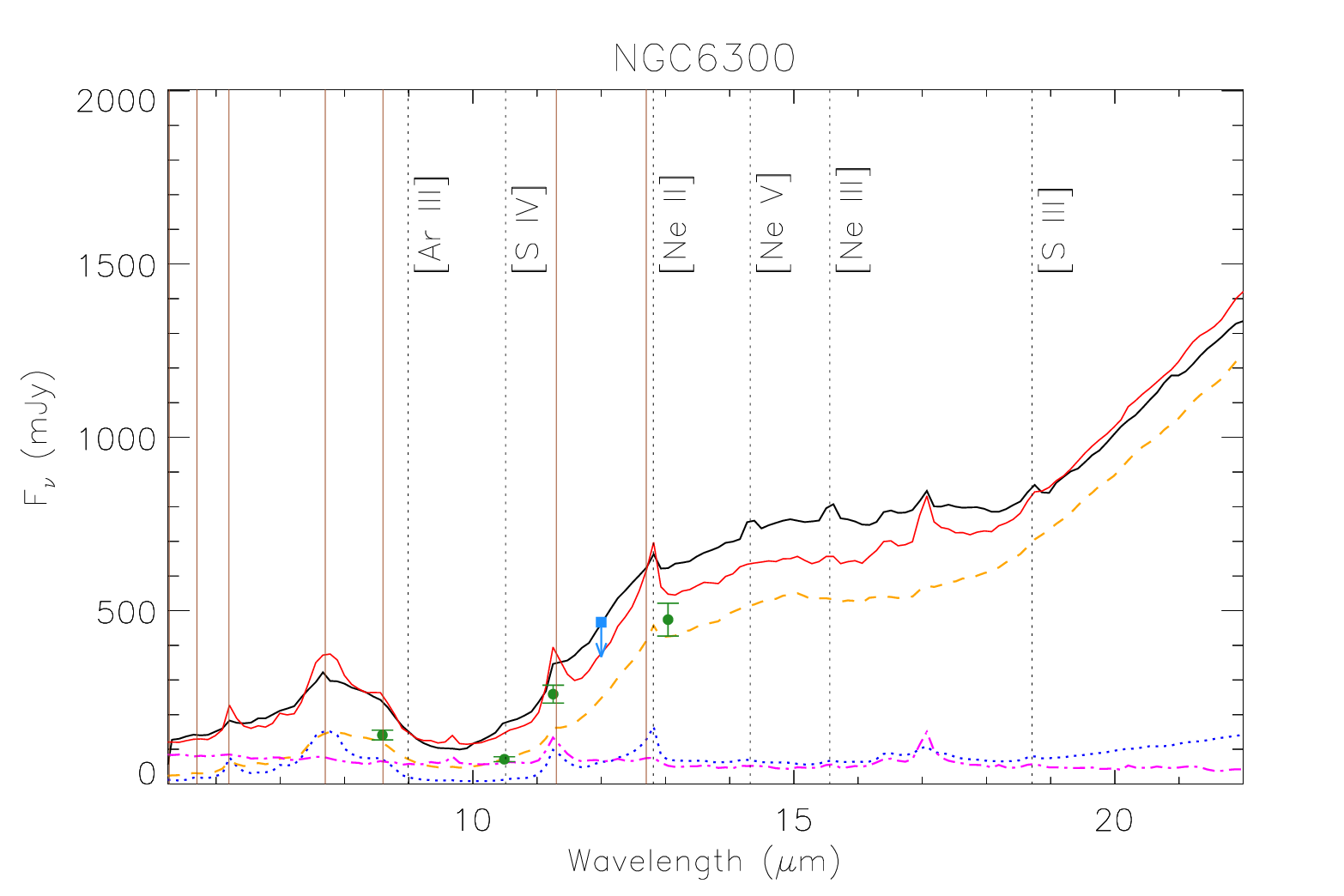}
\includegraphics[width=8.0cm]{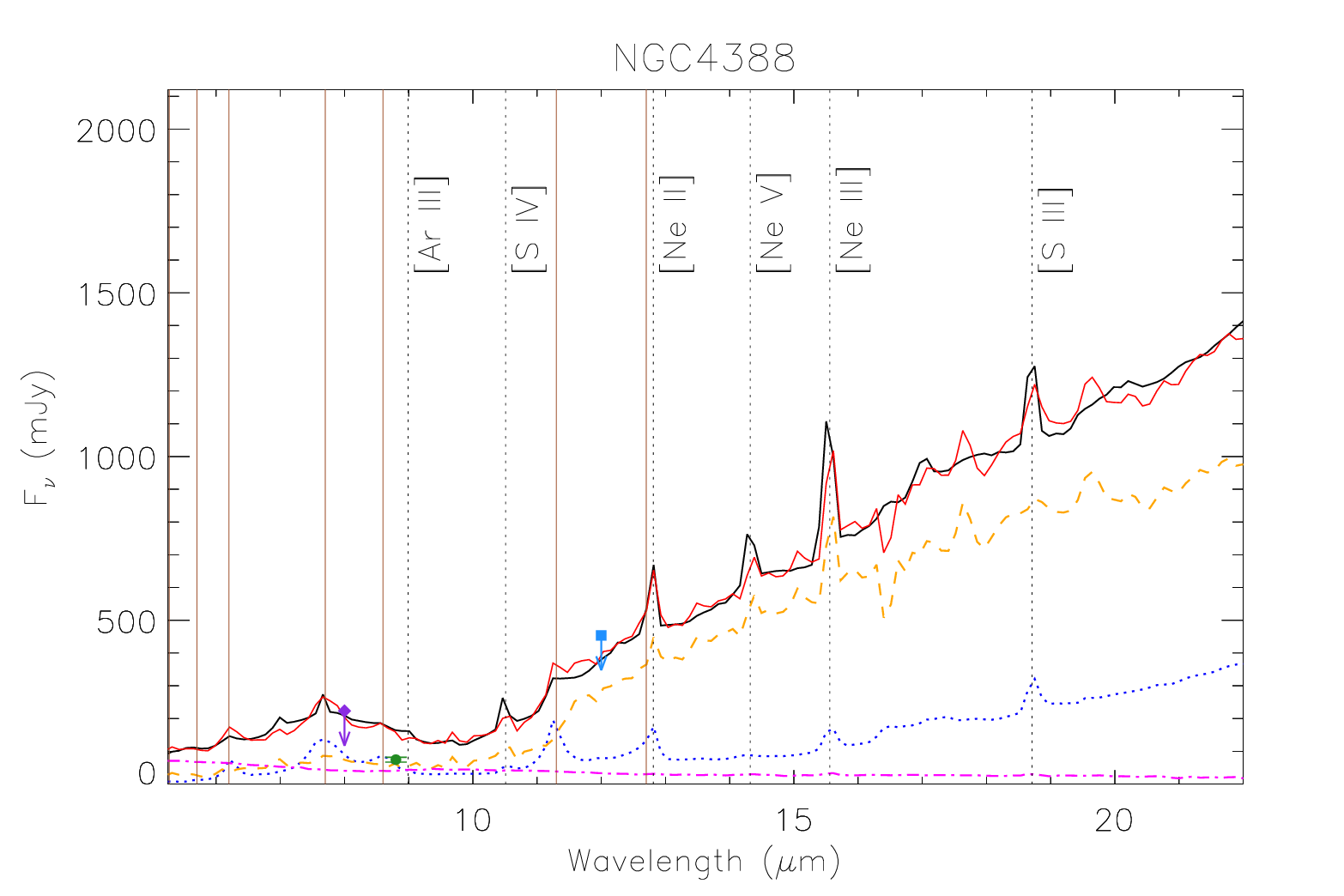}
\includegraphics[width=8.0cm]{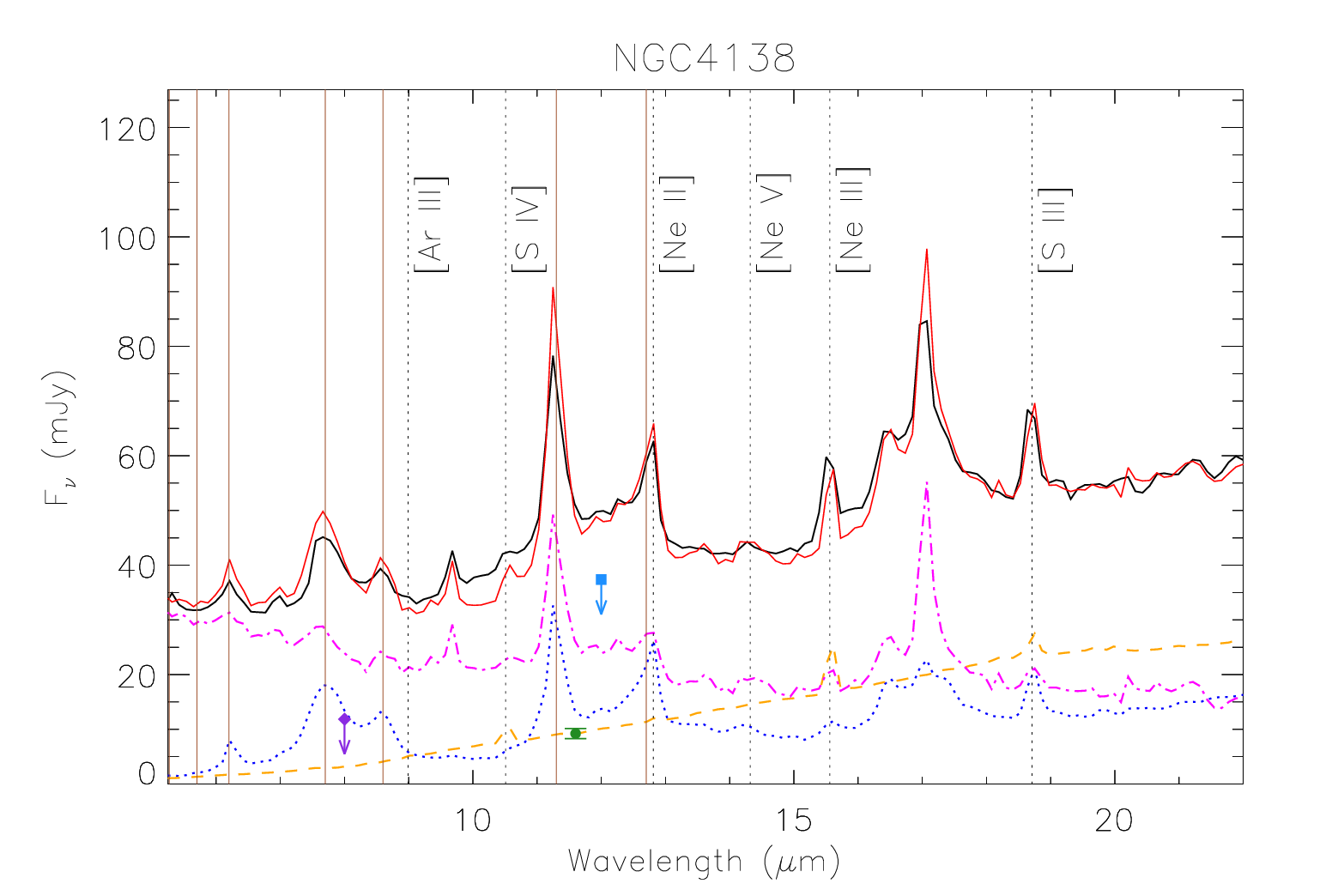}
\includegraphics[width=8.0cm]{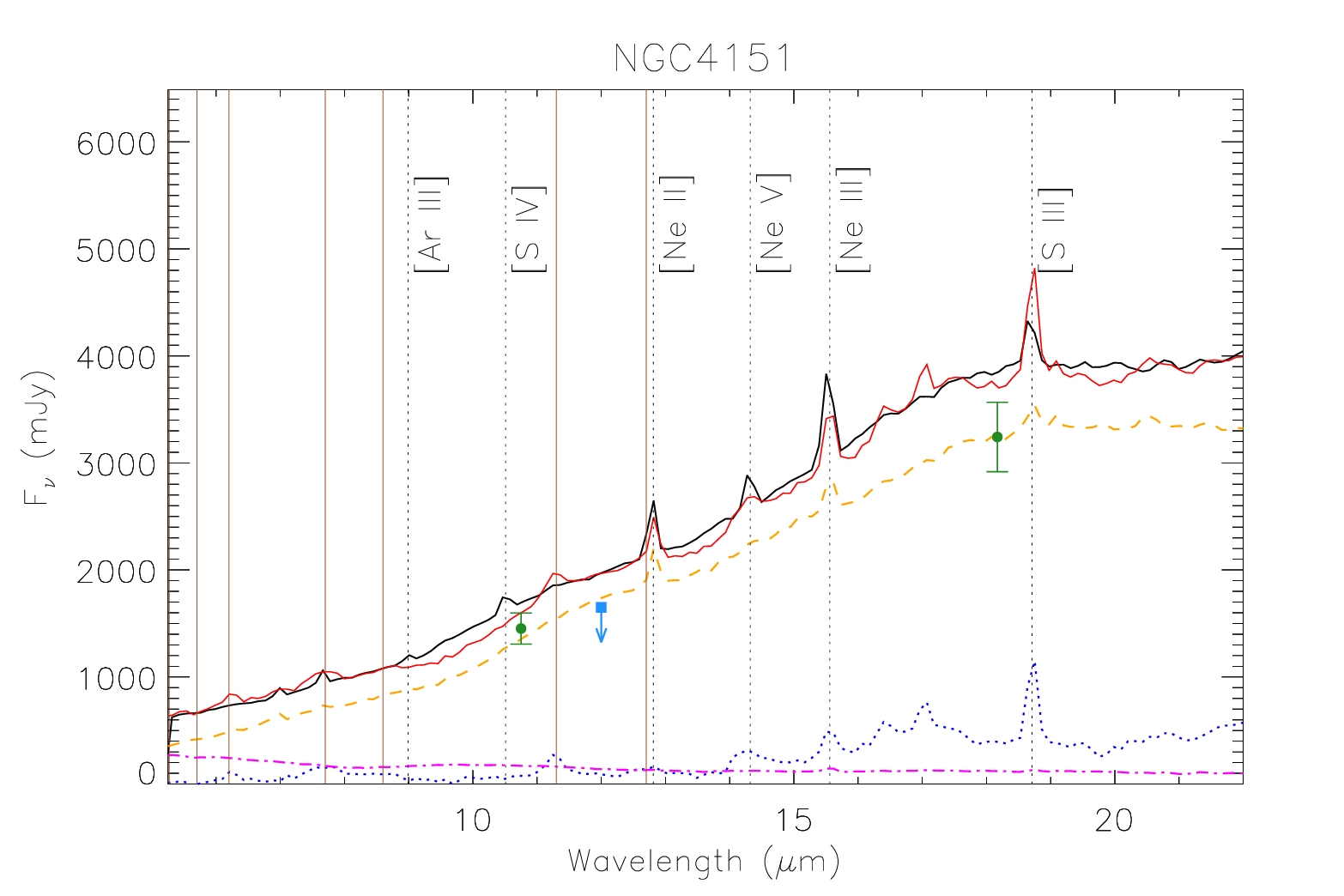}
\par} 
\caption{Spectral decomposition of the {\textit{Spitzer/IRS}} spectra of BCS$_{40}$ sample. We show the {\textit{Spitzer/IRS}} rest-frame spectra (black solid lines), best fits (red solid lines), AGN component (dashed orange lines), PAH component (dotted blue lines) and stellar component (dot-dashed magenta lines). Green circles are the high angular resolution nuclear fluxes used as priors in the fits. Purple diamonds are the arcsecond resolution 8$\mu$m {\textit{IRAC}} nuclear fluxes (as upper limit) and blue squares are the arcsecond resolution 12$\mu$m {\textit{WISE}} fluxes. The brown vertical solid lines correspond to the most important PAH features and the black vertical dotted lines are the main MIR emission lines.}
\label{figB1}
\end{figure*}

\begin{figure*}
\contcaption
\centering
\par{
\includegraphics[width=8.0cm]{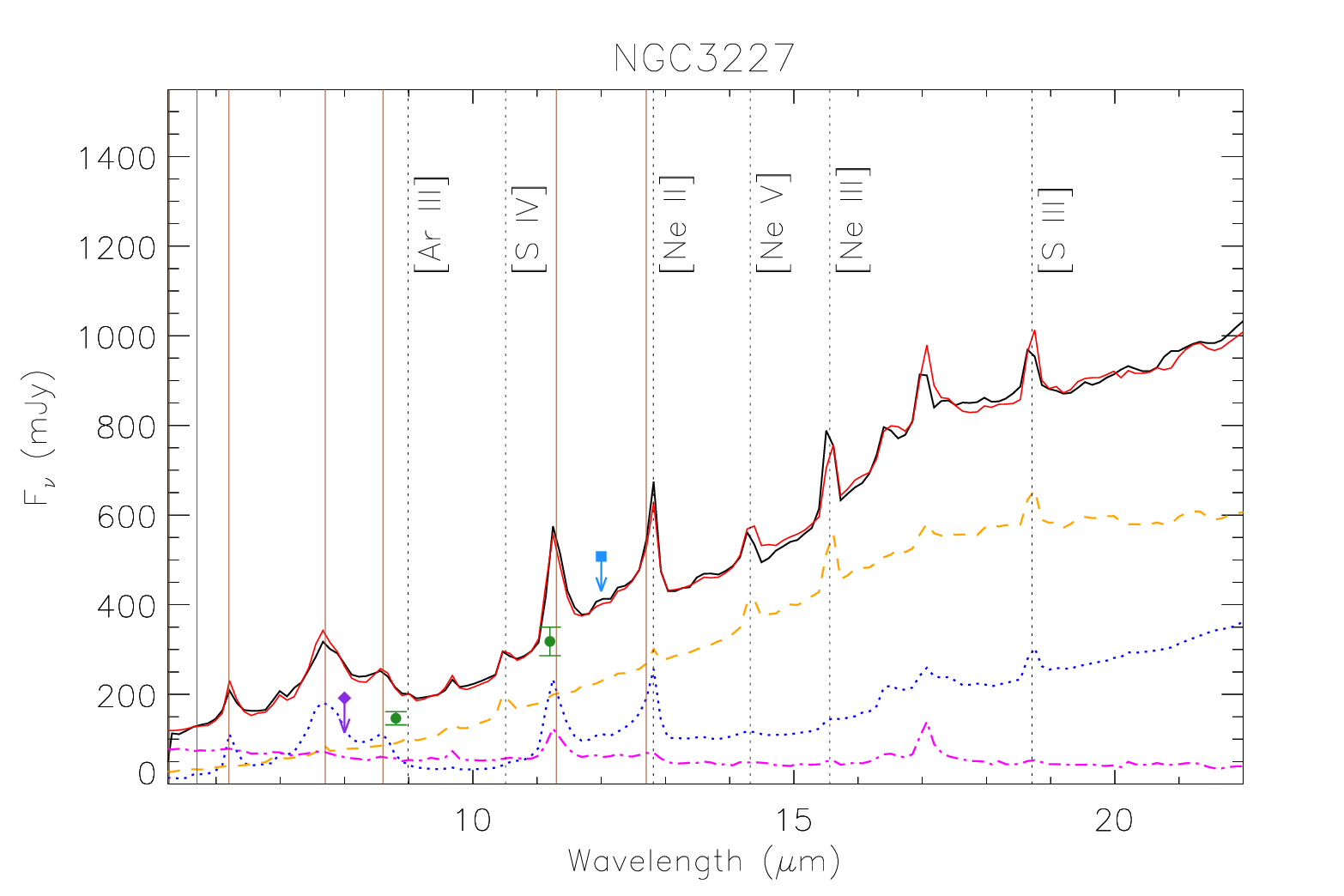}
\includegraphics[width=8.0cm]{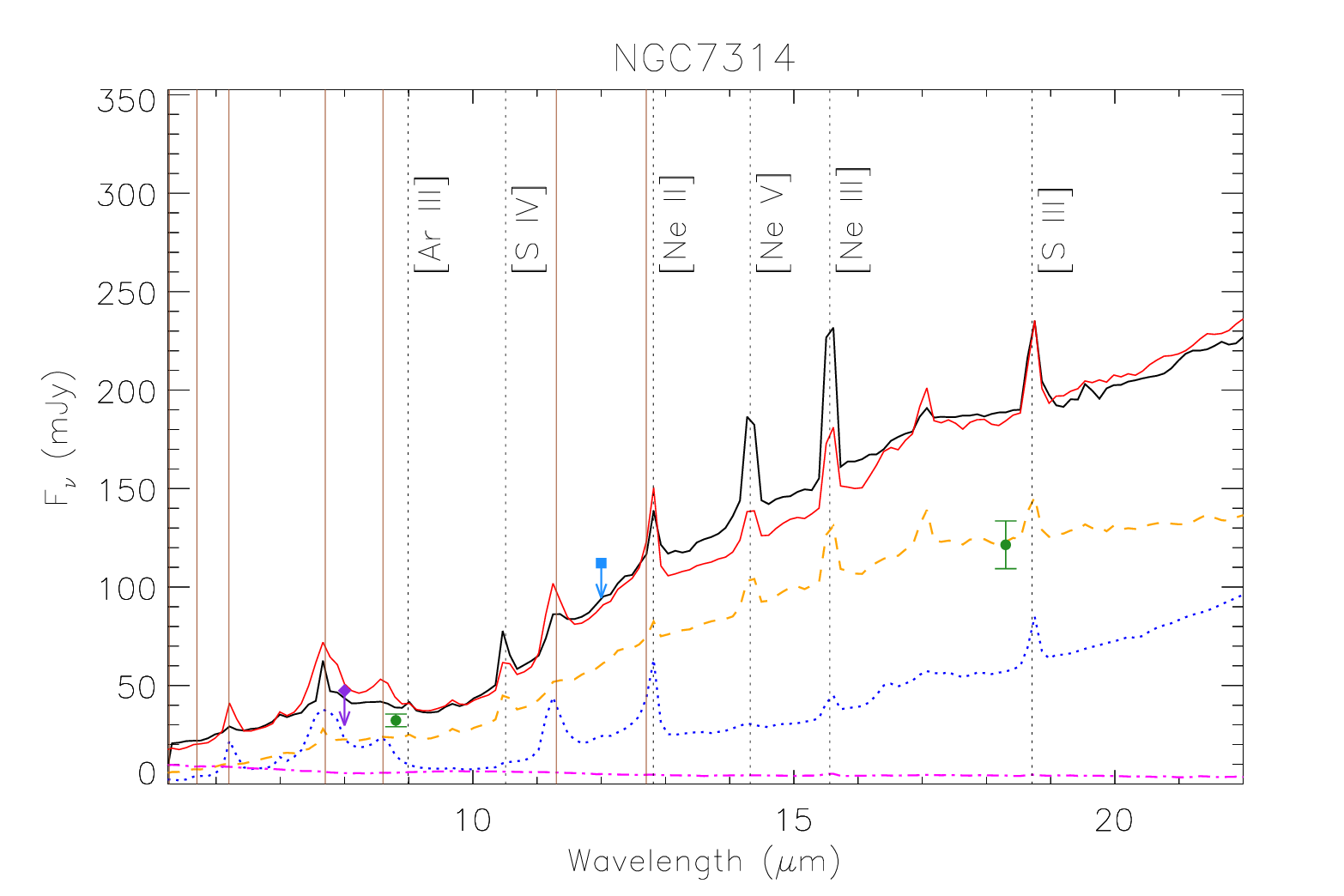}
\includegraphics[width=8.0cm]{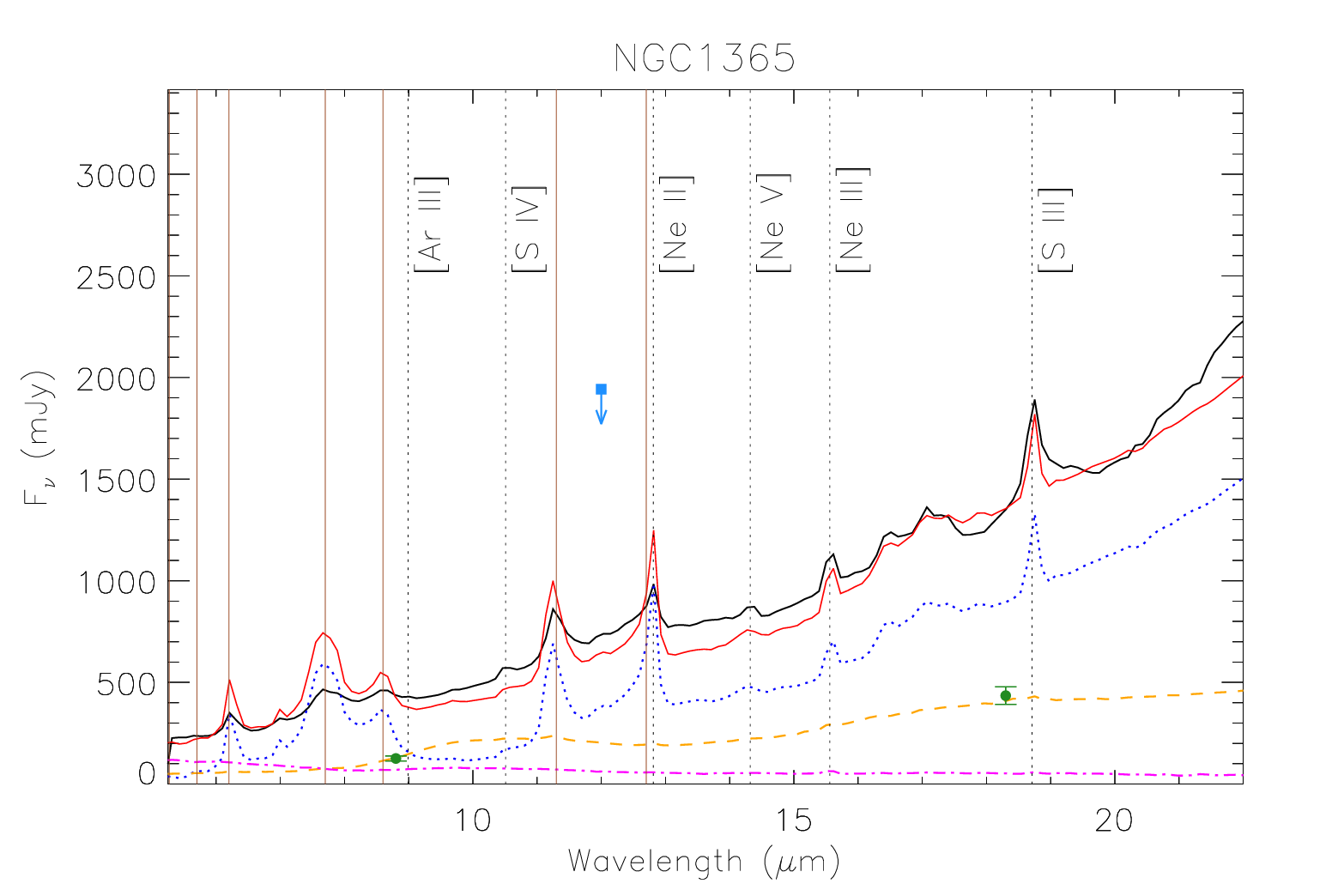}
\includegraphics[width=8.0cm]{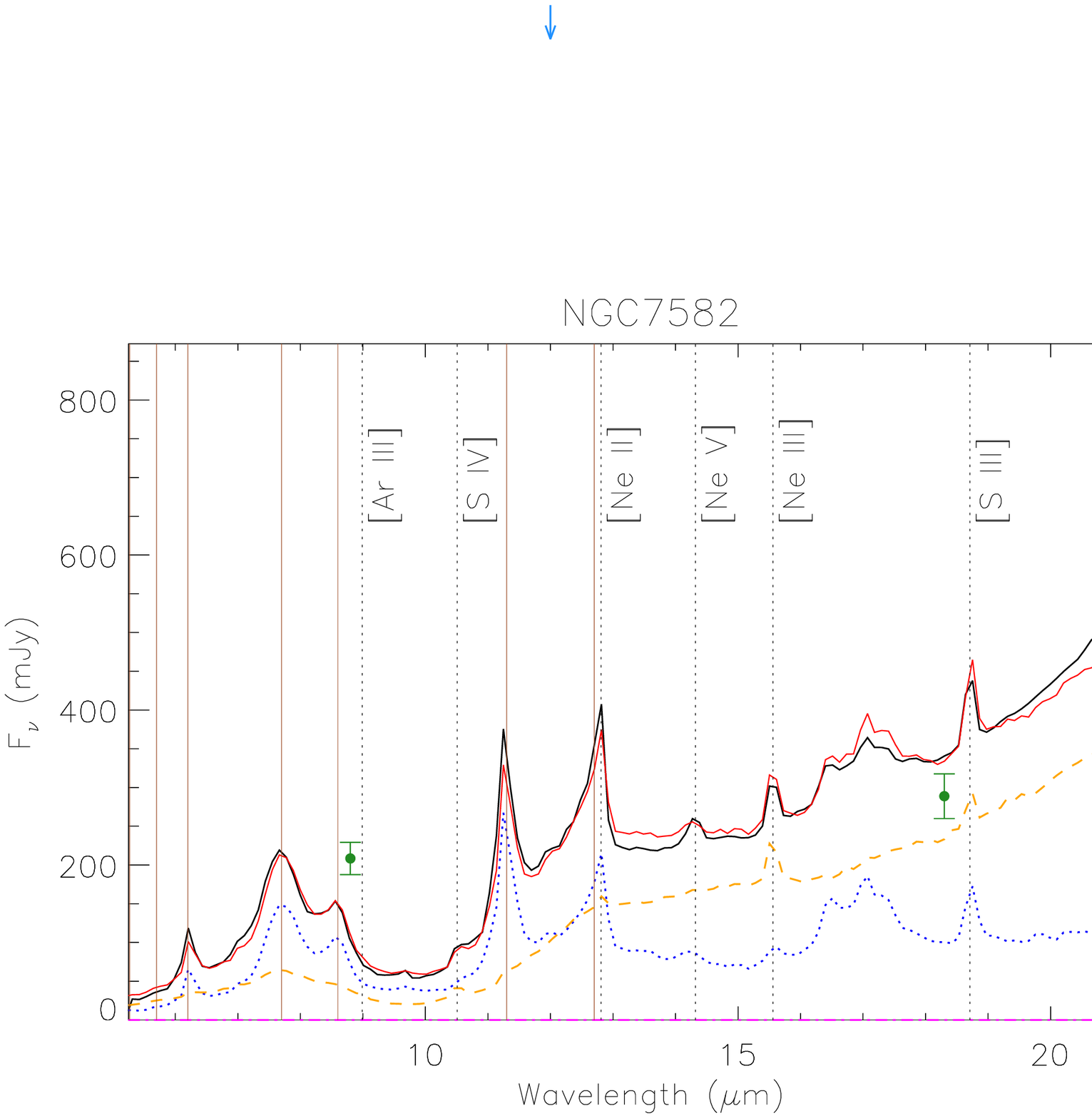}
\includegraphics[width=8.0cm]{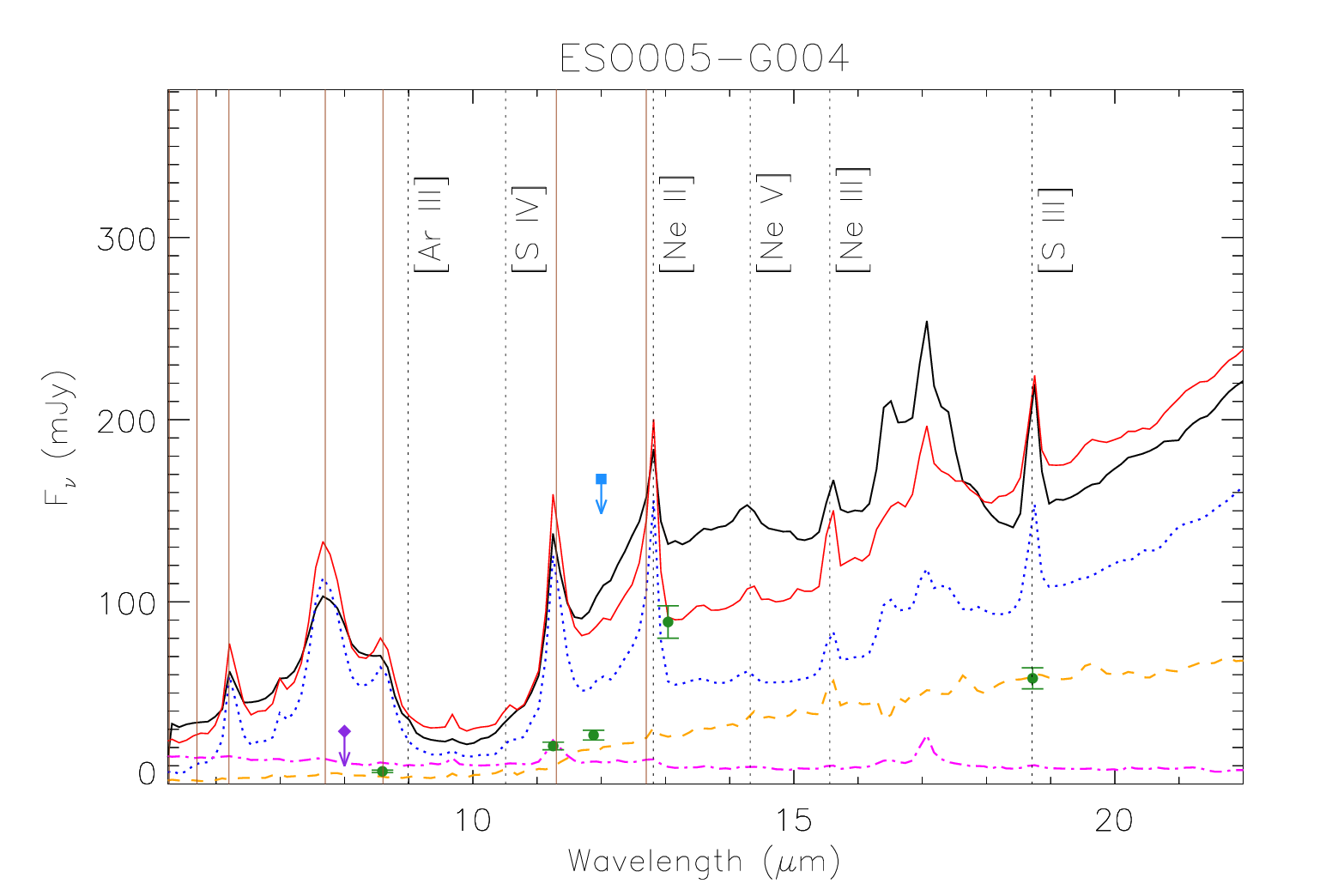}
\includegraphics[width=8.0cm]{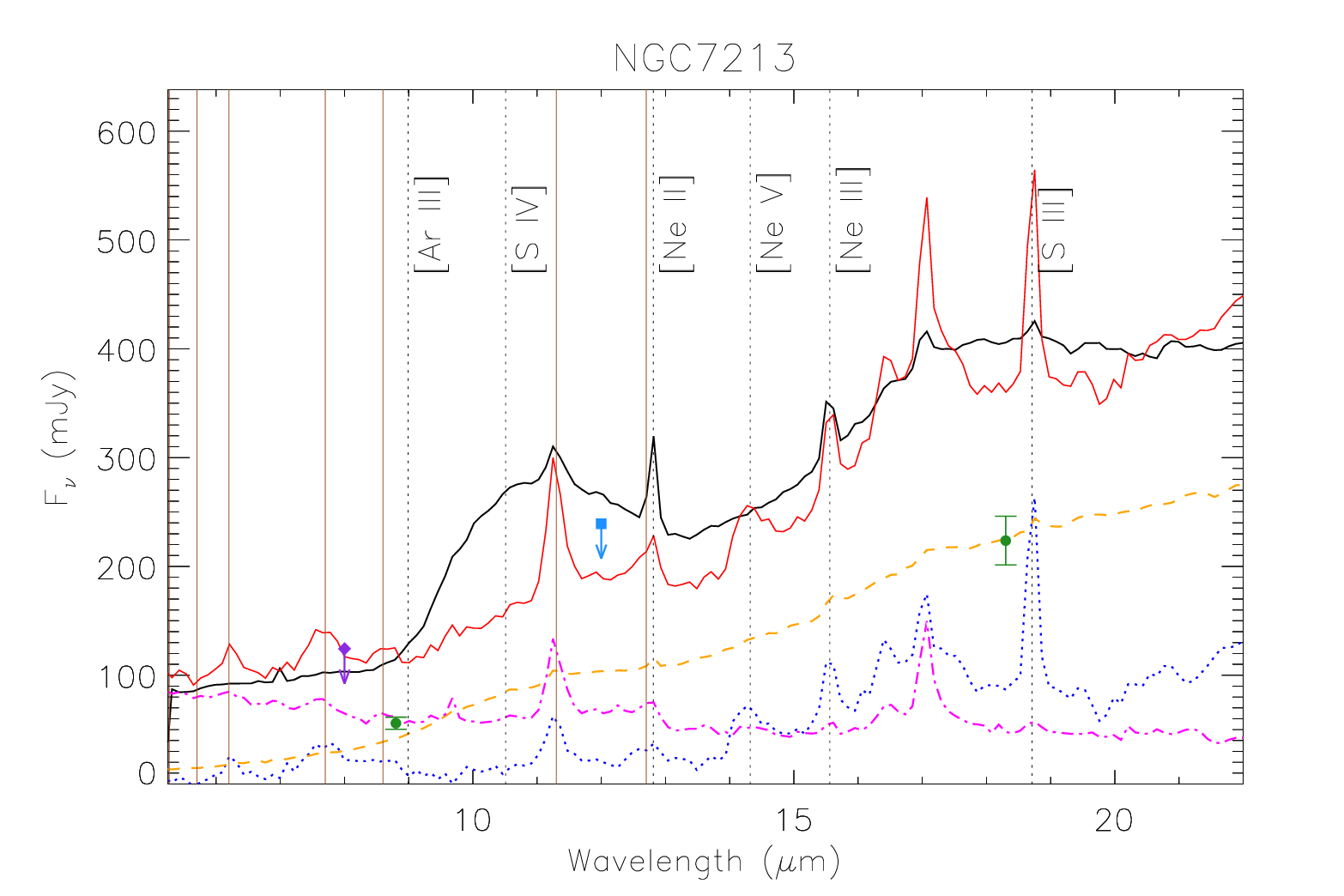}
\includegraphics[width=8.0cm]{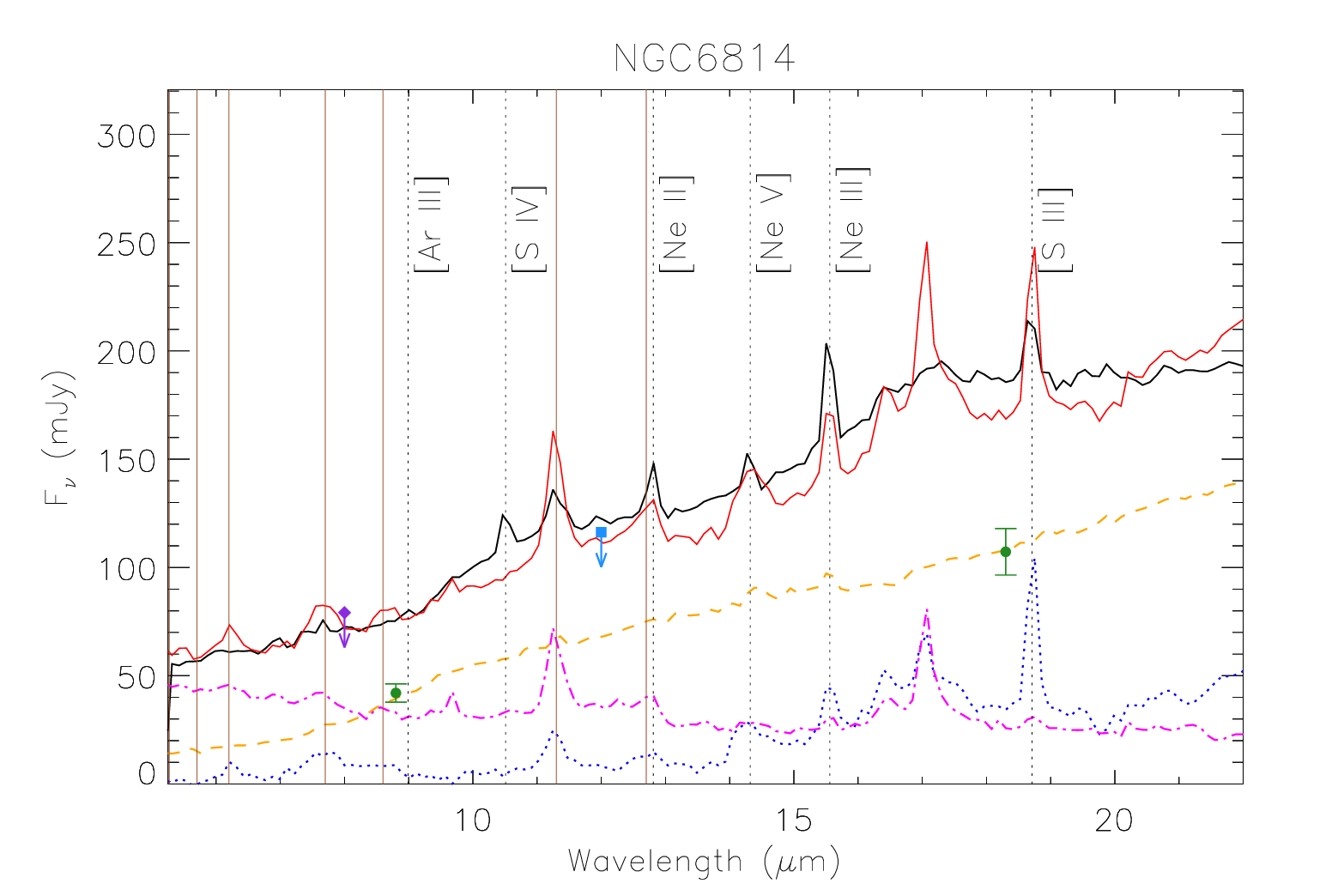}
\includegraphics[width=8.0cm]{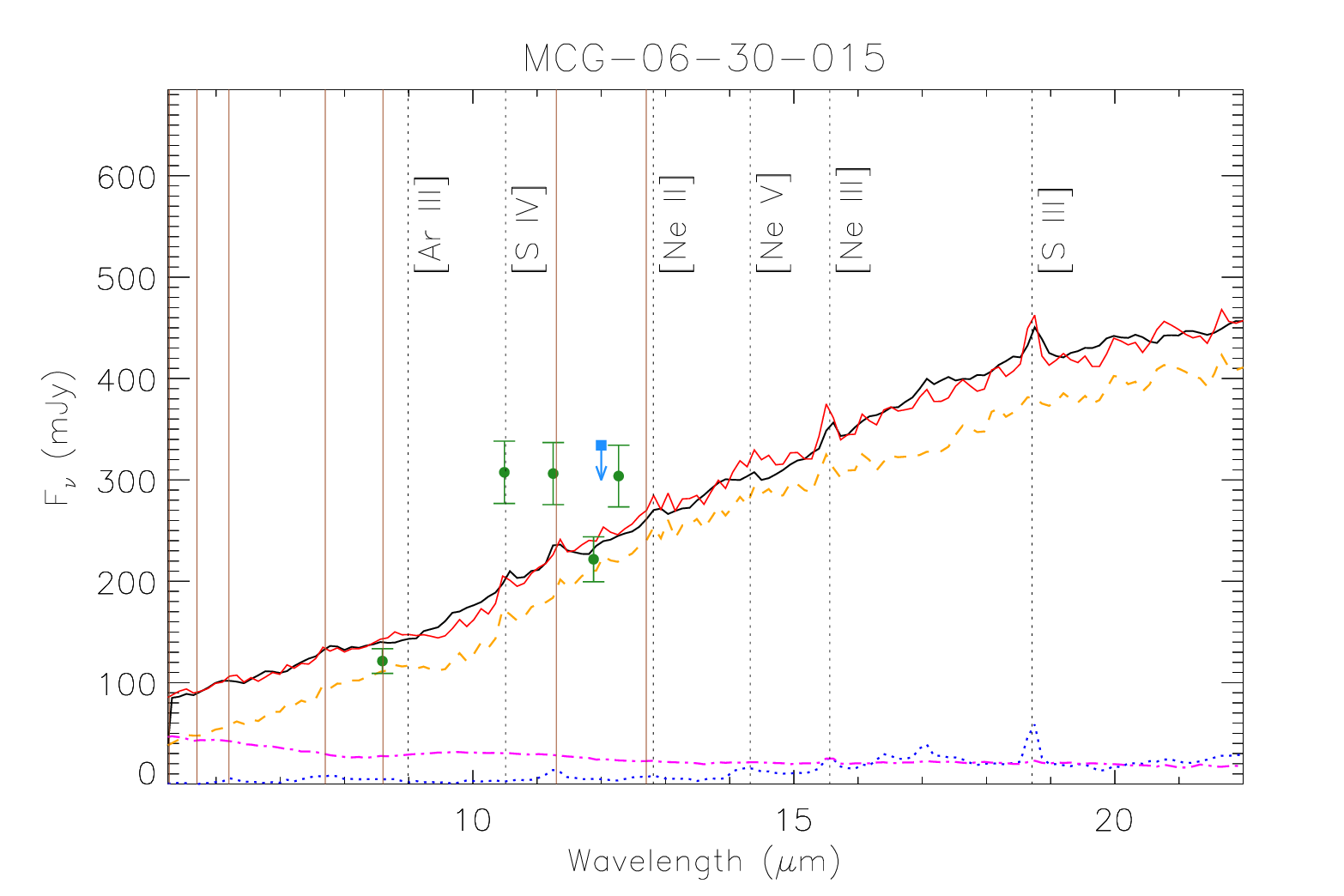}
\par} 
\end{figure*}

\begin{figure*}
\contcaption
\centering
\par{
\includegraphics[width=8.0cm]{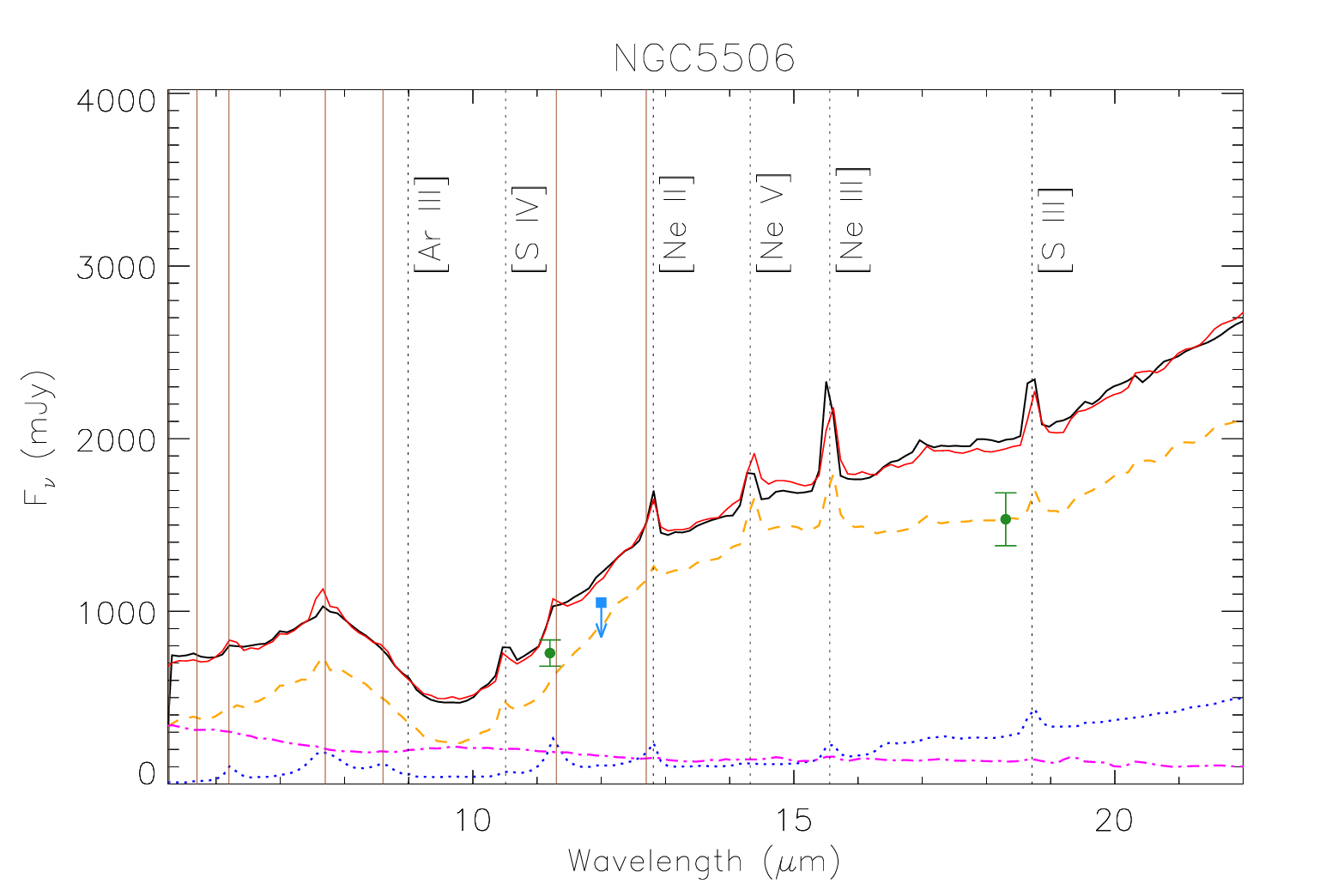}
\includegraphics[width=8.0cm]{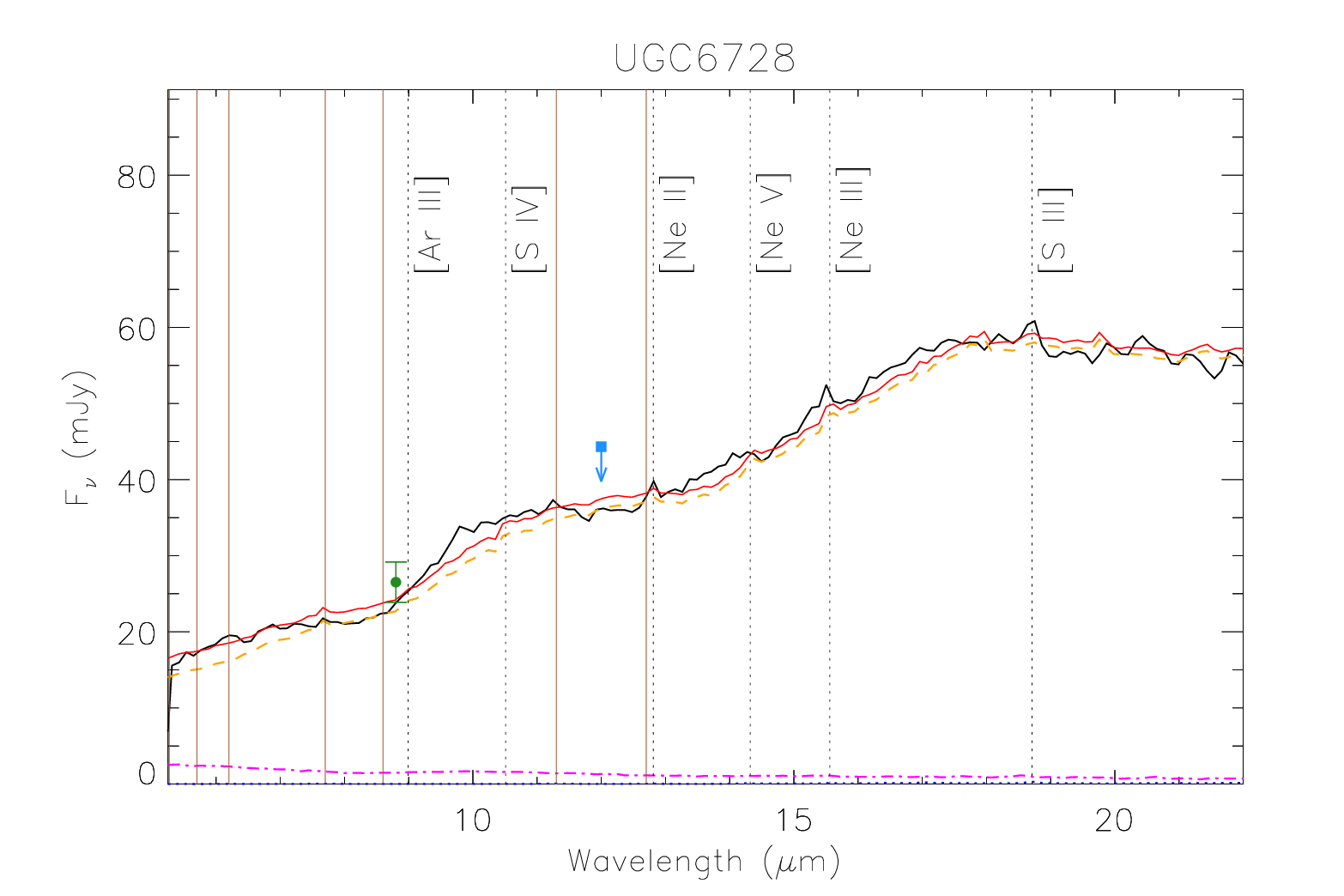}
\includegraphics[width=8.0cm]{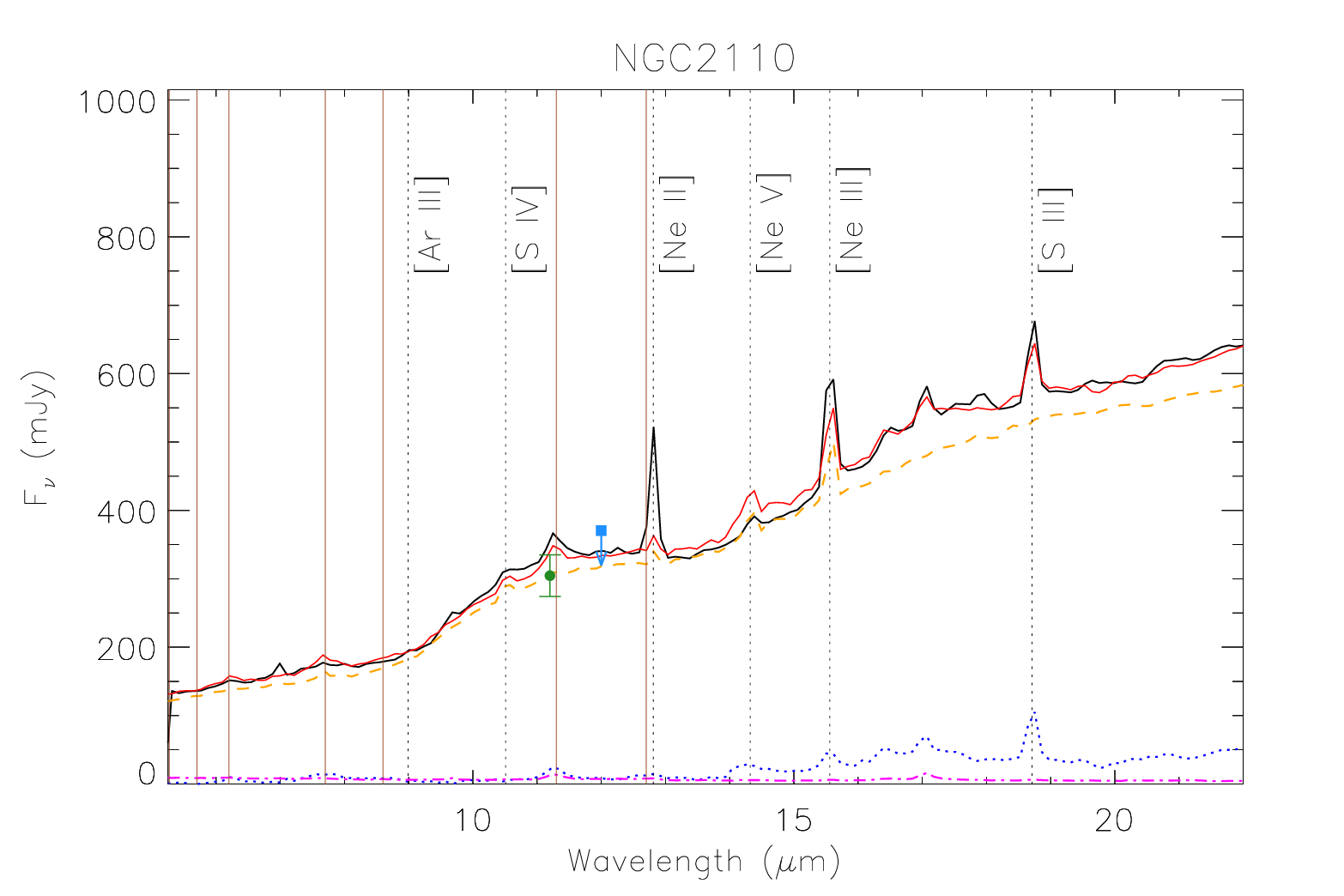}
\includegraphics[width=8.0cm]{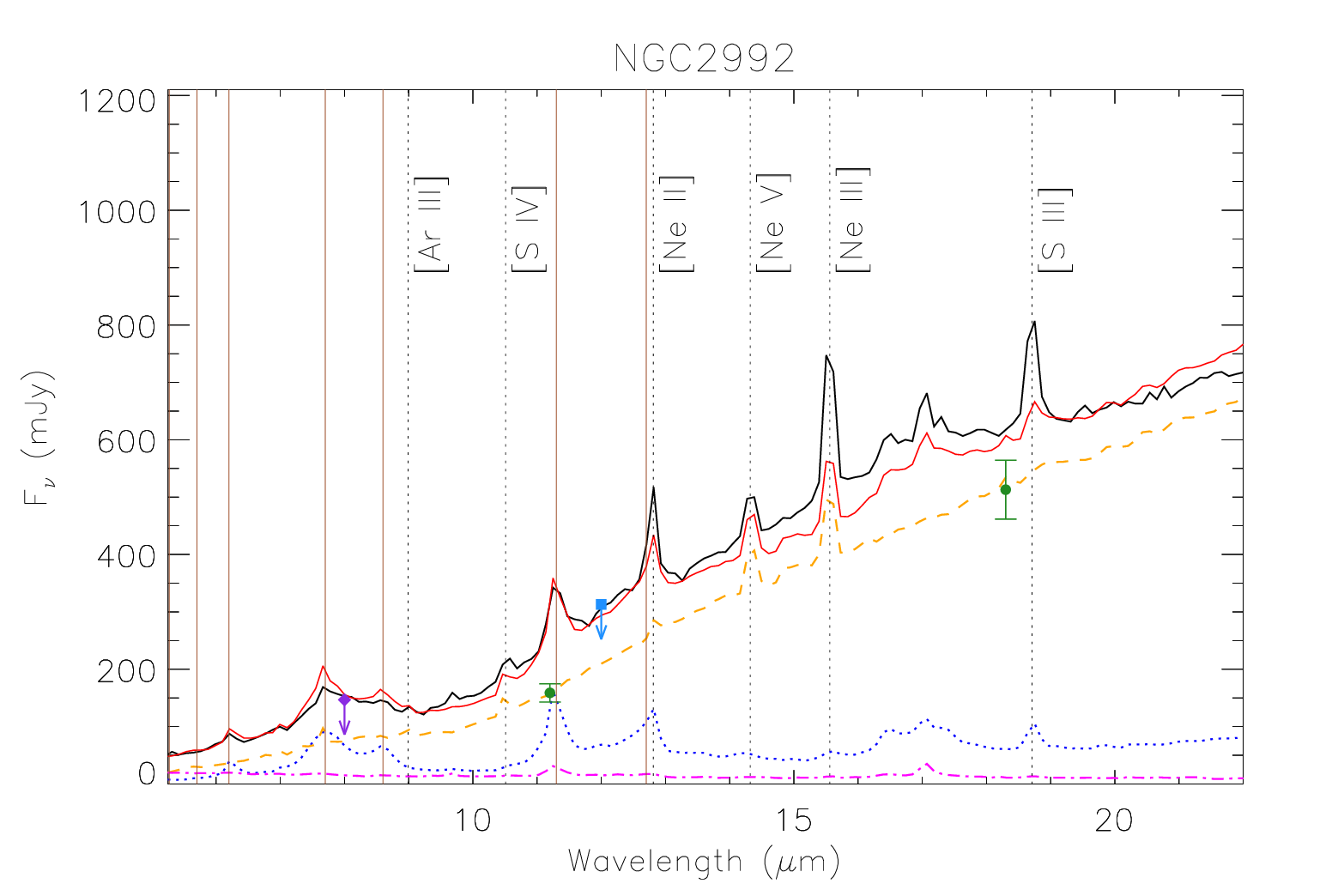}
\includegraphics[width=8.0cm]{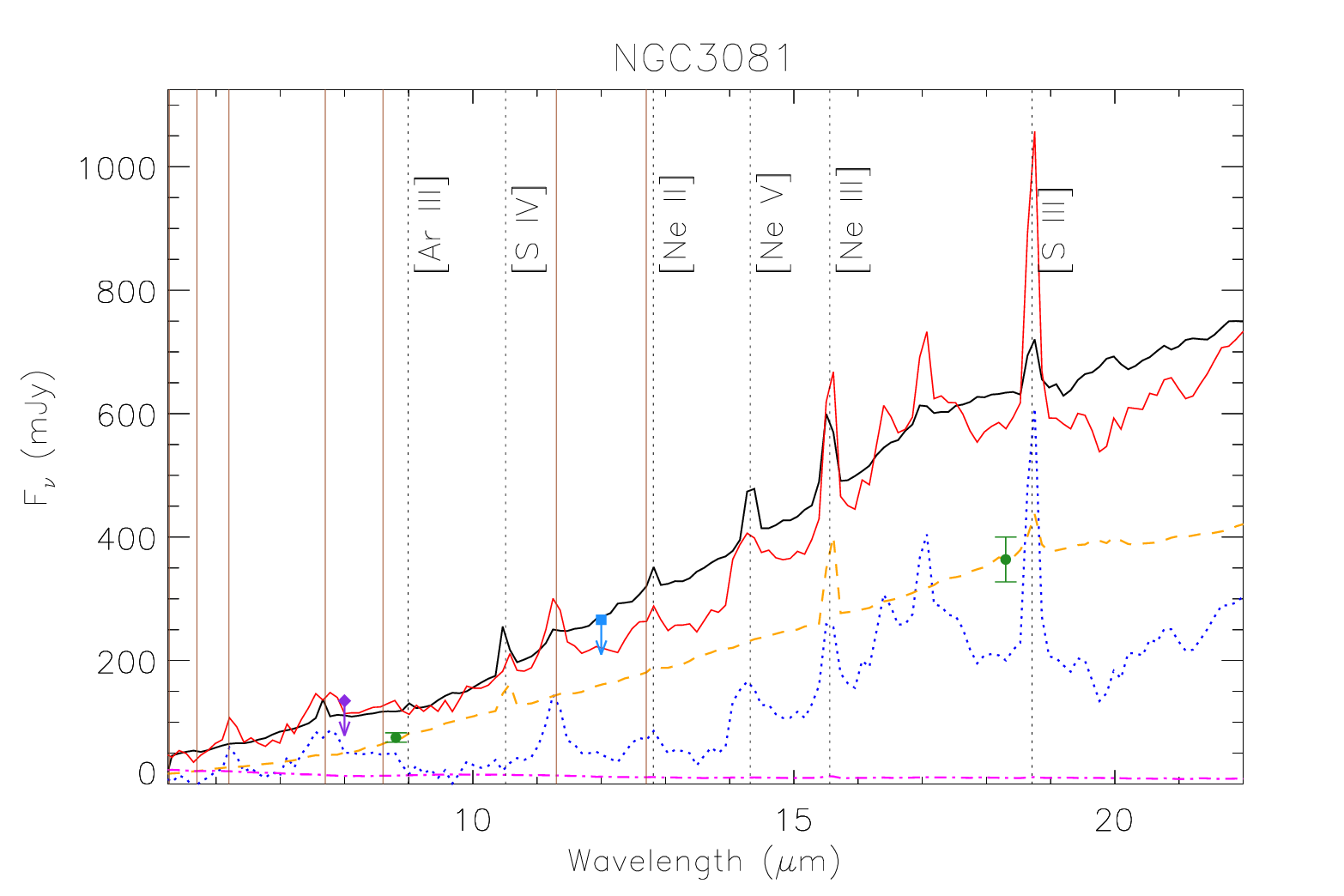}
\includegraphics[width=8.0cm]{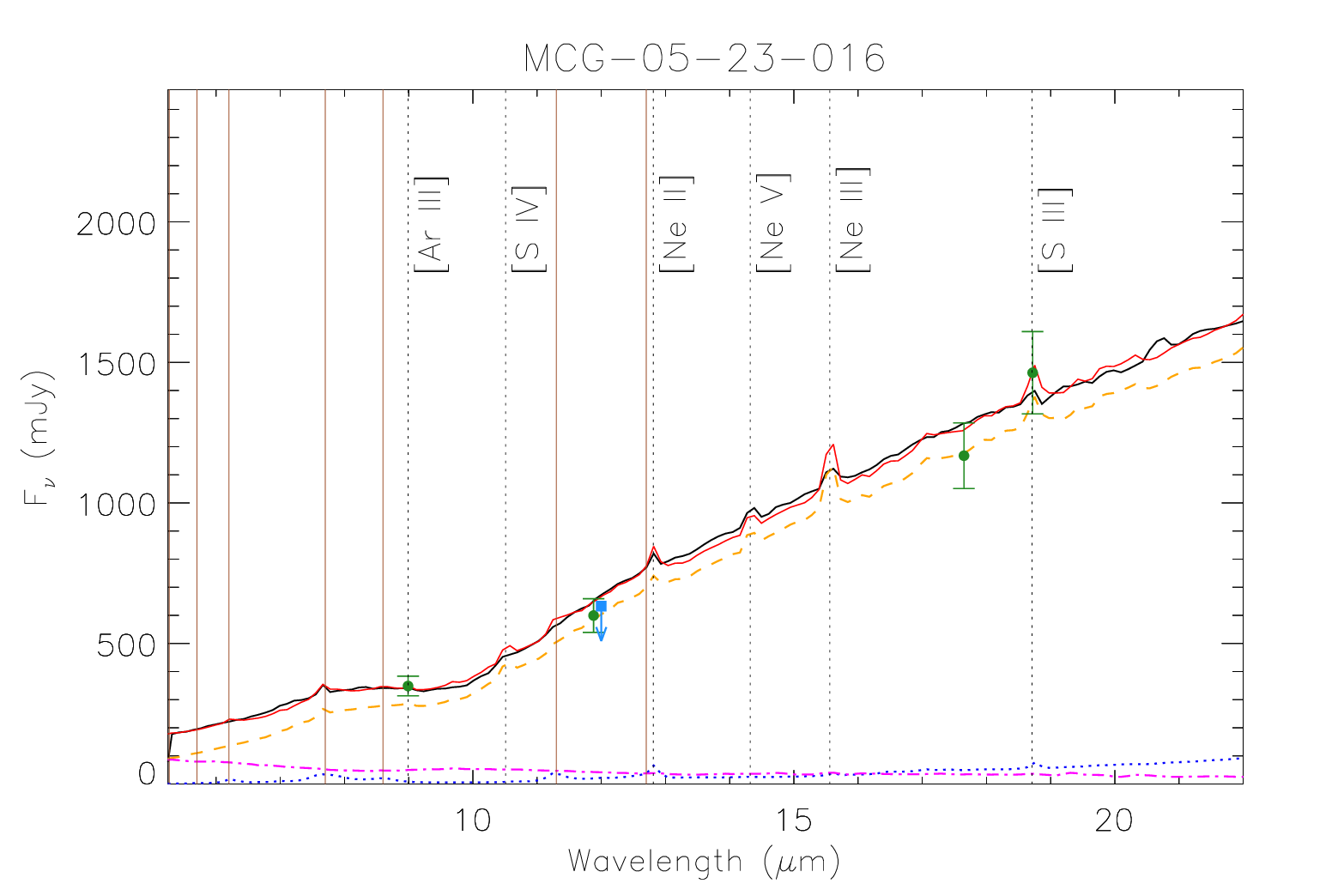}
\includegraphics[width=8.0cm]{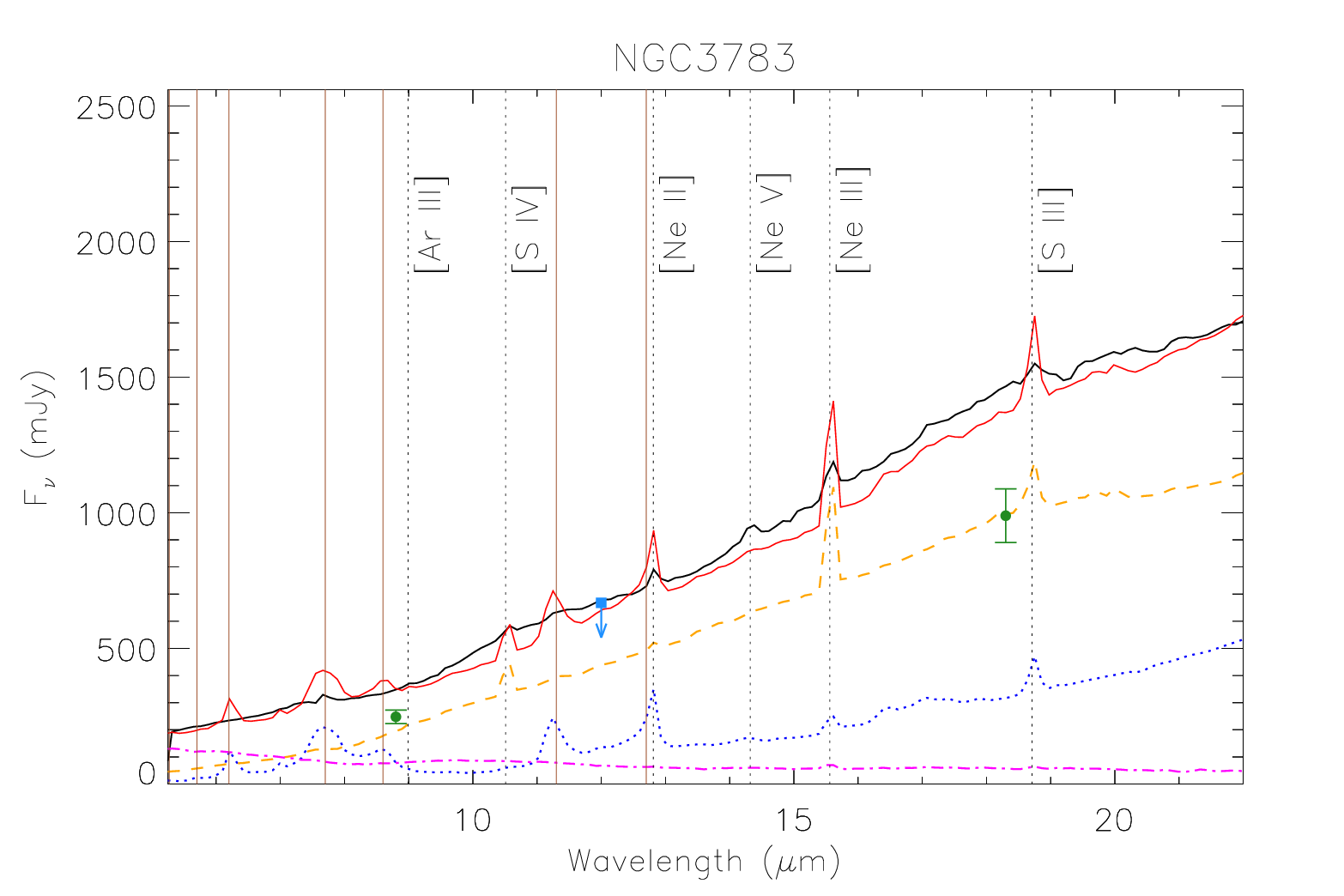}
\includegraphics[width=8.0cm]{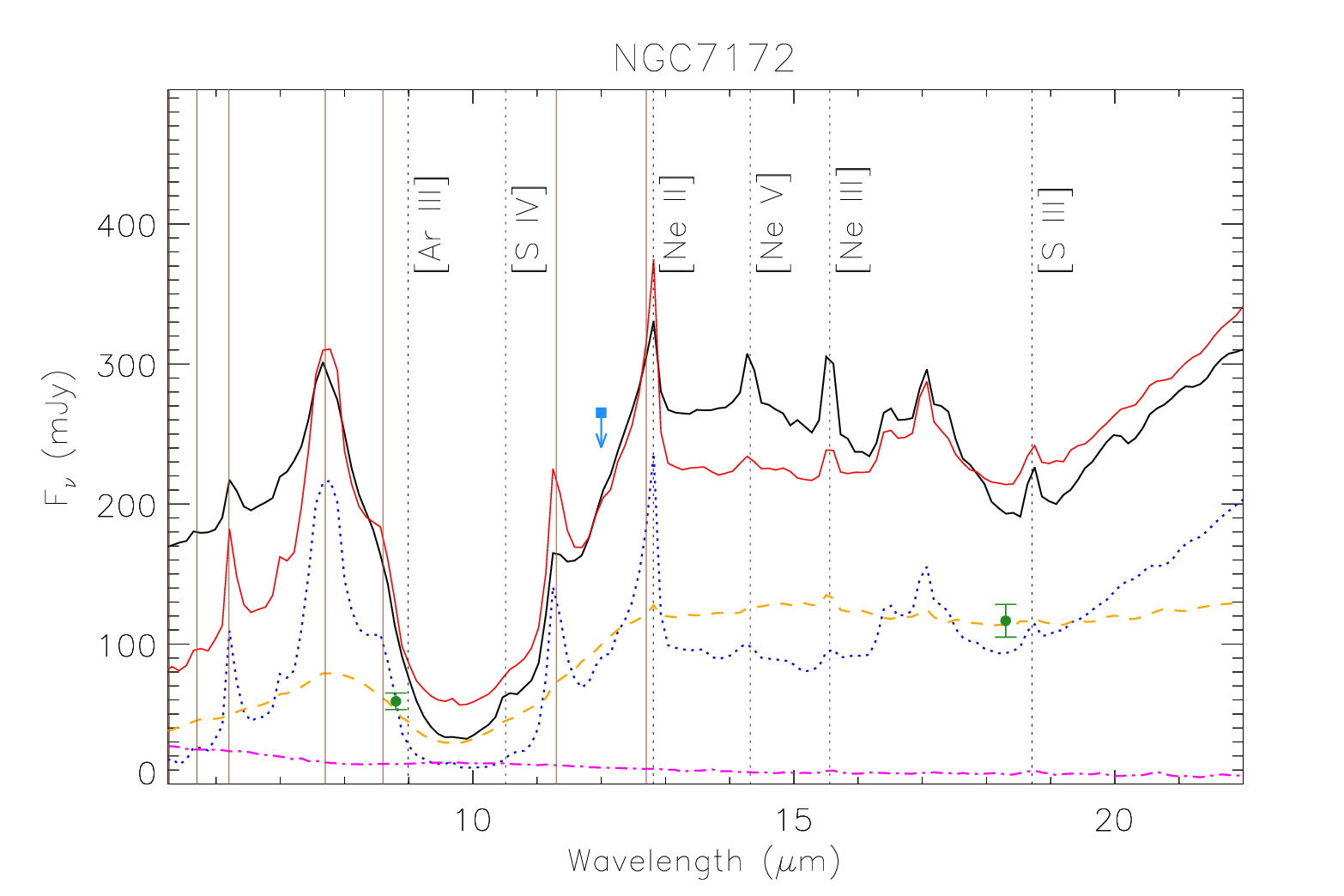}
\par} 
\end{figure*}

\section{Notes on individual objects}
\label{c}

Below, we comment on the possible origin of the heating source of the extended emission for the six objects with extended MIR morphologies at large-scales ($>$400 pc, which is the sample average value) in the high angular resolution data presented in Section \ref{extended}. To do so, we first use the arcsecond and subarcsecond resolution images to compare the extended MIR emission structures of these galaxies (see Fig. \ref{fig7} of Section \ref{extended}). Then, we investigate the origin of the MIR emission using the {\textit{Spitzer/IRS}} spectra to look for AGN and/or SF activity indicators.

We find that for four of these six galaxies the bulk of the extended MIR emission is mainly related to SF. On the other hand, for NGC\,3081 and NGC\,5506, the {\textit{Spitzer/IRS}} spectra show faint PAH features, weak [Ne\,II] emission line and prominent AGN tracers. We also found that the majority of these galaxies have a relatively small contribution of AGN emission ($<$50\%) to the {\textit{Spitzer/IRS}} spectra, except for NGC\,2992, NGC\,3081 and NGC\,5506, which have AGN contributions of 71\%, 63\% and 72\%, respectively (see Appendix \ref{B}).

{\bf{NGC\,4945}} is a practically edge-on spiral galaxy with a highly obscured nucleus containing both an AGN and a starburst. This galaxy presents a bright cluster close to the central region ($\sim$50~pc) and extended emission  along PA$\sim$45 deg out to $\sim$500~pc in the N-band image, being weaker in the Q-band. NGC\,4945 shows the most prominent 6.2~$\mu$m PAH feature of the sample and it also presents the 11.3~$\mu$m PAH feature and strong [Ne\,II] emission. On the other hand, there is no significant [S\,IV] and [Ne\,V] emission. Therefore, the bulk of the extended MIR emission of this galaxy is mainly SF activity. See \citet{Imanishi11} for a detailed study of this galaxy.\\

{\bf{NGC\,1365}} is a barred spiral galaxy which presents bright clusters around the nucleus in the N- and Q-band, which are also visible in the {\textit{IRAC}} image. There is also faint extended emission towards the south (PA$\sim$210 deg; $\sim$600~pc from the nucleus) of the N- and Q-band images, which becomes weaker in the Q-band and matches that of the {\textit{IRAC}} image. The {\textit{Spitzer/IRS}} spectrum shows strong 6.2 and 11.3~$\mu$m PAH features as well as [Ne\,II] and [Ne\,III]. This galaxy also presents weak AGN tracers such as [S\,IV] or [Ne\,V]. Therefore, we conclude that the bulk of this MIR emission is due to SF activity. See \citet{Herrero12} for a detailed study of this galaxy.\\

{\bf{NGC\,7582}} is a highly inclined barred spiral galaxy. This galaxy has an obscured AGN surrounded by a star-forming disk and a dust lane crossing over the nucleus. The subarcsecond resolution Q-band image reveals a bright cluster towards the south, at $\sim$200~pc from the nucleus, which is not present in the N-band image. There is also faint emission extending out to $\sim$630~pc from the southeast to the northwest (PA$\sim$155 deg). The {\textit{Spitzer/IRS}} spectrum shows important 6.2 and 11.3~$\mu$m PAH features and also strong [Ne\,II] emission. On the contrary, the spectrum shows really weak AGN indicators such as [S\,IV] and [Ne\,V]. Therefore, the bulk of this MIR emission is likely related with SF activity. See \citet{Wold06} and \citet{WoldGaliano06} for a detailed study of this galaxy.\\

{\bf{NGC\,5506}} is a practically edge-on spiral galaxy which shows extended MIR emission around the nucleus in the N- and Q-band images. The high angular resolution N-band image reveals faint extended emission from north to south extending out to $\sim$560~pc. However, there is also faint extended emission towards the east in the Q-band image. The {\textit{Spitzer/IRS}} spectrum of this galaxy shows strong [Ne\,III] and [Ne\,V] features and faint PAH features. Therefore, we conclude that the origin of this extended MIR emission is likely nuclear activity. See \citet{Roche07} for a detailed study of this galaxy.\\

{\bf{NGC\,2992}} is an spiral galaxy which is part of the interacting system Arp\,245. The high angular resolution N-band image reveals faint extended emission along PA$\sim$30 deg and out to $\sim$2~kpc, which is not present in the Q-band. This extended emission is also present in the 8~$\mu$m {\textit{IRAC}} image. Considering the strong 6.2 and 11.3~$\mu$m PAH features seen in the {\textit{Spitzer/IRS}} spectrum, the bulk of this extended emission is likely produced by dust heated by SF activity. See \citet{Bernete2015} for a detailed study of this galaxy.\\

{\bf{NGC\,3081}} is a barred spiral galaxy which presents extended emission in the N- and Q-band images, being brighter in the Q-band. The emission is stronger towards the north, extending out to $\sim$450~pc from the southeast to the northwest (PA$\sim$160 deg). The {\textit{Spitzer/IRS}} spectrum of this galaxy shows faint 6.2 and 11.3~$\mu$m PAH emission, weak [Ne\,II] line emission and strong [Ne\,III], [S\,IV] and [Ne\,V] features. Therefore, based on the lack of PAH features in the {\textit{Spitzer/IRS}} spectrum, we can conclude that the MIR emission of NGC\,3081 is mainly produced by nuclear activity. See \citet{Ramos11c} for a detailed study of this galaxy.

\section{Flux-flux correlations}
\label{D}

The luminosity-luminosity correlations might be caused, at least in part, by distance effects. Therefore, we also checked the correlations in flux-flux space. As expected, the correlations using the luminosities are stronger than the flux-flux correlations. However, they are still significant and we confirmed that the results presented in Section \ref{correlations} hold in flux-flux space (see Table \ref{tabD1}).

\input{tabD1.tex}

\label{lastpage}

\end{document}

%% file: tab1.tex
\begin{table*}
\centering
\begin{tabular}{lccccccc}
\hline
Name &	R.A.&	Dec.& Redshift	&Luminosity &Spatial&Seyfert&b/a\\
 	&	(J2000)&	(J2000)& 	&distance &scale& type&\\
 	&				& &		&(Mpc)&(pc arcsec$^{-1}$)&\\

\hline
NGC\,4395		&	12h25m48.8s&	+33d32m49s&	0.0009	&3.84&19&1.8	&0.83	\\
NGC\,5128 (CenA)	&	13h25m27.6s&	-43d01m09s&	0.0010	&4.28&21&2.0	&0.78	\\
NGC\,4945		&	13h05m27.5s&	-49d28m06s&	0.0011	&4.36&21&2.0	&0.19	\\
NGC\,4051		&	12h03m09.6s&	+44d31m53s&	0.0031	&12.9&62&1.2	&0.75	\\
NGC\,6300		&	17h16m59.5s&	-62d49m14s&	0.0034 	&14.0&68&2.0	&0.67	\\
NGC\,4388		&	12h25m46.7s&	+12d39m44s&	0.0101 	&17.0*&82&2.0	&0.19	\\
NGC\,4138		&	12h09m29.8s&	+43d41m07s&	0.0043	&17.7&85&1.9	&0.65	\\
NGC\,4151		&	12h10m32.6s&	+39d24m21s&	0.0049	&20.0&96&1.5	&0.71	\\
NGC\,3227		&	10h23m30.6s&	+19d51m54s&	0.0050	&20.4&98&1.5	&0.67	\\
NGC\,7314		&	22h35m46.2s&	-26d03m02s&	0.0051 	&20.9&100&1.9	&0.46	\\
NGC\,1365		&	03h33m36.4s&	-36d08m25s&	0.0052	&21.5&103&1.8	&0.55	\\
NGC\,7582		&	23h18m23.5s&	-42d22m14s&	0.0054 	&22.1&106&2.0	&0.42	\\
ESO\,005-G004		&	06h05m41.6s&	-86d37m55s&	0.0058	&24.1&116&2.0	&0.21	\\
NGC\,7213		&	22h09m16.3s&	-47d10m00s&	0.0061	&25.1&120&1.5	&0.90	\\
NGC\,6814		&	19h42n40.6s&	-10d19m25s&	0.0062 	&25.8&123&1.5	&0.93	\\
MCG-06-30-015		&	13h35m53.7s&	-34d17m44s&	0.0065	&26.8&128&1.2	&0.60	\\
NGC\,5506		&	14h13m14.9s&	-03d12m27s&	0.0073 	&30.1&144&1.9	&0.24	\\
UGC\,6728		&	11h45m16.0s&	+79d40m53s&	0.0078	&32.1&153&1.2	&0.63	\\
NGC\,2110		&	05h52m11.4s&	-07d27m22s&	0.0078	&32.4&155&2.0	&0.76	\\
NGC\,2992		&	09h45m42.0s&	-14d19m35s&	0.0083	&34.4&164&1.9	&0.31	\\
NGC\,3081		&	09h59m29.5s&	-22d49m35s&	0.0083	&34.5&164&2.0	&0.76	\\
MCG-05-23-016		&	09h47m40.1s&	-30h56m55s&	0.0087	&35.8&171&2.0	&0.45	\\
NGC\,3783		&	11h39m01.7s&	-37d44m19s&	0.0088	&36.4&173&1.2	&0.89	\\
NGC\,7172		&	22h02m01.9s&	-31d52m11s&	0.0092	&37.9&180&2.0	&0.56	\\
%IRAS07145-2914&07h16m31.2s&	-29d19m29s&0.005646	&23.3&112&Sy 2.0\\

\hline
\end{tabular}						 
\caption{BCS$_{40}$ sample sorted by luminosity distance. Right ascension (R.A.), declination (Dec.), Seyfert type and galaxy inclination (b/a) were taken from the NASA/IPAC Extragalactic Database (NED). The redshift, D$_{L}$ and spatial scale were calculated using a cosmology with H$_0$=73 km~s$^{-1}$~Mpc$^{-1}$, $\Omega_m$=0.27, $\Omega_{\Lambda}$=0.73 and velocity-field corrected using the \citet{Mould00} model, which includes the influence of the Virgo cluster, the Great Attractor, and the Shapley supercluster. *This galaxy is part of the Virgo Cluster \citep{Binggeli85}.}
\label{tab1}
\end{table*}

%% file: tab2.tex
\begin{landscape}
\begin{table}
\scriptsize 
\centering
\begin{tabular}{lcccccccccccccc}
\hline
Name		&Wavel. & Filter&Observation	& Telescope /& Exposure&	FWHM&	e$_{gal}$&	Stand.&	FWHM&	e$_{PSF}$&Resol.&Nuclear&Best&Total\\
			&$\lambda_{c}$/$\Delta\lambda$&name&Date	& Instrument& time&	galaxy&	&	dev.&	PSF&	&Nuc.&flux&PSF &flux\\
			&($\mu$m)&	&	& & (s)&	(arcsec)&	&	$\sigma$&	(arcsec)&		&&(mJy)& \%&(mJy)\\
\hline
NGC\,4395	&12.5/1.2	&Si6&05/02/2010&Gemini/MICHELLE	&1011	&0.369&-&		0.06&	0.369	&	0.03&		B.R.	&	7&100&9\\
NGC\,5128	&8.74/0.78	&Si2&03/05/2004	&Gemini/T-ReCS	&608	&0.311&	0.07&		0.06&	0.304	&	0.05&		B.R.	&511&90&611\\
		&18.33/1.5	&Qa&03/05/2004	&Gemini/T-ReCS	&2042	&0.442&	0.13&		0.87&	0.533	&	0.13&		$\times$	&1473&80&2139\\
NGC\,4945	&8.74/0.78	&Si2&17/03/2006	&Gemini/T-ReCS	&261	&3.855&	0.73&		0.08&	0.322	&	0.05&		\checkmark	&47&$\cdots ^{\dagger}$&1241\\
		&18.33/1.5	&Qa&17/03/2006	&Gemini/T-ReCS	&261	&-&	-&		1.17&	0.524	&	1.57&			 -	&78&$\cdots ^{\dagger}$&1765\\
NGC\,4051	&11.6/1.1	&Si5&02/02/2010&Gemini/MICHELLE	&156	&0.372&	0.07&		0.10&	0.355	&	0.07&		B.R.	&408&100&438\\
NGC\,6300	&8.59/0.84	&PAH1&22/03/2008&VLT/VISIR	&362	&0.320&	0.25&		0.09&	0.246	&	0.08&		\checkmark	&141&90&292\\
		&10.49/0.32	&SIV&22/03/2008&VLT/VISIR	&721	&0.383&	0.23&		0.17&	0.281	&	0.15&		\checkmark	&71&70&167\\
		&11.25/1.18	&PAH2&22/03/2008&VLT/VISIR	&362	&0.352&	0.21&		0.09&	0.346	&	0.09&		B.R.	&259&90&347\\
		&13.04/0.44&NeII$\_$2&22/03/2008&VLT/VISIR	&361	&0.398&	0.24&		0.31&	0.365	&	0.10&		B.R.	&474&90&792\\
NGC\,4388	&8.7/1.1	&Si2&01/02/2015	&GTC/CC		&1044	&0.366& 0.39&		0.10&	0.380   &	0.13&		$\times$&74&70&108\\
NGC\,4138	&11.6/1.1	&Si5&02/02/2010&Gemini/MICHELLE	&259	&0.409&-&		0.08&	0.355	&	0.07&		\checkmark	&9&70&12\\
NGC\,4151	&10.75/5.2	&N&07/05/2001	&Gemini/OSCIR	&360	&0.573&	0.13&		0.09&	0.536	&	0.06&		B.R.	&1452&90&1719\\
		&18.17/1.7	&IHW18&07/05/2001&Gemini/OSCIR	&480	&0.640&	0.25&		0.63&	0.577	&	0.02&		B.R.	&3243&100&4242\\
NGC\,3227	&8.7/1.1	&Si2&17/03/2014	&GTC/CC		&627	&0.321&0.08	    	&0.08   &0.287	&0.02	    &	B.R.	&147&80&260 \\
		&11.2/2.4&	Np&04/07/2006	&Gemini/MICHELLE&376	&0.401&	0.07&		0.04&	0.424	&	0.15&		$\times$	&318&90&436\\
NGC\,7314	&8.74/0.78	&Si2&20/08/2010	&Gemini/T-ReCS	&319	&0.402&	0.11&		0.08&	0.354	&	0.07&		\checkmark	&32&80&48\\
		&18.33/1.5	&Qa&25/09/2010	&Gemini/T-ReCS	&811	&0.545&	0.11&		0.25&	0.555	&	0.12&		$\times$	&122&80&162\\
NGC\,1365	&8.74/0.78	&Si2&08/09/2011	&Gemini/T-ReCS	&145	&0.391&	0.19&	    	0.10& 	0.322 	&  	0.06&	 	\checkmark	&126&60&1111\\
		&18.33/1.5&	Qa&08/09/2011	&Gemini/T-ReCS	&521	&0.579&	0.26&		0.54&	0.554	&	0.10&		B.R.	&435&60&1896\\
NGC\,7582	&8.74/0.78	&Si2&07/09/2011	&Gemini/T-ReCS	&296	&0.359&	0.08&		0.07&	0.409	&	0.12&		$\times$	&208&60&863\\
		&18.33/1.5	&Qa&07/09/2011	&Gemini/T-ReCS	&515	&0.692&	0.35&		0.33&	0.513	&	0.11&		\checkmark	&289&60&2214\\
ESO\,005-G004 	&8.59/0.84&	PAH1&13/11/2010	&VLT/VISIR	&181	&0.536&	0.41&		0.07 &	0.255	&	0.05&		\checkmark	&7&100&16\\
		&11.25/1.18&	PAH2&22/11/2010	&VLT/VISIR	&1994	&0.392&	0.22&		0.04 &	0.384	&	0.17&		B.R.	&21&90&27\\
		&11.88/0.74&PAH2$\_$2&22/11/2010&VLT/VISIR	&1986	&0.381&	0.37&		0.05&	0.365	&	0.13&		B.R.	&27&90&38	\\
		&13.04/0.44&NeII$\_$2&09/10/2007&VLT/VISIR	&181	&0.942&	-&		0.38&	0.534	&	0.16&		\checkmark	&89&60&162\\
		&18.72/1.76&	Q2&16/11/2010	&VLT/VISIR	&497	&1.696&	-&		0.62&	0.536	&	0.29&		\checkmark	&58&30&69	\\
NGC\,7213	&8.74/0.78	&Si2&17/12/2010	&Gemini/T-ReCS	&145	&0.399&	0.26&		0.12&	0.375	&	0.05&		B.R.	&56&90&76\\
		&18.33/1.5	&Qa&07/06/2007	&Gemini/T-ReCS	&319	&0.607&	0.06&		0.74&	0.534	&	0.05&		\checkmark	&224&80&300\\
NGC\,6814	&8.74/0.78	&Si2&28/08/2009	&Gemini/T-ReCS	&145	&0.284&	0.17&		0.11&	0.280	&	0.04&		B.R.	&42&90&55\\
		&18.33/1.5	&Qa&28/08/2009	&Gemini/T-ReCS	&203	&0.627&-&		0.98&	0.534	&	0.11&		\checkmark	&107&40&153\\
MCG-06-30-015	&8.59/0.84	&PAH1&10/03/2010&VLT/VISIR	&362	&0.281&	0.25&		0.27&	0.216	&	0.07&		\checkmark	&121&80&291\\
		&10.49/0.32&SIV&14/04/2006	&VLT/VISIR	&177	&0.314&	0.23&		0.18&	0.296	&	0.18&		B.R.	&308&90&378\\
		&11.25/1.18&PAH2&14/04/2006	&VLT/VISIR	&181	&0.330&	0.25&		0.14&	0.317	&	0.21&		B.R.	&306&80&474\\
		&11.88/0.74&PAH2$\_$2&10/03/2010&VLT/VISIR	&362	&0.336&	0.11&		0.35&	0.308	&	0.12&		B.R.	&222&70&418\\
		&12.27/0.36&NeII$\_$1&14/04/2006&VLT/VISIR	&180	&0.350&	0.12&		0.29&	0.336	&	0.16&		B.R.	&304&80&460\\
NGC\,5506	&11.2/2.4&	Np&06/04/2006	&Gemini/MICHELLE&141	&0.378&	0.04&		0.06&	0.399	&	0.12&		$\times$	&758&80&885\\
		&18.1/1.9&	Qa&06/04/2006	&Gemini/MICHELLE&109	&0.562&	0.10&		0.63&	0.534	&	0.07&		B.R.	&1533&90&2126\\
UGC\,6728	&8.7/1.1	&Si2&01/08/2015&GTC/CC		&695	&0.843&	-&		0.10&	0.606	&  	0.21&		\checkmark	&27&$\cdots ^{\ddagger}$&41\\	
NGC\,2110	&11.2/2.4&	Np&18/03/2007	&Gemini/MICHELLE&376	&0.422&	0.06&		0.07&	0.481	&	0.10&		$\times$	&305&80&308\\
NGC\,2992	&11.2/2.4	&Np&12/05/2006	&Gemini/MICHELLE&188	&0.372&	0.17&		0.07&	0.325	&	0.23&		\checkmark	&159&90&347\\
		&18.1/1.9&	Qa&12/05/2006	&Gemini/MICHELLE&163	&0.561&	0.22&		0.68&	0.537	&	1.67&		B.R.	&513&90&647\\
NGC\,3081	&8.74/0.78	&Si2&25/01/2006	&Gemini/T-ReCS	&130	&0.310&	0.23&		0.13&	0.296	&	0.15&		B.R.	&75&90&99\\
		&18.33/1.5	&Qa&25/01/2006	&Gemini/T-ReCS	&304	&0.623&	1.11&		0.83&	0.561	&	0.16&		B.R.	&364&80&613\\
MCG-05-23-016	&8.99/0.28	&ArIII&30/01/2007&VLT/VISIR	&699	&0.265&	0.14&		0.09&	0.293	&	0.10&		$\times$	&349&60&472\\
		&11.88/0.74&PAH2$\_$2&12/03/2006&VLT/VISIR	&361	&0.338&	0.08&		0.19&	0.335	&	0.15&		B.R.	&599&90&752\\
		&17.65/1.66	&Q1&30/01/2007	&VLT/VISIR	&684	&0.483&	0.13&		0.42&	0.461	&	0.10&		B.R.	&1168&80&1640\\
		&18.72/1.76	&Q2&18/12/2005	&VLT/VISIR	&855	&0.504&	0.22&		0.37&	0.477	&	0.27&		B.R.	&1463&100&1706\\
NGC\,3783	&8.74/0.78	&Si2&01/05/2012	&Gemini/T-ReCS	&145	&0.312&	0.03&		0.12&	0.295	&	0.11&		B.R.	&248&90&347\\
		&18.33/1.5	&Qa&01/05/2012	&Gemini/T-ReCS	&319	&0.521&	0.10&		0.59&	0.532	&	0.13&		$\times$	&989&90&1156\\
NGC\,7172	&8.74/0.78	&Si2&20/09/2011	&Gemini/T-ReCS	&145	&0.409&	0.27&		0.10&	0.389	&	0.06&		B.R.	&59&80&96\\
		&18.33/1.5	&Qa&20/09/2001	&Gemini/T-ReCS	&348	&0.540&-&		0.59&	0.600	&	0.09&		$\times$	&117&70&145\\
\hline
\end{tabular}						 
\caption{Summary of the ground-based MIR imaging observations. Columns from 1 to 6 list the galaxy name, the central wavelength and filter width ($\mu$m), the filter name, the observation date, the telescope/instrument and the on-source exposure time. Columns 7 and 8 correspond to the FWHM and ellipticity of the nucleus of the galaxy. Column 9 lists the standard deviation of the sky background in mJy pixel$^{-1}$ units. Columns 10 and 11 are the same as columns 7 and 8, but for the PSF standard star. Column 12 indicates if the galaxy nucleus is resolved or not. Finally, columns 13, 14 and 15 correspond to the nuclear fluxes, the corresponding percentages of PSF subtraction used for determining each nuclear flux and the total fluxes. B.R. corresponds to barely resolved nucleus. Note that $\dagger$ and $\ddagger$ correspond to nuclear fluxes calculated using aperture photometry and aperture corrections, when the nucleus is not well-defined or the structure of the PSF is not symmetric, respectively.}
\label{tab2}
\end{table}
\end{landscape}

%% file: tab3.tex
\begin{table*}
\scriptsize
\centering
\begin{tabular}{lrrcrrcrcrrc}
\hline
Name&F$_{Nuc}^{Sub}$&F$_{Nuc}^{Arc}$&Best &F$_{WISE}$&L$_{2-10\,keV}$	&2-10	&L$_{14-195\,keV}$	& log N$_{H}$&L$_{[Ne\,II]}$          &L$_{[O\,IV]}$		&MIR		\\
    & 8~$\mu$m	& 8~$\mu$m 	&ArcRes &12~$\mu$m&&keV	&				& 	&  (12.81~$\mu$m) &  (28.89~$\mu$m) 	&lines		\\
    &(mJy)		&(mJy) 		&\% subt.&(mJy)&(10$^{41}$~erg s$^{-1}$)	& Ref.	&(10$^{41}$~erg s$^{-1}$)			&(cm$^{-2}$)&(10$^{39}$~erg s$^{-1}$)&(10$^{39}$~erg s$^{-1}$)	&	Ref.	\\

\hline
NGC\,4395	&2&	14	&90	&11			&0.07	&a&0.46&	22.30	&0.08$\pm$0.01	&  0.14$\pm$0.01     &	j\\
NGC\,5128	&564&	861	&IRS&2833	&2.45	&b&16.40&	22.74	&4.23			&	2.88                &	j\\
NGC\,4945	&12&1856&IRS	&2831	&1.40	&b&4.41&	24.60	&15.90$\pm$1.40	&	1.12$\pm$0.30    &	k\\
NGC\,4051	&183&	247	&80	&468	&2.36	&c&9.16&	20.47	&3.92$\pm$0.18	&	7.36$\pm$0.41    &	j\\
NGC\,6300	&140&	285	&IRS&467	&9.17	&b&21.34&	23.34	&2.70$\pm$0.14	&	6.91$\pm$0.53    &	k\\
NGC\,4388	&74&	223	&70	&454	&102.40	&b&87.48&	23.63	&25.80$\pm$1.60	&	108.00$\pm$9.00     &	j\\
NGC\,4138	&3&	12	&90	&38			&1.85	&d&7.87&	22.90	&1.21$\pm$0.19	&	0.76$\pm$0.07     &	j\\
NGC\,4151	&746&	1005&IRS&1650	&1039.00&e&179.00&	22.48	&64.10			&	113.00                &	j\\
NGC\,3227	&79&	192	&60	&508	&20.49	&e&64.23&	22.80	&32.40$\pm$3.40	&	32.3$\pm$1.20     &	j\\
NGC\,7314	&23&	47	&50	&112	&36.78	&d&29.79&	21.79	&4.69$\pm$0.39	&	36.40$\pm$4.10     &	j\\
NGC\,1365	&84&	433	&IRS&1942	&32.52	&f&39.82&	23.60	&89.40$\pm$9.70	&	80.40$\pm$4.90     &	j\\
NGC\,7582	&57&	171	&IRS &1309	&64.84	&g&39.15&	22.98	&147.00$\pm$2.00&   133.00$\pm$1.00     &	j\\
ESO\,005-G004	&5&	29	&50	&167	&11.07	&g&29.19&	23.88	&11.60$\pm$0.10	&	3.12$\pm$0.13     &	j\\
NGC\,7213	&31&	124	&60	&239	&14.47	&d&39.20&	20.60	&20.70$\pm$1.00	&	2.07$\pm$0.44     &	j\\
NGC\,6814	&29&	79	&90	&116	&26.52	&d&49.38&	20.76	&5.73			&	21.50                 &	l\\
MCG-06-30-015	&99&134	&IRS&334	&33.06	&h&64.45&	21.67	&3.61$\pm$0.10	&	20.20$\pm$0.10     &	j\\
NGC\,5506	&638&	936	&IRS&1051	&161.10	&e&255.80&	22.53	&99.50$\pm$3.60	&	274.00$\pm$1.00     &	j\\
UGC\,6728	&21&	21	&IRS&44		&7.23	&d&45.62&	20.65	&1.73$\pm$0.44	&	5.68$\pm$1.00   &	j\\
NGC\,2110	&160&	174	&IRS&370	&48.87	&h&321.50&	22.57	&75.60$\pm$6.70	&	57.40$\pm$4.30     &	j\\
NGC\,2992	&79&	147	&50	&313	&7.03	&d&93.45&	22.00	&76.00$\pm$5.20	&	162.00$\pm$1.00     &	j\\
NGC\,3081	&51&	135	&80	&266	&231.20	&i&125.30&	23.52	&18.00$\pm$1.70	&	171.00$\pm$1.20     &	j\\
MCG-05-23-016&262&	336	&IRS&633	&105.00	&b&335.80&	22.47	&27.80$\pm$0.10	&	42.90$\pm$11.70     &	j\\
NGC\,3783	&140&	317	&IRS&668	&84.81	&e&255.20&	22.47	&31.40$\pm$1.30	&	62.20$\pm$0.10     &	j\\
NGC\,7172	&75&	245	&IRS&265	&128.00	&b&213.10&	22.89	&55.00$\pm$4.20	&	73.20$\pm$5.90     &	j\\
\hline
\end{tabular}					 
\caption{Details on the MIR and X-ray data. The 8$\mu$m nuclear fluxes were calculated by using a 1~$\mu$m window centered at 8~$\mu$m in the scaled AGN component (see Section \ref{nuclear_fluxes}). Column 1 lists the galaxy name. Columns 2, 3, 4 and 5 correspond to the subarcsecond resolution nuclear fluxes, the arcsecond resolution nuclear fluxes, the best percentage of PSF subtraction used with the arcsecond resolution data and the 12~$\mu$m WISE fluxes. Columns 6 and 7 list the intrinsic 2-10~kev X-ray luminosities and their references. Column 8 and 9 corresponds to the 14-195~keV X-ray luminosities and the column densities, respectively, taken from the {\textit{Swift/BAT}} catalog \citep{Tueller2008}. Finally, columns 10 and 11 list the [Ne\,II] and [O\,IV] emission line luminosities and column 12 corresponds to their references. References: a) \citet{Nardini2011}; b) \citet{DeRosa2012}; c) \citet{Vaughan2011}; d) \citet{Liu2014}; e) \citet{Rivers2011}; f) \citet{Brightman2011}; g) \citet{Winter2009}; h) \citet{Shu2010}; i) \citet{Eguchi2011}; j) \citet{Weaver2010}; k) \citet{Goulding2009}; l) \citet{Pereira-Santaella2010}.}
\label{tab3}
\end{table*}

%% file: tab4.tex
\begin{table}
\centering
\begin{tabular}{lccccc}
\hline
Name&Visual 	&Nuclear	 	&Res.	  &Quantitative\\
    &class.	&vs. total		&Sub	  &class.\\
    &		&flux		&3$\sigma$& \\

\hline
NGC\,4395		&Pos. Ext.	&0.76	&$\times$	& Point-like\\    
NGC\,5128		&Pos. Ext.	&0.75	&\checkmark	& Pos. Ext.\\    
NGC\,4945		&Extended	&0.04	&-			& Extended\\    
NGC\,4051		&Extended	&0.93	&\checkmark	& Pos. Ext.\\    
NGC\,6300		&Extended	&0.54	&\checkmark	& Pos. Ext.\\    
NGC\,4388		&Pos. Ext.	&0.69	&\checkmark	& Pos. Ext.\\ 
NGC\,4138		&Point-like	&0.75	&\checkmark	& Pos. Ext.\\    
NGC\,4151		&Pos. Ext.	&0.80	&\checkmark	& Pos. Ext.\\    
NGC\,3227		&Extended	&0.65	&\checkmark	& Pos. Ext.\\    
NGC\,7314		&Point-like	&0.71	&$\times$	& Point-like\\    
NGC\,1365		&Extended	&0.15	&\checkmark	& Extended\\    
NGC\,7582		&Extended	&0.17	&\checkmark	& Extended\\    
ESO\,005-G004		&Pos. Ext.	&0.66	&$\times$	&Point-like \\    
NGC\,7213		&Pos. Ext.	&0.74	&$\times$	&Point-like \\    
NGC\,6814		&Point-like	&0.73	&$\times$	&Point-like \\    
MCG-06-30-015		&Point-like	&0.58	&\checkmark	&Pos. Ext. \\    
NGC\,5506		&Extended	&0.78	&\checkmark	&Pos. Ext.\\    
UGC\,6728		&Pos. Ext.	&0.64		&-	&Pos. Ext.\\    
NGC\,2110		&Pos. Ext.	&0.99	&\checkmark	& Pos. Ext.\\    
NGC\,2992		&Extended	&0.58	&\checkmark	& Pos. Ext.\\    
NGC\,3081		&Extended	&0.67	&\checkmark	& Pos. Ext.\\    
MCG-05-23-016		&Extended	&0.77	&\checkmark	& Pos. Ext.\\    
NGC\,3783		&Pos .Ext.	&0.78	&\checkmark	& Pos. Ext.\\    
NGC\,7172		&Pos. Ext. &0.70	&$\times$	& Point-like\\    
\hline
\end{tabular}					 
\caption{MIR morphological classification. Columns 1, 2, 3 and 4 list the galaxy name, the visual classification, the ratio between nuclear and total fluxes, and the presence or not of  PSF subtraction residuals at $\geq$3$\sigma$. Column 5 lists the strength of the extended emission, which was calculated by using the ratio between nuclear and total fluxes, and the PSF subtraction residuals. See Section \ref{nuclear_morph} for details. }
\label{tab4}
\end{table}

%% file: tab5.tex
\begin{table}
\centering
\begin{tabular}{lcc}
\hline
Name			&Visual 	&Compact\\
			&morph.		&nucleus\\
			&		&	\\

\hline
NGC\,4395		&Irregular	&\checkmark	\\    
NGC\,5128		&Spiral		&$\times$	\\    
NGC\,4945		&Spiral		&$\times$	\\    
NGC\,4051		&Spiral		&\checkmark	\\    
NGC\,6300		&Spiral		&\checkmark	\\    
NGC\,4388		&Spiral		&\checkmark	\\ 
NGC\,4138		&Spiral		&\checkmark	\\    
NGC\,4151		&Spiral		&\checkmark	\\    
NGC\,3227		&Spiral		&$\times$	\\    
NGC\,7314		&Spiral		&\checkmark	\\    
NGC\,1365		&Spiral		&$\times$	\\    
NGC\,7582		&Elliptical	&$\times$	\\    
ESO\,005-G004	&Elliptical	&$\times$	\\    
NGC\,7213		&Spiral		&\checkmark	\\    
NGC\,6814		&Spiral		&\checkmark	\\    
MCG-06-30-015	&Point-like	&\checkmark	\\    
NGC\,5506		&Point-like	&\checkmark	\\    
UGC\,6728		&Irregular	&\checkmark	\\    
NGC\,2110		&Irregular	&$\times$	\\    
NGC\,2992		&Elliptical	&$\times$	\\    
NGC\,3081		&Spiral		&\checkmark	\\    
MCG-05-23-016	&Point-like	&\checkmark	\\    
NGC\,3783		&Spiral		&\checkmark	\\    
NGC\,7172		&Elliptical	&$\times$	\\    
\hline
\end{tabular}					 
\caption{Arcsecond resolution MIR morphological classification. Columns 1, 2 and 3 list the galaxy name, the visual classification and if the galaxy show a compact nucleus or not.}
\label{tab5}
\end{table}

%% file: tab6.tex
\begin{table*}
%\scriptsize
%\tiny
\centering
\begin{tabular}{llrrrrrr}
\hline
\hline
$X$		&Sample			&N	&R	&P$_{null}$&$\sigma$&a	&b  \\
\hline
L$_{2\mbox{-}10~KeV}$	& BCS$_{40}$		&24	&0.83	&$<$1.0$\times$10$^{-6}$	&0.51&0.75$\pm$0.11	&10.89  \\	
L$_{2\mbox{-}10~KeV}$	& BCS$_{40}^\bigstar$	&23	&0.74	&6.4$\times$10$^{-5}$		&0.51&0.68$\pm$0.14	&13.68  \\
L$_{2\mbox{-}10~KeV}$ & Type 1			&8  	&0.83	&1.1$\times$10$^{-2}$		&0.48&1.41$\pm$0.39	&-17.10  \\
L$_{2\mbox{-}10~KeV}$	& Type 2		&16 	&0.86	&1.7$\times$10$^{-5}$		&0.50&0.72$\pm$0.11	&12.24  \\
\hline                                                                                 	
L$_{14\mbox{-}195~KeV}$& BCS$_{40}$		&24	&0.93	&$<$1.0$\times$10$^{-6}$	&0.26&0.62$\pm$0.05	&16.41  \\
L$_{14\mbox{-}195~KeV}$& BCS$_{40}^\bigstar$	&23	&0.88	&$<$1.0$\times$10$^{-6}$	&0.26&0.58$\pm$0.07	&18.21  \\
L$_{14\mbox{-}195~KeV}$& Type 1			&8  	&0.75	&3.4$\times$10$^{-2}$		&0.32&0.71$\pm$0.26	&12.82  \\
L$_{14\mbox{-}195~KeV}$& Type 2			&16 	&0.96	&$<$1.0$\times$10$^{-6}$	&0.24&0.63$\pm$0.05	&16.03  \\
\hline
L$_{[O\,IV]}$	& BCS$_{40}$			&24	&0.84	&$<$1.0$\times$10$^{-6}$	&0.48&0.74$\pm$0.10	&9.23  \\
L$_{[O\,IV]}$	& BCS$_{40}^\bigstar$		&23	&0.78	&1.2$\times$10$^{-5}$		&0.49&0.74$\pm$0.13	&9.08  \\
L$_{[O\,IV]}$	& Type 1			&8  	&0.81	&1.4$\times$10$^{-2}$		&0.36&1.01$\pm$0.29	&-2.35  \\
L$_{[O\,IV]}$	& Type 2			&16 	&0.89	&5.0$\times$10$^{-6}$		&0.48&0.77$\pm$0.11	&8.07  \\
\hline
L$_{[Ne\,II]SF}$	& BCS$_{40}^{\dagger}$			&16	&0.39	&1.4$\times$10$^{-1}$		&0.68&0.29$\pm$0.19	&27.83  \\
\hline
\end{tabular}					 
\caption{Correlation properties of the BCS$_{40}$ sample. Column 1 corresponds to the quantities given the abscissa of Figs. \ref{fig8}~\& \ref{fig9} whereas the ordinate is the nuclear L$_{8\mu m}$ emission. Columns 2 and 3 list the samples considered and the number of galaxies included, respectively. R, P$_{null}$ and $\sigma$ correspond to the Pearson's correlation coefficient, the null probability and the standard deviation. a and b are the fitting parameters of log($Y$)=a$\cdot$log($X$)+b. Note that we used the Bucley-James least squares linear regression method implemented in the ASURV survival analysis package \citep{Feigelson85, Isobe86}. $\bigstar$ and $\dagger$ correspond to the BCS$_{40}$ sample, but excluding NGC\,4395 and the AGN-dominated sources, respectively.}
\label{tab6}
\end{table*}

%% file: tabB1.tex
\begin{table*}
\centering
\begin{tabular}{lccccrccccc}
\hline
Name 		&	F$_{AGN}$&	F$_{PAH}$&	F$_{Stellar}$&	$\alpha_{AGN}$&	S$_{Si}$ &	L$_{6\mu m}^{\,AGN}$ &	L$_{12\mu m}^{\,AGN}$ &	L$_{12\mu m}^{\,PAH}$& Classification \\
		&	         &               &               &                    &               &               &               &               &\\
		&	         &               &               &                    &               &               &               &               &\\
\hline
NGC\,4395	&0.64	&0.36	&0.00	&-2.83	&-0.09	&0.62	&0.66	&0.34&composite\\
NGC\,5128 (CenA)&0.62	&0.23	&0.15	&-1.71	&-1.06	&0.47	&0.67	&0.23&composite\\
NGC\,4945	&0.05	&0.95	&0.00	&-2.48	&-1.36	&0.04	&0.05	&0.95&SF-dominated\\
NGC\,4051	&0.85	&0.15	&0.00	&-2.09	&0.24	&0.83	&0.86	&0.14&AGN-dominated\\
NGC\,6300	&0.62	&0.17	&0.21	&-2.31	&-1.50	&0.26	&0.66	&0.17&composite\\
NGC\,4388	&0.63	&0.23	&0.14	&-3.71	&-1.04	&0.27	&0.71	&0.20&composite\\
NGC\,4138	&0.17	&0.24	&0.59	&-2.79	&0.20	&0.05	&0.21	&0.28&SF-dominated\\
NGC\,4151	&0.83	&0.06	&0.11	&-2.04	&-0.08	&0.61	&0.88	&0.05&AGN-dominated\\
NGC\,3227	&0.53	&0.28	&0.19	&-2.68	&-0.09	&0.26	&0.57	&0.27&composite\\
NGC\,7314	&0.62	&0.29	&0.09	&-2.62	&-0.33	&0.37	&0.67	&0.27&composite\\
NGC\,1365	&0.29	&0.57	&0.14	&-1.92	&0.60	&0.21	&0.32	&0.59&SF-dominated\\
NGC\,7582	&0.48	&0.52	&0.00	&-2.18	&-1.36	&0.52	&0.48	&0.52&SF-dominated\\
ESO\,005-G004	&0.18	&0.66	&0.16	&-3.71	&-1.04	&0.07	&0.22	&0.65&SF-dominated\\
NGC\,7213	&0.45	&0.14	&0.41	&-2.74	&0.32	&0.15	&0.54	&0.11&composite\\
NGC\,6814	&0.53	&0.10	&0.37	&-2.09	&0.24	&0.26	&0.61	&0.07&composite\\
MCG-06-30-015	&0.82	&0.03	&0.15	&-1.91	&-0.19	&0.54	&0.88	&0.02&AGN-dominated\\
NGC\,5506	&0.72	&0.09	&0.19	&-1.37	&-1.19	&0.53	&0.77	&0.09&AGN-dominated\\
UGC\,6728	&0.95	&0.00	&0.05	&-1.24	&0.12	&0.87	&0.96	&0.00&AGN-dominated\\
NGC\,2110	&0.94	&0.03	&0.03	&-1.62	&0.18	&0.92	&0.95	&0.02&AGN-dominated\\
NGC\,2992	&0.71	&0.22	&0.07	&-2.58	&-0.27	&0.49	&0.71	&0.23&composite\\
NGC\,3081	&0.63	&0.29	&0.08 	&-2.79	&0.20	&0.38	&0.73	&0.22&composite\\
MCG-05-23-016	&0.87	&0.03	&0.10	&-2.14	&-0.26	&0.60	&0.90	&0.03&AGN-dominated\\
NGC\,3783	&0.63	&0.21	&0.16	&-2.79	&0.20	&0.31	&0.68	&0.21&composite\\	
NGC\,7172	&0.45	&0.46	&0.09	&-1.01	&-1.06	&0.45	&0.50	&0.45&SF-dominated\\
\hline
\end{tabular}						 
\caption{Properties derived from the spectral decomposition of the {\textit{Spitzer/IRS}} spectra of BCS$_{40}$ sample. Columns 2, 3 and 4 correspond to the fractional contribution of the AGN component, the PAH component and the stellar component to the MIR spectrum, respectively. Columns 5 and 6 list the spectral index of the AGN component and the silicate strength, respectively (positive and negative values correspond to emission and absorption features, respectively). Columns 7 and 8 correspond to the fractional contribution of the AGN component to the rest-frame spectrum at 6~$\mu$m and 12~$\mu$m, respectively. Column 9 is the same as column 8, but for the PAH component. Finally, column 10 lists the systems classification (see Appendix \ref{B} for details).}
\label{tabB1}
\end{table*}

%% file: tabD1.tex
\begin{table*}
\centering
\begin{tabular}{llrrrrrr}
\hline
\hline
$X$		&Sample			&N	&R	&P$_{null}$&$\sigma$&a	&b  \\
\hline
F$_{2\mbox{-}10~KeV}$	& BCS$_{40}$		&24	&0.64	&8.7$\times$10$^{-4}$		&0.49&0.59$\pm$0.15	&-4.06  \\	
F$_{2\mbox{-}10~KeV}$	& BCS$_{40}^\bigstar$	&23	&0.57	&4.7$\times$10$^{-3}$		&0.50&0.56$\pm$0.18	&-4.40  \\
F$_{2\mbox{-}10~KeV}$ & Type 1			&8  	&0.79	&1.9$\times$10$^{-2}$		&0.51&1.19$\pm$0.38&2.09  \\
F$_{2\mbox{-}10~KeV}$	& Type 2		&16 	&0.63	&8.8$\times$10$^{-3}$		&0.46&0.48$\pm$0.16&-5.15  \\
\hline                                                       
F$_{14\mbox{-}195~KeV}$& BCS$_{40}$		&24	&0.77   &1.3$\times$10$^{-5}$		&0.25&0.44$\pm$0.08&-5.36  \\
F$_{14\mbox{-}195~KeV}$& BCS$_{40}^\bigstar$	&23	&0.74   &6.2$\times$10$^{-5}$		&0.26&0.45$\pm$0.09&-5.24  \\
F$_{14\mbox{-}195~KeV}$& Type 1			&8  	&0.79   &2.1$\times$10$^{-2}$		&0.23&0.52$\pm$0.17&-4.62  \\
F$_{14\mbox{-}195~KeV}$& Type 2			&16 	&0.80   &1.8$\times$10$^{-4}$		&0.25&0.44$\pm$0.09&-5.21  \\
\hline
F$_{[O\,IV]}$	& BCS$_{40}$			&24	&0.67   &3.8$\times$10$^{-4}$		&0.48&0.63$\pm$0.15	&-5.72  \\
F$_{[O\,IV]}$	& BCS$_{40}^\bigstar$		&23	&0.65   &8.4$\times$10$^{-4}$		&0.49&0.67$\pm$0.17	&-5.28  \\
F$_{[O\,IV]}$	& Type 1			&8  	&0.84   &8.8$\times$10$^{-3}$		&0.36&1.02$\pm$0.27	&-1.93  \\
F$_{[O\,IV]}$	& Type 2			&16 	&0.70   &2.2$\times$10$^{-3}$		&0.46&0.61$\pm$0.16	&-5.81  \\
\hline
F$_{[Ne\,II]SF}$	& BCS$_{40}^{\dagger}$			&16	&0.12	&6.7$\times$10$^{-1}$		&0.73&0.12$\pm$0.27	&-11.30  \\
\hline
\end{tabular}					 
\caption{Same as in Table \ref{tab6}, but in flux-flux space.}
\label{tabD1}
\end{table*}

%% file: ms.bbl
\begin{thebibliography}{}

\bibitem[\protect\citeauthoryear{Adams}{1977}]{Adams77} Adams 
T.~F., 1977, ApJS, 33, 19 


\bibitem[\protect\citeauthoryear{Alonso-Herrero et 
al.}{2016}]{Herrero15} Alonso-Herrero A., et al., 2016, MNRAS, 
455, 563 


\bibitem[\protect\citeauthoryear{Alonso-Herrero et 
al.}{2014}]{Herrero14} Alonso-Herrero A., et al., 2014, MNRAS, 
443, 2766 


\bibitem[\protect\citeauthoryear{Alonso-Herrero et 
al.}{2012}]{Herrero12} Alonso-Herrero A., et al., 2012, MNRAS, 
425, 311 

\bibitem[\protect\citeauthoryear{Antonucci}{1993}]{Antonucci1993} Antonucci R., 1993, ARA\&A, 31, 473

\bibitem[\protect\citeauthoryear{Asmus et al.}{2015}]{Asmus2015} 
Asmus D., Gandhi P., H{\"o}nig S.~F., Smette A., Duschl W.~J., 2015, MNRAS, 
454, 766

\bibitem[\protect\citeauthoryear{Asmus et al.}{2014}]{Asmus2014} 
Asmus D., H{\"o}nig S.~F., Gandhi P., Smette A., Duschl W.~J., 2014, MNRAS, 
439, 1648 

\bibitem[\protect\citeauthoryear{Binggeli, Sandage, 
\& Tammann}{1985}]{Binggeli85} Binggeli B., Sandage A., Tammann G.~A., 1985, AJ, 90, 1681 

\bibitem[\protect\citeauthoryear{Brightman 
\& Nandra}{2011}]{Brightman2011} Brightman M., Nandra K., 2011, MNRAS, 413, 1206 

\bibitem[\protect\citeauthoryear{Burtscher et 
al.}{2013}]{Burtscher13} Burtscher L., et al., 2013, A\&A, 558, A149 


\bibitem[\protect\citeauthoryear{Clavel et 
al.}{2000}]{Clavel00} Clavel J., et al., 2000, A\&A, 357, 839 

\bibitem[\protect\citeauthoryear{Dasyra et al.}{2011}]{Dasyra11} 
Dasyra K.~M., Ho L.~C., Netzer H., Combes F., Trakhtenbrot B., Sturm E., 
Armus L., Elbaz D., 2011, ApJ, 740, 94 

\bibitem[\protect\citeauthoryear{de Rosa et 
al.}{2012}]{DeRosa2012} de Rosa A., et al., 2012, MNRAS, 420, 2087 

\bibitem[\protect\citeauthoryear{Diamond-Stanic 
\& Rieke}{2012}]{Diamond12} Diamond-Stanic A.~M., Rieke G.~H., 2012, ApJ, 746, 168 

\bibitem[\protect\citeauthoryear{Diamond-Stanic, Rieke, 
\& Rigby}{2009}]{Diamond09} Diamond-Stanic A.~M., Rieke G.~H., Rigby J.~R., 2009, ApJ, 698, 623 

\bibitem[\protect\citeauthoryear{D{\'{\i}}az-Santos et al.}{2011}]{Diaz-santos11} D{\'{\i}}az-Santos T., et al., 2011, ApJ, 741, 32 


\bibitem[\protect\citeauthoryear{D{\'{\i}}az-Santos et al.}{2010}]{Diaz-santos10} D{\'{\i}}az-Santos T., et al., 2010, ApJ, 723, 993 

\bibitem[\protect\citeauthoryear{Efstathiou 
\& Rowan-Robinson}{1995}]{Efstathiou95} Efstathiou A., Rowan-Robinson M., 1995, MNRAS, 273, 649 

\bibitem[\protect\citeauthoryear{Eguchi et al.}{2011}]{Eguchi2011} 
Eguchi S., Ueda Y., Awaki H., Aird J., Terashima Y., Mushotzky R., 2011, 
ApJ, 729, 31 

\bibitem[\protect\citeauthoryear{Esquej et al.}{2014}]{Esquej14} 
Esquej P., et al., 2014, ApJ, 780, 86 


\bibitem[\protect\citeauthoryear{Fazio et al.}{2004}]{Fazio04} 
Fazio G.~G., et al., 2004, ApJS, 154, 10 

\bibitem[\protect\citeauthoryear{Feigelson 
\& Nelson}{1985}]{Feigelson85} Feigelson E.~D., Nelson P.~I., 1985, ApJ, 293, 192 

\bibitem[\protect\citeauthoryear{Gandhi et 
al.}{2009}]{Gandhi09} Gandhi P., Horst H., Smette A., H{\"o}nig S., Comastri A., Gilli R., Vignali C., Duschl W., 2009, A\&A, 502, 457 


\bibitem[\protect\citeauthoryear{Garc{\'{\i}}a-Bernete et 
al.}{2015}]{Bernete2015} Garc{\'{\i}}a-Bernete I., et al., 2015, 
MNRAS, 449, 1309 

\bibitem[\protect\citeauthoryear{Garc{\'{\i}}a-Gonz{\'a}lez et al.}{2015}]{Garcia-Gonzalez15} Garc{\'{\i}}a-Gonz{\'a}lez J., Alonso-Herrero A., P{\'e}rez-Gonz{\'a}lez P.~G., Hern{\'a}n-Caballero A., Sarajedini V.~L., Villar V., 2015, MNRAS, 446, 3199 

\bibitem[\protect\citeauthoryear{Glasse, Atad-Ettedgui, 
\& Harris}{1997}]{Glasse97} Glasse A.~C., Atad-Ettedgui E.~I., Harris J.~W., 1997, SPIE, 2871, 1197 

\bibitem[\protect\citeauthoryear{Gonz{\'a}lez-Mart{\'{\i}}n et 
al.}{2013}]{Gonzalez-Martin2013} Gonz{\'a}lez-Mart{\'{\i}}n O., et al., 2013, A\&A, 553, A35 

\bibitem[\protect\citeauthoryear{Goulding 
\& Alexander}{2009}]{Goulding2009} Goulding A.~D., Alexander D.~M., 2009, MNRAS, 398, 1165 

\bibitem[\protect\citeauthoryear{Hern{\'a}n-Caballero et 
al.}{2015}]{Hernan-caballero2015} Hern{\'a}n-Caballero A., et al., 2015, 
ApJ, 803, 109 

\bibitem[\protect\citeauthoryear{Hern{\'a}n-Caballero 
\& Hatziminaoglou}{2011}]{Hernan-caballero2011} Hern{\'a}n-Caballero A., Hatziminaoglou E., 2011, MNRAS, 414, 500 

\bibitem[\protect\citeauthoryear{Ho 
\& Keto}{2007}]{Ho07} Ho L.~C., Keto E., 2007, ApJ, 658, 314 

\bibitem[\protect\citeauthoryear{H{\"o}nig \& Kishimoto}{2011}]{Honig11} H{\"o}nig S.~F., Kishimoto M., 2011, A\&A, 534, A121 

\bibitem[\protect\citeauthoryear{H{\"o}nig et al.}{2006}]{Honig06} H{\"o}nig S.~F., Beckert T., Ohnaka K., Weigelt G., 2006, A\&A, 452, 459 

\bibitem[\protect\citeauthoryear{Hopkins 
\& Quataert}{2010}]{Hopkins10} Hopkins P.~F., Quataert E., 2010, MNRAS, 407, 1529 


\bibitem[\protect\citeauthoryear{Horst et 
al.}{2008}]{Horst08} Horst H., Gandhi P., Smette A., Duschl W.~J., 2008, A\&A, 479, 389 


\bibitem[\protect\citeauthoryear{Houck et al.}{2004}]{Houck04} 
Houck J.~R., et al., 2004, ApJS, 154, 18 

\bibitem[\protect\citeauthoryear{Ichikawa et 
al.}{2012}]{Ichikawa2012} Ichikawa K., Ueda Y., Terashima Y., Oyabu 
S., Gandhi P., Matsuta K., Nakagawa T., 2012, ApJ, 754, 45 

\bibitem[\protect\citeauthoryear{Imanishi et 
al.}{2011}]{Imanishi11} Imanishi M., Imase K., Oi N., Ichikawa K., 
2011, AJ, 141, 156 

\bibitem[\protect\citeauthoryear{Isobe, Feigelson, 
\& Nelson}{1986}]{Isobe86} Isobe T., Feigelson E.~D., Nelson P.~I., 1986, ApJ, 306, 490 

\bibitem[\protect\citeauthoryear{Kennicutt et al.}{2003}]{Kennicutt03} Kennicutt R.~C., Jr., et al., 2003, PASP, 115, 928 

\bibitem[\protect\citeauthoryear{Khachikian 
\& Weedman}{1974}]{Khachikian74} Khachikian E.~Y., Weedman D.~W., 1974, ApJ, 192, 581 

\bibitem[\protect\citeauthoryear{Khachikian 
\& Weedman}{1971}]{Khachikian71} Khachikian E.~E., Weedman D.~W., 1971, Afz, 7, 389 

\bibitem[\protect\citeauthoryear{Lagage et al.}{2004}]{Lagage2004} 
Lagage P.~O., et al., 2004, Msngr, 117, 12 

\bibitem[\protect\citeauthoryear{Lebouteiller et 
al.}{2011}]{Lebouteiller11} Lebouteiller V., Barry D.~J., Spoon 
H.~W.~W., Bernard-Salas J., Sloan G.~C., Houck J.~R., Weedman D.~W., 2011, 
ApJS, 196, 8 

\bibitem[\protect\citeauthoryear{Levenson et 
al.}{2009}]{Levenson09} Levenson N.~A., Radomski J.~T., Packham 
C., Mason R.~E., Schaefer J.~J., Telesco C.~M., 2009, ApJ, 703, 390 

\bibitem[\protect\citeauthoryear{Liu et al.}{2014}]{Liu2014} 
Liu T., Wang J.-X., Yang H., Zhu F.-F., Zhou Y.-Y., 2014, ApJ, 783, 106 

\bibitem[\protect\citeauthoryear{Lutz et 
al.}{2004}]{Lutz04} Lutz D., Maiolino R., Spoon H.~W.~W., Moorwood A.~F.~M., 2004, A\&A, 418, 465 

\bibitem[\protect\citeauthoryear{Mart{\'{\i}}nez-Paredes et al.}{2015}]{Martinez-Paredes2015} Mart{\'{\i}}nez-Paredes M., et al., 2015, MNRAS, 454, 3577 

\bibitem[\protect\citeauthoryear{Mason et al.}{2012}]{Mason12} 
Mason R.~E., et al., 2012, AJ, 144, 11 


\bibitem[\protect\citeauthoryear{Mason et al.}{2006}]{Mason06} Mason R.~E., Geballe T.~R., Packham C., Levenson N.~A., Elitzur M., Fisher R.~S., Perlman E., 2006, ApJ, 640, 612 


\bibitem[\protect\citeauthoryear{Mateos et al.}{2015}]{Mateos15} 
Mateos S., et al., 2015, MNRAS, 449, 1422 

\bibitem[\protect\citeauthoryear{Matsuta et 
al.}{2012}]{Matsuta2012} Matsuta K., et al., 2012, ApJ, 753, 104 

\bibitem[\protect\citeauthoryear{Mel{\'e}ndez et al.}{2008b}]{Melendez08b} Mel{\'e}ndez M., Kraemer S.~B., Schmitt H.~R., Crenshaw D.~M., Deo R.~P., Mushotzky R.~F., Bruhweiler F.~C., 2008b, ApJ, 689, 95-107 

\bibitem[\protect\citeauthoryear{Mel{\'e}ndez et 
al.}{2008a}]{Melendez08} Mel{\'e}ndez M., et al., 2008a, ApJ, 682, 
94 

\bibitem[\protect\citeauthoryear{Mould et al.}{2000}]{Mould00} 
Mould J.~R., et al., 2000, ApJ, 529, 786 

\bibitem[\protect\citeauthoryear{Mulchaey et 
al.}{1994}]{Mulchaey94} Mulchaey J.~S., Koratkar A., Ward M.~J., 
Wilson A.~S., Whittle M., Antonucci R.~R.~J., Kinney A.~L., Hurt T., 1994, 
ApJ, 436, 586 

\bibitem[\protect\citeauthoryear{Mu{\~n}oz-Mateos et al.}{2009}]{Munoz-Mateos09} Mu{\~n}oz-Mateos J.~C., et al., 2009, ApJ, 703, 1569-1596 

\bibitem[\protect\citeauthoryear{Nardini 
\& Risaliti}{2011}]{Nardini2011} Nardini E., Risaliti G., 2011, MNRAS, 417, 2571 


\bibitem[\protect\citeauthoryear{Nenkova et 
al.}{2008a}]{Nenkova08} Nenkova M., Sirocky M.~M., Ivezi{\'c} {\v Z}., 
Elitzur M., 2008, ApJ, 685, 147 

\bibitem[\protect\citeauthoryear{Nenkova et 
al.}{2008b}]{bNenkova08} Nenkova M., Sirocky M.~M., Nikutta R., 
Ivezi{\'c} {\v Z}., Elitzur M., 2008, ApJ, 685, 160 

\bibitem[\protect\citeauthoryear{Packham et 
al.}{2005}]{Packham05} Packham C., Radomski J.~T., Roche P.~F., 
Aitken D.~K., Perlman E., Alonso-Herrero A., Colina L., Telesco C.~M., 
2005, ApJ, 618, L17 


\bibitem[\protect\citeauthoryear{Peeters, Spoon, 
\& Tielens}{2004}]{Peeters04} Peeters E., Spoon H.~W.~W., Tielens A.~G.~G.~M., 2004, ApJ, 613, 986 

\bibitem[\protect\citeauthoryear{Pereira-Santaella et 
al.}{2010}]{Pereira-Santaella2010} Pereira-Santaella M., Diamond-Stanic 
A.~M., Alonso-Herrero A., Rieke G.~H., 2010, ApJ, 725, 2270 

\bibitem[\protect\citeauthoryear{Pier 
\& Krolik}{1992}]{Pier92} Pier E.~A., Krolik J.~H., 1992, ApJ, 401, 99 

\bibitem[\protect\citeauthoryear{Prieto, P{\'e}rez Garc{\'{\i}}a, 
\& Rodr{\'{\i}}guez Espinosa}{2002}]{Prieto02} Prieto M.~A., P{\'e}rez Garc{\'{\i}}a A.~M., Rodr{\'{\i}}guez Espinosa J.~M., 2002, MNRAS, 329, 309 


\bibitem[\protect\citeauthoryear{Radomski et 
al.}{2003}]{Radomski2003} Radomski J.~T., Pi{\~n}a R.~K., Packham 
C., Telesco C.~M., De Buizer J.~M., Fisher R.~S., Robinson A., 2003, ApJ, 
587, 117 

\bibitem[\protect\citeauthoryear{Radomski et 
al.}{2002}]{Radomski2002} Radomski J.~T., Pi{\~n}a R.~K., Packham 
C., Telesco C.~M., Tadhunter C.~N., 2002, ApJ, 566, 675 

\bibitem[\protect\citeauthoryear{Ramos Almeida et 
al.}{2014}]{Ramos14} Ramos Almeida C., et al., 2014, MNRAS, 
445, 1130 

\bibitem[\protect\citeauthoryear{Ramos Almeida et 
al.}{2011a}]{Ramos2011} Ramos Almeida C., et al., 2011, ApJ, 731, 
92 

\bibitem[\protect\citeauthoryear{Ramos Almeida et 
al.}{2011c}]{Ramos11c} Ramos Almeida C., et al., 2011, MNRAS, 
417, L46 

\bibitem[\protect\citeauthoryear{Ramos Almeida et 
al.}{2009}]{Ramos09} Ramos Almeida C., et al., 2009, ApJ, 702, 
1127 

\bibitem[\protect\citeauthoryear{Ramos Almeida et 
al.}{2007}]{Ramos07} Ramos Almeida C., P{\'e}rez Garc{\'{\i}}a 
A.~M., Acosta-Pulido J.~A., Rodr{\'{\i}}guez Espinosa J.~M., 2007, AJ, 134, 
2006 


\bibitem[\protect\citeauthoryear{Rigby, Diamond-Stanic, 
\& Aniano}{2009}]{Rigby09} Rigby J.~R., Diamond-Stanic A.~M., Aniano G., 2009, ApJ, 700, 1878 

\bibitem[\protect\citeauthoryear{Rivers, Markowitz, 
\& Rothschild}{2011}]{Rivers2011} Rivers E., Markowitz A., Rothschild R., 2011, ApJS, 193, 3 

\bibitem[\protect\citeauthoryear{Roche et al.}{2007}]{Roche07} Roche P.~F., Packham C., Aitken D.~K., Mason R.~E., 2007, MNRAS, 375, 99 

\bibitem[\protect\citeauthoryear{Roche et al.}{1991}]{Roche91} 
Roche P.~F., Aitken D.~K., Smith C.~H., Ward M.~J., 1991, MNRAS, 248, 606 

\bibitem[\protect\citeauthoryear{Roche 
\& Aitken}{1985}]{Roche85} Roche P.~F., Aitken D.~K., 1985, MNRAS, 213, 789 


\bibitem[\protect\citeauthoryear{Ruschel-Dutra et 
al.}{2014}]{Ruschel14} Ruschel-Dutra D., Pastoriza M., Riffel R., 
Sales D.~A., Winge C., 2014, MNRAS, 438, 3434 

\bibitem[\protect\citeauthoryear{Sales et al.}{2013}]{Sales13} 
Sales D.~A., Pastoriza M.~G., Riffel R., Winge C., 2013, MNRAS, 429, 2634 

\bibitem[\protect\citeauthoryear{Sazonov et 
al.}{2012}]{Sazonov12} Sazonov S., et al., 2012, ApJ, 757, 181 

\bibitem[\protect\citeauthoryear{Schartmann et 
al.}{2005}]{Schartmann05} Schartmann M., Meisenheimer K., Camenzind M., Wolf S., Henning T., 2005, A\&A, 437, 861 


\bibitem[\protect\citeauthoryear{Shi et al.}{2006}]{Shi06} 
Shi Y., et al., 2006, ApJ, 653, 127 

\bibitem[\protect\citeauthoryear{Shu, Yaqoob, 
\& Wang}{2010}]{Shu2010} Shu X.~W., Yaqoob T., Wang J.~X., 2010, ApJS, 187, 581 

\bibitem[\protect\citeauthoryear{Siebenmorgen, Heymann, 
\& Efstathiou}{2015}]{Siebenmorgen2015} Siebenmorgen R., Heymann F., Efstathiou A., 2015, A\&A, 583, A120 

\bibitem[\protect\citeauthoryear{Siebenmorgen et 
al.}{2005}]{Siebenmorgen05} Siebenmorgen R., Haas M., Kr{\"u}gel E., Schulz B., 2005, A\&A, 436, L5 

\bibitem[\protect\citeauthoryear{Soifer et al.}{2001}]{Soifer01} Soifer B.~T., et al., 2001, AJ, 122, 1213 


\bibitem[\protect\citeauthoryear{Soifer et al.}{2000}]{Soifer00} Soifer B.~T., et al., 2000, AJ, 119, 509 


\bibitem[\protect\citeauthoryear{Spinoglio et 
al.}{2012}]{Spinoglio12} Spinoglio L., Dasyra K.~M., Franceschini 
A., Gruppioni C., Valiante E., Isaak K., 2012, ApJ, 745, 171 

\bibitem[\protect\citeauthoryear{Spinoglio 
\& Malkan}{1992}]{Spinoglio92} Spinoglio L., Malkan M.~A., 1992, ApJ, 399, 504 


\bibitem[\protect\citeauthoryear{Stalevski et 
al.}{2012}]{Stalevski2012} Stalevski M., Fritz J., Baes M., Nakos T., 
Popovi{\'c} L.~{\v C}., 2012, MNRAS, 420, 2756 

\bibitem[\protect\citeauthoryear{Sturm et al.}{2005}]{Sturm05} 
Sturm E., et al., 2005, ApJ, 629, L21 

\bibitem[\protect\citeauthoryear{Telesco et 
al.}{2003}]{Telesco03} Telesco C.~M., et al., 2003, SPIE, 4841, 
913 

\bibitem[\protect\citeauthoryear{Telesco et 
al.}{1998}]{Telesco1998} Telesco C.~M., Pina R.~K., Hanna K.~T., 
Julian J.~A., Hon D.~B., Kisko T.~M., 1998, SPIE, 3354, 534 

\bibitem[\protect\citeauthoryear{Tody}{1986}]{tody86} Tody D., 
1986, SPIE, 627, 733 

\bibitem[Tristram et al. (2009)]{Tristram09} Tristram, K. R. W. et al. 2009, A\&A, 502, 67

\bibitem[\protect\citeauthoryear{Tueller et 
al.}{2008}]{Tueller2008} Tueller J., Mushotzky R.~F., Barthelmy S., 
Cannizzo J.~K., Gehrels N., Markwardt C.~B., Skinner G.~K., Winter L.~M., 
2008, ApJ, 681, 113 

\bibitem[\protect\citeauthoryear{Ueda et al.}{2015}]{Ueda15} 
Ueda Y., et al., 2015, ApJ, 815, 1 

\bibitem[\protect\citeauthoryear{Vaughan et 
al.}{2011}]{Vaughan2011} Vaughan S., Uttley P., Pounds K.~A., 
Nandra K., Strohmayer T.~E., 2011, MNRAS, 413, 2489 

\bibitem[\protect\citeauthoryear{Weaver et al.}{2010}]{Weaver2010} 
Weaver K.~A., et al., 2010, ApJ, 716, 1151 


\bibitem[\protect\citeauthoryear{Winter et al.}{2010}]{Winter2010} 
Winter L.~M., Lewis K.~T., Koss M., Veilleux S., Keeney B., Mushotzky 
R.~F., 2010, ApJ, 710, 503 


\bibitem[\protect\citeauthoryear{Winter et al.}{2009}]{Winter2009} 
Winter L.~M., Mushotzky R.~F., Reynolds C.~S., Tueller J., 2009, ApJ, 690, 
1322 

\bibitem[\protect\citeauthoryear{Wold et 
al.}{2006}]{Wold06} Wold M., Lacy M., K{\"a}ufl H.~U., Siebenmorgen R., 2006, A\&A, 460, 449 


\bibitem[\protect\citeauthoryear{Wold 
\& Galliano}{2006}]{WoldGaliano06} Wold M., Galliano E., 2006, MNRAS, 369, L47 


\bibitem[\protect\citeauthoryear{Wright et al.}{2010}]{Wright10} 
Wright E.~L., et al., 2010, AJ, 140, 1868 


\bibitem[\protect\citeauthoryear{Wu et al.}{2005}]{Wu05} Wu 
H., Cao C., Hao C.-N., Liu F.-S., Wang J.-L., Xia X.-Y., Deng Z.-G., Young 
C.~K.-S., 2005, ApJ, 632, L79 


\end{thebibliography}
